\documentclass[11pt,a4paper,twoside,nofootinbib,british]{book}
\usepackage[UKenglish]{babel}
\usepackage[T1]{fontenc}

\usepackage[math]{anttor}
\usepackage[utf8]{inputenc}
\usepackage{babel}
\usepackage{latexsym,amsmath,amssymb}
\usepackage{graphicx,overpic}

\usepackage{bm}
\usepackage{textcomp}
\usepackage{graphicx}
\usepackage{enumerate}
\usepackage{float} 
\usepackage{bm}
\usepackage{verbatim}
\usepackage{enumerate}
\usepackage{color}
\usepackage{dcolumn}
\usepackage{geometry}
\usepackage{textcomp}
\usepackage{mathrsfs}
\usepackage{subfigure}
\usepackage{cite}
\usepackage{hyperref}
\usepackage[stable]{footmisc}
\usepackage[T1]{fontenc}
\usepackage[footnotesize,bf,it]{caption}

\def\b{\begin{equation}}
\def\e{\end{equation}}

\def\b{\begin{equation}}
\def\e{\end{equation}}

\def\openone{\leavevmode\hbox{\small$1$\normalsize\kern-.37em$1$}}

\makeatletter
\def\cleardoublepage{\clearpage\if@twoside \ifodd\c@page\else
    \hbox{}
    \thispagestyle{empty}
    \newpage
    \if@twocolumn\hbox{}\newpage\fi\fi\fi}
\makeatother \clearpage{\pagestyle{empty}\cleardoublepage}

\numberwithin{equation}{section}

\newcommand{\qed}{\nobreak \ifvmode \relax \else
      \ifdim\lastskip<1.5em \hskip-\lastskip
      \hskip1.5em plus0em minus0.5em \fi \nobreak
      \vrule height0.75em width0.5em depth0.25em\fi}

      \usepackage{fancyhdr}
\begin{document}
\thispagestyle{empty}

{\Large \noindent {\bf \centerline{Department of Optics}
\\[0.05cm]
\centerline{Faculty of Natural Sciences}
\\[0.05cm]
\centerline{Palack\'y University}
\\[0.05cm]
\centerline{Olomouc, Czech Republic}
\\[2cm]

{\Huge {\bf \centerline{Quantum properties of
\\[0.1cm]}
\centerline{ superposition states,
\\[0.1cm]}
\centerline{squeezed states, and of some
\\[0.1cm]}
\centerline{  parametric processes
\\[1cm]}}}

{\huge \centerline{Author: {\bf Faisal Aly Aly El-Orany
\\[1cm]}}}

{\Large {\bf \centerline{ Doctoral Thesis
\\[.1cm]}}
\centerline{Branch: {\bf Optics and Optoelectronics
\\[1cm]}}

{\Large \centerline{Supervisor: {\bf Prof. RNDr. Jan Pe\v{r}ina,
DrSc.
\\[.5cm] }}}

\centerline{{\bf Olomouc, 2001 }}}}}

\newpage
\begin{center}
{\Huge Acknowledgments}
\end{center}
\vspace{.5cm}

First of all, I would like to start this thesis by passing my thanks to my
supervisor, Professor Jan Pe\v{r}ina, under his supervision I worked
more than three years; I thank him for collaboration in
 the most important and beautiful period of my life, I will never
forget how much he was kind to me. I am very gratitude to
him for  long discussions and suggestions of interesting problems
in nonlinear optics.
This work could  not be done in the present form without the support of
all the members
of the Department of Optics of Palack\'y University.

My warm thanks come next to Professor Mohamed Sebawe Abdalla
from the Mathematics Department, College of Science,
King Saud University, Riyadh, Saudi Arabia, for his suggestions of several
problems and helping in some cases in the calculations.
 I wish to thank him for everything he did for me
from the earlier days when I started to be engaged with  quantum optics.

I would like to thank sincerely Professor Vlasta Pe\v{r}inov\'a for many
discussions, support, advise as well as supply with some articles.
My acknowledge extends to  Dr. Jarom\'{i}r K\v{r}epelka  for many  numerical advises.

I am obliged much to my mother, brothers and sister,
I missed them  too much, although they are always with me,
and to my wife  and son  who  give  beautiful meaning to my life,
and   support me and are patient with me.

\setcounter{page}{1} \thispagestyle{empty} \tableofcontents

\newpage

  \chapter{Introduction}

The failure of classical mechanics to account for many experimental results
such as the stability of atoms and matter, blackbody radiation, etc. led
physicists to the realization of new tool to deal with these problems and
thus  quantum mechanics came to the vicinity of life at the beginning of
the last century.
Indeed, quantum mechanics is the one of the crowning achievements of modern
physics where the quantized electromagnetic field
is supported  by the experimental observations of nonclassical states of the
radiation field, e.g. squeezed states, sub-Poissonian photon statistics
and photon antibunching.
Nowadays, quantum optics, the union of quantum field theory and physical optics,
is undergoing a time of revolutionary change. The subject has evolved from
early studies on the coherence properties of radiation to the laser in the
modern areas of study involving, e.g.,
the role of squeezed states of the radiation field and atomic coherence
in quenching quantum noise in interferometry and optical amplifiers.
Furthermore, quantum optics provides a powerful new probe for addressing
fundamental issues of quantum mechanics such as complementarity, hidden
variables, and other aspects central to the foundations of quantum physics
and philosophy.

For the description of a quantum system, the concept of state is used,
which is the same, a wave function, a state vector or a density matrix
containing the information about the possible results of measurement on
the system. Quantum optics has statistical origin and therefore
the state of a quantum system contains all information necessary to
completely determine its statistics (the probabilistic nature of a
quantum system).
Among the wide variety of possible radiation-field states, there are some
fundamental types that play a special role in quantum optics.
Strictly speaking, there are three types of these states which are widely
used, namely, coherent, number and squeezed states.
Following the development of the quantum theory of radiation and with
the advent of the laser, the coherent states of the field, that mostly
describe a classical electromagnetic field, were widely studied.
Indeed, these states are more appropriate basis for many optical fields.
 However, number states are purely nonclassical states and
they are a useful representation of high-energy photons. However, there are
experimental difficulties which have prevented the generation of photon number
states with more than a small number of photons. Despite this the number
states of the electromagnetic field have been used as a basis for several
problems in quantum optics including some laser theories.
The  squeezed states
(i.e. the states of light with reduced fluctuations in one quadrature below
the level associated with the vacuum state)  are very important owing
to their potential  applications, e.g. in the optical communication systems,
interferometric techniques, and in an optical waveguide tap.
Furthermore, squeezed states have been seen in several laboratories.
In addition to these states, several quantum states in the literature have
 appeared, in particular those which are connected with the
superposition principle. Using such principle, in the first part of this
thesis, we will  develop new class of states  which are superposition
of  squeezed and displaced number states.
In fact, this class of states   generalizes
some considerable results given in the literature.
We close this part by giving an example of multidimensional squeeze
operator which is more general than usually used.

On the other hand, nonlinear optics (NLO) has become a very important
subfield of optics since
its inception over 40 years ago. The origin of this branch is the
study of the phenomena that occur as a consequence of the modification of the
optical properties of a material system by the presence of light.
The impact of NLO on science and technology has been twofold. First, it has
enhanced our understanding of fundamental light-matter interactions. Second,
it has been a deriving force in the rejuvenation of optical technology
for several areas of science and engineering. For example, NLO provides the
key to many developments, e.g. the advent of telecommunications using optical
fibers,  carrying  information on a laser beam in the process of communication,
storage, retrieval, printing or sensing, and there are increasing efforts to
achieve ever greater data-processing capabilities.
One of the most promising devices in NLO is the nonlinear
directional coupler that consists of two or more parallel optical waveguides
fabricated from some nonlinear material. Both waveguides are placed close enough
to permit flux-dependent transfer of energy between them.
 This flux transfer can be controlled by the device design and
 the input flux as well.
This device is
  experimentally implemented. Several models of this device
will be investigated  in the second part of this thesis.

As is well-known the quantum statistical properties of radiation present an important
branch of modern physics with rapidly increasing applications in
spectroscopy, quantum generators of radiation, optical communication, etc.
Using these methods, quantum optics was able to predict many nonclassical
phenomena, such as squeezing of vacuum fluctuations in field quadratures,
antibunching of photons, sub-Poisson photon statistics exhibiting reduced
photon-number fluctuations below the Poisson level of ideal laser light,
quantum oscillations in photocount distributions, violation of various
classical statistical inequalities, collapse and revival of atomic
inversion, etc.

This doctoral thesis deals with investigating various quantum statistical
aspects for new quantum states and  for nonlinear
 directional couplers.
The first part will be devoted to the static regime where we
introduce the superposition of squeezed
and displaced number states and discuss their quantum properties,
phase properties and the influence of thermal noise on their behaviour.
 We conclude this part by giving an example of
 new type of multidimensional squeeze operator model which is
more general than usually used and which includes two
different squeezing mechanisms. The dynamical regime will be
considered in the second part where we study the evolution
 of quantum states like coherent, number and thermal states
in some nonlinear medium, such as nonlinear couplers.
To achieve this goal in a systematic way we give first a survey of
previous works in the literature related to our topic in chapter 2 and
in chapter 3 we write down all mathematical relations controlling the
quantum statistical properties of the models under discussion.

\chapter{Contemporary state of the problem}
\vskip0.5cm
In this chapter, we display the most important relevant results
 in quantum optics which are used in our work. This will be done by
 throwing the
light on the results of both squeezed light and  the propagation of  light
in the nonlinear media.
\section{Squeezed light}
\subsection{Squeezed states and squeezed superposition states}

Squeezed states of the electromagnetic field are purely quantum
states.  They are defined through the relation
\begin{equation}
|\xi_{1},\alpha\rangle =\hat{D}(\alpha )\hat{S}(\xi_{1})|0\rangle ,\label{1}
\end{equation}

\noindent where $|0\rangle$ is the vacuum state,
$\hat{D}(\alpha )$ ~\cite{ex[1]} and $\hat{S}(\xi_{1})$ ~\cite{[5]}
are displacement and squeeze operators
 which  are given, respectively, by

\begin{equation}
\hat{D}(\alpha )=\exp (\hat{a}^{\dagger}\alpha-\hat{a}\alpha^{*}),\label{3}
\end{equation}

\begin{equation}
\hat{S}(\xi_{1}) = \exp (\xi^{*}_{1} \hat{a}^{2} - \xi_{1} \hat{a}^{\dagger 2} )
,\label{2}
\end{equation}

\noindent where $\hat{a}$ and $\hat{a}^{\dagger }$ are annihilation and
creation operators, respectively, and $\xi_{1}$ and $\alpha$ are complex
parameters.

The significant feature for these states is that they can have less uncertainty
in one quadrature than a coherent state.
These states
 can exhibit a number of distinctly quantum features,
such as sub-Poissonian statistics ~\cite{[1],{[2]},{[3]}},
 as well as they have no nonsingular representation in terms of
the Glauber-Sudarshan $P$ distribution ~\cite{[4]}.
 It is important to refer that the concept of squeezed coherent states (SCS)
have been applied
to other quantum mechanical systems. For instance, they may play a role
in increasing the sensitivity of a gravitational wave detector
~\cite{[4]}.
SCS have
appeared at the first time ~\cite{[5]} as a simple generalization of the well known
minimum-uncertainty wavepackets. Authors of ~\cite{[4],{[6]}} have demonstrated
further details concerning
the properties of such states, or the so-called two-photon
coherent states ~\cite{[6]}.

 A lot of efforts have been directed towards the methods of generating SCS.
For example,
 we can mention, authors of ~\cite{[7],{[8]},{[9]}} have shown that squeezing may be generated
 in an optical four-wave mixing. Further progress has been done
in papers ~\cite{[10],{[11]}},
where the
first experimental observations of squeezing were given in four-wave mixing
in an atomic beam of Sodium vapor ~\cite{[10]} as well as in optical
fibers ~\cite{[11]}.
Particular attention was given to generate squeezed light in parametric
amplifiers ~\cite{[3],{[13]},{[14]},{[15]},{[16]}}, where the treatment of
parametric oscillation and intracavity
second harmonic generation provides the basis for subsequent calculations of
squeezing in these systems.  Furthermore, more activities have been focused
on studying generation of SCS in: nondegenerate parametric oscillator
~\cite{[17]}, optical
bistability ~\cite{[18],{[19]}}, and resonance fluorescence
~\cite{[20],{[21]},{[22]}}. For more complete
information about SCS the reader can consult papers
~\cite{[23],{[24]},{[25]}}.

The concept of superposition of quantum states has been extended by several
authors to include SCS ~\cite{[26],{[27]}}.  Authors of ~\cite{[26]} have
introduced  even and odd displaced squeezed states as a superposition of two
vacuum squeezed displaced states. For such superposition
 the authors studied the higher-order squeezing with respect
to the definition given by
Hong and Mandel ~\cite{[28]}. Also it is interesting to refer to
~\cite{[27]} where extensive
efforts have been done to introduce a superposition of set of SCS
and the authors managed
to calculate and discuss their quasiprobabality distribution
functions as well as the generation scheme of such superposition states.

Squeezed and displaced number state (SDNS) ~\cite{[29]} is an energy
eigenstate of a quantum harmonic oscillator which is displaced
and then squeezed
 (it may also be called
the generalized squeezed coherent state (GSCS)).
For SDNS bunching and antibunching properties have been investigated
~\cite{[30]}.
For more details about its properties, discussion as well as
 new methods for analytical investigation one can see paper ~\cite{[31]}.
Further, for such a state, the most general
form as well as the time-dependent expectation values, uncertainties of
wave-functions and probability densities have been  given in ~\cite{[32]}
 using the
functional form for the squeeze and the time-displacement operators
~\cite{[33]}. Also we can mention that a special case of SDNS has been
discussed in ~\cite{[61]} where these states have been treated from the
point of view of non-diagonal $P$-representation.
The authors have discussed the properties of the squeezed and displaced
Fock states as generalized states and their discussion has been extended
to the Glauber $R$-representation for the density operator as well as to
the phase distribution.
The physical interpretation of SDNS has been considered similarly to SCS
~\cite{[29]}, in other words,
 SDNS is the coherent state formed due to two excitations on
a particular number state.
  Also we can mention paper ~\cite{[34]} where squeezed (but not displaced)
number state was produced.

  We can also refer to several applications of SCS.  For example, SCS have
several potential applications, one, for instance, is in optical
communication systems ~\cite{[35]}.  Also the interferometric techniques
may provide ways to detect very weak forces such as caused by gravitational
radiation and may experience limitations on sensitivity due to quantum noise
arising from photon counting and radiation pressure fluctuations ~\cite{[36]}.
Another application is in an optical waveguide tap ~\cite{[37]} where it has
been shown that a high signal-to-noise ratio may be obtained using SCS in an
optical waveguide to tap a signal carrying waveguide.
\subsection{Entangled squeezed states}
Entangled states gain their feature from the quantum correlation between
different quantum mechanical
systems where
the individual operator of the single
system does not exhibit nonclassical effects, however, the compound system
can exhibit such effects. To be more specific,  if the measurement of an
observable of the first system (say), for correlated system, is performed,
this projects the other system into new states; otherwise the systems are
uncorrelated.
Squeezing property is the important phenomenon distinguishing well
mechanism of correlation of systems, where squeezing can occur in
combination of the quantum mechanical systems
 even if single systems are not themselves squeezed.

The most significant example for compound squeezing is related to
the two-mode  squeeze operator defined by
an effective unitary operator as

\begin{equation}
\hat{S}(\xi_{2}) = \exp (\xi^{*}_{2}\hat{a}^{\dagger} \hat{b}^{\dagger}
- \xi_{2}\hat{a}\hat{b}), \label{4}
\end{equation}

\noindent where $\hat{a}$ and $\hat{b}$ are the annihilation  operators
of first and second modes respectively, and $\xi_{2}$ is a complex
parameters. Expression (\ref{4}) represent
the evolution operator of  the nondegenerate parametric amplifier with
classical pumping and this operator can produce perfect squeezing
only in the correlated states of two field modes.
There are several models  which have been given in the literature, e.g.
~\cite{entan1}--~\cite{entan5}.  The aim of these articles
is to examine the statistical properties for single- and compound-modes.
Moreover, the ideas that quantum correlation can give rise to squeezing in
the combination of system operators has been shown true for multimode
squeezed
states of light ~\cite{[2],{[3]}, {[10]},{entan1},{entan6},{entan7}}
 and for dipole
fluctuations in multimode squeezed states ~\cite{entan2}.
\subsection{Phase properties of squeezed states}
In classical optics, the concepts of the intensity and  phase of optical
fields have a well-defined meaning. That is the electromagnetic field ($E$)
associated with one mode, $E=A\exp (i\theta)$, has  a well defined
amplitude ($A$) and phase ($\theta$). This is not so simple in quantum optics
where the mean photon number and the phase are represented by noncommuting
operators and consequently they cannot be
defined well simultaneously. In fact, the concept of phase is a controversional
problem from the earlier days of quantum optics ~\cite{phasx1,{phas15},{phas16}}. In general
there are three methods of treating this issue ~\cite{phasx2}.
The first one considers the phase as a physical quantity in analogy to
position or momentum by representing it with a linear Hermitian phase
operator. The second one involves c-number variables (real or complex) in
phase spaces or  their associated distribution functions, or ensembles
of trajectories. The third one is the operational phase approach in which
the phase information is inferred from the experimental
data by analogy with the classical analysis of the experiment.
Each approach has advantage and disadvantage points.

Squeezed states have phase  sensitive noise
properties and therefore several works have been devoted to follow
such properties. We can mention that the authors of ~\cite{phas1}
have investigated the fluctuation properties of squeezed states
using a phase-operator formalism defined by Susskind and Glogower
~\cite{phas2}. They have shown that similarly as coherent states of high
intensity
approach a semiclassical number-phase uncertainty product, the squeezed
states retain their quantum properties and their number-phase uncertainty
relations are not minimized.
However, the exact general results of the same technique have been obtained
in ~\cite{phas3,{phas4}} for single- and two-mode squeezed states. Indeed,
these are mathematical treatment for the problem. Exact phase calculations
for different definitions have been presented in ~\cite{phas5} showing that
the measured-phase-operator formalism leads to contrasting behaviour compared
with that based on the Susskind-Glogower or Hermitian-phase-operator
formalisms. In the framework of Pegg-Barnett formalism
~\cite{phas6,{phas7},{phas8},{phas9}} several works have been done
treating not only  single-mode squeezed states ~\cite{phas10,{phas11}}
but also two-mode squeezed states ~\cite{phas12}.
The main results of these articles are: for single-mode squeezed states
with non-zero displacement coherent amplitude, the phase distribution exhibits
the bifurcation phenomenon or single peak-structure under certain conditions.
However, it has been shown that the joint phase distribution for the
two-mode squeezed vacuum depends only on the sum of the phases of the
two modes, and that the sum of the two phases is locked to a certain
value as the squeeze parameter increases ~\cite{phas14}.

We refer to ~\cite{phas15,{phas16}} where the investigation of phase
properties of various quantum mechanical models in greater
details have been given.

\subsection{Squeezed states with thermal noise}
It is worthwhile mentioning that squeezing of thermal radiation field
has been already produced in a microwave Josephson-junction parametric
amplifier ~\cite{ther1} where a thermal input field has been considered in a
squeezing device and the generated field exhibits substantial noise reduction.
The aim of such work is to generate nonclassical fields for interaction
studies with Rydberg
atoms in high-Q microwave cavities, where thermal noise in input fields
is
always large. Further, it has been suggested ~\cite{[36]} that interferometers
for
detection of gravitational waves could employ squeezing techniques in
order
to improve their resolution. Practically, these systems will inevitably
experience thermal noise, so it is essential to be aware of the
various
representations of squeezed states and their physical interpretation
~\cite{ther3}.

As a result of the fact that signal beams are usually accompanied by thermal
noise, many authors
concentrated on the studying of the influence of thermal noise on the
behaviour of quantum states ~\cite{ther4}--~\cite{ther14}.
Some of these studies give particular attention to the calculation
of photon-counting distribution ~\cite{ther4}--~\cite{ther8}. This calculation
basically  depends on the Gaussian form of Wigner function. It has been shown
that for large values of the squeeze parameter  the  photon-counting
distribution is oscillating even for strong thermal noise. The quantum
statistical properties of squeezed thermal light ~\cite{ther9}--~\cite{ther13},
e.g. quadrature squeezing, second- and higher-order correlation functions,
different representation for the density operator, and
quasiprobability distribution functions, reveal that
the degree of purity of the input thermal light is left unchanged by the
subsequent squeezing and displacement processes. Moreover, an
effective squeezing
could be recognized under particular choice of the parameters and this fact
is ensured by the behaviour of Glauber $P$-function which it no longer
exists.
For strong squeezing both squeezing properties and normalized
correlation functions to all orders do not depend on the initial intensity
of thermal field.
The  quantum phase distributions and variances of strong coherent and
phase-squeezed states mixed with thermal light have been  considered
in ~\cite{ther14} showing  that the effect of thermal noise
on the coherent phase distribution becomes important only when the number of
thermal photons is of the order of one-half ($(1/2)e^{-r}$, $r$ is the real
squeeze parameter).

Here we may refer also to the squeezed thermal reservoir which has been
studied in detail ~\cite{ther15}--~\cite{ther18}. One can see that the authors of
 ref. ~\cite{ther17} have shown for
atom radiation in a squeezed thermal reservoir that the two quadratures of
the atomic polarization are damped at different rates, which is
consistent
with the case of a squeezed vacuum reservoir. Furthermore, when the atom is
driven by a coherent field, it has been found that the steady-state
polarization
depends on the relative phase of the squeezing and  driving field.
This phase dependence becomes less pronounced when the number of thermal
photons increases. This behaviour suggests a number of novel applications,
such as new schemes for optical bistability ~\cite{ther19}. Nevertheless,
an objection against the reservoir technique can be given when an exact
solution of the density matrix
equation becomes unavailable even in the steady-state regime.
\section{Light propagation}
Classical optics describes quite successfully the propagation of laser
light, both in free-space and inside a transparent medium. The reason
is that in a coherent state of radiation, the electric and magnetic fields
may be written in terms of their expectation values, and thus their
propagation may be treated classically through the macroscopic Maxwell equations.
The classical Maxwell equations permit the calculation of both the spatial
progression and the temporal evolution of a propagating electromagnetic
field, and treat the interaction of the field in a medium phenomenologically
through the induced polarization. However, in quantum optics the
simultaneous use of the Hamiltonian and the momentum operators yields
operatorial spatial-temporal equations of motion for the electric and magnetic fields,
having a form equivalent to that of the classical Maxwell equations and thus
the quantum propagative phenomena can be rigorously described
~\cite{prop1,{mook}}.

In this section we throw the light on two examples representing the
 propagation of light in media and which are frequently
used in this thesis; namely, parametric processes and nonlinear directional
 coupler.
\subsection{Parametric processes}
There has been a great interest in the field of nonlinear optics for
both the practical applications and the theoretical aspects of the
nonlinear effects.
 Experiments in this field were made possible by
the fact that the lasers with a sufficiently high output ($10^5$ to
$10^6$ Watt/cm$^2$) had become easily available ~\cite{grah1}. At this power
level the nonlinear susceptibilities of certain media were producing
observable effects ~\cite{bloe1},  e.g. such as the phenomena of parametric
fluorescence and
parametric oscillation ~\cite{gior1}. Indeed, linearity or nonlinearity is a
property of the medium through which the light travels, rather than the
property of the light itself. Light interacts with light via the medium.
More precisely, the presence of an optical field modifies the properties
of the medium which, in turn, modify another optical field or even the
original field itself ~\cite{salh1}.
For example, the study of the structure of crystals and
molecules has long utilized the phenomena of light scattering from atoms
or molecules having two energy levels. The frequency of the incident
beam  may then be shifted up or down, by an amount equal to the difference
in the two energy levels of the scatterer. The resulting lower- and
higher- frequency scattered waves are the Stokes and anti-Stokes
components, respectively ~\cite{mish1}. However, in the coherent Raman
effect the presence of a monochromatic light wave in a Raman active
medium gives rise to parametric coupling between an optical vibrational
mode and a mode of the radiation field which represents the scattered
(Stokes) wave. In the case of Brillouin scattering a similar form of
coupling holds, with the vibrational mode oscillating at an acoustic
rather than an optical frequency ~\cite{[3]}.

As  was known the nonlinear processes in the quantum mechanical domain
has led to the prediction and the observation of many quantum phenomena,
e.g. squeezing of vacuum fluctuations and photon antibunching.
In the heart of nonlinear  optics there are two
 significant  processes  which have been attracted
amount of study; namely, parametric frequency converter (PFC) and
parametric amplifier (PA).  PFC can be described by a process of
exchanging photons between two optical fields of different frequencies:
signal mode at frequency $\omega_{1}$ and idler mode at frequency
$\omega_{2}$. This model  can be  applied to describe
various optical phenomena, e.g. to find analogies between PFC
and  beam splitter ~\cite{perin}, two-level atom driven by a single mode
of electromagnetic field ~\cite{miel}, and Raman scattering ~\cite{perin,{orlov}}.
The quantum properties of PFC are discussed in
~\cite{louis}. Further, some authors studied this model as the lossless
linear coupler, e.g. ~\cite{jansz1}-~\cite{jansz4}. In this situation the model is
considered to be represented by two electromagnetic waves which
are guided  inside the structure consisting of
two adjacent and parallel waveguides; the linear exchange of energy between
these two waveguides
is established via the evanescent field ~\cite{marc}.

On the other hand, PA is designed in the most familiar form   to
amplify an oscillating signal by means of a particular coupling of the
mode  to a second mode of oscillation, the idler
mode. The coupling parameter is made to oscillate with time in a way
which gives rise to a steady increase of the energy in both the signal
and idler modes ~\cite{[3]}. The importance of PA
is related to the fact that it is the source for  squeezed light ~\cite{[6]}.
For example, degenerate  and non-degenerate PA  are  sources  for
single-mode ~\cite{[6]} and two-mode ~\cite{entan1,{entan2},{entan3}}
squeezing of vacuum fluctuations, respectively.

The parametric process has been employed in experiments. For example, the
fourth-order interference effects arise when pairs of photons produced in
parametric down-conversion are injected into Michelson interferometers ~\cite{kwia}.
However, the second-order interference is observed in the superposition of
signal photons from two coherently pumped parametric down-conversions when the
paths of the idler photons are aligned ~\cite{zou1}.  Further, squeezed states
of the electromagnetic field are generated by degenerate parametric
down-conversion in optical cavity ~\cite{[16]} where noise reductions greater than
$50\%$ relative to the vacuum noise level are observed in a balanced homodyne
detector.  Also, the observation of high-intensity sub-Poissonian
light using the correlated  "twin"  beams generated by
an optical parametric oscillator has been demonstrated ~\cite{mertz}.

The optical processes  involving the competition between PFC and
 PA are of interest from theoretical and experimental points of view,
 e.g. in three-mode interaction ~\cite{mish1,{martin}}.
For these processes  the quantum theory can be briefly reported as
follows ~\cite{shen}: The nonlinear interaction couples different
photon modes and leads to energy transfer between modes.  Photons in
some modes may be annihilated, while those in the other modes created,
and hence the photon distribution is disturbed. In every time the rate of
energy transfer between the modes depends on the statistical properties
of the light fields. Statistics is particularly important in this case
for analyzing the results of experiments. As is expected the output
field depends on the statistical nature of both the incident beams and
the fluctuations in the medium.
Moreover, the measurements of the statistical properties of the output field
could yield information on the  properties of the medium
if those of the incident radiation are known.

\subsection{Nonlinear directional coupler}
In quantum optics many simple quantum systems have been examined from the
point of view of completely quantum statistical description including not
only amplitude and intensity (energy) development of such systems, but also
higher-order moments and complete statistical behaviour. Such results  have
fundamental physical meaning for interpretation of quantum theory
~\cite{cou1} and
they are useful for applications in optoelectronics and photonics as
well. These results can be successfully transferred to more complicated and
more practical systems, such as optical couplers composed of two or more
waveguides connected linearly by means of evanescent waves.
The waveguides
 can be linear or nonlinear employing various nonlinear optical
processes, such as optical parametric processes, Kerr effect, Raman or
Brillouin scattering, etc.
Such devices play important role in
optics, optoelectronics and photonics as switching and memory elements for
all-optical devices (optical processors and computers). When one linear and
the other nonlinear waveguides are employed, we have a nonlinear optical
coupler producing nonclassical light in the nonlinear waveguide which can be
controlled from the linear waveguide, i.e. one can control light by light. The
generation and transmission of nonclassical light exhibiting squeezed vacuum
fluctuations and/or sub-Poissonian photon statistics in nonlinear optical
couplers can further be supported when all the waveguides are nonlinear. The
possibility to generate and to transmit effectively nonclassical light in
this way is interesting especially in optical communication and
high-precision measurements where the reduction of quantum noise increases
the precision.

Since the pioneering work on nonlinear couplers which has been done by
Jensen ~\cite{jen1}, a series of articles have been devoted to
the study of this important optical device from both classical
~\cite{clas1}--~\cite{clas11} (using the coupled-mode theory) and quantal
viewpoints ~\cite{qu1}--~\cite{faisal2}.
In the framework of quantum mechanics particular attention has been
paid to quantum statistical properties in relation
to quantum noise and generation and transmission of nonclassical light.
Such generation and transmission of
nonclassical light can be very effective as a consequence of using
evanescent waves involved in the interaction.
 For a review of role of quantum statistical
properties in nonlinear couplers, see ~\cite{qu20}.

\chapter{Goals of the thesis}
The main goal of the doctoral thesis is a study of  quantum statistics
for some static and dynamic regimes of nonlinear processes in quantum optics.

In the first part (static regime):
1- we want to develop a general class of quantum
states as a superposition of displaced and squeezed number states.
We study the quantum statistics for this class rigorously.
We also investigate the effect of thermal noise on the properties of
 such class of states. Also we report the methods of generation
 for such superposition.\newline
2- we  want to give an example of
 new type of multidimensional squeeze operator model which is
more general than usually used and which includes two
different squeezing mechanisms. All basic properties related to
this operator are discussed in greater details.

In the second part of this thesis  we concentrate on studying the statistical
properties of an optical field propagating within a nonlinear directional coupler.
Our starting point is the Hamiltonian, which represents a nonlinear directional
coupler.
We assume one-passage propagation, so that losses in the beams have been
neglected. In other cases they can be described in the standard quantum
way in the form of interaction of light beams with reservoirs, as for
instance described in ~\cite{goal}.
Moreover, we  treat the problems of propagation in the Hamiltonian formalism
assuming the energy of the system does not have directionality. However, in the
case
that all waves are propagating with the same velocity, time and space
relate by the velocity of propagation $v$, $z=vt$.
For all these models we investigate the effect of switching between the input
modes and the outgoing fields from the coupler.
We can consider three problems of light propagation in this device:
1- a symmetric directional coupler operating by
nondegenerate parametric amplification.\newline
2-  nonlinear optical couplers composed of two nonlinear
waveguides operating by the second subharmonic generation, which are
coupled linearly through evanescent waves and nonlinearly through nondegenerate
optical parametric interaction.\newline
3-
a nonlinear asymmetric directional coupler
composed of a linear waveguide and a nonlinear waveguide operating by
nondegenerate parametric amplification.

\chapter{ Methods and tools of   quantum theory used}
In this chapter we review the quantum methods and tools for controlling  the
nonclassical phenomena, such as correlation function, quadratures squeezing,
quasiprobability distribution
function, photon-number distribution and phase distribution.
Further, we do not describe a complete details for these methods and
tools  when they
can be found in the standard text book.  Moreover, we  write down the
formulae related to the dynamical regime where those of the static
regime can be obtained by simply setting $t=0$ as a main tool the
coherent-state technique is used.

\section{ Correlation functions}
Antibunched and/or sub-Poissonian light is an example of nonclassical
light and can be determined
from a photocounting-correlation measurement.
Starting with the experiment of Hanbury Brown and Twiss ~\cite{chs1},
strong interest in
the photon-counting statistics of optical fields began. Traditional
diffraction and interference experiments and spectral measurements may be
considered as being performed in the domain of one photon or linear optics.
The theory of higher-order optical phenomena, described by higher-order
correlation functions of the electromagnetic field, was founded by Glauber
~\cite{chs2}, who introduced the measure of super-Poissonian statistics
 (classical phenomenon) and
sub-Poissonian statistics (nonclassical phenomenon) of photons in any state.
A state (of a single mode for convenience) which displays sub-Poisson
statistics is
characterized by the fact that  the variance of the photon number
$\langle (\triangle \hat{n}_{j}(t))^{2}\rangle$
is less than the average photon number
$\langle \hat{n}_{j}(t)\rangle =\langle \hat{A}^{\dagger}_{j}(t)
\hat{A}_{j}(t)\rangle$. This  can be written by means of
 the normalized normal second-order
correlation function  as

\begin{equation}
g_{j}^{(2)}(t) = \frac{\langle \hat{a}_{j}^{\dagger 2}(t)
\hat{a}_{j}^{2}(t)\rangle } {\langle \hat{a}_{j}^{\dagger}(t) \hat{a}
_{j}(t)\rangle^{2}} \\
 =1+\frac{\langle (\triangle\hat{n}
_{j}(t))^{2}\rangle - \langle \hat{a}_{j}^{\dagger}(t)
\hat{a}_{j}(t)\rangle}{\langle \hat{a}_{j}^{\dagger}(t)
\hat{a}_{j}(t)\rangle^{2}}, \label{5}
\end{equation}

\noindent where the subscript $j$ relates to the $j$th mode, and
the photon number variances have the form

\begin{equation}
\langle (\triangle\hat{n}_{j}(t))^{2}\rangle =
\langle(\hat{a}_{j}^{\dagger}(t) \hat{a}_{j}(t))^{2}\rangle -\langle \hat{a}
_{j}^{\dagger}(t) \hat{a}_{j}(t)\rangle^{2}.  \label{6}
\end{equation}
 Then it holds that $g_{j}^{(2)}(t)<1$ for
sub-Poissonian distribution of photons, $g_{j}^{(2)}(t)>1$ for
super-Poissonian distribution of photons and when $g_{j}^{(2)}(t) =1$
Poissonian distribution occurs corresponding to a coherent state.
Furthermore, for instance, the generation of sub-Poissonian light has been established in
a semiconductor laser ~\cite{chs3} and in the microwave region using masers
operating in the microscopic regime ~\cite{chs4}.
An application of radiation exhibiting the sub-Poissonian
statistics to optical communications has been considered in ~\cite{chs5}.

On the other hand, it has been shown explicitly in ~\cite{chs6,{chs7}}
that sub-Poissonian photon
statistics need not be associated with antibunching, but can be
accompanied by bunching. However, within the framework of the
classical theory, light cannot be antibunched, i.e. antibunched
light is a manifestation of a quantum effect.
The basic formula to study this phenomenon
is the two-time normalized intensity  correlation function ~\cite{chs7,
{chs8}}.
 For the
$j$th mode, this function is defined by

\begin{equation}
g^{(2)}_{j}(t,t+\tau)=\frac{\langle
\hat{a}^{\dagger}_{j}(t)\hat{a}^{\dagger}_{j}(t+\tau)\hat{a}_{j}(t+\tau)
\hat{a}_{j}(t)
\rangle}{\langle \hat{a}^{\dagger}_{j}(t)\hat{a}_{j}(t)\rangle
\langle \hat{a}^{\dagger}_{j}(t+\tau)\hat{a}_{j}(t+\tau)\rangle}.\label{7}
\end{equation}

\noindent The importance of this function in the analysis of photon
antibunching comes from the direct relation between this function and the
joint detection probability of  two photons, one at time $t$
and another at time $t+\tau$. It is clear that using (\ref{7}) for
 $\tau\rightarrow 0$
as a definition of bunching properties, then
 the bunching/antibunching and super-/sub-Poissonian statistics
 are in one-to-one correspondence.
More general definition of photon antibunching can be adopted
~\cite{chs7,{chs8}} if $g^{(2)}_{j}(t,t+\tau)$ increases
from its initial value at $\tau=0$. This can be represented in
equivalent differential form, assuming that $g^{(2)}_{j}(t,t+\tau)$ is
a well behaved function in $\tau$, as

\begin{equation}
K_{j}(t)=\frac{\partial g^{(2)}_{j}(t,t+\tau)}{\partial
\tau}|_{\tau=0}>0; \label{8}
\end{equation}
\noindent photon bunching is given by the opposite condition ($K_{j}(t)<0$),
otherwise the photons are unbunched. It is reasonable  pointing
out that antibunching and sub-Poissonian behaviour always accompany each
other for the single-mode time-independent fields ~\cite{chs9}.

Finally, we turn our attention to discuss the effect of intermodal
correlation in terms of anticorrelations between different modes in the model.
This can be done by two means. The
first mean is given by introducing the photon-number operator
$\hat{n}_{j,k}=\hat{A}^{\dagger}_{j}\hat{A}_{j}+
\hat{A}^{\dagger}_{k}\hat{A}_{k}$ and calculating the quantity $\langle
(\triangle W_{j,k})^{2}\rangle$=\\$\langle
:(\hat{n}_{j,k})^{2}:\rangle -\langle \hat{n}_{j,k}\rangle^{2}$, where :
: denotes the normally ordered operator, i.e. creation operators
$\hat{A}^{\dagger}_{j} $ are to
the left of annihilation operators $\hat{A}_{j}$.
The quantum anticorrelation effect is then characterized in terms of
the variance of
the photon number, which is less than the average of the photon number
for nonclassical light, by negative values
of $\langle (\triangle W_{j,k})^{2}\rangle$, i.e. negative
cross-correlation taken two times is stronger than the sum of quantum
noise in single modes ~\cite{chs10}.
The second way is based on
 violation of Cauchy-Schwarz
inequality.  The violation of Cauchy-Schwarz inequality can be
observed in a two-photon interference experiment ~\cite{chs11}.
Classically, Cauchy-Schwarz inequality has the form ~\cite{chs12}

\begin{equation}
\langle I_{1}I_{2}\rangle^{2}\leq \langle I^{2}_{1}\rangle \langle
I^{2}_2\rangle, \label{9}
\end{equation}

\noindent where $I_{j},j=1,2$ are classical intensities of light
measured by different detectors in a
double-beam experiment.
In quantum theory,
 the deviation from this classical inequality can be represented
by the factor ~\cite{chs13}

\begin{equation}
I_{j,k}=\frac{[\langle \hat{A}^{\dagger 2}_{j}\hat{A}^{2}_{j}
\rangle\langle \hat{A}^{\dagger 2}_{k}\hat{A}^{2}_{k}\rangle]^{\frac{1}{2}}}{
\langle
\hat{A}^{\dagger}_{j}\hat{A}_{j}\hat{A}^{\dagger}_{k}
\hat{A}_{k}\rangle}-1. \label{10}
\end{equation}

\noindent The negative values for the quantity $I_{j,k}$
mean that the intermodal correlation is larger than the correlation between
the photons in the same mode ~\cite{entan5} and this indicates strong
violation of the Cauchy-Schwarz inequality. Finally, anticorrelation
between modes can be measured by detecting single modes separately by
two photodetectors and correlating their outputs.
\section{Quadrature squeezing}
As we mentioned earlier, squeezed light  possesses less noise than a coherent light
in one of the field quadratures and can exhibit a number of features
having no classical analogue. This light can be measured by
homodyne detection where the signal is superimposed on a strong
coherent beam of the local oscillator.

There are several definitions for squeezing, e.g. standard squeezing
~\cite{chs14}, amplitude-squared squeezing ~\cite{chs15}, two-mode squeezing
~\cite{chs16}, higher-order squeezing ~\cite{chs17},
principal squeezing ~\cite{chs18}, etc.
Of course,  the  quantum mechanical systems can exhibit different types
of squeezing at the same time.
It has been shown that when a beam of light propagates through a
nonlinear crystal, in the process of generation of the second harmonic, the
fundamental mode becomes squeezed in the sense of standard ~\cite{chs14} as
well as amplitude-squared squeezing ~\cite{chs15}. Further,
the parametric amplifier in a cavity can ideally produce squeezed light
with characteristics akin to the single and two modes when operating in
degenerate and non-degenerate regimes, respectively ~\cite{chs16}.

In this thesis
 we investigate single-mode, two-mode and three-mode
squeezing on the basis of the two quadratures $\hat{X}_{n}$ and
$\hat{Y}_{n}$ (where the subscript $n$ takes on 1,2,3 associated with
the single-, two- and three-mode squeezing) which are
related to the conjugate electric and magnetic field operators
$\hat{E}$ and $\hat{H}$. They are defined in the standard way.
Assuming that these two  quadrature operators satisfy the following
commutation relation

\begin{equation}
\left[ \hat{X}_{n},\hat{Y}_{n}\right] =C, \label{9b}
\end{equation}

\noindent  where $C$ is a c-number  specified later,  the following
uncertainty relation holds

\begin{equation}
\langle (\triangle \hat{X}_{n})^{2}\rangle \langle (\triangle \hat{Y}_{n}
)^{2}\rangle \geq \frac{|C |^{2}}{4}, \label{10b}
\end{equation}

\noindent  where $\langle (\triangle \hat{X}_{n})^{2}\rangle=\langle
\hat{X}_{n}^{2}\rangle -\langle \hat{X}_{n}\rangle ^{2}$ is the variance.
Therefore, we can say that the model possesses $X_{n}$-quadrature squeezing
if the $S_{n}$-factor ~\cite{chs19},

\[
S_{n}=\langle (\triangle \hat{X}_{n}(t))^2\rangle -
0.5|C |,  \]
\begin{equation}
=\frac{\langle (\triangle \hat{X}_{n}(t))^2\rangle -
0.5|C |}{0.5|C |} ,   \label{11}
\end{equation}

\noindent satisfies the inequality $-1\leq S_{n}<0$. Similar expression
 for the $Y_{n}$-quadrature  ($Q_{n}$-parameter) can be obtained.

For example, the two quadratures of three-mode squeezing are defined  as

\begin{equation}
\hat{X}_{3}(t)=\frac{1}{2}[ \hat{A}_{1}(t)+\hat{A}_{2}(t)
+ \hat{A}_{3}(t)+\hat{A}_{1}^{\dagger}(t)+\hat{A}_{2}^{\dagger}(t)+
\hat{A}_{3}^{\dagger}(t)], \label{12}
\end{equation}

\begin{equation}
\hat{Y}_{3}(t)=\frac{1}{2i}[ \hat{A}_{1}(t)+\hat{A}_{2}(t)+
\hat{A}_{3}(t)-\hat{A}_{1}^{\dagger}(t)-\hat{A}_{2}^{\dagger}(t)-
\hat{A}_{3}^{\dagger}(t)]. \label{r13}
\end{equation}
The expressions for the single-mode and
two-mode squeezing can be
obtained easily from (\ref{12}) and (\ref{r13})
 by dropping the operators of absent mode,
e.g. for the 1st mode, single-mode squeezing can be obtained by setting
the operators of 2nd and 3rd modes equal zero. It should be taken into
account that $C=\frac{1}{2},1,\frac{3}{2}$ corresponding to the single-mode,
two-mode and three-mode squeezing, respectively.
\section{Quasiprobability functions}
Evaluation of various time-dependent mode observables is most
conveniently achieved with the aid of corresponding time-dependent
characteristic functions, their normal, antinormal and symmetric
forms, and the Fourier transforms of these characteristic functions
(quasiprobability functions). All of these are related to the density
matrix which provides a complete statistical description of the system.
There are three types of quasiprobability functions:
Wigner $W$-, Glauber $P$-, and Husimi $Q$-functions.
These functions could be used also as  crucial to
describe the nonclassical effects of the system, e.g. one can employ the
negative values of $W$-function, stretching of $Q$-function and high
singularities in $P$-function. Furthermore, these functions are now
accessible from measurements ~\cite{wig}.

Indeed, the detailed statistics of the  coupled field modes can be
obtained from several photon-counting experiments. Most often we are
interested in the quantum statistics of either one mode which
determine the ensemble averages of the observable of this mode, or the
composite statistics of the compound modes which reflect their mutually
correlated properties. That is why we consider in this thesis phase
space distributions  for the single- and compound-modes
for different types of initial states.

The $s$-parametrized joint characteristic function
for the compound system ($n$ modes) defined by

\begin{equation}
C^{(n)}(\underline{\zeta},t,s)={\rm Tr} \left\{
\hat{\rho}(0)\exp \left[\sum_{j=1}^{n}(\zeta_{ j} \hat{A}_{j}^{\dagger}(t)-
\zeta_{j}^{*} \hat{A}_{j}(t)+\frac{s}{2}|\zeta_{j}|^{2})\right] \right\},
\label{13}
\end{equation}

\noindent where $\hat{\rho}(0)$ is the initial density matrix  for the
system  under consideration, $\underline{\zeta}=(\zeta_{1},\zeta_{2},...,
\zeta_{n})$ is a set of parameters, and $s$ takes on values $1, 0$ and $-1$ corresponding to
normally, symmetrically and antinormally ordered characteristic functions,
respectively.

Thus the $s$-parametrized joint quasiprobability functions
are given by

\begin{equation}
W^{(n)}(\underline{\beta},t,s)=\frac{1}{\pi^{2n}}
\underline{\int} C^{(n)}(\underline{\zeta},t,s)
\prod_{j=1}^{n}\exp(\beta_{j}\zeta_{j}^{*}-
\beta_{j}^{*}\zeta_{j}) d^{2}\zeta_{j},
\label{14}
\end{equation}

\noindent where
$\underline{\beta}=(\beta_{1},\beta_{2},...,\beta_{n})$  are complex
amplitudes and
$\underline{\int}=\int...\int$ $n$-fold integral. When
$s=1, 0, -1$, equation (\ref{14}) gives formally Glauber
$P$-function, Wigner $W$-function and Husimi $Q$-function, respectively.

On the other hand, the $s$-parametrized quasiprobability  functions
for the single-mode can be attained by means of integrating $(n-1)$
 times in the corresponding
joint quasiprobability functions or by using the single-mode
$s$-parametrized characteristic function.
For instance, the characteristic function $C^{(1)}(\zeta_{1},t,s)$ for mode
$\hat{A}_{1}$ (say) can be
obtained from $C^{(n)}(\underline{\zeta},t,s)$ by simply setting
$\zeta_{k}=0, \quad k=2,3,...,n$.
Hence the $s$-parametrized single-mode  quasiprobability  can be derived by

\begin{equation}
W^{(1)}(\beta_{1},t,s)=
\underline{\int} W^{(n)}(\underline{\beta},t,s)
\prod_{j=2}^{n}d\beta_{j},
\label{15}
\end{equation}

\noindent etc., or

\begin{equation}
W^{(1)}(\beta_{j},t,s)=\frac{1}{\pi^{2}}
\int
C^{(1)}(\zeta_{j},t,s)
\exp(\beta_{j}\zeta_{j}^{*}-
\beta_{j}^{*}\zeta_{j}) d^{2}\zeta_{j}.
\label{16}
\end{equation}

\noindent  $C^{(1)}(\zeta_{j},t,s)$ in (\ref{16}) is the $s$-parametrized
 characteristic  function for single-mode.
The superscripts (1) and (n) in the above equations
stand for single-mode case and $n$-mode case, respectively.

The various moments of the bosonic operators for
the system, using the characteristic functions and quasiprobability
functions, in the normal
form (N), antinormal form (A) and symmetrical form (S), corresponding to
$s=1,-1,0$, respectively,
 can be obtained by

\[
\langle \prod _{j=1}^{n}\hat{A}_{j}^{\dagger m_{j}}(t)\hat{A}_{j}^{n_{j}}(t)
\rangle_{N,A,S}= \prod _{j=1}^{n}\frac{\partial^{m_{j}+n_{j}}}
                           {\partial \zeta _{j}^{m_{j}}
                            \partial (-\zeta _{j}^{*})^{n_{j}}}C^{(n)}
                             (\underline{\zeta },s,t)_{s=1,-1,0}
                             |_{\underline{\zeta }=\underline{\zeta}^{*}=0}\]

\begin{equation}
=\underline{\int} W^{(n)}(\underline{\beta },s,t)_{(s=1,-1,0)}\prod
_{j=1}^{n}
\beta_{j}^{*m_{j}}\beta _{j}^{n_{j}}d^{2}\beta _{j},
\label{17}
\end{equation}

\noindent where $n_{j}, m_{j}$ are positive integers.
The formulae (\ref{17}) are valid for the single mode and compound
modes as well.

There are several applications of the quasiprobability function.
We restrict ourselves to those of the single-mode case because the generalization
to the multi-mode case is  straightforward.

First, the degree of purity ${\rm Tr}\rho ^{2}$ for any state
can be evaluated with aid of the symmetrical characteristic function
given by the relation

\begin{equation}
{\rm Tr}(\rho ^{2})=\frac{1}{\pi }\int |C(\zeta
,s=0)|^{2}d^{2}\zeta
.\label{scf36}
\end{equation}
One can find for a pure state ${\rm Tr}(\rho ^{2})=1$ and for a
mixed state ${\rm Tr}(\rho ^{2})<1$.
Also, the quasiprobability function can be used to calculate
the matrix elements of $\hat{\rho}_{1}(t)$  in the
$n$-quantum representation at time $t$  by means of
 the relation ~\cite{[3]}

\begin{equation}
\langle m|\hat{\rho}_{1}(t)|n_{1}\rangle=
\int\frac{\beta^{m}_{1}\beta^{*n_{1}}_{1}}{\sqrt{n_{1}!m!}}
W^{(1)}(\beta_{1},s=1,t)\exp (-|\beta_{1}|^{2}) d^{2}\beta_{1}.
\label{18}
\end{equation}
It is clear that this formula reduces to the photon-number distribution
when $m=n_{1}$.
Further, the photon-number distribution can be obtained also from
 the integral relation in terms
of Wigner function of the specified mode and Laguerre polynomials as

\begin{equation}
P^{(1)}(n_{j},t)=\frac{2(-1)^{n_{j}}}{n_{j}!}\int W^{(1)}(\beta
_{j},t)\exp(-2|\beta _{j}|^{2})L_{n_{j}}(4|\beta _{j}|^{2})d^{2}\beta
_{j},   \label{19}
\end{equation}

\noindent where $j$ denotes the mode under consideration and $
W^{(1)}(\beta_{j} ,t)$ is the single mode Wigner function for
the specified mode. Also quasiprobability
functions can be used to investigate the phase
distribution for the structure by integrating these functions over
the radial variable ~\cite{phasx1,{phas15},{phas16}}

\begin{equation}
P(\theta,s,t)=\int_{0}^{\infty}W^{(1)}(\beta_{j},s,t) |\beta_{j}| d |\beta_{j}|,
\label{20}
\end{equation}
where $\beta_{j}=|\beta_{j}|\exp(i\theta)$.

More details on the phase properties in terms of Pegg-Barnett formalism
will be given in the following section.
\section{Phase properties-Pegg-Barnett formalism}
Here we give essential background for Pegg-Barnett ~\cite{phas6}--~\cite{phas9}
phase formalism.
This formalism is based on introducing a finite $(s+1)$-dimensional space
$\Psi$ spanned by the number states $|0\rangle,|1\rangle,...,|s\rangle$.
The physical variables (expectation values of Hermitian operators)
are evaluated in the finite dimensional space $\Psi$ and  at
the final stage the limit $s\rightarrow \infty$ is taken.
A complete orthonormal basis of $s+1$ states is defined on $\Psi$ as

\begin{equation}
|\Theta_{m}\rangle =\frac{1}{\sqrt{s+1}}\sum^{s}_{k=0}\exp (ik\Theta_{m})
|k\rangle, \label{21}
\end{equation}
where

\begin{equation}
\Theta_{m}=\Theta_{0}+\frac{2\pi m}{s+1}, \qquad m=0,1,...,s.
\label{22}
\end{equation}
The value of $\Theta_{0}$ is arbitrary and defines a particular basis
of $s+1$ mutually orthogonal states. The Hermitian phase operator is
defined as
\begin{equation}
\hat{\Phi}_{\theta}=\sum^{s}_{m=0}\Theta_{m} |\Theta_{m}\rangle
\langle \Theta_{m}|,
\label{23}
\end{equation}
where the subscript shows the dependence on the choice of $\Theta_{0}$.
The phase states (\ref{21}) are eigenstates of the phase operator (\ref{23})
with the eigenvalues $\Theta_{m}$ restricted to lie within a phase window
between $\Theta_{0}$ and $2\pi+\Theta_{0}$.
The expectation value of the phase operator (\ref{23}) in a pure state
$|\psi\rangle=\sum^{\infty}_{m=0}C_{m}|m\rangle$, where $C_{m}$ is the
 weighting coefficient including the normalization constant,
is given by
\begin{equation}
\langle \psi |\hat{\Phi}_{\theta}|\psi\rangle =\sum^{s}_{m=0}\Theta_{m}
|\langle \psi|\Theta_{m}\rangle|^{2}.
\label{24}
\end{equation}

The density of phase states is $(s+1)/(2\pi)$, so the continuum phase
distribution as $s$ tends to infinity is

\begin{eqnarray}
\begin{array}{rl}
P(\Theta)={\rm lim}_{s\rightarrow \infty}\frac{s+1}{2\pi}|\langle
\Theta_{m}|\psi\rangle|^{2} \\
\\
=\frac{1}{2\pi} \sum_{m,m^{'}=0}^{\infty} C_{m}C^{*}_{m^{'}}
\exp[i(m-m^{'})\Theta],
\end{array} \label{25}
\end{eqnarray}
where $\Theta_{m}$ has been replaced by the continuous phase variable
$\Theta$. As soon as the phase distribution $P(\Theta)$ is known, all
the quantum-mechanical phase moments can be obtained as a classical
integral  over $\Theta$.
One of the particular interesting quantities in the description of the phase
is the phase variance determined by

\begin{equation}
\langle (\triangle \hat{\Phi})^{2}\rangle =\int \Theta^{2} P(\Theta)d
\Theta -\left(\int \Theta P(\Theta)d\Theta \right)^{2}.\label{27}
\end{equation}
As is well known the mean photon number and the phase are conjugate
quantities in this approach and consequently they obey the
following uncertainty relation

\begin{equation}
\langle (\triangle \hat{\Phi})^{2}\rangle
\langle (\triangle \hat{n})^{2}\rangle \geq \frac{1}{4}
|\langle [\hat{n},\hat{\Phi}]\rangle|^{2}. \label{28}
\end{equation}

The number--phase commutator appearing on the right-hand side of (\ref{28})
can be calculated for any physical state ~\cite{phas6}--~\cite{phas9} as
\begin{equation}
\langle [\hat{n},\hat{\Phi}]\rangle=i[1-2\pi P(\Theta_{0})].\label{29}
\end{equation}
In relation to (\ref{28}), we can give  the notion of the number and phase
squeezing ~\cite{wod,{buz2}} through the relation

\begin{equation}
S_{n}=\frac{\langle (\triangle \hat{n})^{2}\rangle}
{\frac{1}{2}|\langle [\hat{n},\hat{\Phi}]\rangle|}-1,\label{30}
\end{equation}

\begin{equation}
S_{\theta}=\frac{\langle (\triangle \hat{\Phi})^{2}\rangle}
{\frac{1}{2}|\langle [\hat{n},\hat{\Phi}]\rangle|}-1.\label{31}
\end{equation}
The values of $-1$ in these equations means maximum squeezing of the
photon number or the phase.

In this thesis we calculate the quantities given in this chapter for the models under
discussion and then  numerical simulations are performed
using, e.g. Surfer or Matlab program.

\chapter{Scientific results and their analysis: part I (static regime)}
In this part we investigate the quantum properties for a
superposition of squeezed and displaced number states (SSDNS)
without and with thermal noise. We suggest also a new type of
multidimensional squeeze operator and discuss its properties in details.
\section{ Quantum statistical properties of
superposition of squeezed and displaced number states}
This section is devoted to study properties of
superposition of squeezed and displaced number states (SSDNS) such as
orthogonality property, wave function, photon-number distribution,
quasiprobability functions and phase properties in terms of Pegg-Barnett
formalism.
\subsection{ State formalism and some of its properties}
Here we develop  a general class of quantum states, as a result of a
superposition between two quantum states, described as a single mode vibration
of electromagnetic field which is on a sudden squeezed-plus-displaced by a collection
    of two displacements $180^{0}$ out of phase given as

\begin{equation}
 | r ,\alpha ,n\rangle _{\epsilon}= \lambda _{\epsilon }
 [ \hat{D}(\alpha ) +\epsilon \hat{D}(-\alpha )]\hat{S}(r)|n\rangle ,
\label{scf1}
\end{equation}
where $\hat{D}(\alpha )$ and $\hat{S}(r)$ are the displacement and squeeze
operators, given by (\ref{3}) and (\ref{2}), respectively, with $\xi_{1}=r$
and  $\alpha$ are real;
    $\epsilon =|
\epsilon |e ^{{\rm i}\phi}$ is a parameter specified later on,
$\phi$ is a phase; $|n\rangle$ is the
number (Fock) state, and $\lambda _{\epsilon }$ is a normalization constant
 given as
\begin{equation}
 | \lambda_{\epsilon }^{2}| ^{-1}= 1+|\epsilon |^{2}+2|\epsilon |
\exp (-2t^{2} ) {\rm L}_{n}(4t^2)
\cos \phi ,    \label{scf2}
\end{equation}
where $t=\alpha e^{r}$, ${\rm L}_{n}(.)$ is the Laguerre polynomial.
During the derivation of (\ref{scf2}), we used the relations
\begin{eqnarray}
\begin{array}{lr}
\hat{D}^{\dagger}(\alpha ) \hat{a} \hat{D}(\alpha )= \hat{a} +\alpha ,\qquad
\hat{S}^{\dagger}(r)\hat{a}\hat{S}(r)=\hat{a} C_{r} -\hat{a}^{\dagger}
S_{r},\\
\hat{D}(\zeta )\hat{D}(\beta )=\hat{D}(\zeta +\beta )\exp [\frac{1}{2}(\zeta
\beta^{*} -\zeta^{*}\beta )] ,\quad
\hat{S}^{\dagger}(r)\hat{D}(\beta )\hat{S}(r)=\hat{D}(\beta C_{r} +\beta ^{*}
S_{r}),
\end{array}
\label{scf3}
\end{eqnarray}
and the relation ~\cite{scfr1}
\begin{equation}    
\langle n| \hat{D}(t) | k\rangle  =
\left\{
\begin{array}{rl}
\sqrt{\frac{k!}{n!}} \exp (-\frac{t^{2}}{2}) t^{n-k} {\rm L}^{n-k}_{k}(t^{2})
\;\;&{\rm for}\;n\geq k ,\\
\sqrt{\frac{n!}{k!}} \exp (-\frac{t^{2}}{2}) (-t)^{k-n} {\rm L}^{k-n}_{n}(t^{2})
\;\;&{\rm for}\;k\geq n ,
\end{array}
\right.
\label{scf4}
\end{equation}
where $C_{r}=\cosh r, S_{r}=\sinh r$; ${\rm L}^{\gamma }_{n}(.)$
is associated Laguerre  polynomial. The same states (\ref{scf1}) were investigated
independently in \cite{ALkad}.

The density matrix $\hat{\rho}$ of this state can be written as
\begin{equation}
\hat{\rho}=
| \lambda_{\epsilon }|^{2}(\hat{\rho}_{M}+\hat{\rho}_{I}).  \label{scf5}
\end{equation}
The part of the density matrix corresponding to the statistical mixture of
two squeezed displaced number states is:
\begin{equation}
\hat{\rho}_{M}=
 \hat{D}(\alpha )\hat{S}(r)|n\rangle
\langle n| \hat{S}^{\dagger}(r)\hat{D}^{\dagger}(\alpha )
+|\epsilon|^{2} \hat{D}(-\alpha )\hat{S}(r)|n\rangle
\langle n| \hat{S}^{\dagger}(r)\hat{D}^{\dagger}(-\alpha ),  \label{scf6}
\end{equation}
while the quantum interference part has the form
\begin{equation}
\hat{\rho}_{I}=
 \epsilon^{*}
 \hat{D}(\alpha )\hat{S}(r)|n\rangle
\langle n| \hat{S}^{\dagger}(r)\hat{D}^{\dagger}(-\alpha )
+\epsilon \hat{D}(-\alpha )\hat{S}(r)|n\rangle
\langle n| \hat{S}^{\dagger}(r)\hat{D}^{\dagger}(\alpha ).  \label{scf7}
\end{equation}
This quantum interference part of the density matrix contains information
about the quantum interference between component states
$\hat{D}(\pm\alpha )\hat{S}(r)|n\rangle$ and this will be responsible for
some interesting behaviour of the phase distribution, as we will see.
For completeness, the physical interpretation of such states can be related
to a superposition
of coherent states formed due to two excitations on  particularly excited
harmonic oscillators ~\cite{[29]}.
 It is clear that these states enable us to
obtain generalizations of some results given in the literature,
e.g. ~\cite{ther9,{ther10},{ther12},{perin},{scfr2},{scfr3},
{scfr4},{scfra4},{scfraa4}}. Firstly, the wavefunction of state
(\ref{scf1}) can be determined through ~\cite{[31]} as
\begin{equation}
\Psi^{(\epsilon)}_{n}(x,r,\alpha)=
\langle x|r,\alpha,n\rangle_{\epsilon}
= \lambda _{\epsilon } \frac{\partial ^{n}}{\partial\mu^{n}} \int
\langle x|\mu\rangle\langle \mu|
 [ \hat{D}(\alpha ) +\epsilon \hat{D}(-\alpha )]\hat{S}(r)
 \exp (\mu \hat{a}^{\dagger})|0\rangle \frac{d^{2}\mu}{2\pi}, \label{scf8}
\end{equation}
where we have used the overcompleteness relation for coherent states;
 after substituting for the element $\langle x|\mu\rangle$ from
\cite{ther15} into (\ref{scf8}) and using similar technique as in ~\cite{[31]},
we arrive at (with $\alpha$ is real)
\begin{eqnarray}
\begin{array}{rl}
\Psi^{(\epsilon)}_{n}(x,r,\alpha)=\frac{\lambda_{\epsilon}e
^{\frac{r}{2}}}
{\sqrt{2^{n}n!}}(\frac{\omega}{\pi \hbar})^{\frac{1}{4}}\left\{
\exp\left[
-\frac{e^{2r}}{2}(x\sqrt{\frac{\omega}{\hbar}}-\sqrt{2}\alpha)^{2}
\right]
{\rm H}_{n}[e^{r}(x\sqrt{\frac{\omega}{\hbar}}-\sqrt{2}\alpha)]
\right.
\\
\\
 +\epsilon
\exp\left[ -\frac{e^{2r}}{2}(x\sqrt{\frac{\omega}{\hbar}}+\sqrt{2}\alpha)^{2}
\right]
\left. {\rm H}_{n}[e^{r}(x\sqrt{\frac{\omega}{\hbar}}+\sqrt{2}\alpha)]\right\},
\end{array}
\label{scf9}
\end{eqnarray}
where $\omega$ and $\hbar$ are frequency of the harmonic oscillator and
 Planck's constant divided by $2\pi$; they appeared here as
a result of using the wavefunction $\langle x|\beta \rangle$
when describing a coherent  state of the harmonic oscillator in the
coordinate representation in the derivation of (\ref{scf9}).

The inner product of the ket $| r, \alpha , n \rangle_{\epsilon }$
with the bra $_{\epsilon '}\langle r ', \alpha ', m|$ can be calculated
using the wavefunction
   (\ref{scf9}) through the relation
\begin{equation}
_{\epsilon '}\langle  m, \alpha ' ,r '|n ,\alpha
,r\rangle_{\epsilon}=\int_{-\infty}^{+\infty}dx
\Psi^{(\epsilon ')*}_{m}(x,r',\alpha')
\Psi^{(\epsilon)}_{n}(x,r,\alpha)
, \label{scf10}
\end{equation}
together with the identity
 ~\cite{scfr5}

\begin{eqnarray}
\begin{array}{rl}
\sqrt{\frac{M}{\pi}}\int_{-\infty}^{+\infty}dx
{\rm H}_{m}(x){\rm H}_{n}(\Lambda x+d) \exp(-Mx^{2}+cx)=
\exp (\frac{c^2}{4M})
\\
\\
\times (\sqrt{\frac{M-1}{M}})^{m}
 (\sqrt{\frac{M-\Lambda ^{2}}{M}})^{n}
 \sum_{j=0}^{min(m,n)}  \frac{n! m!}
{j!(n-j)!(m-j)!}
\\
\\
\times \left(\frac{2\Lambda}{\sqrt{(M-1)(M-\Lambda^{2})}}\right)^{j}
 {\rm H}_{m-j}\left(\frac{c}{2\sqrt{(M-1)M}}\right)
{\rm H}_{n-j}\left(\frac{c\Lambda+2dM}{2\sqrt{M(M-\Lambda)}}\right),
\end{array}
\label{scf11}
\end{eqnarray}
as

\begin{eqnarray}
\begin{array}{rl}
 _{\epsilon '}\langle  m, \alpha ' ,r '|n ,\alpha
,r\rangle_{\epsilon}=\frac{\lambda_{\epsilon ' }
\lambda_{\epsilon }(\frac{\tanh R}{2})^{\frac{m}{2}}}{\sqrt{n!
m!\cosh R}}\sum_{j=0}^{min(m,n)}  \frac{n! m![\frac{-\tanh R}{2}]^{\frac{(n-j)}{2}}}
{j!(n-j)!(m-j)!}[\frac{2}{\sqrt{\sinh 2R}}]^{j}
\\
\\
\times
\left\{ \exp [\frac{\tau ^{2}_{1}}{2}(\tanh R-1)]
 {\rm H}_{n-j}(X_{1})
{\rm H}_{m-j}(X_{2})[1+(-1)^{(n+m)}\epsilon '^{*}\epsilon]
\right.
\\
\\
\left. + \exp [\frac{\tau^{2}_{2}}{2}(\tanh R-1)]
{\rm H}_{n-j}(Y_{1}) {\rm H}_{m-j}(Y_{2})
[ \epsilon '^{*}
+(-1)^{(n+m)}\epsilon]\right\}
\end{array}
\label{scf12}
\end{eqnarray}

\noindent and
\begin{eqnarray}
\begin{array}{rl}
R=r-r',\quad
\tau_{1}=(\alpha -\alpha ') e^{r},\quad
\tau_{2}=(\alpha +\alpha ' ) e^{r},\quad
X_{1}=\frac{{\rm i}\tau_{1} e^{-R}}{\sqrt{\sinh 2R}},
\\
X_{2}=\frac{\tau_{1} }{\sqrt{\sinh 2R}},\quad
Y_{1}=\frac{{\rm i}\tau_{2} e^{-R}}{\sqrt{\sinh 2R}},\quad
Y_{2}=\frac{\tau_{2} }{\sqrt{\sinh 2R}},
\end{array}
\label{scf14}
\end{eqnarray}
where ${\rm H}_{m}(.)$ is the Hermite polynomial of order m.
It is clear that for  $\epsilon '=-1, \epsilon =1$ and $n=m$ equation
(\ref{scf12}) vanishes, that is
the inner product of two states, one from even subspace and the other
from odd subspace, is zero. Here we assume that $\epsilon '=\epsilon $.
It is clear from (\ref{scf12}) that SSDNS are not orthogonal
similarly as coherent states,
but under a certain constrains they can be approximately orthogonal,
e.g. when  $\tau_{1}$ and $\tau_{2}$ are very large such that
$\tanh R<1$.
It is easy to prove that for $r '=r, \alpha =\alpha '$ and $n=m$
equation (\ref{scf12})
reduces to unity.

 When $\epsilon^{'}=0,\alpha^{'}=0,r^{'}=0$ and $r\rightarrow 0$,
 (\ref{scf12}) reduces to the distribution
coefficient for the superposition of displaced Fock states
as

\begin{eqnarray}
\begin{array}{rl}
\langle m | \alpha,n,\rangle_{\epsilon}=
\lambda_{\epsilon}\sqrt{n!m!}\exp(-\frac{\alpha^{2}}{2})
\sum_{j=0}^{min(m,n)}  \frac{ (-1)^{n-j}\alpha^{n+m-2j}}
{j!(n-j)!(m-j)!}[1+(-1)^{n+m}\epsilon ]
\\
\\
=\lambda_{\epsilon}\left[\frac{p!}{q!}\right]^{\frac{1}{2}}
(-1)^{n-p}\alpha^{n+m-2p}\exp(-\frac{\alpha^2}{2})[1+(-1)^{n+m}\epsilon]
{\rm L}^{q-p}_{p}(\alpha^{2}),
\end{array}
\label{scfa14}
\end{eqnarray}
where $p={\rm min}(n,m)$ and $q={\rm max}(n,m)$.

Further, taking $ \epsilon '=0, \alpha '=0$, and $r '=0$ in (\ref{scf12}) and using
the well-known definition of the photon-number distribution
$P(m)=|\langle m|r, \alpha ,n\rangle_{\epsilon} |^{2}$,
we get

\begin{equation}
\begin{array}{rl}
P^{(\epsilon )}(m)=\frac{|\lambda_{\epsilon }|^{2} (\frac{\tanh r}{2})^{m}}{n! m!
\cosh r}\exp [t^{2}(\tanh r -1)]
\left|\sum_{j=0}^{min(n,m)}  \frac{n! m!}{(n-j)! (m-j)!
 j!}\right.
\\
\\
\times \left. [\frac{2} {\sqrt{\sinh 2r}}]^{j}[-\frac{\tanh
r}{2}]^{\frac{(n-j)}{2}}
\left\{  {\rm H}_{m-j}(\frac{t}{\sqrt{\sinh 2r}}) {\rm H}_{n-j}
(\frac{{\rm i}te^{-r}}{\sqrt{\sinh 2r}}) [1 +(-1)^{n+m}\epsilon]
 \right\}\right|^{2}.
\label{scf15}
\end{array}
\end{equation}
When $r\rightarrow 0$, equation (\ref{scf15}) reduces to
\begin{equation}
\begin{array}{rl}
P^{(\epsilon )}(m)=\frac{n! m!  e^{-\alpha^{2}}}{1+|\epsilon|^{2}
+|\epsilon|\cos\phi e^{-\alpha^{2}}}\left|\sum_{j=0}^{min(n,m)}
\frac{\alpha^{n+m-2j}}{(n-j)! (m-j)! j!}\right.
\\
\\
\times \left.\left[ (-1)^{n-j} +(-1)^{m-j}|\epsilon|\cos \phi +
{\rm i}|\epsilon |(-1)^{m-j}\sin\phi \right]\right|^{2}.
\label{scf16}
\end{array}
\end{equation}
It is clear that for $\epsilon =0$ and $\alpha =0$,
$P^{(\epsilon)}(m)=\delta_{n,m}$, as expected.
Indeed, the oscillations in the photon-number distribution of squeezed states
~\cite{scfr6} are important effect which emphasizes the deviation from the
 Poissonian distribution.
 These oscillations are remarkable for large squeezing parameter
  and can be interpreted using the Bohr-Sommerfeld band picture in phase space.
Here we study the photon-number distribution for SSDNS with particular attention
to the effect of the overlap between  oscillators.
During our numerical analysis  we use only three values
for $\epsilon $, namely, $ 1, -1$ and ${\rm i}$, respectively in
correspondence to
even squeezed and displaced number states (ESDNS), odd squeezed and
displaced number states
(OSDNS) and Yurke squeezed and displaced number states (YSDNS).
These states have been considered in connection with, for
$r=0,n=0$, even coherent, odd coherent and Yurke-Stoler states.

\begin{figure}[h]%
 \centering
    \includegraphics[width=6cm]{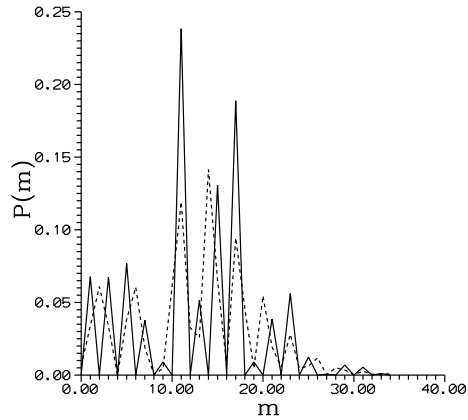}
   \caption{
Photon-number distribution $P(m)$ of ESDNS (solid curve) and
YSDNS (dashed curve)
 for $\alpha =2$, $n=3$ and $r=0.8$. }
  \end{figure}


From Fig. 5.1 it is clear  the oscillatory behaviour of squeezed states
as expected.
Comparison of the  solid and dashed curves
shows that interference part (the overlap between different oscillators, i.e.
$\hat{\rho}_{I}$) increases the oscillations in the
 photon-number distribution.
Intuitively, when  $n$  increases
the oscillations in $P(m)$ become more pronounced
 as the zeros  of the Hermite polynomial increase.
Furthermore, this situation is still valid with increasing $r$
~\cite{ther7}.
It is convenient to point out that the numerical analysis of $P(m)$
for OSDNS has a similar behaviour as for ESDNS, whereas for the statistical
mixture it is similar to that of YSDNS, i.e.
it does not depend on the quantum interference between the components of
the states $\hat{D}(\pm\alpha)\hat{S}(r)|n\rangle$.
 We conclude this subsection by discussing  the
sub-Poissonian  behaviours of SSDNS with the help of $g^{(2)}(0)$.
This can be done by calculating the expectation values of
$\hat{a}^{\dagger}\hat{a}$ and $\hat{a}^{\dagger2}\hat{a}^{2}$
in terms of the state (\ref{scf1}) and inserting the results into
(\ref{5}) (with $t=0$).
The calculations of the previous moments  are rather
lengthy but  straightforward, where the relations
(\ref{scf3}) and (\ref{scf4})
should be frequently used.

\begin{figure}[h]%
  \centering
  \subfigure[]{\includegraphics[width=5cm]{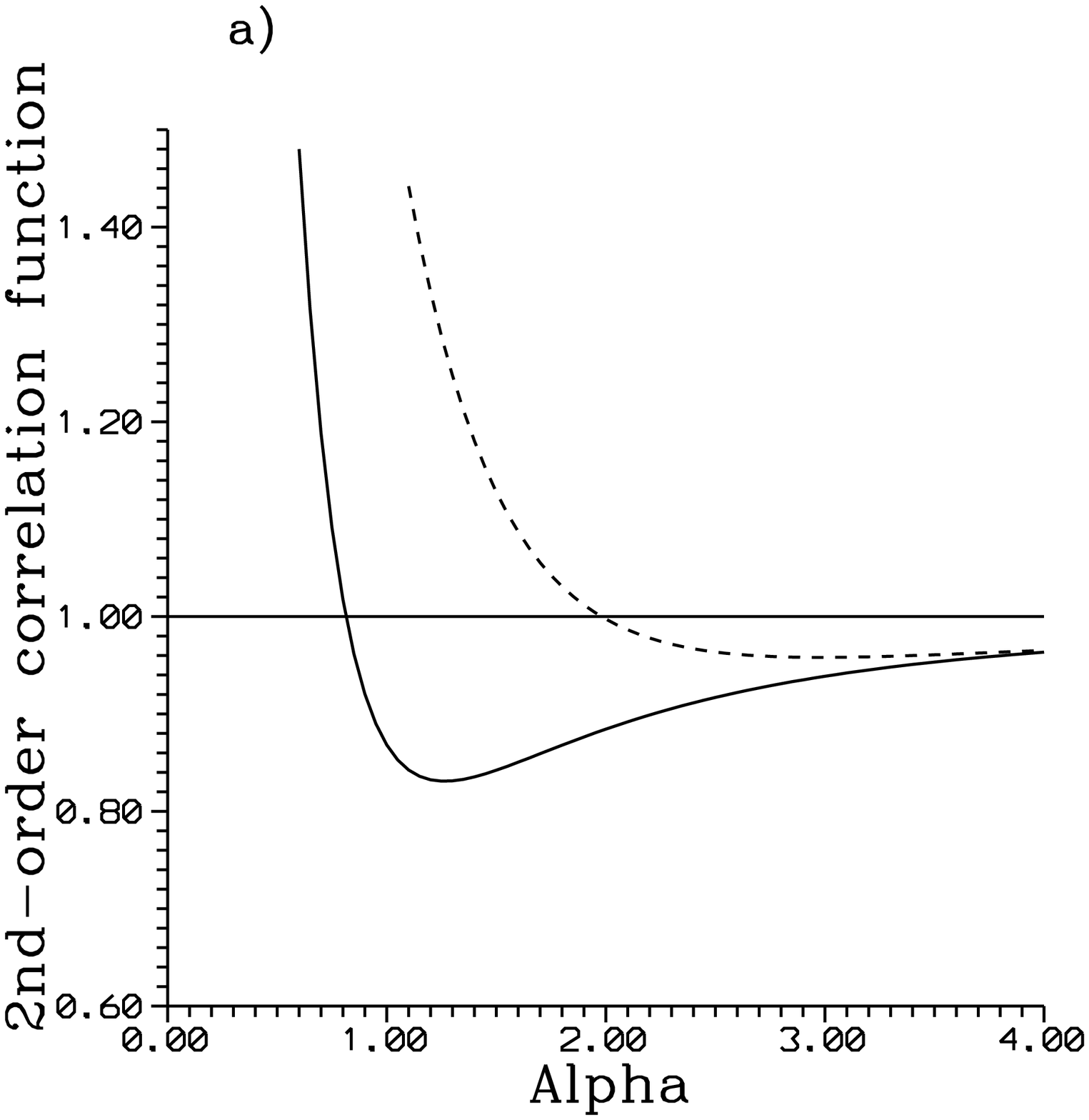}}
 \subfigure[]{\includegraphics[width=5cm]{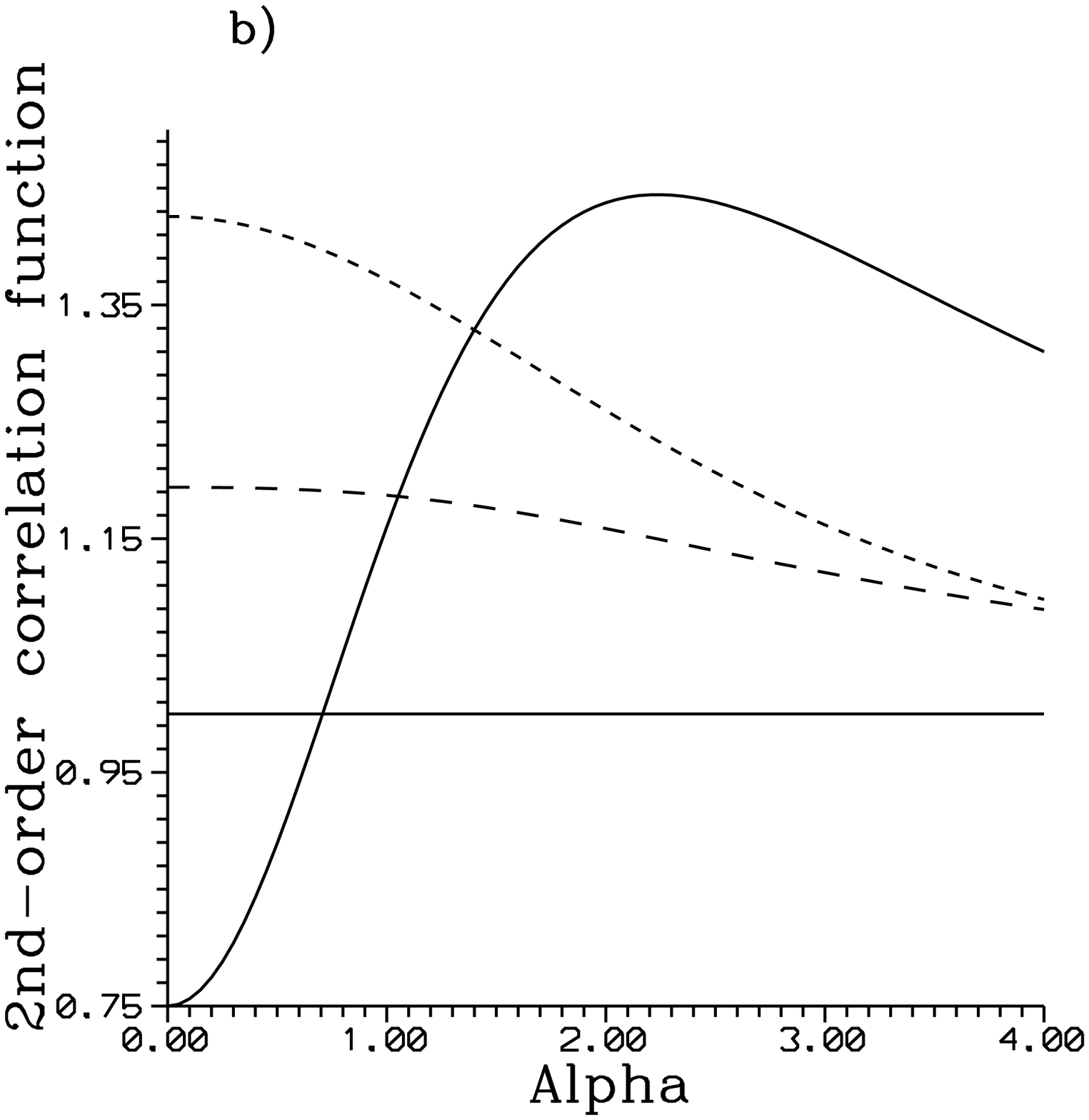}}
    \caption{
Normalized normal second-order correlation function $g^{(2)}(0)$
a) for YSDNS, $n=0$ and $r=0.5$ (solid curve), $0.9$ (dashed
curve), the straight line is corresponding to $g^{(2)}(0)$ for the
coherent state; b) for YSDNS, $n=4$ and $r=0$ (solid curve), $0.5$
(long-dashed curve), $0.9$ (short-dashed curve), the straight line
is corresponding to $g^{(2)}(0)$ for the coherent state. }
  \end{figure}
Here, we do not write down their expressions.
Now we start our numerical analysis with particular attention paid to
Yurke squeezed and displaced number states.
In fact, Yurke and Stoler ~\cite{scfr3} have shown that a coherent state
propagating through an amplitude dispersive medium, under specific
conditions of parameters, can evolve into a superposition of two coherent
states $180^{0}$ out of phase.
For such  Yurke-Stoler states $g^{(2)}(0)=1$, which corresponds to
Poissonian distribution. As soon as $r$ increases ($r=0.5,0.9$),
the Poissonian distribution disappears and $g^{(2)}(0)$ takes on values
corresponding to super-Poissonian statistics in the short starting interval
of $\alpha$ and sub-Poissonian behaviour
elsewhere (Fig. 5.2a). Also it is clear that  the length of the
sub-Poissonian interval decreases as the value of $r$  increases.
On the other hand, when the number of quanta increases (e.g. $n=4$) for
the same values of $r$,
we see that $g^{(2)}(0)$ takes on always super-Poissonian
values for large value of $r$ ($r=0.5,0.9$) and it has
sub-Poissonian behaviour when $r=0$ in the short
starting interval of $\alpha $ (Fig. 5.2b).

In general we noted  $g^{(2)}(0)$ for ESDNS and OSDNS has the same
behaviour as for
YSDNS. In other words, the known behaviours of $g^{(2)}(0)$ for even coherent states
 and odd coherent
states are changed by increasing the values of $r$ and $n$, and
the sub-Poissonian interval decreases as $r$ increases.


\subsection{ Quasiprobability distribution functions}
Here we  examine the quasiprobability distribution functions
for SSDNS. We can adopt two types of these functions which are
  $Q$-function (Husimi) and $W$-function (Wigner).
We use the $s$-parametrized characteristic function $C(\zeta,s)$ as
defined in section 4.3 to calculate
such functions.
Using  the definition of $\hat{\rho}$ given by (\ref{scf5})
together with the relations
(\ref{scf3})-(\ref{scf4})
 in a straightforward way, we get
\begin{equation}
\begin{array}{rl}
C^{(\epsilon )}(\zeta,s )=| \lambda_{\epsilon} | ^{2} \exp (-| k
^{2}|/2+\frac{s}{2}|\zeta|^{2})
{\rm L}_{n}(|k^{2}|)\left[ \exp [(\zeta -\zeta^{*})\alpha ]\right.
\\
\\
+\left. \exp [(\zeta ^{*}
-\zeta )\alpha ] |\epsilon| ^{2} \right]
+ |\epsilon| \lambda_{\epsilon} | ^{2} \left[
\exp (-{\rm i}\phi -\frac{| k ^{2}_{+}|}{2}+\frac{s}{2}|\zeta|^{2}) {\rm
L}_{n}(|k^{2}_{+}|)\right.
\\
\\
+\left.
\exp ( {\rm i}\phi -\frac{| k ^{2}_{-}|}{2}+\frac{s}{2}|\zeta|^{2}) {\rm L}_{n}(|k^{2}_{-}|)\right],
 \label{scf17}
\end{array}
\end{equation}
where
\begin{eqnarray}
\begin{array}{lr}
k=\zeta C_{r}+\zeta^{*} S_{r}, \quad
k_{+}=(\zeta + 2\alpha ) C_{r}+(\zeta ^{*} + 2\alpha)S_{r}, \\
k_{-}=(\zeta - 2\alpha ) C_{r}+(\zeta ^{*} - 2\alpha)S_{r}.
\label{scf18}
\end{array}
\end{eqnarray}

As soon as the characteristic function is obtained, the $W-$function
and $Q-$function  can be calculated easily
by inserting (\ref{scf17}) into  (\ref{16}) and carrying out the integration
(see Appendix A for the details), we get

\begin{eqnarray}
\begin{array}{lr}
W^{(\epsilon )}(x,y )=\frac{2(-1)^{n}| \lambda_{\epsilon} | ^{2}}{\pi} \left\{
|\epsilon |^{2}\exp [-2(y^{2} e^{-2r}+e^{2r}(x+\alpha)^{2})]
 {\rm L}_{n}[4(y^{2}
 e^{-2r}+e^{2r}(x+\alpha)^{2})]\right.
\\
\\
+\exp [-2(y^{2} e^{-2r}+e^{2r}(x-\alpha)^{2})]
{\rm L}_{n}[4(y^{2}
e^{-2r}+e^{2r}(x-\alpha )^{2})]
\\
\\
 \left.+ 2|\epsilon | \exp [-2(y^{2} e^{-2r}+e^{2r} x ^{2})]
  {\rm L}_{n}[4(y^{2}
e^{-2r}+e^{2r} x^{2})]\cos(4y\alpha -\phi )\right\},
\label{scf19}
\end{array}
\end{eqnarray}

\begin{equation}
\begin{array}{rl}
Q^{(\epsilon )}(\beta)=\frac{| \lambda_{\epsilon} | ^{2}(\frac{\tanh r}{2})^{n}}
{\pi n! \cosh r} \left\{ \exp [-|\beta-\alpha |^{2}- {\rm Re}(\beta-\alpha)^{2}
\tanh r]
\right.
\\
\\
\times
|{\rm H}_{n}[\frac{{\rm i}(\beta^{*}-\alpha)}{\sqrt{\sinh 2r}}]|^{2}
+|\epsilon| ^{2} \exp [-|\beta+\alpha |^{2}-{\rm Re}(\beta+\alpha )^{2}\tanh r]
\\
\\
\times |{\rm H}_{n}[\frac{{\rm i}(\beta^{*}+\alpha )}{\sqrt{\sinh 2r}}]|^{2}
+2|\epsilon |
\exp [-\frac{1}{2}(|\beta-\alpha |^{2}+|\beta+\alpha |^{2})]
\\
\\
{\rm Re} \left[ \exp [-\frac{\tanh r}{2}[(\beta-\alpha )^{2}
+(\beta+\alpha ) ^{2}]
+2{\rm i}\alpha {\rm Im} \beta+{\rm i}\phi ]\right.
\\
\\
\times \left. \left. {\rm
H}_{n}[\frac{{\rm i}(\beta^{*}
+\alpha)}{\sqrt{\sinh 2r}}]
{\rm H}_{n}[\frac{-{\rm i}(\beta-\alpha )}{\sqrt{\sinh 2r}}] \right] \right\},
\label{scf20}
\end{array}
\end{equation}
where $\beta$ in (\ref{scf19}) has been taken as
$x+{\rm i}y$ and $x$ and $y$ which are corresponding to the quadratures.

The interference in phase space
which is representative to the superposition states may be seen clearly
in the behaviour of the $W$-function.
In general we find that there are two regimes controlling  the behaviour
of $W$-function as well as the phase distribution  for the states
(\ref{scf1}) which are
$\alpha>>1$ and $\alpha \leq 1$ provided that $n$ is finite. For the
first regime, the even- and
odd-cases (i.e. $\epsilon=1$ and $-1$) provide  similar behaviours and
there is one-to-one correspondence  between the components of density matrix
(\ref{scf5}) and that corresponding to the curves.
Nevertheless, all these features are washed out
in the second regime and the behaviour  dramatically
change.

\begin{figure}[h]%
  \centering
  \subfigure[]{\includegraphics[width=5cm]{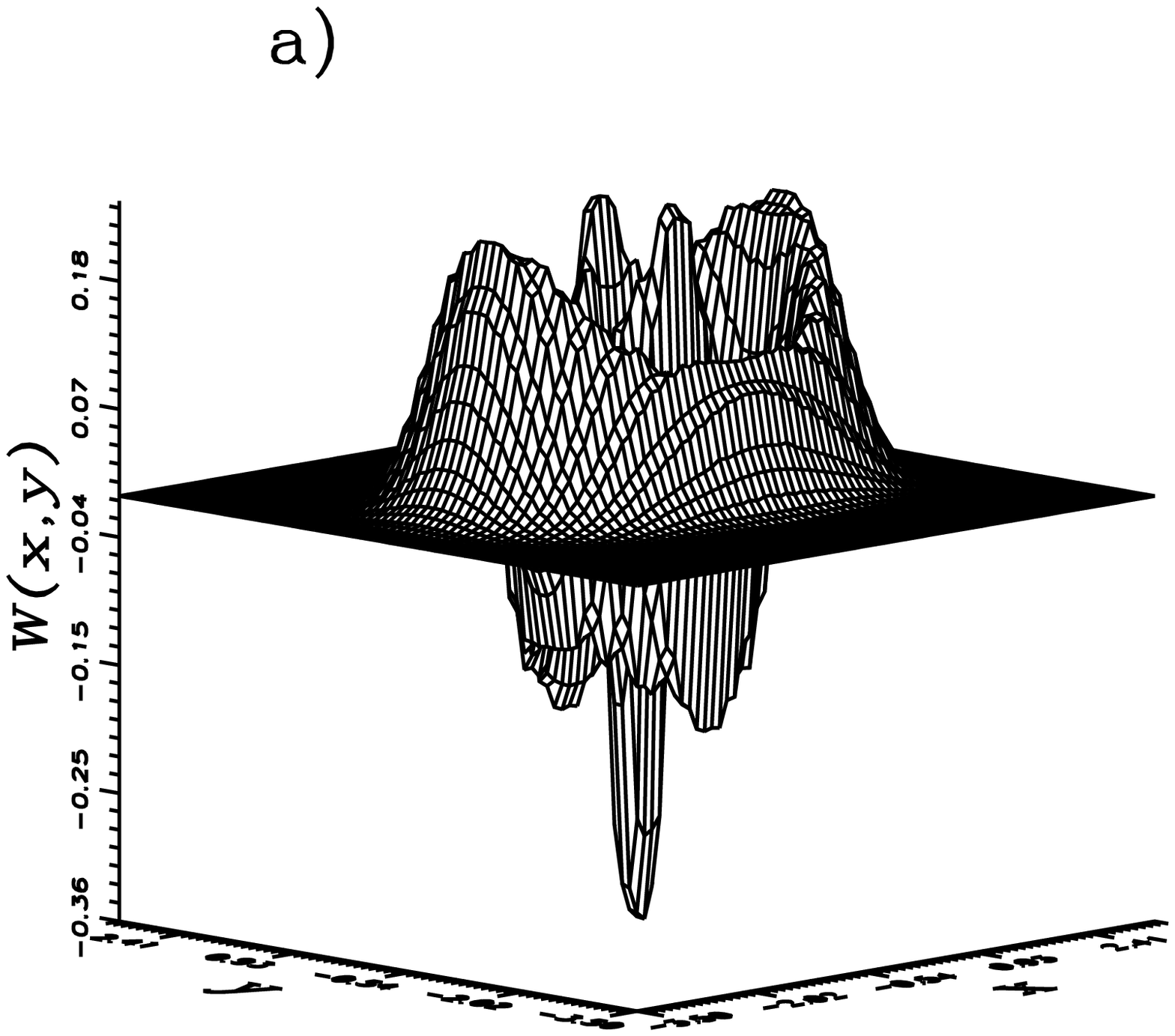}}
 \subfigure[]{\includegraphics[width=5cm]{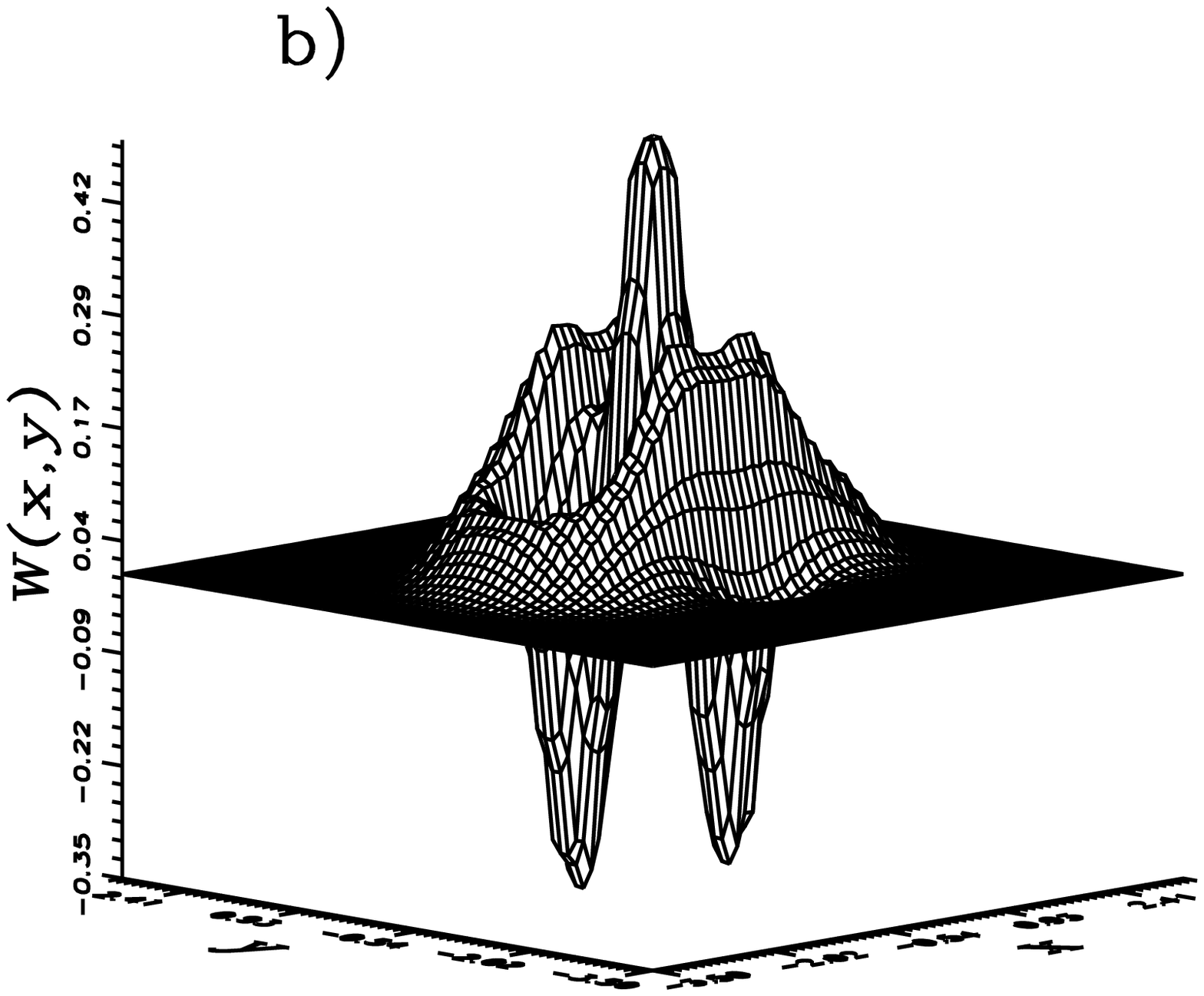}}
    \caption{
 $W$-function against $x$ and $y$ for $n =1,\alpha
=1$ and for a) $\epsilon=1$; b) $\epsilon =-1$. }
  \end{figure}


More illustratively,  $W$-function (\ref{scf19}) includes three terms, the first
two terms representing  the statistical mixture of squeezed displaced
number states and the third one is
the interference part. When $\alpha$  is small (as shown in Fig. 5.3a and b for
superposition of displaced number states) the contributions of these
components are comparable so that even- and odd-cases
distributions have different shapes.
On the other hand, when $\alpha$ is large the structures of the $W$-function
for  even- and odd-cases
are almost similar, i.e. they include two symmetrical  peaks originated
 at $\pm\alpha$ and interference fringes are in between.
Of course, such situation is still valid if
squeezing in the displaced superimposed number state in optical cavity is
considered, however, the peaks will be then stretched.
This  behaviour is reflected in the behaviour of the phase
distribution, as we will see.

The quantum-interference term is not visible in  the Q-function ~\cite{scfr8}.
For ESDNS the Q-function is shown in Figs. 5.4a-d for shown values of the
parameters.
By numerical analysis we obtained
for $r=0$ and $n=0$  two identical
 Gaussian shapes which are stretched by increasing the value of $r$.
 As $n$  increases ($n=1$),
 having $r=0$, the peaks for $r=0,n=0$ are replaced by two identical
closed top hole peaks, each of them represents the Q-function of
the displaced Fock state $|1\rangle$, and  there is evidence
for splitting in the internal
edges of the holes in Fig. 5.4a. Increasing $r (r=0.9)$ for $n=1$ as before,
the two peaks for $r=0$ are completely separated from each
other, by the action of squeezing, and each of them is split to two
identical Gaussian-like shapes with the same
root (Fig. 5.4b).
 However, when $n=4$ and $r=0.5,0.9$, we observe that for $r=0.5$ the
 two peaks occurred for $n=1$
are replaced by two  symmetrical stretched circular holes
(Fig. 5.4c). But for $r=0.9$ each hole
is converted to two symmetric peaks separated by four spikes (Fig. 5.4d).


\begin{figure}[h]%
  \centering
  \subfigure[]{\includegraphics[width=5cm]{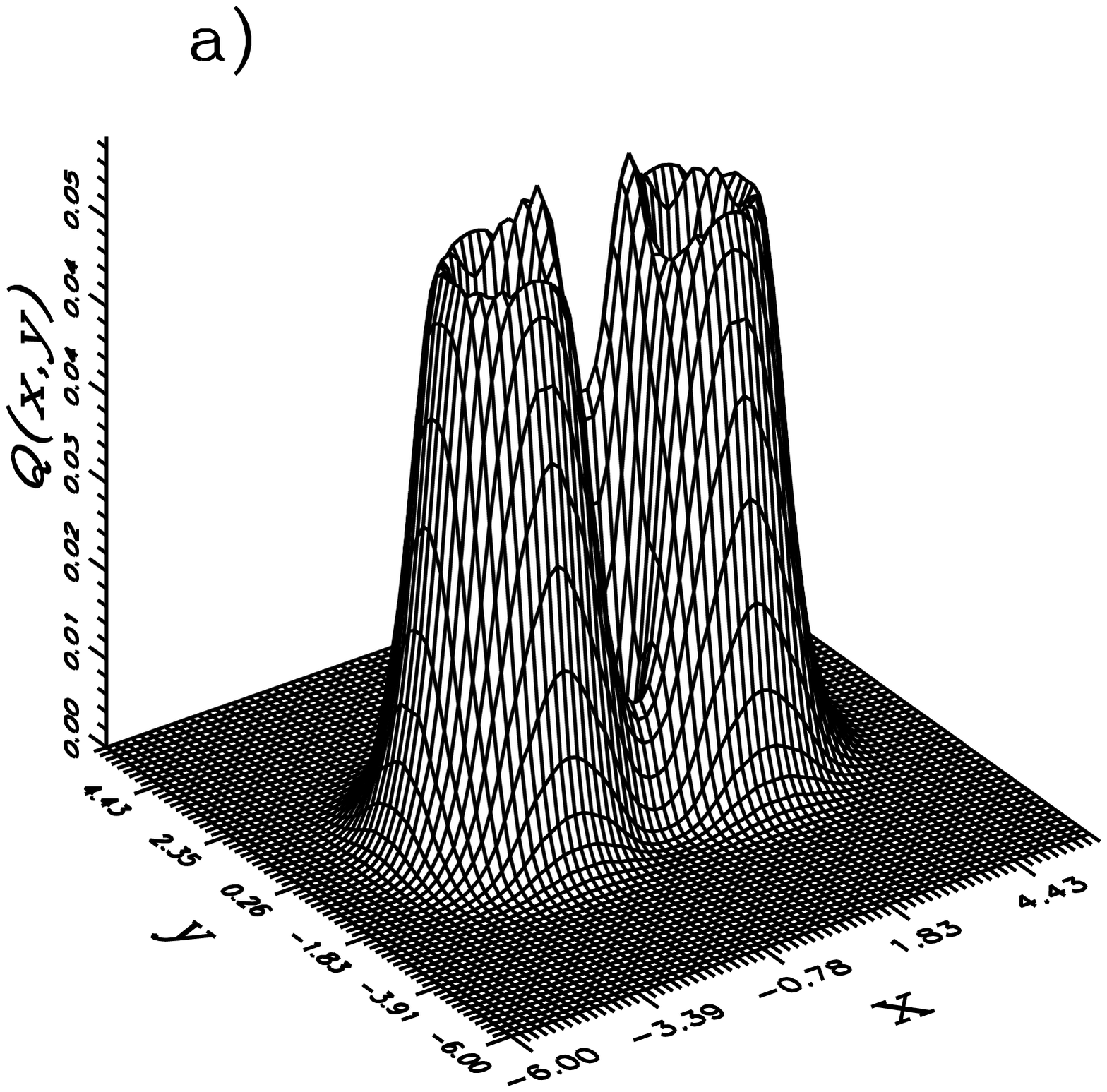}}
 \subfigure[]{\includegraphics[width=5cm]{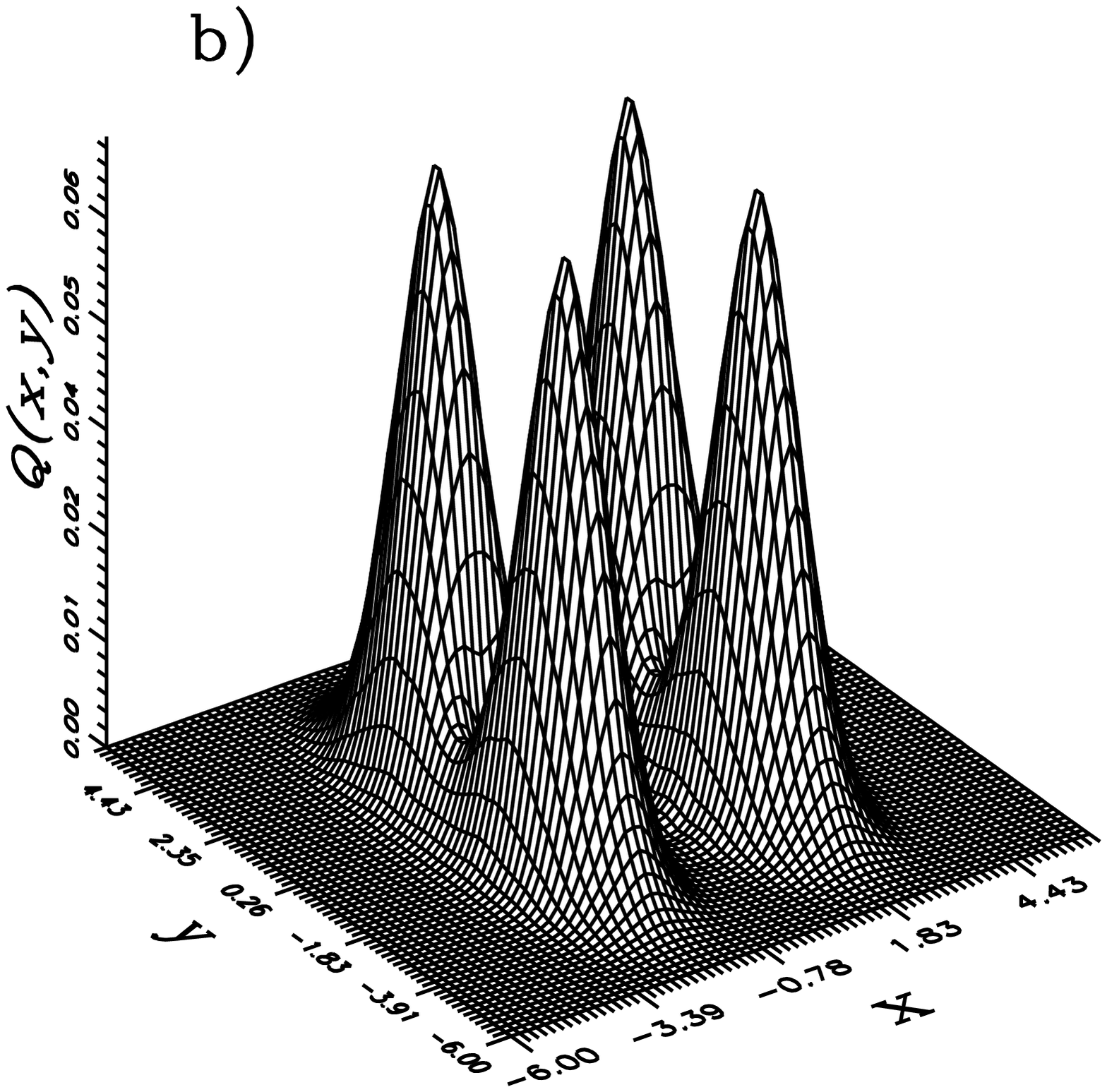}}
 \subfigure[]{\includegraphics[width=5cm]{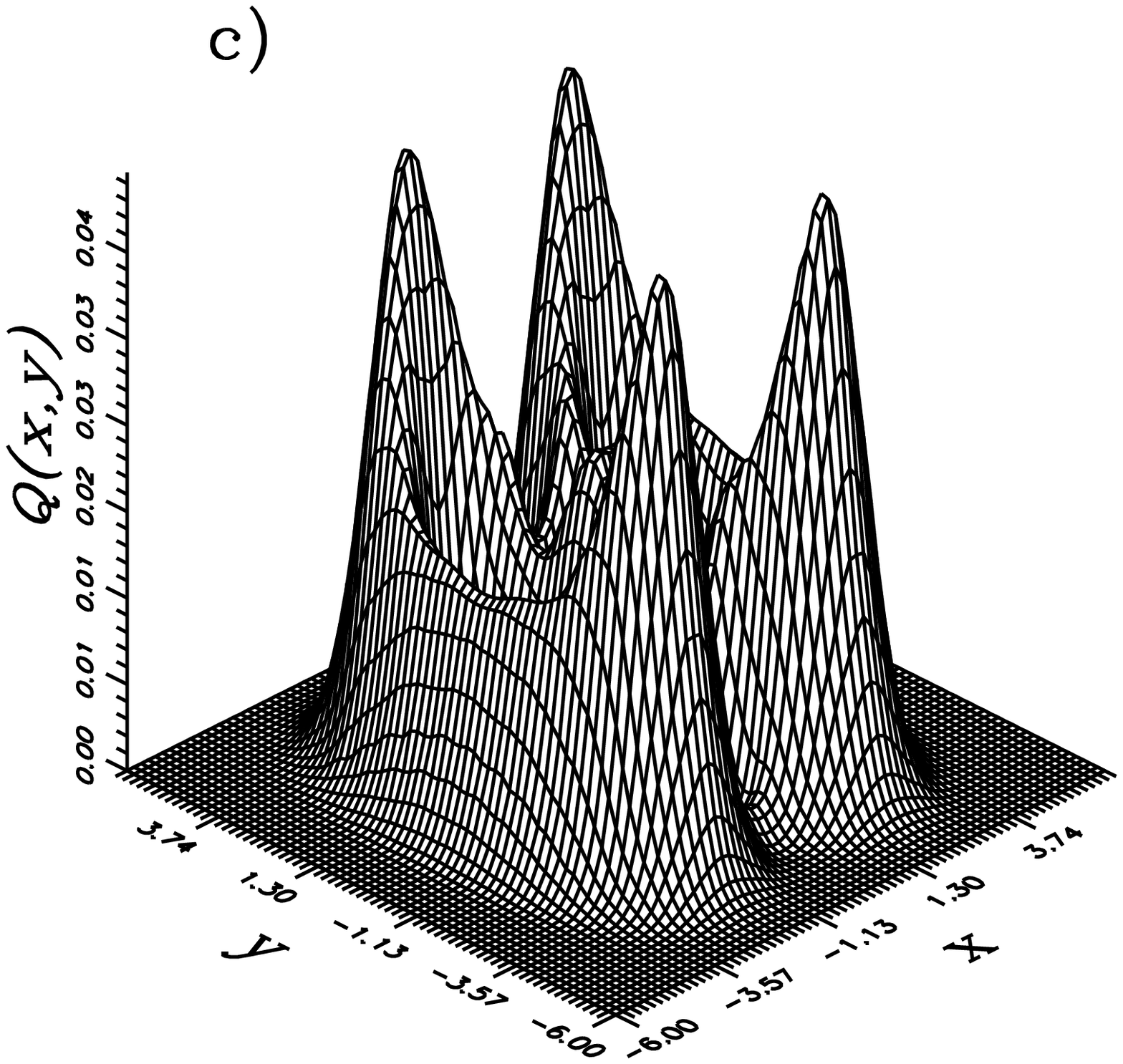}}
 \subfigure[]{\includegraphics[width=5cm]{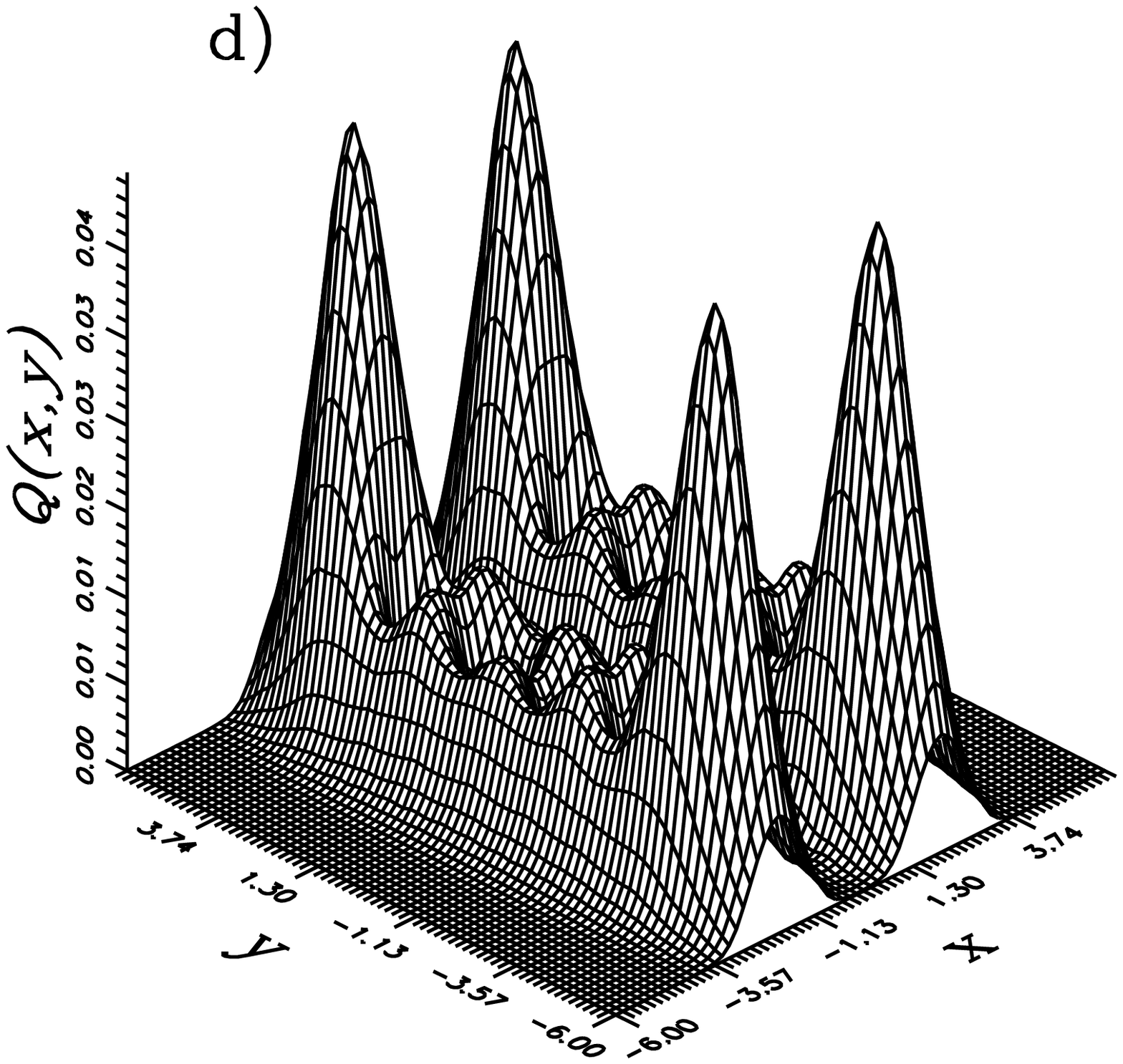}}
    \caption{
Q-function  for ESDNS and $\alpha =2$ for a) $n=1$ and $r=0$;
 b) $n=1$ and $r=0.9$;
 c) $n=4$ and $r=0.5$; d)  $n=4$ and $r=0.9$.} \label{fig4}
  \end{figure}

%

The Q-function for OSDNS and  YSDNS can be
represented by
similar figures. This shows that the interference term plays negligible
role in the Q-function.
\subsection{Phase properties}
We shall use the relations of the Pegg-Barnett formalism given in
section 4.4 to study the phase distribution for the
superposition of displaced and squeezed number states (\ref{scf1}). In this case
$C_{m}$ in (\ref{25}) is given by $C_{m}=C_{m}(r,\alpha,n,\epsilon)=
\langle m|r,\alpha,n\rangle_{\epsilon}$ and this quantity
can be deduced from (\ref{scf12}) by suitable choice of the parameters.

The photon-number variance  is defined as

\begin{eqnarray}
\begin{array}{rl}
\langle (\triangle \hat{n})^{2}\rangle
=\langle \hat{n}^{2}\rangle-
\langle \hat{n}\rangle^{2}\\
\\=\langle \hat{a}^{\dagger 2}\hat{a}^{ 2}\rangle
+\langle \hat{n}\rangle -\langle \hat{n}\rangle^{2},
\end{array}   \label{scf21}
\end{eqnarray}
where the number operator $\hat{n}=\hat{a}^{\dagger}\hat{a}$.
In terms of (\ref{scf1})
the quantities $\langle \hat{a}^{\dagger 2}
\hat{a}^{ 2}\rangle$ and $\langle \hat{n}\rangle$
 can be straightforwardly calculated.
For completeness, the phase variance of (\ref{scf1}) reads
\begin{eqnarray}
\begin{array}{rl}
\langle (\triangle \hat{\Phi})^{2}\rangle=\frac{\pi^{2}}{3}+4{\rm Re}
\sum_{m>m'}C_{m}(r,\alpha,n,\epsilon)C^{*}_{m'}(r,\alpha,n,\epsilon)
\frac{(-1)^{m-m'}}{(m-m')^{2}}
\\
\\
  -4\left[{\rm Re} i\sum_{m>m'}C_{m}(r,\alpha,n,\epsilon)C^{*}_{m'}
  (r,\alpha,n,\epsilon) \frac{(-1)^{m-m'}}{(m-m')}
\right]^{2}.
\end{array}\label{scf22}
\end{eqnarray}
The value $\pi^2/3$ is the phase variance for a state with uniformly
distributed phase, e.g. the vacuum state.

At the mean time, for understanding the behaviour of the phase distribution
of the states  (\ref{scf1}) it is more
convenient firstly to study such properties for two  subsidiary states which
are the superposition of displaced number states ~\cite{oba}, and
squeezed and displaced number states ~\cite{[29]}.
These two states are of interest because it has been shown for the former
states that they can exhibit strong sub-Poissonian character as well as
quadrature squeezing. The  bunching and antibunching properties for the
latter states have been discussed in ~\cite{[30]}.  Investigation
of such properties for these two kinds of states most clearly illustrates
the role of the different parameters in the state (\ref{scf1}) in the
behaviour of the distribution.
In the following  we discuss the phase distribution determined
by (\ref{25}), and the phase variance and phase squeezing, respectively.
For simplicity  we restrict our investigation to real
values of $\alpha$ and $\epsilon$.
\\
\\
\noindent {\bf (a) Phase probability distribution}
\\
\\
As we mentioned before, there are two regimes controlling  the behaviour
of the phase distribution for the states under discussion
similar to the $W$-function.
We start our discussion by focusing the
attention on the behaviour of the superposition of displaced number state.
It is important to mention that the properties of displaced number states
$\hat{D}(\alpha)|n\rangle$  have been
given in ~\cite{scfraa4} and their phase properties in ~\cite{tan}. In fact the
phase properties of such states are  interesting since they connect
the number state, which has no phase information,
and coherent states, which play the boundary role between the classical
and nonclassical
states and  always exhibit single peak structure of phase distribution.
The phase distribution of displaced number states exhibits nonclassical
oscillations with the number of peaks which is equal to $n+1$. This result is
interpreted in terms of the area of overlap in phase space ~\cite{tan}.
Unfortunately, the situation is completely different for the superposition
of displaced number state (see Figs. 5.5a and b for shown values of the
parameters).
From Fig. 5.5a it
is clear that the quantum interference between component states
$\hat{D}(\alpha)|n\rangle$ and $\hat{D}(-\alpha)|n\rangle$ leads to the
lateral nonclassical oscillations which become more pronounced and narrower
as $n$  increases. However, the statistical mixture of displaced number
states shows $n+1$ peaks around
 $\Theta =0$. We have not presented  the phase distribution for
 the odd-case since it is similar to that of the even-case.
 Comparing this behaviour with that of even (or odd) coherent states,
we conclude that  the  number states in the superposition make
the phase information more significant ~\cite{phasx1}.
Turning our attention to the Fig. 5.5b where $\alpha=1$, we can see that
the distribution
of displaced number states exhibits two-peak structure as expected for
$n=1$ ~\cite{tan}, however, the distribution of even-case displays one
central peak
at $\Theta=0$ and two wings as $\Theta\rightarrow \pm \pi$, and finally
the distribution of odd-case provides four peaks. That is the
distribution here is
irregular, however, more smoothed than before and the structure of
the statistical
mixture (central part) is modified by the action of the interference term
arising from  the superposition of the states; this was the case for
$W$-function.
\begin{figure}[h]%
  \centering
  \subfigure[]{\includegraphics[width=5cm]{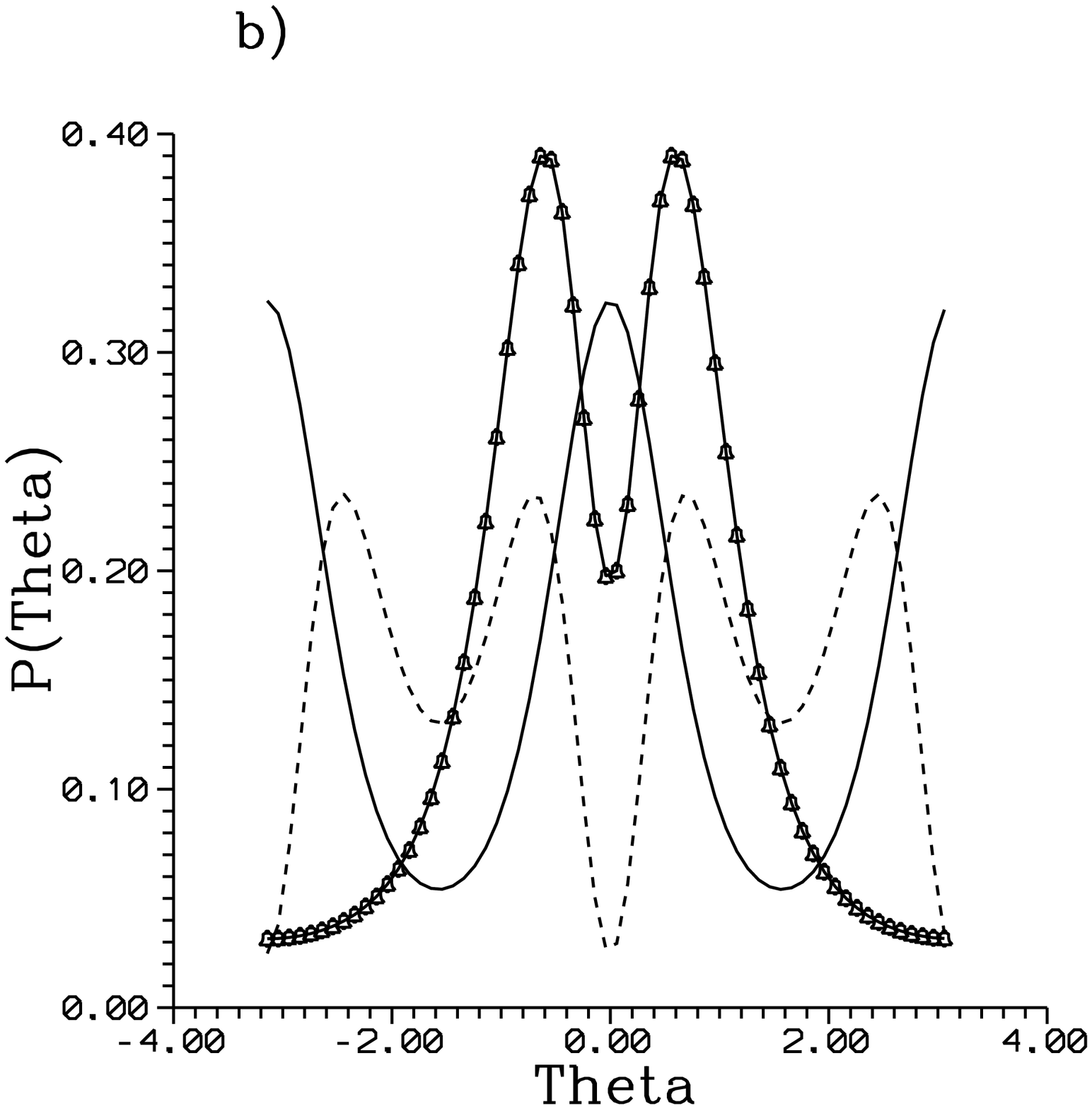}}
 \subfigure[]{\includegraphics[width=5cm]{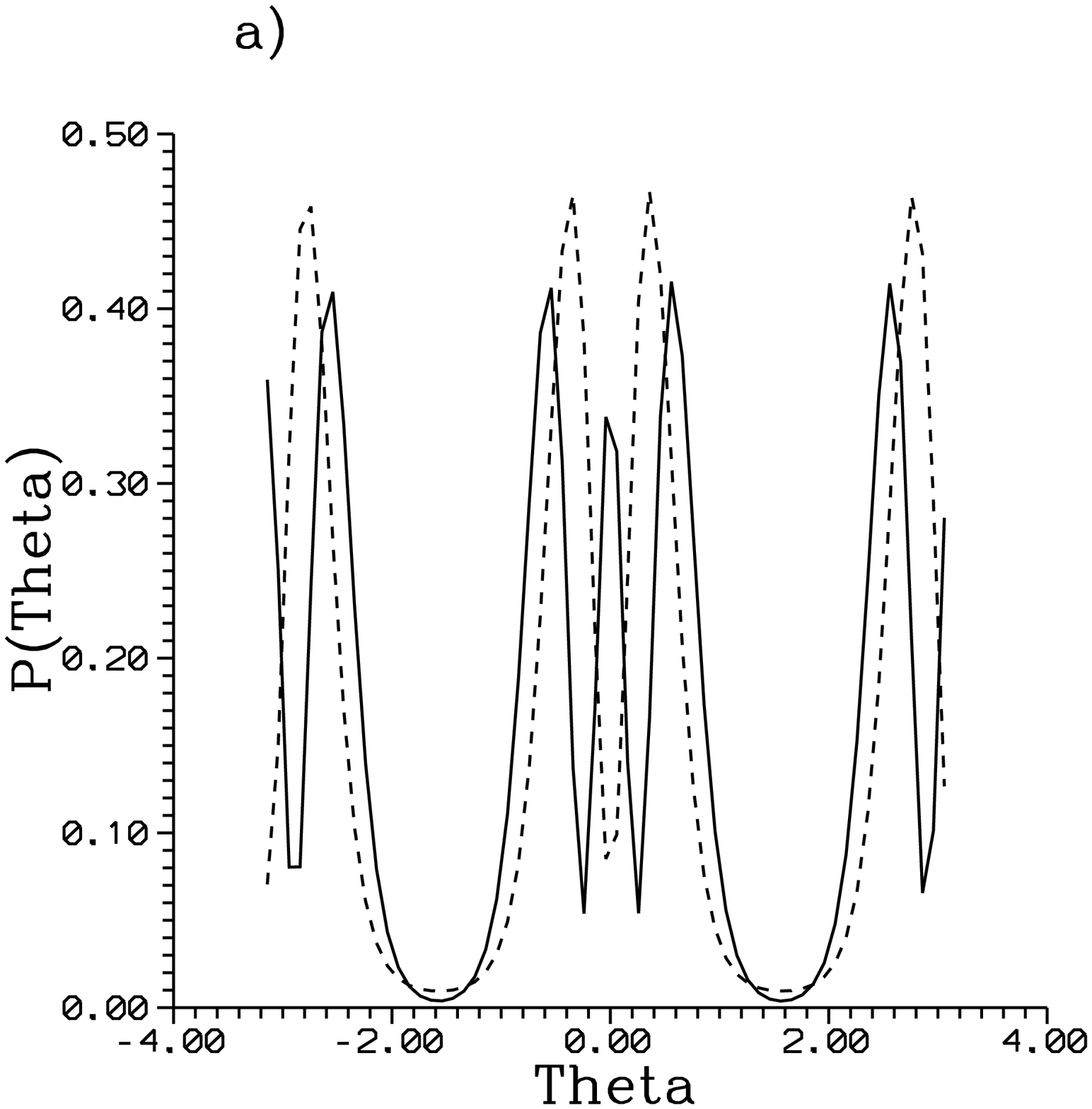}}
     \caption{
Phase distribution $P(\Theta)$ for superposition of displaced
number states and for a) $\alpha=2,\epsilon =1$ , $n=1$ (dashed
curve), and $2$ (solid curve); b) $ \alpha=1,n=1$ and $\epsilon
=0$ (bell-centered curve), $1$ (solid curve), and $-1$ (dashed
curve).}
  \end{figure}

%

Before discussing the phase properties of the displaced and squeezed
number states it is reasonable to  remind the
behaviour of the well known squeezed states. As known for squeezed states
with non-zero displacement coherent amplitude, the phase distribution exhibits
the bifurcation phenomenon. In this phenomenon the single peak structure of
the coherent component is evolved into two peaks structure with respect to
both $\alpha $ (for large fixed value of squeezing parameter $r$) and $r$
(when $\alpha $ takes fixed value) ~\cite{phas11}.
 This phenomenon has been
recognized as a result of the competition between the two peaks structure of
the squeezed vacuum state and the single-peak structure of the coherent state.
For squeezed and displaced number states such phenomenon cannot occur
due to the effect of the Fock state which replaces the initial peak ($r=0$)
for coherent state by a multi-peak structure, i.e. by $n+1$ peaks (see Fig. 5.6a
for shown values of the parameters).   From this figure one can observe that
there is a three-peak structure corresponding to the case $n=2$ of
displaced
Fock state. The height of the central peak (i.e. at $\Theta=0$) is almost
the same and equals $(1/2\pi)|\sum_{m=0}^{\infty}C_{m}|^{2}$. That is
the central value of the phase distribution $P(\Theta=0)$ is insensitive
to squeezing provided that $r$ is finite.  However, the lateral  peaks
undergo phase squeezing as $r$ increases, i.e. the peaks become narrower.

Now  we can investigate  the behaviour of the
superposition of displaced and squeezed number states (see Figs. 5.6b-d for
shown parameters). Figs. 5.6b and c are given for the second regime
for even- and odd-cases, respectively. From Fig. 5.6b  we can see that the
initial oscillations are increased compared with Fig. 5.6a as a result of the
interference in phase space. Further, the initial lateral peaks can evolve
in the course of increasing $r$
to provide bifurcation shape,
i.e. the distribution curve undergoes a transition from single- to a
double-peaked form with increasing $r$; however, the central peak is almostly
unchanged with increasing $r$. This peak splitting is connected with the
squeezed
states ~\cite{phas11}. Further, as the number of peaks  increases for $r>0$, the
distribution becomes more and more narrower. On the other hand, the phase
distribution of the odd-case is quite different as we have shown before
where the initial peaks are not significantly changing as in the
even-case. More precisely, initial distribution becomes broader for a while
and suddenly (at $r\simeq 0.5$) breaks off to start to be a narrower
distribution for later $r$.
It is clear that  in this regime the state (\ref{scf1}) becomes more and more
nonclassical  and  the even- and odd-cases are distinguishable.
 Further, the phase distribution of the even-case is
more sensitive with respect to squeezing than the odd-case.
Nevertheless, for the first regime where $\alpha$  is large we noted that the
phase distribution carries
 at least the same initial information regardless  of the value of $r$
(see Fig. 5.6d for the shown values of the parameters).

\begin{figure}[h]%
  \centering
  \subfigure[]{\includegraphics[width=5cm]{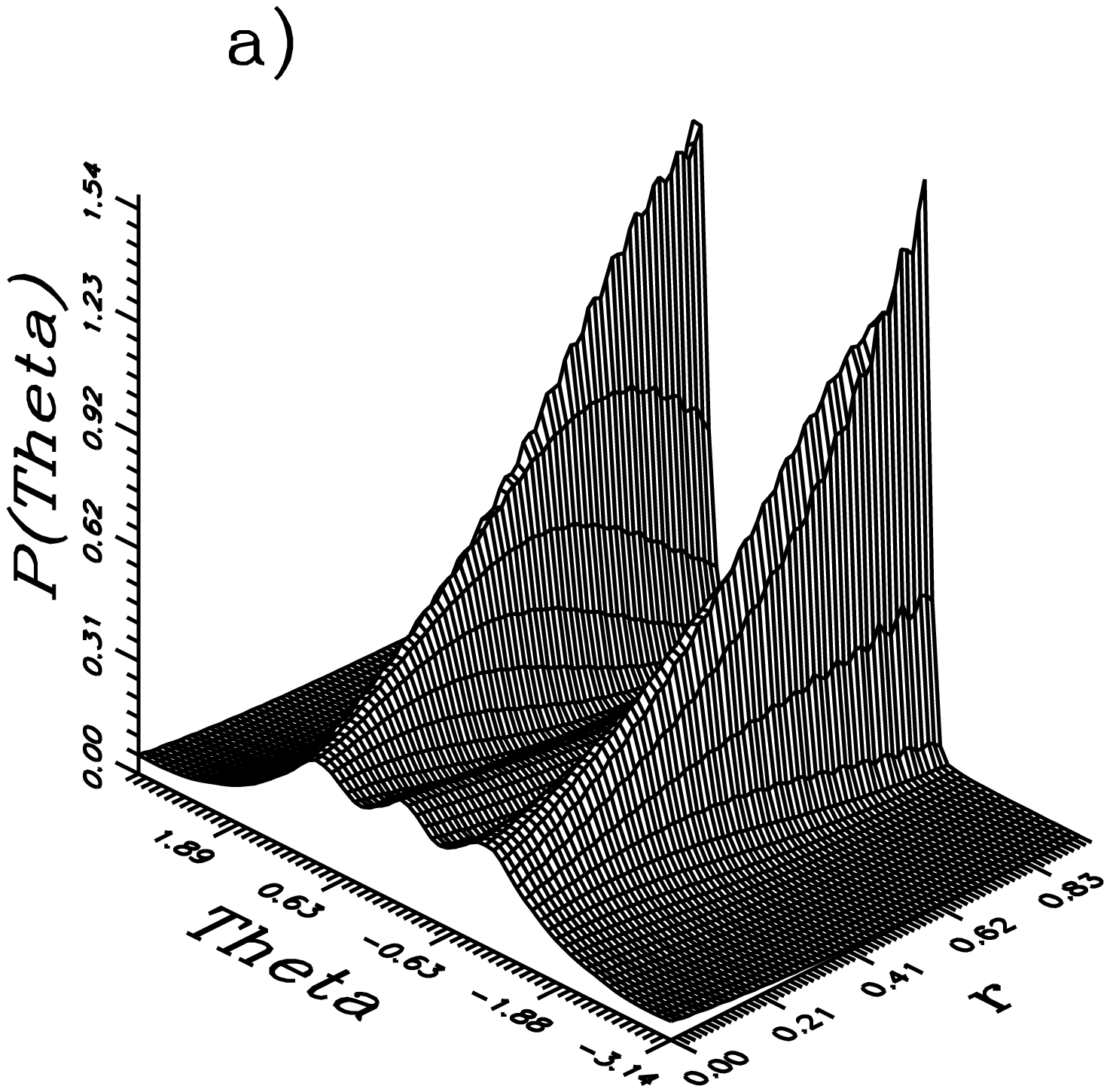}}
 \subfigure[]{\includegraphics[width=5cm]{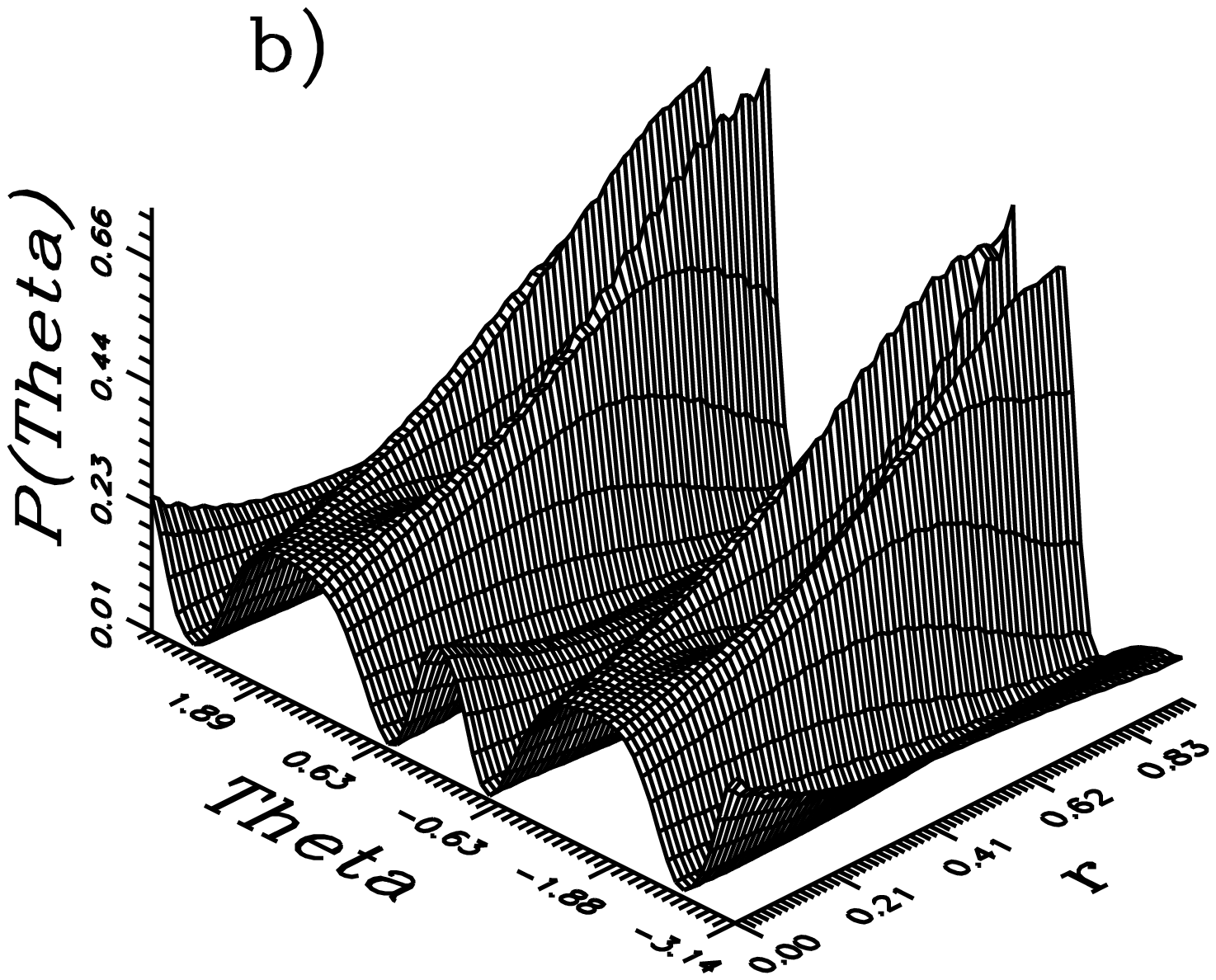}}
 \subfigure[]{\includegraphics[width=5cm]{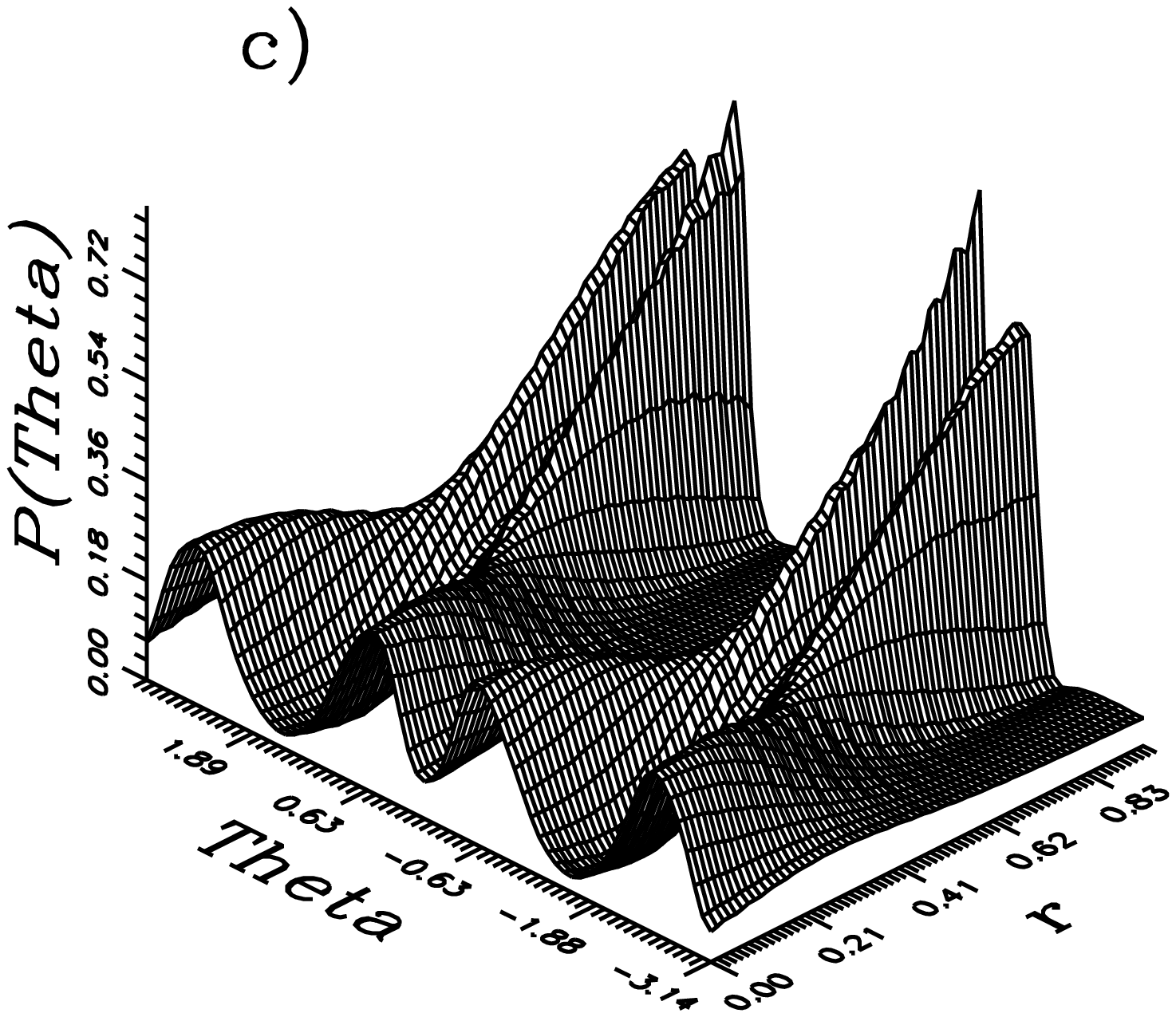}}
 \subfigure[]{\includegraphics[width=5cm]{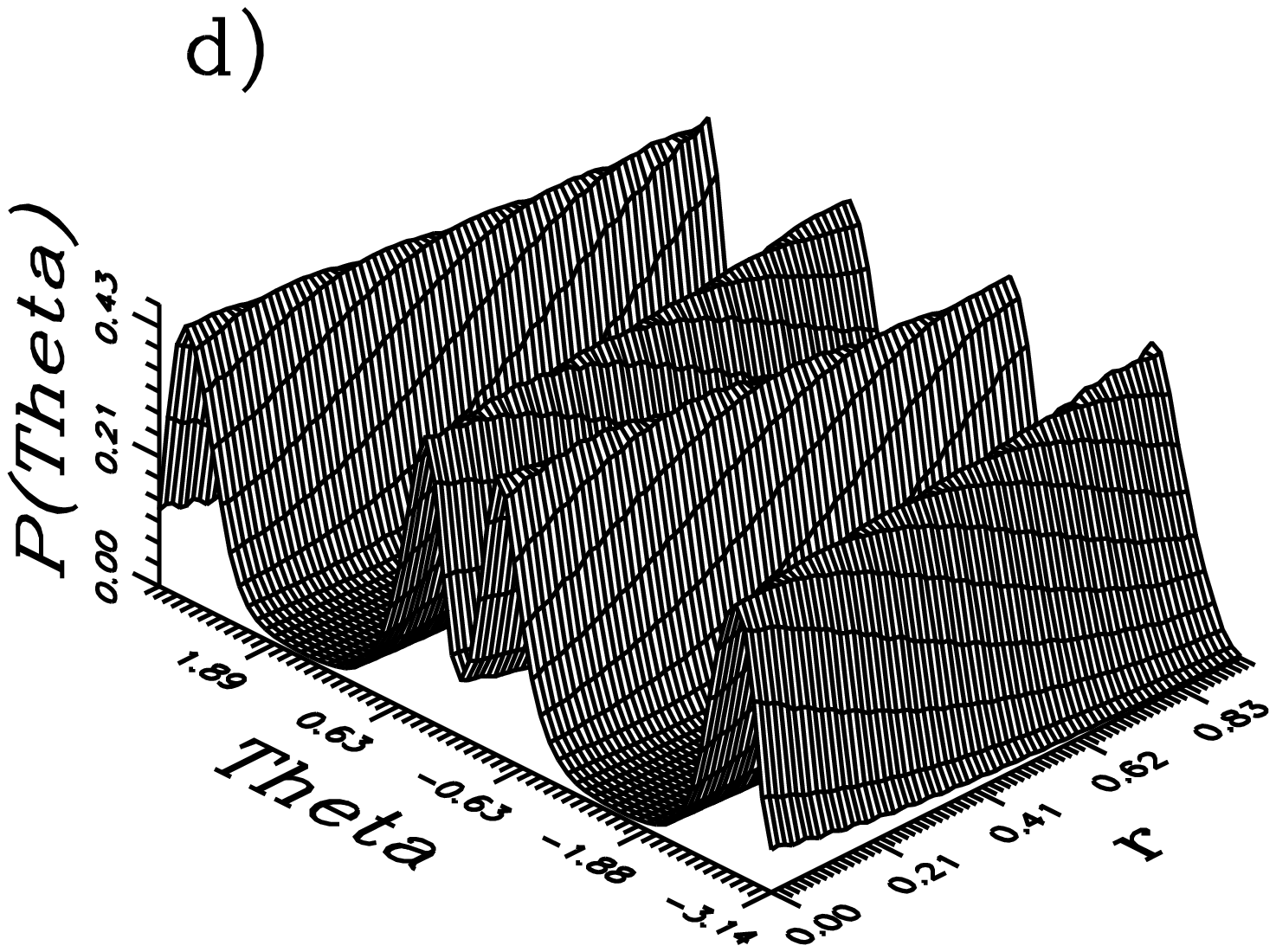}}
    \caption{
Phase distribution $P(\Theta)$ for superposition of displaced and
squeezed number states for $
(\epsilon,\alpha,n)=(0,1,2),(1,1,2),(-1,1,2)$ and $(1,2,2)$
corresponding to the cases a,b,c and d, respectively.}\label{fig6}
  \end{figure}

   \noindent{\bf (b) Variance, amplitude and phase squeezing}
\\
\\
\noindent Here we investigate the behaviour of
the phase variance, and amplitude and phase fluctuations
following (\ref{scf22}), (\ref{30}) and (\ref{31}), respectively.
We start our discussion by analyzing the behaviour of
the superposition of displaced number states. For this purpose Figs. 5.7a
and 7b are  shown for the phase variance, and amplitude and phase
fluctuations, respectively. It is seen that  in general the phase
variance starts from the value $\pi^2/3$ (the vacuum state value)
and returns back  to it when $\alpha$ is large, but through different routes.
To be more specific, the phase variance of displaced number states starts
from the value for the vacuum, goes to a minimum, and then comes again
to $\pi^2/3$. However, the behaviour of  the superposition of
displaced number states takes different ways to arrive at the same result,
i.e. it starts from $\pi^2/3$ as before, goes to the
maximum value and eventually comes back to  the value of vacuum.
The comparison of the two cases shows the role
 of the quantum mechanical interference between state components.
Further, as $n$ increases, the oscillations in the variance  become
more pronounced.
Comparison of the behaviour of even- and odd-cases shows that
they are different
only over  the initial short interval of $\alpha$, i.e. when $\alpha$ is
small, and this agrees with what we have discussed earlier. So we can
conclude that for intensities high enough  of the coherent field, the
variance of the phase is approximately randomized. The route to this
randomization is dependent on the choice of $\epsilon$.
With respect to the amplitude and phase fluctuations, we can note
from (\ref{29}) that these quantities depend not only on the intensity of
the field, but also on the choice of the reference angle $\Theta_{0}$.
We have chosen here $\Theta_{0}=-\pi$, where the mean value  $|\langle
[\hat{N},\hat{\Phi}_{\theta}]\rangle|$ of squeezed displaced number
state approaches unity. Fig. 5.7b has been obtained to illustrate
 the parameters $S_{N}$ and $S_{\theta}$ which provide information
about the degree of squeezing in $\hat{n}$ and $\hat{\Phi}_{\theta}$.
One can observe from this figure that when $\epsilon =0$ (displaced number
state) and $\alpha\rightarrow 0$, the parameter $S_{N}$ tends to $-1$,
which means that the number state is $100\%$ squeezed with respect
to the operator $\hat{n}$. This situation is
expected since $\langle (\triangle \hat{n})^{2}\rangle=0$ for the number
state. Further, the larger the number of quanta is
the shorter is the interval  over which $S_{N}$
is squeezed. Also when $\alpha>>1$, squeezing in  $S_{\theta}$ is
remarkable, whereas $S_{N}$ becomes unsqueezed.
This result can be deduced from the behaviour of the phase variance
(see Fig. 5.7a), where $\langle (\triangle \hat{\Phi})^{2}\rangle\simeq 0$ at
$\alpha\simeq 4$ and this should be connected with maximum squeezing in
$S_{\theta}$ at this point.
Such behaviour of $S_{N}$ and $S_{\theta}$ confirms the fact that
the number of photons and phase are conjugate quantities in this approach.
As is known displaced
number states are not minimum uncertainty states and  the variances for
the quadrature operators never go below the standard quantum limit.
Moreover, they may exhibit sub-Poissonian statistics for the range
$\alpha^{2}\leq 1/2$ ~\cite{scfraa4}.
However, there is no relation between the sub-Poissonian statistics and
the fluctuation in the amplitude or the phase. This fact has been shown
before
for the down-conversion process with quantum pump where the signal mode
can exhibit amplitude squeezing and at the same time it is
super-Poissonian ~\cite{tann}.

We proceed by discussing the behaviour of the superposition states
(long-dashed and circle-centered curves) in Fig. 5.7b where the interference
 in phase space starts
to play a role. We noted (from our numerical analysis) that only
the even case can provide squeezing in $S_{N}$ with maximum value at the
origin and squeezing interval larger than that discussed before. Indeed,
this maximum value is related also to that of number state at $\alpha=0$.
It should be stressed here that $\alpha=0$ for the odd-case may lead to
a singularity.
As we mentioned before the superposition of displaced number states
 can exhibit strong sub-Poissonian character as well as
quadrature squeezing ~\cite{oba}.
Now we can illustrate the role of the squeezing in the superimposed
displaced number states optical cavity with respect to variance,
amplitude and phase fluctuations. It is obvious, when squeezing is
considered, that the initial
value (at $\alpha=0$) of the phase variance is shifted  since we
have initially  squeezed number state which is providing phase information.
However, when $\alpha$ is large and $r$ is finite or also  $r$ is large,
it can be proved simply that the coefficient $C_{m}(r,\alpha,n,\epsilon)$
vanishes  and consequently the phase variance tends to $\pi^{2}/3$
(becomes randomized). Moreover,  the routs are here similar to those of
Fig. 5.7a. On the other hand, squeezing could be seen
only in $S_{\theta}$ for  $\epsilon=0$ (see squared-centered curve in
Fig. 5.7b). Furthermore, comparison of the short-bell-centered curve (of
displaced Fock state) and squared-centered
curve reveals that squeezing parameter reduces the amount of squeezing in
$S_{\theta}$, too.
This means that the superposition of
displaced and squeezed number states  provides quadratures
squeezing which possesses less information about amplitude and phase fluctuations.
\begin{figure}[h]%
  \centering
  \subfigure[]{\includegraphics[width=5cm]{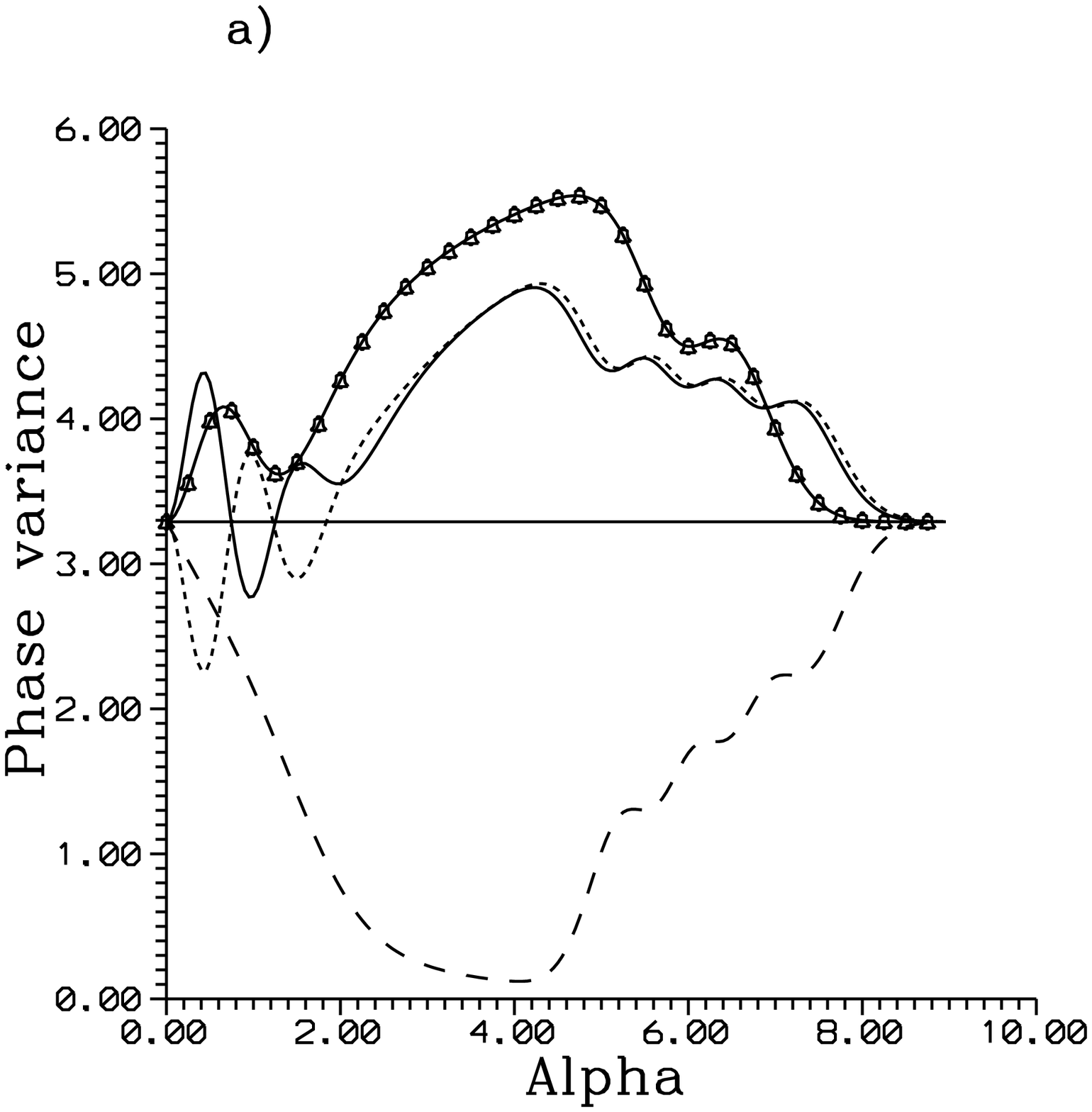}}
 \subfigure[]{\includegraphics[width=5cm]{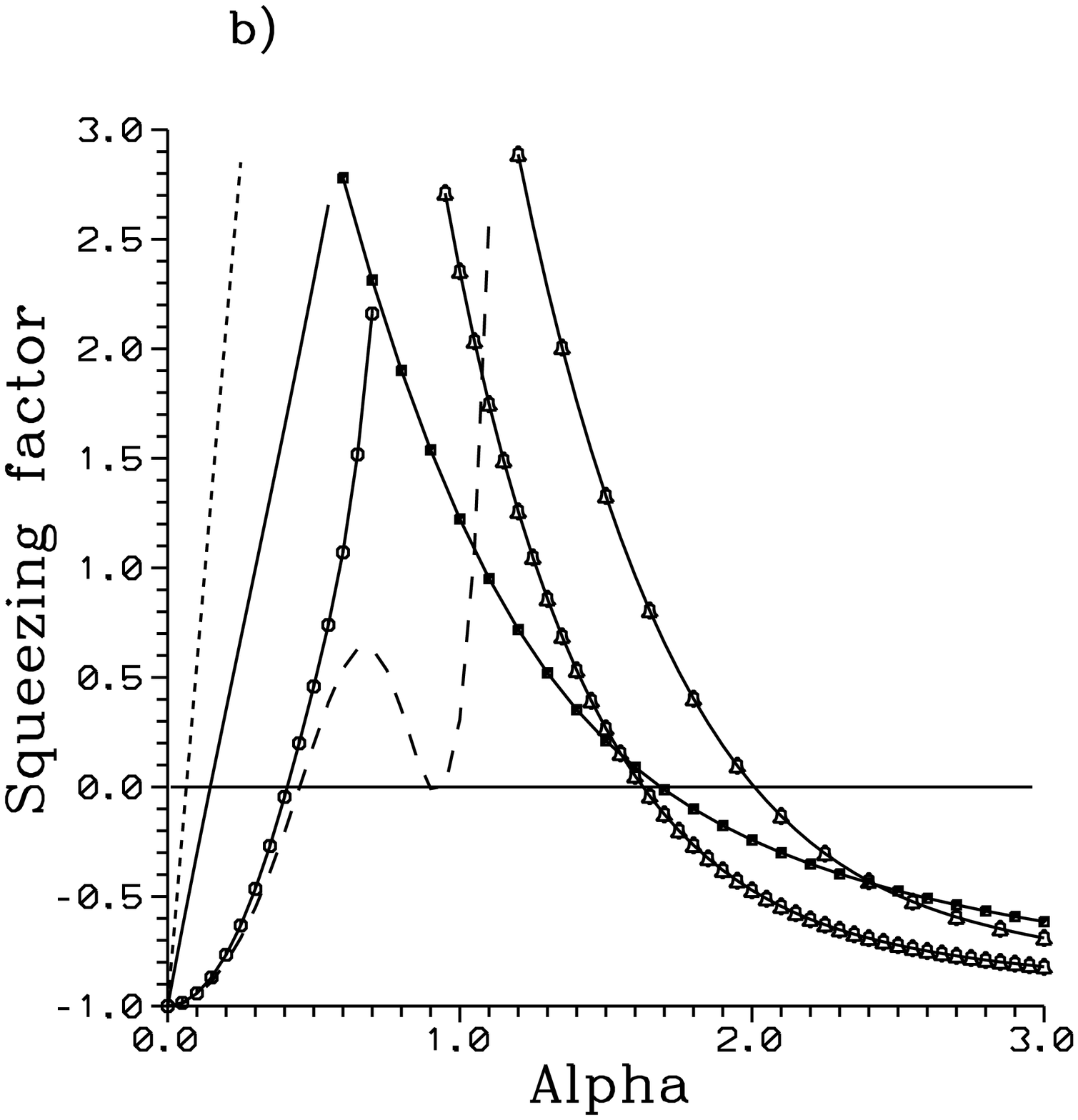}}
    \caption{
a) Phase variance of a superposition of displaced number states
against $\alpha$ for $(\epsilon,n)=(0,3)$ (long-dashed curve),
$(1,1)$ (bell-centered  curve), $(1,3)$ (solid  curve) and
$(-1,3)$ (short-dashed curve). The solid straight line is
corresponding to the phase variance of vacuum. b) Amplitude and
phase fluctuations of a superposition of displaced and squeezed
number states against $\alpha$ for $(\epsilon,n,r)=(0,1,0)$ (
$S_{N}$ solid curve and  $S_{\theta}$ short-bell-centered curve),
$(0,2,0)$
 ( $S_{N}$ short-dashed curve and  $S_{\theta}$ long-bell-centered curve),
$(1,1,0)$ ($S_{N}$  long-dashed curve), $(1,2,0)$ ($S_{N}$
circle-centered curve) and $(1,2,0.5)$ ($S_{\theta}$
squared-centered curve). The solid straight line is the bound of
squeezing. }
 \end{figure}

\subsection{ Scheme of production}
Nonclassical states of light in cavity are of increasing importance in
quantum optics. There are two principal approaches how to generate them
~\cite{[34],{scfr9}}: {\bf (i)} Find
 the appropriate Hamiltonian which transforms via unitary time evolution
 the initial states to the desired final state, e.g. Schr\"odinger cat
states. {\bf (ii)} Make a measurement on one of two entangled quantum systems
 and obtain the state of the other system by the corresponding state reduction.

We use here,  to suggest generation of superposition of such states, quantum state
engineering, which is based on injecting $N$ two-level excited identical
atoms propagating through vacuum cavity field.  The atoms passing the cavity one
by one in such a way that each injected atom increases the number of Fock
states, are building up the cavity field state by one.
In  microwave experiments this is achieved by using Rydberg atoms and very high-Q
superconducting cavities so that spontaneous emission and cavity
damping are quite negligible on the time scale of the atom-field interaction.
The $k$th atom enters the cavity in a given coherent superposition
state $C_{g}^{k}|g\rangle_{k}+ C_{e}^{k}|e\rangle_{k}$, where
$|g\rangle$ ($|e\rangle$) denotes ground (excited) atom state and
$C^{k}_{g}$ and $C^{k}_{e}$ are the corresponding superposition weights, and
it can  be detected in the ground state. The atom-field resonant interaction is
 described by Jaynes-Cummings model via the interaction Hamiltonian
 $\hat{H}=\hbar \chi (\hat{a}\sigma_{+}+\hat{a}^{\dagger}\sigma_{-})$, where
 $\sigma_{+}$ and $\sigma_{-}$ are atom raising and lowering operators and $\chi$
is atom-field coupling constant. Once the $N$ atoms crossed the cavity, leaving
their photon in the cavity and decaying to the ground state, the pure
field state in the cavity is ~\cite{scfr10}
\begin{equation}
|\psi\rangle =\lambda^{'}\sum_{m=0}^{N}\Lambda_{m}^{N}|m\rangle, \label{g1}
\end{equation}
where $\lambda^{'}$ is a normalization constant and the coefficients
$\Lambda_{m}^{N}$ come from the old coefficients $\Lambda_{m}^{N-1}$
and $\Lambda_{m-1}^{N-1}$ via
 the recurrence formula
 \begin{equation}
 \Lambda_{m}^{N}=(1-\delta_{m,0}) \Lambda_{m-1}^{N-1} C_{e}^{(N)}\sin (\chi T_{N}\sqrt{m})
+(1-\delta_{m,N}) \Lambda_{m}^{N-1} C_{g}^{(N)}\cos (\chi T_{N}\sqrt{m}), \label{g2}
\end{equation}
where $\Lambda_{0}^{0}=1$ corresponds to the initial vacuum state and
 $T_{N} $ is the interaction time of the $N$th atom with the field.

Now SSDNS can be represented in the Fock state basis as
\begin{equation}
|r,\alpha,n\rangle_{\epsilon}=\sum_{m}^{\infty}C_{m}|m\rangle, \label{g3}
\end{equation}
where $C_{m}=\langle m|r,\alpha,n\rangle_{\epsilon}$ is the expansion
 coefficient, transition amplitude, and  can be obtained
 from (\ref{scf12}). Then by controlling  $C_{e},C_{g}$, the atomic superposition
 coefficients, we can prepare each atom, in principle, before entering the cavity
  in such a way that $\Lambda^{N}_{m}=C_{m}$. It must be held
$N>\bar{n}$, where $\bar{n}$ is the mean number of photons for the desired
state. In this way the desired state may  be  obtained.
The  solution of the recurrence relations (\ref{g2}), that give the unknown
coefficients $C_{e},C_{g}$ in terms of the known coefficients $C_{m}$,
is available ~\cite{[34],{scfr9}}.

Recently,  a great progress has been done in quantum states generation using
trapping ions and laser cooling ~\cite{[34]}. An ion trapped in a harmonic
potential can be regarded as a particle with a quantized center-of-mass
motion. One can  consider an ion trapped in a harmonic potential and driven
by two laser beams tuned  to the first lower and upper vibrational sidebands,
respectively ~\cite{scfr11}.
  The author showed that, in the rotating-wave approximation, considering
sideband limit, i.e. the vibrational frequency is much larger than the other
characteristic frequency of the problem, and the behaviour of the ion in
the Lamb-Dicke regime, i.e. $\eta\ll1$, with choosing the amplitudes and
phases of the lasers in the appropriate
 forms, the interaction Hamiltonian  describing the problem takes the form
\begin{equation}
\hat{H}_{I}=-{\rm i}g (\hat{a}+\hat{a}^{\dagger})
( S_{-}+S_{+}),
\end{equation}
where $g=\eta\Omega\exp(-\frac{\eta^{2}}{2})$, and $S_{-},S_{+}$ are
the raising, lowering operators for the two-level ion. Further, $\Omega $
is the common Rabi frequency of the respective lasers
 and $\eta$ is the Lamb-Dicke parameter which is connected with the
 vibrational frequency $\nu$, the wave vector of the driven field $k$
 and the mass of the trapped ion $M$ are related by $\eta=\frac{k}{\sqrt{2M\nu}}$.
Assuming the ion is initially in the ground  state
$|g\rangle$ and the motion is in the arbitrary state $|\phi\rangle$, then
 after an interaction time   $ \tau $  the system is in the entangled state
\begin{equation}
|\psi ( \tau)\rangle =\cos (g\tau\hat{O})|\phi\rangle|g\rangle -{\rm i}
\sin (g\tau\hat{O})|\phi\rangle |e\rangle
, \label{g10}
\end{equation}
where $\hat{O}=-{\rm i}( \hat{a}-\hat{a}^{\dagger})$.
Detecting the internal state of the ion and considering the ion in the
ground state, then the vibrational motion collapses to

\begin{equation}
|\psi \rangle_{g}= N _{g }
 [ \hat{D}(\alpha )+ \hat{D}(-\alpha )]|\phi\rangle,\label{g11}
\end{equation}
where $ \alpha $
has been considered equal to $   g\tau, \hat{D}( \alpha) $ is the displacement
operator as before and $N_{g}$ is a normalization factor. If $|\phi\rangle$ is a
squeezed number state ~\cite{[6],{scfr3}} we get the motional even squeezed and
displaced number state of ion. On the other hand, if we find the ion is in
the excited state similarly we can get odd squeezed and displaced number state.

Furthermore, the quantum states
 discussed in this section can be generated  from a micromaser,
using quantum superposition
  of squeezed and displaced states from nonlinear optical processes,
  such as optical parametric generation and amplification,
to which a radiation in the Fock state is initially
introduced.
\subsection{ Conclusions}
Superposition of quantum states controls the behaviour of single states
by making it more or less pronounced, as a consequence of
the interference between different probability amplitudes.
We have introduced a superposition of two squeezed displaced number states and
discussed the quantum statistical properties of the
resulting state. We concentrated  on three special sub-states,
ESDNS, OSDNS and YSDNS corresponding to $\epsilon =1, -1$ and ${\rm i}$,
respectively. For these states we studied scalar product,
photon-number distribution $P(m)$,
normalized second-order correlation function $g^{(2)}(0)$, $W$-function, $Q$-function,
and phase properties. By means of the scalar product
of SSDNS we have demonstrated that the states
which belong to the same subspace are
not orthogonal in general, but if $\tau_{1}$ and $\tau_{2}$ are large
such that $\tanh R <1$, they become  approximately orthogonal.
The oscillations in $P(m)$ have been demonstrated.
These oscillations are more pronounced
as $r$ and $n$ increases.  Such oscillations have been recognized as a striking
 feature of  highly nonclassical  states.
 The normalized intensity correlation
function $g^{(2)}(0)$ for this superposition
shows that the sub-Poissonian interval decreases as $r$ increases.
For quasiprobability distribution functions for $n\neq 0$ but finite,
 we have noted that for  $W$-function there are two regimes controlling
its  behaviour depending on whether
the superposition is macroscopic ($\alpha >> 1$) or microscopic
($\alpha \leq 1$).
In the first regime,  the even- and odd-cases (i.e. $\epsilon=1$ and $-1$)
give similar behaviours and   the structure of the density matrix is
remarkable in figures.
In other words, for this case Wigner function includes two symmetrical  peaks originated
 at $\pm\alpha$ and interference fringes are in between.
However, for the second regime  the structures of the $W$-function
for  even- and odd-cases are quite different.
Further, in general we noted that there are negative values
 of the $W$-function for different $\epsilon$, which are  signature of
their quantum mechanical features. For all cases
 squeeze parameter is responsible for peaks stretching, which
 increases as $r$ increases, and the occupation number $n$ is responsible
 for "chaotic" behaviour, which is more pronounced as $n$ increases.
The $Q$-functions exhibit a number of interesting quantum effects. For example,
the Q-function of shifted ground state, e.g. $\alpha\neq 0,r=0,n=0$, is
squeezed by
increasing the value of $r$ and has several peaks and spikes by increasing the
number of quanta $n$.
Also it is noted that the quantum-interference term
is more visible in the W-function but not in the Q-function \cite{scfr8}.
For the phase distribution
we noted that in the first regime,  the even- and odd-cases
give similar behaviours and   the structure of the density matrix is
remarkable in figures similar to $W$-function.
All these facts are washed out
in the second regime where the behaviour  becomes irregular, however,
smooth.  In general we noted  that the higher the number of quanta is,
the more peaks the distribution possesses. Influence of the squeezing parameter
could be recognized in the second regime where the distribution exhibits
peak-splitting and peak-narrowing and this is in contrast with the fist
regime.
For  the phase variance we conclude that it asymptotically goes to
the value $\pi^2/3$ of the uniform distribution when either $\alpha$ or
$r$ is large, but through different routes. We have shown also that
this superposition can exhibit photon-number fluctuations and phase
fluctuations.

Finally, based on recent results ~\cite{scfr10,{scfr11}} we have suggested ways for generation of
such superposition in the framework of micromaser ~\cite{scfr10} and trapped ions
~\cite{scfr11}.

\section{ Quantum statistical properties of superposition of squeezed and
displaced  states with thermal noise}
In this section we  investigate the influence of thermal noise
on the properties of the previous states (\ref{scf1}) using density operator
formalism.

\subsection{Density operator for the states}

Sometimes we have not enough information to specify completely the state of
the system and hence we cannot form its wave function and consequently the
density operator must be used to describe the state of the
quantum-mechanical system. The well-known example for such a state is
provided by the thermal excitation of photons in a cavity mode
maintained at
the temperature $T$, which is described by the density operator,
$\hat{\rho}_{\rm ch}$, as \cite{scfr12}

\begin{equation}
\hat{\rho}_{\rm ch}=\frac{1}{1+\bar{n}}\sum_{n=0}^{\infty
}z^{n}|n\rangle\langle n|,  \label{scf23}
\end{equation}
where $z=\frac{\bar{n}}{\bar{n}+1}$ is the quotient of Bose-Einstein
(geometric) distribution and mean number of photons $\bar{n}=\left(
\exp (\frac{\hbar \nu }{K_{B}T})-1\right) ^{-1}$ for thermal equilibrium at
temperature $T$, $\nu $ being radiation frequency, $K_{B}$ is
the Boltzmann constant and $\hbar$ has the same meaning as before.
It is clear that the thermal distribution has a diagonal expansion
in terms of the Fock state $|n\rangle $. This diagonality causes the
electric field expectation value to vanish in thermal equilibrium. Such
thermal field can be generated by a thermal source composed of many
independent atomic radiators and consists of the superposition of waves of
many different frequencies within some continuous range. These waves can
be regarded as independent waves with random phases \cite{perin}.

In the following we define the thermal superposition squeezed and displaced
states (TSDS) similar to that of squeezed thermal state \cite{ther9}-
\cite{ther12}, displaced
squeezed thermal state \cite{ther13} and a superposition of coherent state with
thermal noise \cite{scfr13}, as a sum of states (\ref{scf1}) weighted by the Bose-Einstein
distribution, which can be represented by the density operator in the form

\begin{equation}
\hat{\rho}_{T}=\frac{\lambda _{T}^{(\epsilon )}}{1+\bar{n}}%
\sum_{n=0}^{\infty }z^{n}[\hat{D}(\alpha )+\epsilon \hat{D}(-\alpha
)]\hat{S}(r)|n\rangle \langle n|\hat{S}(-r)[\hat{D}(-\alpha )+\epsilon
^{*}\hat{D}(\alpha )],  \label{scf24}
\end{equation}
where all the notations have the same meaning as before
and $\lambda _{T}^{(\epsilon )}$ is the normalization constant, obtained
with the aid of the normalization relation ${\rm Tr}\hat{\rho}=1$ to
take
the form

\begin{equation}
\lambda_{T}^{(\epsilon )}=\left[ 1+ |\epsilon |^{2} +2|\epsilon |\cos
\phi
\exp [-2(1+2\bar{n})t^{2}]\right]^{-1}.  \label{scf25}
\end{equation}

It is noted from equation (\ref{scf24}) that the thermal superposition of
displaced and squeezed number states which has been considered in optical
cavity, is not a product of the states for independent oscillators. In
fact we can find  interesting nonclassical effects which can occur as a
result of existing correlations between oscillators;  this will be seen later.

As before the parameter $\epsilon $ can take either the
value $1,$ or $-1$ or $i$  corresponding to thermal squeezed even
displaced state (TSEDS), thermal squeezed odd displaced state (TSODS)
and thermal squeezed Yurke displaced state (TSYDS), respectively.

In the following we calculate the quasiprobability distribution
functions which will be used to derive the second-order
correlation function,  photon-number distribution,
purity, and  phase distribution.

\subsection{Quasiprobability distribution functions}

Using the $s$-parametrized characteristic function given in section 4.3
together with (\ref{scf24}) we straightforwardly get

\begin{eqnarray}
\begin{array}{lr}
C(\zeta ,s)=\lambda _{T}^{(\epsilon )}\exp \left[-(A+\frac{1
}{2}[1-s])|\zeta |^{2}-\frac{E}{2}(\zeta ^{2}+\zeta ^{*2})\right]
\\
\times \biggl\{\exp (\zeta \alpha ^{*}-\zeta ^{*}\alpha )+|\epsilon
|^{2}\exp (\zeta ^{*}\alpha -\zeta \alpha ^{*})
\\
+|\epsilon |\exp \left[
-i\phi -G-(\zeta C^{*}+\zeta ^{*}C)\right]
+|\epsilon |\exp \left[ i\phi -G+(\zeta C^{*}+\zeta ^{*}C)\right]
\biggr\},
\end{array}
\label{scf26}
\end{eqnarray}

\noindent where

\begin{eqnarray}
\begin{array}{rl}
G=(1+2\bar{n})[2|\alpha |^{2}\cosh 2r+(\alpha
^{2}+\alpha^{*2})\sinh 2r], \\
 C=(1+2\bar{n})\left[ \alpha \cosh 2r+\alpha ^{*}\sinh
2r\right] ,\\
 A=\bar{n}+(1+2\bar{n})S_{r}^{2},\quad
 E=(1+2\bar{n})S_{r}C_{r}.\label{scf27}
\end{array}
\end{eqnarray}
During the derivation of (\ref{scf26}), relations
(\ref{scf4}) and (A.2) in appendix A have been used.
Consequently, using (\ref{16}) together with the identity
(A.4) in appendix A, the $s$-parametrized quasiprobability distribution functions are

\begin{eqnarray}
\begin{array}{lr}
W(\beta ,s)=\frac{\lambda _{T}^{(\epsilon )}}{\pi
\sqrt{[A+\frac{1}{2}(1-s)]^{2}-E^{2}}}\\
\\
\times\biggl\{ \exp \left[ -\frac{[A+\frac{1}{2}(1-s)]|\beta -\alpha|^{2}+\frac{
1}{2}E[(\beta -\alpha )^{2}+(\beta ^{*}-\alpha ^{*})^{2}]}{(A+\frac{1}{2}[1-s%
])^{2}-E^{2}}\right] \\
\\
+|\epsilon |^{2}\exp \left[ -\frac{[A+\frac{1}{2}(1-s)]|\beta +\alpha
|^{2}+\frac{1}{2}E[(\beta +\alpha )^{2}+(\beta ^{*}+\alpha
^{*})^{2}]}{(A+%
\frac{1}{2}[1-s])^{2}-E^{2}}\right] \\
\\
+|\epsilon |\left[ \exp \left( i\phi -G+K\right)
+\exp\left( -i\phi -G+K^{*}\medskip \right) \right]\biggr\},
\end{array}
\label{scf28}
\end{eqnarray}

\noindent where $K$ is given by

\begin{equation}
K=\frac{[A+\frac{1}{2}(1-s)](C-\beta )(C^{*}+\beta
^{*})-E[({\rm Re}C-i{\rm Im}\beta )^{2}+(i{\rm Im}C-{\rm Re}\beta
)^{2}]}{(A+\frac{1}{
2}[1-s])^{2}-E^{2}},\label{scf29}
\end{equation}
where Re and Im mean real and imginary parts.
As we mentioned before when   $s=1,0,$ and $-1$, expression (\ref{scf28}) gives
Glauber P-function, Wigner W-function and Husimi Q-function, respectively.

By taking the appropriate values of the parameters $\epsilon ,\alpha
,r$, one can easily use expression (\ref{scf28}) to obtain the corresponding
ones for some special states, such as thermal squeezed coherent state
\cite{ther13}, and squeezed thermal state \cite{ther9,{ther10}}, also
for thermal  odd- and even-coherent states \cite{scfr13}.

\begin{figure}[h]%
 \centering
    \includegraphics[width=6cm]{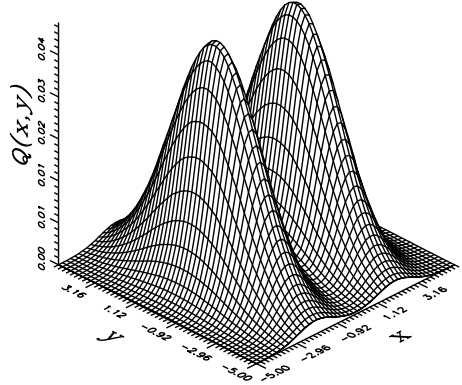}
   \caption{
 Q-function for TSYDS and $\alpha =2$ and $r=0.5$ for
 $\bar{n}=2.5$.} \label{fig8}
\end{figure}

In the following we  examine the Q-function ($s=-1$), and the
W-function ($s=0$) for TSYDS given by  (\ref{scf28}). It should be
noted that for all the cases we  assume the parameter $\alpha $ is real
and equal to 2. In Fig. 5.8 we have plotted the Q-function against
both $x={\rm Re}\beta $ and $y={\rm Im}\beta$, for $r=0.5$, and
$\bar{n}=2.5$. From this figure we
can observe two identical peaks, both of them are stretched under the
action of the squeezing parameter $r$. Further, these peaks are broader
than those of the statistical mixture of coherent states and this is connected
with mean number of thermal photons, i.e. for large $\bar{n}$ the peaks become
more and more broader.

\begin{figure}[h]%
  \centering
  \subfigure[]{\includegraphics[width=5cm]{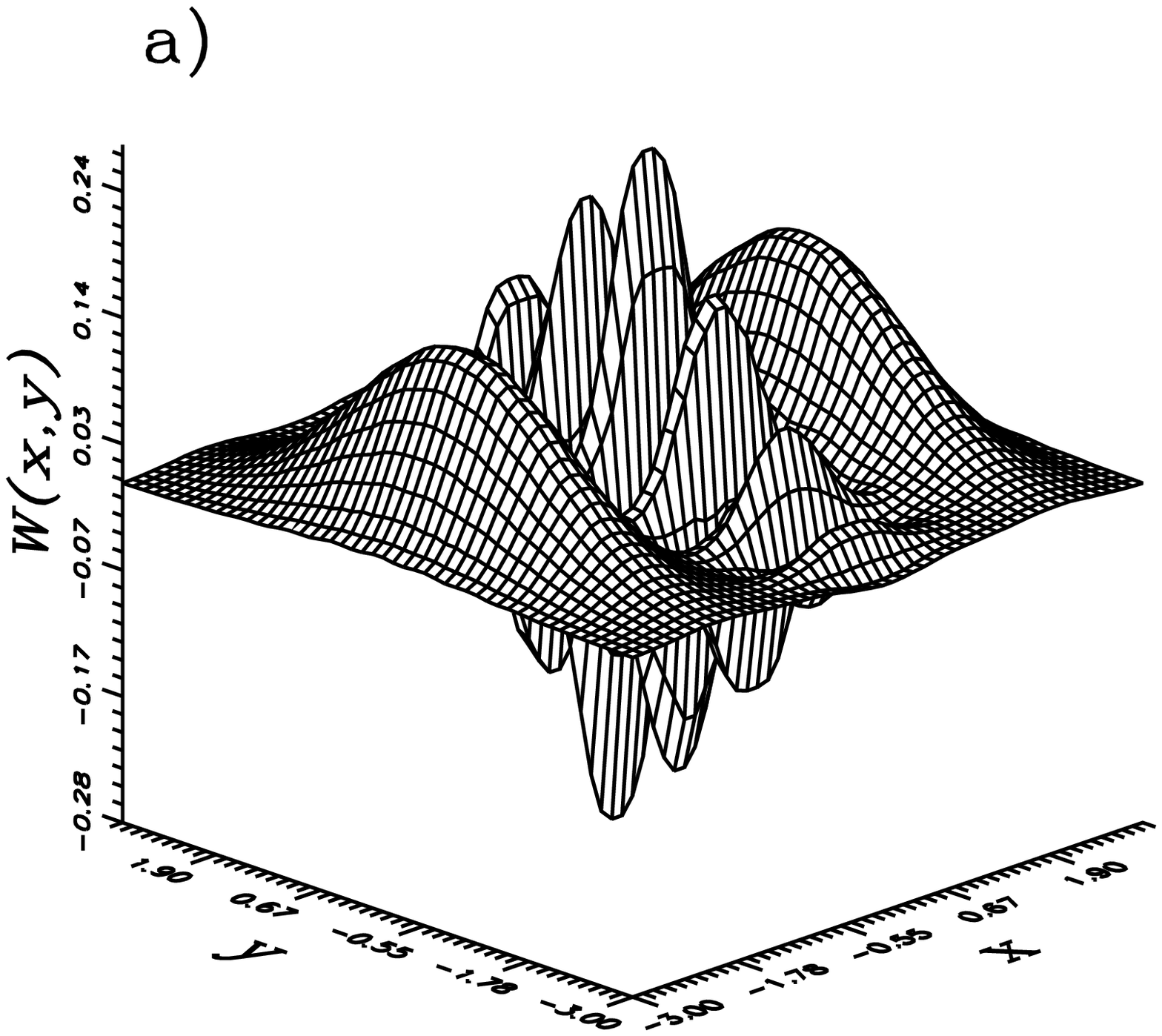}}
 \subfigure[]{\includegraphics[width=5cm]{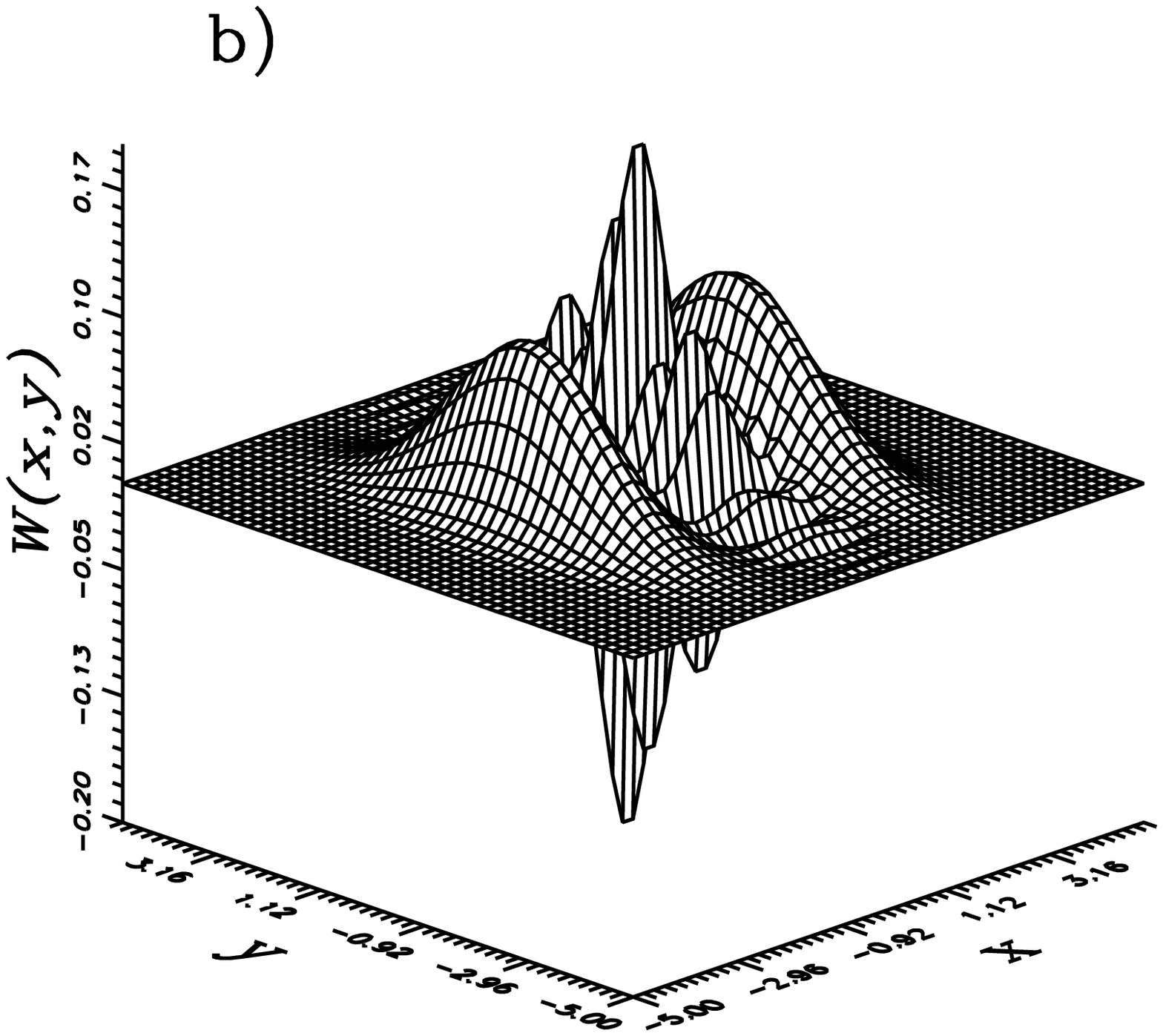}}
    \caption{
 W-function for TSYDS and $\alpha =2$ and $r=0.5$ for
a) $\bar{n}=0.5$ ; b) $\bar{n}=1$. } \label{fig9}
 \end{figure}

In Figs. 5.9a and b we have plotted the W-function against $x={\rm
Re}\beta $ and $y={\rm Im}\beta $ for $r=0.5,\bar{n}=0.5$
 and for $\bar{n}=1$, respectively. In Fig. 5.9a we can see the
W-function in $x-y$ plane, showing that the oscillator phase space
consists of two Gaussian bells, corresponding to statistical mixture of
individual composite states; also we
can see interference fringes inbetween arising as a result of the
contribution originating from the quantum superposition. Furthermore, we
find that the $W$-function takes some negative values, which is indicating
nonclassical effect,  due to the interference of the
states in phase space. Increasing thermal photon number $\bar{n}$
and keeping the value of $r$ as before, we can realize that the two
Gaussian bells are stretched in
the $y$-direction and the interference fringes become more sharper than
before. In the mean time the negative values in W-function start to
smooth out reflecting the classical effect of the thermal noise in the system
(see Fig. 5.9b).

\subsection{Second-order correlation function}

Here we study the sub-Poissonian statistics for the state under consideration.
So that the relation (\ref{17})
has been used to deduce  the expectation values  $\langle \hat{a}^{\dagger }
\hat{a}\rangle $ and $%
\langle \hat{a}^{\dagger 2}\hat{a}^{2}\rangle $
as

\begin{equation}
\langle \hat{a}^{\dagger }\hat{a}\rangle =\lambda
_{T}^{(\epsilon
)}\left[ (A+|\alpha |^{2})(1+|\epsilon |^{2})+2|\epsilon
|e^{-G}(A-|C|^{2})\cos \phi \right] , \label{scf30}
\end{equation}

\begin{eqnarray}
\begin{array}{rl}
\langle \hat{a}^{\dagger 2}\hat{a}^{2}\rangle =\lambda
_{T}^{(\epsilon )}\{\left[ 2A^{2}+4|\alpha |^{2}A+E^{2}+|\alpha
|^{4}-E(\alpha ^{2}+\alpha ^{*2})\right] (1+|\epsilon |^{2})\\
\\
+2|\epsilon |e^{-G}\left[
2A^{2}+E^{2}-4|C|^{2}A+|C|^{4}-E(C^{2}+C^{*2})\right] \cos \phi\}.
\label{scf31}
\end{array}
\end{eqnarray}
\begin{figure}[h]%
 \centering
    \includegraphics[width=6cm]{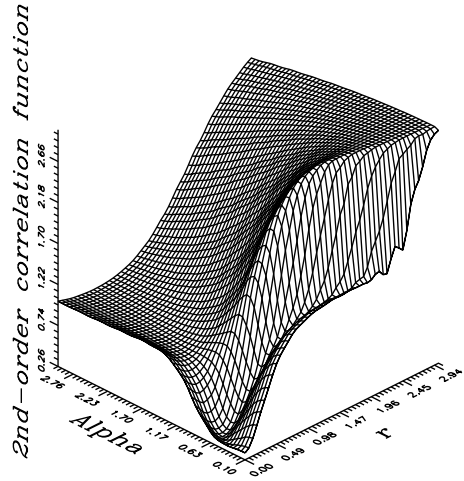}
   \caption{
 Normalized normal second-order correlation function $
g^{(2)}(0)$ for TSODS for $\bar{n}=0.1$. } \label{fig10}
\end{figure}

As is well-known in the thermal optical cavity, photons have tendency to
bunch each other, where thermal light can be described as a classical
light.
In the thermal coherent squeezed optical field in cavity the loss of
nonclassical effects is much faster than at zero temperature. This is noticed
from the numerical investigations of the function $g^{(2)}(0)$, for both
TSEDS and TSYDS; for example when we take $\bar{n}\neq 0$, we find the
values of $g^{(2)}(0)$ are always greater than one showing super-Poissonian
behaviour whatever the values of $r$ and $\alpha $ are. On the other hand,
for TSODS, and for small value of $\bar{n}$ we realized that there is
sub-Poissonian behaviour with maximum value when $\alpha $ and $r$
are small, which indicates that the
states are closed to the odd coherent states, see Fig. 5.10. Finally we would
like to point out that for $\bar{n}\geq 1$ sub-Poissonian behaviour  entirely
 disappeared.

\subsection{Photon-number distribution}
Here we discuss the photon-number distribution $P(n)$
for the states given by  (\ref{scf24}). As we mentioned before,
this function can be
deduced using the relation between Laguerre polynomial and Wigner
function given by (\ref{19}).
In this case the integration may be
carried out applying the same technique as in Appendix A
 and   the photon-number distribution reads

\begin{equation}
\begin{array}{rl}
 P(n)  = \frac{\lambda _{T}^{(\epsilon )}}{\sqrt{(A+1)^{2}-E^{2}}}%
\sum_{m=0}^{n}(1-\nu _{-})^{m}(1-\nu _{+})^{n-m}   \\
  \\
 \times \left\{ (1+|\epsilon |^{2})L_{m}^{-\frac{1}{2}%
}(-X_{-})L_{n-m}^{-\frac{1}{2}}(-X_{+})\exp \left[
-\frac{[(A+1)|\alpha
|^{2}+E(\alpha _{1}^{2}-\alpha _{2}^{2})]}{(A+1)^{2}-E^{2}}\right]
\right. \\
 \\
 \left. +2|\epsilon
|L_{m}^{-\frac{1}{2}}(C_{2}^{2}Y_{-})L_{n-m}^{-%
\frac{1}{2}}(C_{1}^{2}Y_{+})\exp \left[ \frac{%
[(A+1)|C|^{2}-E(C_{1}^{2}-C_{2}^{2})]}{(A+1)^{2}-E^{2}}-G\right] \cos
\phi
\right\} , \label{scf32}
\end{array}
\end{equation}
where

\begin{eqnarray}
\begin{array}{rl}
\nu _{\pm }=[A+1\pm E]^{-1},\quad
 X_{+}=\alpha _{2}^{2}[(A+E)(A+E+1)]^{-1},\\
\\
 X_{-}=\alpha _{1}^{2}[(A-E)(A-E+1)]^{-1},\quad
 Y_{\pm }=[(A\pm E)(A\pm E+1)]^{-1},
\label{scf33}
\end{array}
\end{eqnarray}

\noindent and $L_{n}^{k}(.)$ is again the associated Laguerre polynomial.

Now  when we take $r\rightarrow 0$ in (\ref{scf32}) and using the
identity \cite{scfr14}
\begin{equation}
\sum_{m}^{n}L_{m}^{k_{1}}(x)L_{n-m}^{k_{2}}(y)=L_{n}^{k_{1}+k_{2}+
1}(x+y),
\label{scf34}
\end{equation}
we arrive at
\begin{equation}
\begin{array}{rl}
P(n)  = \frac{2\lambda _{T}^{(\epsilon )}(r=0)}{\bar{n}+1}\left(
\frac{\bar{n}}{\bar{n}+1}\right) ^{n}\left\{ (1+|\epsilon |^{2})
\exp \left( -\frac{|\alpha |^{2}}{1+\bar{n}}\right) L_{n}
\left( -\frac{|\alpha|^{2}}{1+\bar{n}}\right) \right.   \\
  \\
 \left. +2|\epsilon |\exp \left[ -\frac{(2\bar{n}+1)|\alpha
|^{2}}{1+%
\bar{n}}\right] L_{n}\left( \frac{(2\bar{n}+1)^{2}|\alpha
|^{2}}{\bar{n}(1+%
\bar{n})}\right) \cos \phi \right\} ,\label{scf35}
\end{array}
\end{equation}
which is in agreement with equation (26) of \cite{scfr13} provided that
we set $|\epsilon
|=1$
for even- and odd-thermal coherent states.

\begin{figure}[h]%
 \centering
    \includegraphics[width=6cm]{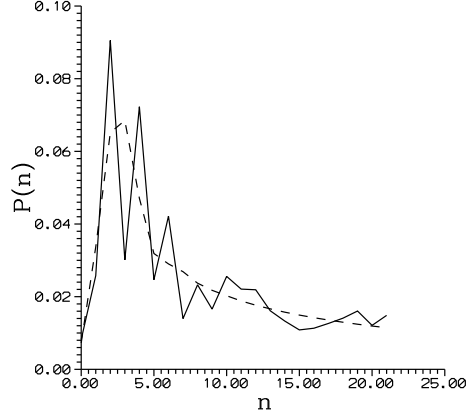}
   \caption{
 $P(n)$ for TSEDS for $r=1.5,\bar{n}=3$ and
$|\alpha|^{2}=4$; the dashed curve is for the statistical mixture.
} \label{fig11}
\end{figure}


In general  for the superposition of coherent thermal
optical fields in the cavity there are large scale macroscopic
oscillations
in the photon-number distribution which disappear for $\bar{n}>>|\alpha
|^{2}$, see \cite{scfr13}. In Fig. 5.11 we have plotted $P(n)$ given by
(\ref{scf32})
for TSEDS against the photon number $n$ for fixed values of $\bar{n}$, $%
\alpha $ and $r$ (solid curve). Comparing our results with those obtained
in \cite{scfr13}, we can say that $P(n)$ in the present case have oscillations
more pronounced than those given in the thermal even coherent state case.
This emphasizes the role of squeezing parameter in the thermal optical cavity.
In fact these oscillations are not only indication to nonclassical
effects, but are also emerging from the quantum interference between
the components of the state in the phase space. This can be seen when
comparing
 dashed curve (corresponding to the statistical mixture) and solid
curve in the figure (these curves are well defined for $n$ positive integer
or zero).

Now let us discuss the degree of purity ${\rm Tr}\rho ^{2}$ for TSDS.
Substituting  (\ref{scf26}) into
(\ref{scf36}), carrying out the integration and putting $|\epsilon |=1$, we
get\bigskip

\begin{eqnarray}
\begin{array}{rl}
{\rm Tr}(\rho^{2}_{T})= \frac{\lambda_{T}}{2\bar{n}+1}\{
2+\exp\left( -\frac{4e^{2r}\alpha^{2}}{2\bar{n}+1 }\right)
 +4\exp \left[ -e^{2r}\alpha^{2}(2\bar{n}+1+\frac{1}{2\bar{n}+1}
)\right]\\
\\
\times \cos (\phi )
 +\exp \left[ -4e^{2r}\alpha ^{2}(2\bar{n}%
+1)\right] \cos (2\phi )\}.\label{scf37}
\end{array}
\end{eqnarray}
As we mentioned before for a pure state ${\rm Tr}(\rho ^{2})=1$ and for a
mixed state ${\rm Tr}(\rho ^{2})<1$.
The authors of \cite{ther13} have shown that the degree of purity of
the input chaotic light states is unchanged by subsequent squeezing or
displacement process. Further, in \cite{scfr13} it has been shown that
the superposition of coherent thermal fields in a cavity is quite
complicated, thermal odd coherent state is more chaotic than the thermal
light, but for thermal even
coherent state it was found that $\frac{1}{1+\bar{n}}\leq {\rm Tr}(\rho
_{TSS}^{2})<1$. In comparing equation (29) in \cite{scfr13} with expression
(\ref{scf37}),
we can say that the degree of purity for thermal odd (even) coherent
state is larger than that of TSODS (TSEDS) owing to squeezing parameter $r$.
In Fig. 5.12 we have plotted ${\rm Tr}(\rho _{T}^{2})$ for TSYDS against $\alpha $
for  fixed value of $\bar{n}=1$ and different values of $r\quad (r=0,0.6)$.
From this
figure it is easy to see that the inequality $\frac{1}{1+\bar{n}}\leq {\rm
Tr}(\rho _{T}^{2})<1$ is satisfied as for thermal even coherent state, where
we find that ${\rm Tr}(\rho _{T}^{2})$ starts to increase its value comparing to
the value for thermal light until it reaches the maximum, then it starts again
to decrease until it takes the value of thermal light, but this is reached after long
period of $\alpha $. Also we noticed that as $r$ increases, the ${\rm
Tr}(\rho _{T}^{2})$ rapidly decreases to the value for chaotic light.

\begin{figure}[h]%
 \centering
    \includegraphics[width=6cm]{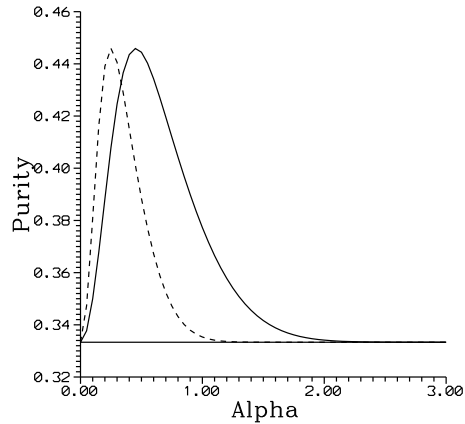}
   \caption{
 $Tr{\rho^{2}}$ for TSYDS for $\bar{n}=1$, $r=0$
(solid curve) and $r=0.6$ (dashed curve). } \label{fig12}
\end{figure}

\subsection{Phase property}

As is well known, the nonclassical properties of electromagnetic waves
are
progressively destroyed by presence of noise and losses. Therefore the
effect of thermal noise on quantum phase measurement, especially of
nonclassical light, has attracted only a little attention \cite{ther14}.
Here we  consider this effect based on the quasiprobability
distribution function technique,  which  is more convenient for
this problem.
So inserting $W(\beta ,s)$  given by (\ref{scf28}) into  equation (\ref{20})
and using the identity
\begin{equation}
\int_{0}^{\infty} \exp(-ax^2-bx-c) dx=
\frac{1}{2}\sqrt{\frac{\pi}{a}}[1- {\rm erf}(\frac{b}{2\sqrt{a}})],
\label{scfa38}
\end{equation}
where ${\rm erf}(.)$ is the Gauss error function defined by
\begin{equation}
{\rm erf}(x)=\frac{2}{\sqrt{\pi }}\int_{0}^{x}\exp (-y^{2})dy, \label{scf40}
\end{equation}
after minor algebra we arrive at

\begin{eqnarray}
\begin{array}{rl}
P^{(\epsilon)}(\theta,s)=\frac{\lambda_{T}}{\pi
\sqrt{A+\frac{1-s}{2}+E\cos
(2\theta)}}\left\{ \left[ \frac{(1+|\epsilon|^{2})}{2\sqrt{h_{1}}}%
\exp(-\tau_{1})\right.\right. \\
\\
+\left.\frac{h_{2}\exp(\tau_{2})\sqrt{\pi}}{2}[1-|\epsilon|^{2}+(1+
|\epsilon|^{2}) {\rm erf}(h_{2}\sqrt{h_{1}}) ]\right] \\
\\
+\left. |\epsilon|\exp(\tau_{3})\left[\frac{\cos \phi
\exp(\frac{h^{2}_{3}}{%
h_{1}}) }{\sqrt{h_{1}}}+\frac{h_{3}\sqrt{\pi}}{h_{1}}[\sin\phi + i {\rm
erf}
(ih_{3}\sqrt{h_{1}})\cos \phi ]\right] \right\},
\label{scf38}
\end{array}
\end{eqnarray}

\noindent where

\begin{eqnarray}
\begin{array}{rl}
h_{1}=\frac{A+\frac{1-s}{2}+E\cos (2\theta
)}{(A+\frac{1-s}{2})^{2}-E^{2}},\quad
h_{2}=\frac{\alpha [A+\frac{1-s}{2}+E]\cos \theta }{(A+\frac{1-s}{2}%
)^{2}-E^{2}\cos (2\theta )},\quad
h_{3}=\frac{C^{2}\sin \theta }{(A+\frac{1-s}{2})+E},\\
\\
\tau_{1}=\frac{\alpha^{2}}{(A+\frac{1-s}{2})-E}, \quad
\tau_{2}=\frac{\alpha^{2}\sin^{2}\theta
}{A+\frac{1-s}{2}+E\cos(2\theta)},\quad
\tau_{3}=\frac{C^{2}\cos^{2}\theta }{A+\frac{1-s}{2}+E\cos(2\theta)}
-G.
\label{scf39}
\end{array}
\end{eqnarray}

The phase distribution functions obtained by means of quasidistribution
functions
may possess some difficulties, such as the singularity  using the $P$-function
and the negative values using $W$-function for some states.
However, these difficulties can be avoided by considering the case
$s=-1$, corresponding to the $Q$-function which is always positive.
\begin{figure}[h]%
  \centering
  \subfigure[]{\includegraphics[width=5cm]{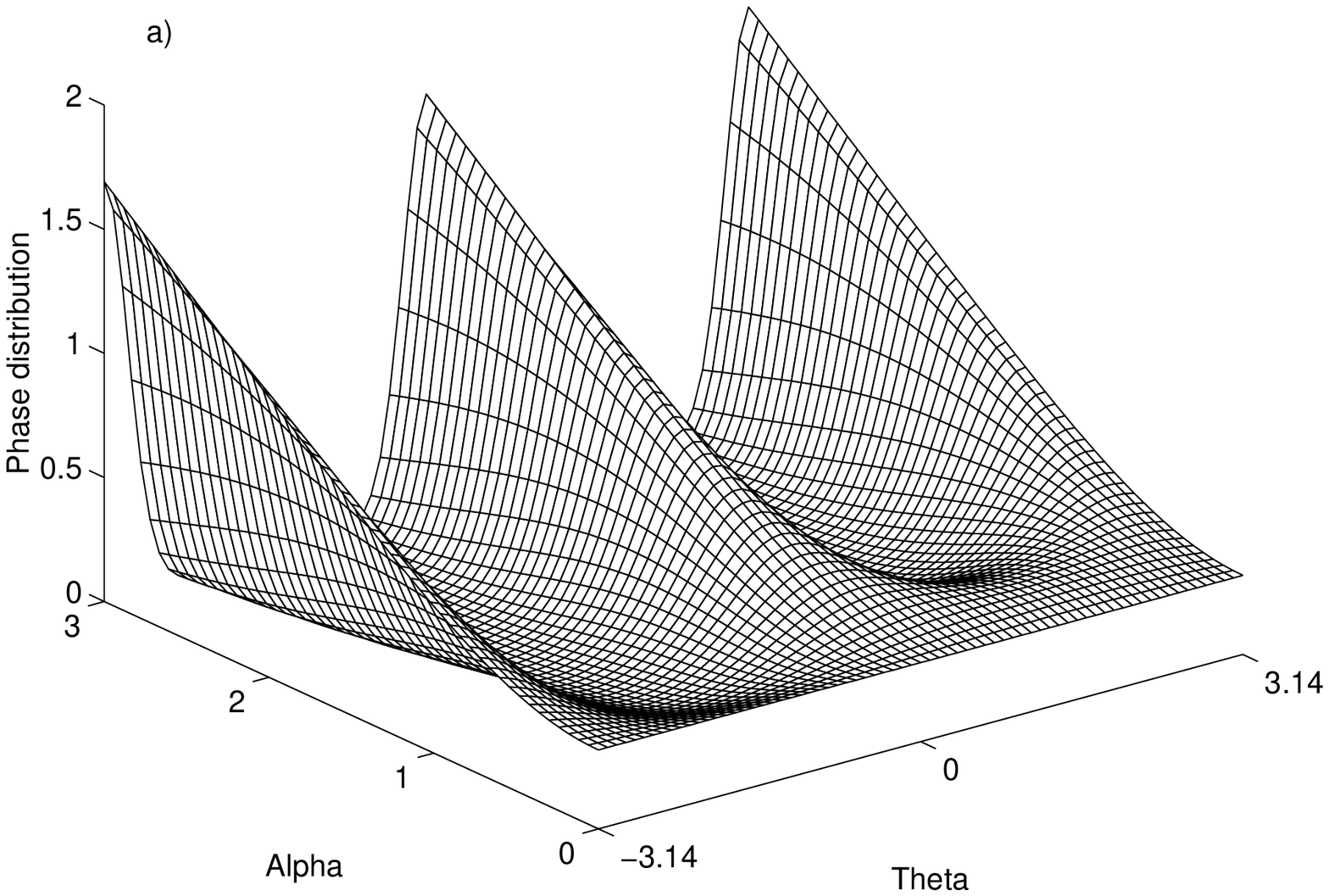}}
 \subfigure[]{\includegraphics[width=5cm]{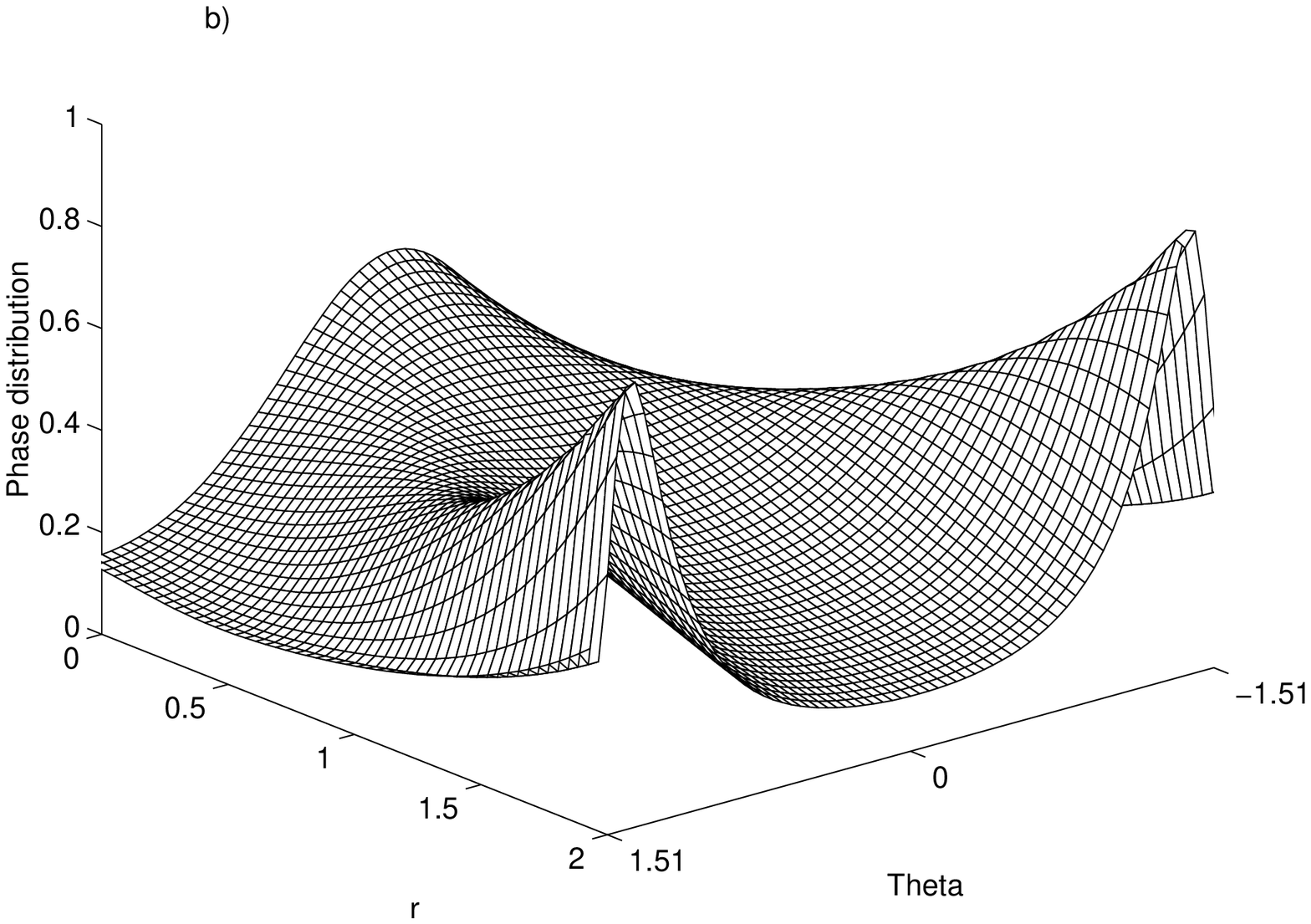}}
\subfigure[]{\includegraphics[width=5cm]{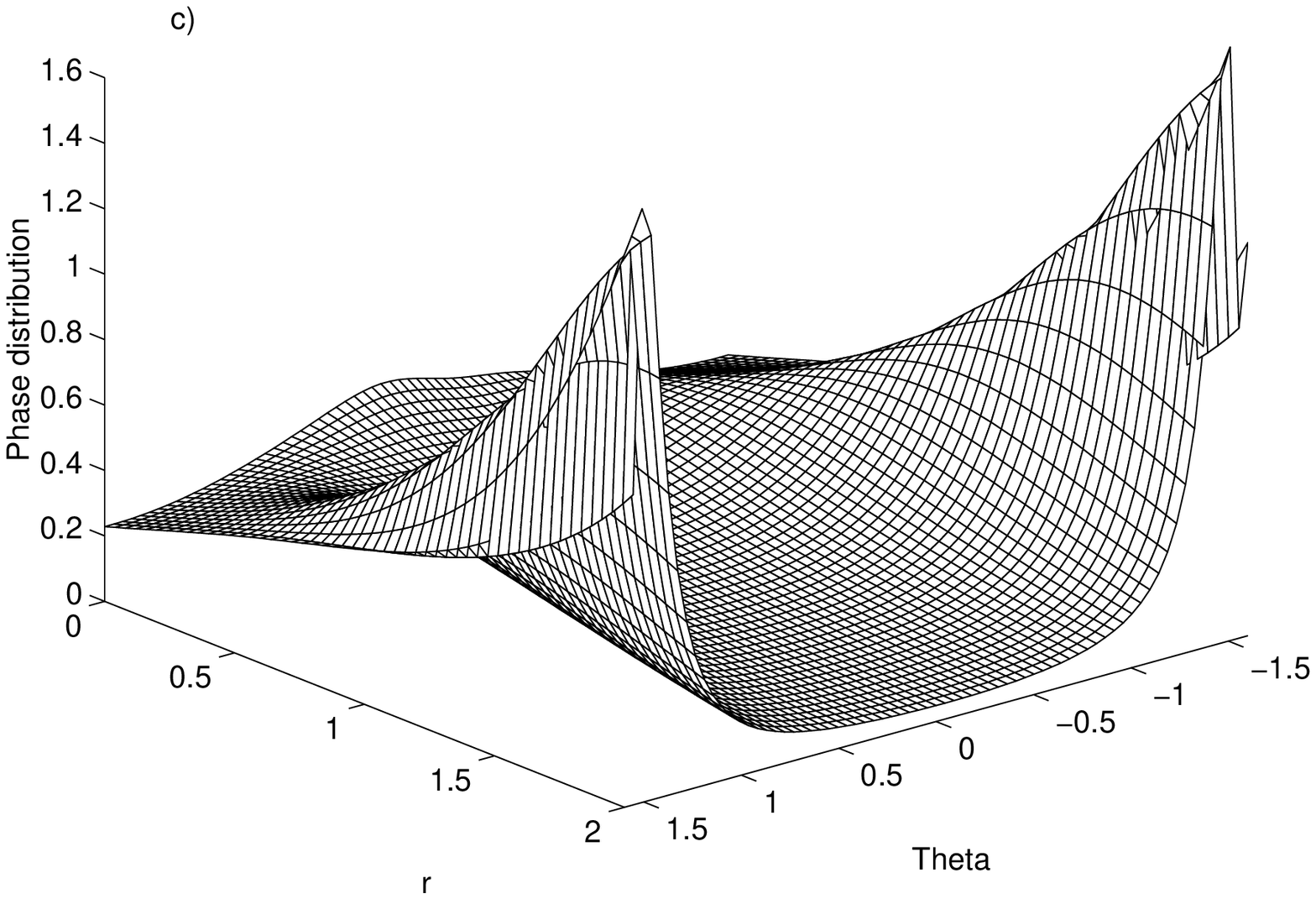}}
    \caption{
 Phase distribution for TSYDS for a) $P(\theta ,\alpha )$
for $r=0,\bar{n}=0$; b) $P(\theta ,r)$ for $\bar{n}=0,\alpha =1$;
c) $P(\theta,r)$ for $\bar{n}=2,\alpha =1$. } \label{fig13}
 \end{figure}

The phase distribution for superposition of quantum states, say even
and/or
odd coherent states, has two-peak structure \cite{scfr15}.
For the present case we have displayed in Figs. 5.13a-c the
phase distribution for TSYDS. In Fig. 5.13a we plotted the phase
distribution
for the Yurke-Stoler state with $r=0,$ and $\bar{n}=0$.
In this figure we can see two-peak structure for even (odd) coherent state,
which has been replaced by single peak at $\theta =0$ and two wings as
$\theta \rightarrow \pm \pi $. This behaviour has been realized in
studying
the odd binomial states, see \cite{scfr16}. The squeezing effect in the
Yurke-Stoler optical field in the cavity, i.e. squeezed Yurke-Stoler
state ($
\bar{n}=0$), has also been displayed in Fig. 5.13b. It is clear in our
case
that the bifurcation phenomenon appears in range of $\theta $ shorter than
that for the squeezed state, where the central peak is at $\theta =0$ and
the lateral peaks are at $\theta =\pm 1.5$. Finally we have plotted in
Fig. 5.13c the effect of thermal noise in addition to squeezing for fixed value of
the parameter $\alpha\quad (\alpha =2)$. From this figure we can realize that
the latter bifurcation has been destroyed, e.g. the central peak is washed
out, which indicates the influence of losses in the optical cavity,
when the nonclassical effect disappeared.

\subsection{Conclusions}

In this section we have considered the
thermal
noise in the distribution of superposition squeezed and displaced number
states. An observation of such a system can be found in a trapped ions,
where a motion of a harmonically bound $^{9}Be^{+}$ ions can be
used, see for example \cite{[34]}. This can be seen when ions are
initially
cooled by a laser close to zero point of motion and the coupling between
their motional and internal states due to applied (classical) radiation
can
be described by the Jaynes--Cummings-type interaction. In this case the
evolution of the internal atomic state will create states having
quantum
character, such as number state. On the other hand, if the ions
are
not in the Fock state, the motional state is characterized by a density
operator whose diagonal elements have a number distribution $P(n)$ given
by

\begin{equation}
P_{\downarrow }(t)=\frac{1}{2}\left[ 1+{\sum_{n=0}^{\infty
}P(n)\cos
(2\Omega _{n,n+1}t)\exp (-\nu _{n}t)}\right] ,\label{scf41}
\end{equation}
 where $P_{\downarrow }$ is the probability of occupation in $%
|\downarrow \rangle $ ($|\downarrow \rangle $ is the hyperfine ground
state
of the ion), $\Omega _{n,n+1}$ is the Rabi frequency and $\nu _{n}$ is the
decoherence rate between levels $|n\rangle $ and $|n+1\rangle $. $P(n)$
in
expression (\ref{scf41}) has been derived analytically for more general case, see
equation (\ref{scf32}), which
 can  be controlled by the detuning Doppler cooling.
Therefore
we have an advantage to measure and control further phenomena by using the
photon-number distribution given by  (\ref{scf41}).

For thermal superposition of squeezed and displaced states, the
nonclassical
effects have been traced via quasiprobability distribution functions,
second-order correlation function, photon-number distribution, and phase
distribution. We have seen that the correlation between different
oscillators is essentially responsible for nonclassical effects, which
has been demonstrated by the behaviour of Wigner function and photon-number
distribution. The thermal light photons have tendency to bunch each
other, this has been reflected in the behaviour of our state, for example the
normalized correlation function $g^{(2)}(0)$ shows sub-Poissonian statistics but
only
for real and negative values of $\epsilon $, provided $\bar{n}$ is
sufficiently small. $Q$-function exhibits always two peaks strucure
provided that $\epsilon\neq 0$.
Nevertheless, for the sake of comparison, it is important to remind that
for the state (\ref{scf1}), as we have shown in section 5.1,
 the $W$-function exhibits negative values for
a wide
range of phase space, i.e. the role of the interference term is less
effective; $g^{(2)}(0)$ shows sub-Poissonian statistics for a different
type of $\epsilon$, i.e. $\epsilon=\pm 1,i$; further $Q$-function displays
multi-peak structures related to the excitation number $n$.

We have also calculated the photon-number
distribution
using the Wigner function and we have shown that the oscillations in $P(n)$
are
increased when the squeezing is considered for superposition of thermal
coherent fields in the cavity. Also we have found that the
purity coefficient for
this superposition which is mixed rather than pure. Finally, we have considered the
phase properties of the optical system and shown that for a short range of
$%
\theta $,
bifurcation distinguishing the squeezed states can occur in the absence
of thermal noise.
This phenomenon disappears with increasing
thermal noise in the cavity.

   \section{ Quantum statistics of a solvable three-boson squeeze operator
model}
In this section  we  introduce
a highly correlated multidimensional squeeze operator. This operator
consists
of three modes in interaction and it represents the time-dependent
evolution operator of the interaction part of the Hamiltonian
\cite{abd1,{abd2}} determined as

\begin{eqnarray}
\begin{array}{lc}
\frac{H}{\hbar}=  \sum_{j=1}^{3} \omega_{j} \hat{a}_{j}^{\dagger}
\hat{a}_{j} - i \lambda_{1} \left\{\hat{a}_{1} \hat{a}_{2} \exp [i (\omega_{1}+
\omega_{2})t] - {\rm h.c.}\right\} \\
\\
 \hspace{1in} -i \lambda_{2} \left\{ \hat{a}_{1} \hat{a}_{3}
\exp [i(\omega_{1}+ \omega_{3}) t] -\mbox{h.c.}\right\} -i \lambda_{3} \left\{
\hat{a}_{2}
\hat{a}_{3}^{\dagger} \exp [i(\omega_{2}- \omega_{3})t] - \mbox{h.c.}\right\},
\label {scf42}
\end{array}
\end{eqnarray}
 where $\hat{a}_{j}$ and $\hat{a}_{j}^{\dagger}$ satisfy the
commutation relations

\begin{equation}
[\hat{a}_{k}, \hat{a}_{j}^{\dagger}]=  \delta_{kj}, \quad \delta_{kj}
=\{_{0 \quad k~\neq j}^{1 \quad k=j},  \label{scf43}
\end{equation}
 and $\omega _{j}$ are the field frequencies and $\lambda_{j}$
are the effective intermodal coupling constants. In fact, Hamiltonian equation (\ref{scf42})
describes a two-photon parametric coupling of modes $1$ and $2$, and $1$
and
$3$ (two photons are simultaneously created or annihilated in both the
quantum modes through the interaction with classical pumping mode), and
linear interaction of modes $2$ and $3$. By introducing the
transformation $%
\hat{A}_{j}=\hat{a}_{j}\exp (i\omega _{j}t)$, the Hamiltonian (\ref{scf42})
may be transformed into the interaction picture,

\begin{equation}
\frac{H}{\hbar}=  - i \lambda_{1} ( \hat{A}_{1}\hat{A}_{2} - \hat{A}
_{1}^{\dagger}\hat{A}_{2}^{\dagger} ) - i \lambda_{2} (\hat{A}_{1}
\hat{A}_{3} -\hat{A}_{1}^{\dagger}\hat{A}_{3}^{\dagger}) - i \lambda_{3}
(\hat{A}_{3}^{\dagger} \hat{A}_{2} - \hat{A}_{3} \hat{A}_{2}^{\dagger} ).
\label{scf44}
\end{equation}

It is important mentioning that  the interaction (\ref{scf44})
can be established in a bulk nonlinear crystal exhibiting
the second-order nonlinear properties in which three dynamical
modes of frequencies $\omega_1, \omega_2, \omega_3$ are induced
by three beams from lasers of these frequencies.
When pumping this crystal by means of the corresponding
strong coherent pump beams, as indicated in the Hamiltonian,
we can approximately fulfil the phase-matching conditions for the
corresponding processes, in particular if the frequencies are close
each other (biaxial crystals may be helpful in such an arrangement).
Also a possible use of quasi-phase matching may help in the realization,
which is, however, more difficult technologically \cite{real}.
Another possibility to realize such interaction  is to use
a nonlinear directional coupler which is composed of two optical
waveguides fabricated from  nonlinear material described by the quadratic
susceptibility $\chi^{(2)}$. Modes 1 and 2 propagate in the first
waveguide and 1 and 3 in the second waveguide (auxiliary device such
as bandgap quantum coupler \cite{qu15} or a set of mirrors can be used
to generate two identical modes (mode 1) in each
waveguide). The interactions between the modes in  the same waveguide are
established by strong pump coherent light. In this case the coupling
constants $\lambda_{1}$ and $\lambda_{2}$ are proportional to the second
order susceptibility $\chi^{(2)}$ of the medium and they also
include the amplitudes of the pump. The linear coupling between modes 2 and
3 is established through the evanescent waves \cite{marc}.

For completeness the time-evolution operator of equation (\ref{scf44}) is

\begin{equation}
\exp (-i \frac{H}{\hbar} t) = \exp [ \lambda_{1} t (\hat{A}
^{\dagger}_{1}
\hat{A}_{2}^{\dagger} - \hat{A}_{1} \hat{A}_{2} ) + \lambda_{2} t (
\hat{A}%
_{1}^{\dagger}\hat{A}_{3}^{\dagger}-\hat{A}_{1} \hat{A }_{3} ) +
\lambda_{3}
t (\hat{A}_{3} \hat{A}_{2}^{\dagger}- \hat{A} _{3}^{\dagger} \hat{A}_{2}
)],
 \label{scf45}
\end{equation}

\noindent which can be identified well with a time-dependent three-mode
squeeze operator

\begin{equation}
\hat{S}(\underline{r})= \exp [ r_{1} (\hat{A}^{\dagger}_{1} \hat{A}
_{2}^{\dagger} - \hat{A}_{1} \hat{A}_{2} ) + r_{2} ( \hat{A}_{1}^{\dagger}
\hat{A}_{3}^{\dagger}-\hat{A}_{1} \hat{A}_{3} ) + r_{3} (\hat{A}_{3}
\hat{A}_{2}^{\dagger}- \hat{A}_{3}^{\dagger} \hat{A}_{2} )],
 \label{scf46}
\end{equation}
 where $r_{j}=\lambda _{j}t$, with $0\leq r_{j}<\infty $,
$j=1,2,3$
and $\underline{r}=(r_{1},r_{2},r_{3})$. It is evident that this
squeeze
operator must involve two different squeezing mechanisms (terms
involving $%
r_{1}$ and $r_{2}$) and then it is more complicated than squeezing
operators that have appeared in the literature earlier
\cite{[5], {[6]},{entan1}}-\cite{entan5}. Now if we set

\begin{equation}
 \hat{A} = (\hat{A}_{1}^{\dagger}\hat{A}_{2}^{\dagger} -\hat{A}_{1}
  \hat{A}_{2}), \quad \hat{B}  =(\hat{A}_{1}^{\dagger}
\hat{A}_{3}^{\dagger}- \hat{
A}_{1} \hat{A}_{3} ), \quad
 \hat{C}=(\hat{A}_{2}\hat{A}_{3}^{\dagger
}-\hat{A}_{3}\hat{A}_{2}^{\dagger }),
 \label{scf47}
\end{equation}
 then we have the following commutation relations

\begin{equation}
 [\hat{A}, \hat{B}]=  - \hat{C}, \quad
\quad [\hat{B}, \hat{C}]=\hat{A}, \quad  \quad [\hat{C},
\hat{A}] =\hat{B}.
 \label{scf48}
\end{equation}
\noindent We may conclude that the squeeze operator (\ref{scf46}) involves
correlations and can be regarded as the exponential of linear combination of
three generators, which are closed under the commutation relations (\ref{scf48}),
and it represents the $su(1,1)$ generalized coherent state.
As we mentioned earlier the quantum correlation between different quantum
mechanical
systems is responsible for the  nonclassical effects for such type
of  operator.

Our plan of studying operator (\ref{scf46})  will be as follows:
We derive the
basic relations for three-mode squeeze operator and then  we
 discuss three-mode
photon-number sum and photon-number difference as well as squeezing
phenomenon. Moreover, we give two examples of three modes
squeezed states related to this operator
which are three-mode squeezed coherent and number states.
For these two states we  discuss sub-Poissonian phenomenon
and quasidistribution functions.

\subsection{ Properties of the correlated quantum systems}

Squeezed state of light is distinguishable by long-axis variance of noise
ellipse for one of its quadratues in phase-space. This property is connected
with the pairwise nature of the unitary operator (\ref{2}) under which the
initial state evolves.
This operator exhibits some well-known basic relations
summarized in the literature
\cite{[5]}. We  introduce similar relations corresponding to
the three-mode squeeze operator (\ref{scf46}), and then we use them
to deduce the three-mode photon-number sum and difference. Also we discuss
squeezing phenomenon related to this operator.

The squeeze operator (\ref{scf46}) provides a Bogoliubov transformation of
the annihilation and creation operators that mixe the three modes as

\begin{eqnarray}
\begin{array}{rl}
\bar{A}_{1} \equiv \hat{S}^{-1} (\underline{r})
\hat{A}_{1}
\hat{S}( \underline{r}) = \hat{A}_{1} f_{1} + \hat{A}_{2}^{\dagger}
f_{2}+
\hat{A}_{3}^{\dagger} f_{3},\\
 \bar{A}_{2} \equiv \hat{S}^{-1} ( \underline{r})
\hat{A}_{2}
\hat{S}( \underline{r}) = \hat{A}_{2} g_{1} + \hat{A}_{3} g_{2}+
\hat{A}
_{1}^{\dagger} g_{3},\\
 \bar{A}_{3} \equiv \hat{S}^{-1} (\underline{r})
\hat{A}
_{3} \hat{S}(\underline{r}) = \hat{A}_{3} h_{1} +
\hat{A}_{1}^{\dagger}
h_{2}+ \hat{A}_{2} h_{3},
\label {scf49}
\end{array}
\end{eqnarray}
 where
\begin{eqnarray}
\begin{array}{lr}
f_{1} = \cosh \mu + \frac{2 r_{3}^{2}}{\mu^{2}}
\sinh^{2}(
\frac{\mu}{2}), \quad
 f_{2} = \frac{r_{1}}{\mu} \sinh \mu - \frac{2 r_{2}
r_{3}}{
\mu^{2}} \sinh^{2} (\frac{\mu}{2}),\\
 f_{3} = \frac{r_{2}}{\mu} \sinh \mu + \frac{2r_{1}
r_{3}}{
\mu^{2}} \sinh^{2}(\frac{\mu}{2}),
\quad g_{1} = \cosh \mu - \frac{2 r_{2}^{2}}{\mu^{2}}
\sinh^{2}(\frac{\mu}{2}), \\
 g_{2} = \frac{r_{3}}{\mu} \sinh \mu + \frac{2 r_{1}
r_{2}}{
\mu^{2}} \sinh^{2}(\frac{\mu}{2}), \quad
 g_{3} = \frac{r_{1}}{\mu} \sinh \mu + \frac{2r_{2}
r_{3}}{%
\mu^{2}} \sinh^{2}(\frac{\mu}{2}), \\
 h_{1} = \cosh \mu - \frac{2 r_{1}^{2}}{\mu^{2}}
\sinh^{2}(
\frac{\mu}{2}), \quad
 h_{2} = \frac{r_{2}}{\mu} \sinh \mu - \frac{2 r_{1}
r_{3}}{
\mu^{2}} \sinh^{2} (\frac{\mu}{2}),\\
 h_{3} = \frac{-r_{3}}{\mu} \sinh \mu + \frac{2r_{1}
r_{2}}{%
\mu^{2}} \sinh^{2} (\frac{\mu}{2}),
\label {scf50}
\end{array}
\end{eqnarray}
 where $\mu = \sqrt{r_{1}^{2}+ r_{2}^{2} -r_{3}^{2} } $ and
$r_{3}^{2} < r_{1}^{2} + r_{2} ^{2}$.  The derivation of these relations
is given in appendix B.

The commutation relations (\ref{scf43}) under the transformations (\ref{scf49})
hold also for the operators $\bar{A}_{j}$. Using these transformations we can
easily calculate the statistical properties for each mode. It is worth to
mention that the corresponding expressions for the single mode squeezed
operator \cite{[5]} and for two-mode squeezed operator \cite{entan1} can be
obtained from (\ref{scf49}) by taking $r_{2}=r_{3}=0,$ and
$\hat{A_{3}}\rightarrow \hat{0}$
together with $\hat{A_{2}}\rightarrow \hat{A}_{1}$ for only single mode case.

We may point out that a strong correlation is built up between the three
modes described by squeeze operator (\ref{scf46}). This is quite obvious for the
case of the parametric amplification when two mode waves are mixed to
generate a third wave via nonlinear medium, e.g. in an optical crystal with
nonlinear second order susceptibility \cite{[5]}. This can be
demonstrated
with the help of three-mode pure squeezed vacuum states
$\hat{S}(\underline{r
})\prod_{j=1}^{3}|0_{j}\rangle $, where $\hat{S}(\underline{r})$ is
the
squeeze operator (\ref{scf46}). In this case the eigenstates of the three-mode
photon-number difference $\hat{A}_{1}^{\dagger }\hat{A}_{1}-\hat{A}
_{2}^{\dagger }\hat{A}_{2}-\hat{A}_{3}^{\dagger }\hat{A}_{3}$ correspond to
zero
eigenvalue, thus

\begin{equation}
 \Delta ( \hat{A}_{1}^{\dagger} \hat{A}_{1} - \hat{A}
_{2}^{\dagger} \hat{A}_{2} -\hat{A}_{3}^{\dagger} \hat{A}_{3} )^{2}
=0.
\label{scf51}
\end{equation}
 However, the situation will be different for three-mode
photon-number sum, using (\ref{scf49}) and
after minor calculations we obtain

\begin{eqnarray}
\begin{array}{rl}
\Delta ( \hat{A}_{1}^{\dagger} \hat{A}_{1} + \hat{A}
_{2}^{\dagger} \hat{A}_{2} +\hat{A}_{3}^{\dagger} \hat{A}_{3}
)^{2}=
f^{2}_{1} (f^{2}_{1}-1)+g^{2}_{3}(1+g^{2}_{3})\\
+h^{2}_{2}(1+h^{2}_{2})+ 2(f^{2}_{1}g^{2}_{3}+ f^{2}_{1}h^{2}_{2}+
h^{2}_{2}g^{2}_{3}).
\label{scf51}
\end{array}
\end{eqnarray}

In order to see the correlation between modes we have to calculate the
fluctuations for both sum and difference operators thus having

\begin{eqnarray}
\begin{array}{rl}
2\left(\langle \hat{A}^{\dagger}_{1}\hat{A}_{1} \hat{A}^{\dagger}_{2}\hat{A}%
_{2}\rangle+ \langle \hat{A}^{\dagger}_{1}\hat{A}_{1}
\hat{A}^{\dagger}_{3}
\hat{A}_{3}\rangle+ \langle \hat{A}^{\dagger}_{2}\hat{A}_{2} \hat{A}%
^{\dagger}_{3}\hat{A}_{3}\rangle -\langle \hat{A}^{\dagger}_{1}\hat{A}%
_{1}\rangle \langle\hat{A}^{\dagger}_{2}\hat{A}_{2}\rangle \right.\\
-\left.\langle \hat{A}^{\dagger}_{1}\hat{A}_{1} \rangle \langle\hat{A}%
^{\dagger}_{3}\hat{A}_{3}\rangle- \langle
\hat{A}^{\dagger}_{2}\hat{A}_{2}
\rangle \langle\hat{A}^{\dagger}_{3}\hat{A}_{3}\rangle\right)
=2(f^{2}_{1}g^{2}_{3}+ f^{2}_{1}h^{2}_{2}+ h^{2}_{2}g^{2}_{3}).
\label{scf52}
\end{array}
\end{eqnarray}
 It is clear that this quantity has non-zero value and this is
the signature of photon-number correlation between modes.

Now we show how quantum correlations between the systems can give rise to
squeezing if operators act in the spaces not only corresponding to three systems
(three-mode squeezing)  but also to two systems (two-mode squeezing),
rather than to the individual systems (single-mode squeezing). This will be done using
 three-mode pure squeezed vacuum states
$\hat{S}(\underline{r})\prod_{j=1}^{3}|0_{j}\rangle $ as before.
For this purpose, using results of section 4.2,
we calculate the squeezing variances
for  three-mode squeezing  quadratures given by (\ref{12}) and
(\ref{r13}) in terms of
 three-mode pure squeezed vacuum states. After minor algebra we get

\begin{eqnarray}
\begin{array}{rl}
\langle (\triangle \hat{X})^{2}\rangle
=\frac{1}{4}\Bigl[
f^{2}_{1}+f^{2}_{2}+f^{2}_{3}
+g^{2}_{1}+g^{2}_{2}+g^{2}_{3} \\
+h^{2}_{1}+h^{2}_{2}+h^{2}_{3}
+4h_{2}g_{3}+ 4f_{1}h_{2}+ 4f_{1}g_{3} \Bigr],
\label{scf53}
\end{array}
\end{eqnarray}

\begin{eqnarray}
\begin{array}{rl}
\langle (\triangle \hat{Y})^{2}\rangle
=\frac{1}{4}\Bigl[ f^{2}_{1}+f^{2}_{2}+f^{2}_{3}
+g^{2}_{1}+g^{2}_{2}+g^{2}_{3} \\
+h^{2}_{1}+h^{2}_{2}+h^{2}_{3}
+4h_{2}g_{3}- 4f_{1}h_{2}- 4f_{1}g_{3} \Bigr].
\label{scf54}
\end{array}
\end{eqnarray}

As we mentioned earlier the expressions for the single-mode and
two-mode squeezing can be
obtained easily from (\ref{scf53}) and (\ref{scf54}).
It should be taken into
account that $C=\frac{1}{2},1,\frac{3}{2}$ corresponding to the single-mode,
two-mode and three-mode squeezing, respectively.

From (\ref{scf53}) and (\ref{scf54}) one can easily
prove that the model cannot exhibit single-mode squeezing, e.g. for the
first mode we have

\begin{equation}
S_{1}=S_{2}=2(f^{2}_{2}+f^{2}_{3}), \label{scf55}
\end{equation}
where the relations between the coefficients $f_{j}$ resulting
from the commutation rules of $\hat{A}_{j}$ have been used to get such
relation.

\begin{figure}[h]%
 \centering
    \includegraphics[width=6cm]{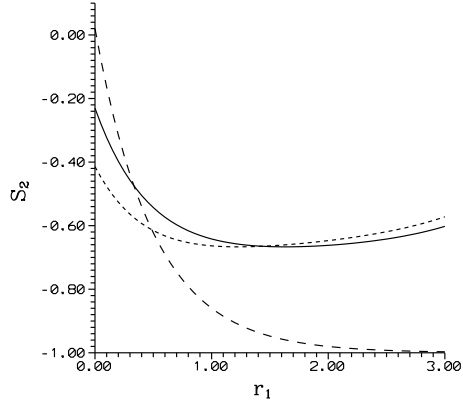}
   \caption{
$S_{2}$ against squeeze parameter $r_{1}$ for three-mode squeezing
and for $r_{3}=0.1$, $r_{2}=0.1$ (solid curve) and $0.2$
(short-dahed curves); and for two-mode squeezing (between modes 1
and 2, long-dashed curve) $(r_{2},r_{3})=(0.2,0.1)$. }
\label{fig14}
\end{figure}

For compound modes the system can provide two-mode (only between
modes (1, 2) and (1,3)) as well as
three-mode squeezing in the $Y$-quadrature  as shown in Fig. 5.14 (for
the shown values of squeeze parameters).
From this figure one can observe that the behaviour
of three-mode squeezing factors is smoothed in such a way that they
are initially squeezed and  their squeezing values
 reach their maximum, then they start again
to decrease into  unsqueezed values for large domain of
$r_{1}$ (which is not shown in the figure).
Also, the initial values of
squeezing are sensitive to the values of the squeeze parameters (compare
solid and  short-dashed curves). Concerning two-mode squeezing factor
(long-dashed curve)  one can see  it is monotonically
decreasing function with lower limit $-1$.
It is important mentioning that the  squeeze operator under discussion
cannot practically provide  maximum squeezing (for three-mode
squeezing), i.e. $S_{2}=-1$, and this, of course, is in contrast with the
single-mode and two-mode squeeze operators (\ref{2}) and (\ref{4})
where they  can display maximum squeezing for large values of squeeze
parameter, however, they cannot provide squeezing initially.
In conclusion it is quite obvious   that  squeezing can occur in the
combined systems
even if the individual systems are not  themselves squeezed. The mechanism
for this process is the correlation  between the systems.

\subsection{ Three-mode squeezed coherent and number states}

Here we  examine the Glauber second-order correlation
function
for squeezed coherent states and squeezed number states related to
the three-mode squeeze operator (\ref{scf46}).
Further we discuss the anticorrelation properties for three-mode squeezed
coherent states.

First  we  define
the three-mode squeezed coherent states as

\begin{equation}
|\underline{\alpha},\underline{r}\rangle \equiv
\hat{S}(\underline{r})\hat{D}
(\underline{\alpha}) |0 \rangle_{1}|0\rangle_{2}|0\rangle_{3},\label{scf56}
\end{equation}
 where $\hat{S}(\underline{r})$ is the three-mode squeeze
operator
(\ref{scf46}) and $\hat{D}(\underline{\alpha})$ is the three-mode Glauber
displacement operator given by

\begin{equation}
\hat{D}(\underline{\alpha}) = \exp \sum_{j=1}^{3} ( \alpha_{j}
\hat{A}_{j}^{\dagger} - \alpha_{j}^{*} \hat{A}_{j} ),
\label{scf57}
\end{equation}
and $\underline{\alpha}=(\alpha_{1},\alpha_{2},\alpha_{3})$.

The unitarity of the operator (\ref{scf46}) provides
that three-mode squeezed coherent states are not orthonormal
but they are complete in the following sense

\begin{equation}
\hat{1}=\frac{1}{\pi^{3}}\int \int \int |\underline{\alpha}, \underline{r}%
\rangle\langle \underline{r},\underline{\alpha}| d^{2}\alpha_{1}
d^{2}\alpha_{2}d^{2}\alpha_{3}.
\label{scf58}
\end{equation}

Using (\ref{scf49}) the mean photon numbers for  various modes in three-mode
squeezed coherent states are given by

\begin{eqnarray}
\begin{array}{rl}
\langle \hat{n}_{1} \rangle_{\rm coh} = |f_{1}\alpha_{1}
+f_{2}\alpha^{*}_{2}+f_{3}\alpha^{*}_{3}|^{2}+f^{2}_{2}+f^{2}_{3}, \\
 \langle\hat{n}_{2}\rangle_{\rm coh} = |g_{1}\alpha_{2}
+g_{2}\alpha_{3}+g_{3}\alpha^{*}_{1}|^{2}+ g^{2}_{3} ,
\label{scf59}
\end{array}
\end{eqnarray}
 and the photon-number variances are

\begin{eqnarray}
\begin{array}{rl}
\langle (\Delta \hat{n}_{1})^{2} \rangle_{\rm coh} =
(2f_{1}^{2}-1)\langle \hat{n}_{1}\rangle_{\rm coh} -(f_{1}^{2}-1)^{2},\\
\langle (\Delta \hat{n}_{2})^{2} \rangle_{\rm coh} =
(2g_{3}^{2}+1)\langle \hat{n}_{2}\rangle_{\rm coh} -g_{3}^{4},
\label{scf60}
\end{array}
\end{eqnarray}
where ${\rm coh}$ stands for squeezed coherent states. Expressions
related with the 3rd mode can be obtained from those of the 2nd mode by
using the following transformation

\begin{equation}
(g_{1},g_{2},g_{3})\rightarrow (h_{3},h_{1},h_{2}). \label{scf61}
\end{equation}

Having
obtained equations (\ref{scf59}) and (\ref{scf60}), we are  ready to
examine
the second-order correlation function given by
(\ref{5}). In the following
we  restrict our discussions to the first mode $1$, because the other
modes would have similar behaviour.

In phase space, squeezed coherent states $|\alpha ,r\rangle $ are
represented by a noise ellipse with the origin at $\alpha $, and
they do not exhibit sub-Poisson distribution, i.e. they exhibit Poisson
distribution  at $r=0$ which
is growing rapidly to superthermal distribution, i.e. $g^{(2)}(0)>2$, and
it persists for a large domain of $r$ \cite{ther9,{ther10}}. In our model of
three-mode squeezed coherent states we  show that they exhibit only
partial coherence behaviour, i.e. $1<g^{(2)}(0)<2$.

 The condition for sub-Poissonian statistics is that the variance
$\langle (\triangle \hat{n}_{j})^{2}\rangle $ must be less than the
mean photon number $\langle \hat{n}_{j}\rangle $. We show that this
condition
will not be fulfilled for all modes, i.e. sub-Poissonian light cannot be
obtained, because  we have, e.g. for the first mode, the inequality

\begin{equation}
2 |f_{1}^{2}\alpha_{1}
+f_{2}\alpha^{*}_{2}+f_{3}\alpha^{*}_{3}|^{2}+f^{2}_{2}+f^{2}_{3}<0,
\label{scf62}
\end{equation}
 which will not be satisfied for any values of the coherent
amplitudes $\alpha_{j}$.

For the first mode $\hat{A}_{1}$ we have

\begin{equation}
g_{1}^{(2)}(0)= 1+ \left[
\frac{2(f_{1}^{2}-1)\langle\hat{n}
_{1}\rangle_{coh}-(f_{1}^{2}-1)^{2}}
{\langle\hat{n}_{1}\rangle^{2}_{\rm coh} }
\right].
\label{scf63}
\end{equation}
 It is clear that when $r_{j}=0$, we recover the normalized second-order
correlation function for coherent states.
According to (\ref{scf63}), to reach Poissonian statistics
or thermal statistics, the expectation value $\langle\hat{n}_{1}\rangle$
of the photon number should have the following values:

\begin{equation}
\langle \hat{n}_{1}\rangle _{\rm p}=\frac{1}{2}(f_{1}^{2}-1),\qquad
 \langle \hat{n}_{1}\rangle_{\rm th}=f^{2}_{1}-1,
\label{scf64}
\end{equation}
 where subscripts ${\rm p}$ and ${\rm th}$ denote  the
corresponding quantities for Poisson and thermal distributions,
respectively. For instance, to obtain thermal field, from (\ref{scf59}) and
 (\ref{scf64}), i.e. $\langle \hat{n}_{1}\rangle_{\rm th}$, we have

\begin{equation}
|f_{1}\alpha_{1}+f_{2}\alpha^{*}_{2}+f_{3}\alpha^{*}_{3}|^{2}=0.
\label{scf65}
\end{equation}
 It is clear that (\ref{scf65}) is satisfied only when $\alpha
_{j}=0$,
i.e. for three-mode squeezed vacuum states, and hence super-thermal
statistics
cannot be obtained. Similar procedures show that Poissonian statistics
cannot be obtained. This is a consequence of intermodal correlations of
three-mode squeezed coherent states, for $\alpha _{j}\neq 0$, light
exhibits
only partially coherence. This can be seen in Fig. 5.15, where we
display the normalized normal
second-order correlation function for the first mode against
$
r_{1}$ for $\alpha _{j}=2\exp (i\frac{\pi }{4}),j=1,2,3$,
$r_{3}=0.2$ with $
r_{2}=0.4,0.8,$ and $1.5$ corresponding to solid, short-dashed,
and long-dashed curves, respectively. Also we have displayed the
corresponding normalized second-order correlation
function for two-mode squeezed coherent state (centered curve)
for the amplitudes $\alpha _{j}=2\exp (i\frac{\pi }{4}),j=1,2$, for sake of
comparison. In this figure it is clear that partial coherence is
dominant, and it  persists for large values of $r_{1}$ and the initial values
of $g_{j}^{(2)}(0)$ are
 sensitive to $r_{2}$. However, the normalized second-order
correlation function for two-mode squeezed coherent state (centered curve) is a
growing function, starting from 1 (Poisson distribution)
 when there is no squeezing $r_{1}=0$, and  becomes stable for large values of
 $r$ displaying partial coherence behaviours. These values exceed
those for three-mode squeezed coherent state for large domain of $r$.
\begin{figure}[h]%
 \centering
    \includegraphics[width=6cm]{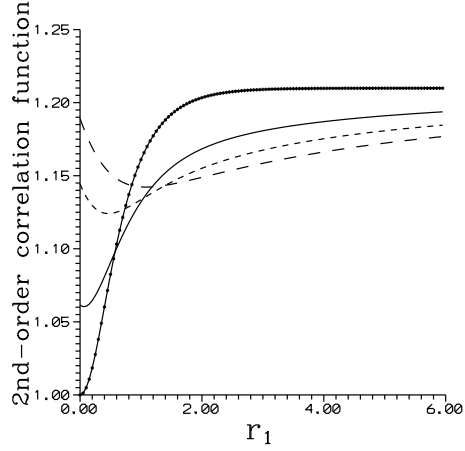}
   \caption{
Normalized normal second-order correlation function $%
g_{1}^{(2)}(0)$ (first mode) for three-mode squeezed coherent
state against $r_{1}$, when $\alpha _{j}=2\exp (i\frac{\pi
}{4}),j=1,2,3$, $r_{3}=0.2$, taking $r_{2}=0.4$ (solid curve),
$0.8$ (short-dashed curve) and $1.5$ (long-dashed curve); centered
curve is corresponding to the second-order correlation function
for two-mode squeezed coherent state and $\alpha _{j}=2\exp
(i\frac{\pi }{4}),j=1,2$. } \label{fig15}
\end{figure}


Second, we  discuss the
sub-Poisson properties of three-mode squeezed number states. This state
can
be written with the aid of three-mode squeeze operator as

\begin{equation}
|\underline{n},\underline{r}\rangle \equiv \hat{S}(\underline{r}) |n_{1}
\rangle|n_{2}\rangle|n_{3}\rangle,
\label{scf66}
\end{equation}
 for simplicity we  set
$\underline{n}=(n_{1},n_{2},n_{3})$.

The mean photon numbers in  three-mode squeezed number state are

\begin{eqnarray}
\begin{array}{lr}
\langle \hat{n}_{1}\rangle_{\rm n}=\bar{n}_{1}f^{2}_{1}+ (\bar{n}
_{2}+1)f^{2}_{2} +(\bar{n}_{3}+1)f^{2}_{3}, \\
\langle \hat{n}_{2}\rangle_{\rm n}=\bar{n}_{2}g^{2}_{1}+
\bar{n}_{3}g^{2}_{2} +(
\bar{n}_{1}+1)g^{2}_{3},
\label{scf67}
\end{array}
\end{eqnarray}

\noindent and the variances of the photon number are

\begin{eqnarray}
\begin{array}{lr}
 \langle (\Delta \hat{n}_{1})^{2} \rangle_{\rm n} =
f_{1}^{2}f_{2}^{2}[\bar{n}_{1}+\bar{n}_{2} +2\bar{n}_{1}\bar{n}_{2}+1] +
f_{1}^{2}f_{3}^{2}[\bar{n}_{1}+\bar{n}_{3} +2\bar{n}_{1}\bar{n}_{3}+1] \\
+f_{2}^{2}f_{3}^{2}[\bar{n}_{2}+\bar{n}_{3} +2\bar{n}_{2}\bar{n}_{3}] ,
\\
\\
 \langle (\Delta \hat{n}_{2})^{2} \rangle_{\rm n} =
g_{1}^{2}g_{2}^{2}[\bar{n}_{3}+\bar{n}_{2} +2\bar{n}_{2}\bar{n}_{3}] +
g_{1}^{2}g_{3}^{2}[\bar{n}_{1}+\bar{n}_{2} +2\bar{n}_{1}\bar{n}_{2}]
\\
+g_{2}^{2}g_{3}^{2}[\bar{n}_{1}+\bar{n}_{3} +2\bar{n}_{1}\bar{n}_{3} +1],
\label{scf68}
\end{array}
\end{eqnarray}
 where $\bar{n}_{j}$ is the mean photon number for the $j$th
mode. The expression for the third mode can be obtained using (\ref{scf61}).
It is worthwhile to refer to \cite{ther9, {ther10},{pau1},{ther12}, {pau3}},
where further discussions related to  squeezed number states are
given. For instance, when the squeezing is  not significant, i.e. $r$
is small,
the normalized second-order correlation function can be less than unity,
which indicates that the light field has sub-Poissonian statistics
\cite{ther10}.
Furthermore, squeezed vacuum state exhibits super-Poisson statistics  (precisely
superthermal statistics) for $r\neq 0$ \cite{[24]}. Here, in contrast to the
latter results, we prove that three-mode squeezed number state can
exhibit thermal statistics when $\bar{n}_{j}=0,j=1,2,3$, i.e. for
three-mode squeezed vacuum states, otherwise sub-Poissonian statistics
or
partially coherence behaviour is dominant. We focus our attention to the
first mode. According to the definition of normalized
second-order correlation function to obtain thermal statistics, we need to
have

\begin{equation}
\langle (\Delta \hat{n}_{1})^{2} \rangle_{\rm n} - \langle
\hat{n}_{1}\rangle_{\rm n} - \langle \hat{n}_{1}\rangle^{2}_{\rm n}=0;
\label{scf69}
\end{equation}
from (\ref{scf67}) and (\ref{scf68}) and
after minor algebra one finds

\begin{equation}
\bar{n}_{1}(\bar{n}_{1}+1)f^{4}_{1} +
\bar{n}_{2}(\bar{n}_{2}+1)f^{4}_{2}+
\bar{n}_{3}(\bar{n}_{3}+1)f^{4}_{3}=0.
\label{scf70}
\end{equation}
 It is clear that the equality sign is only satisfied for
$\bar{n}_{j}=0,j=1,2,3$. In other words, superthermal statistics will never
occur. This is in contrast with the  squeezed number state.
\begin{figure}[h]%
  \centering
  \subfigure[]{\includegraphics[width=5cm]{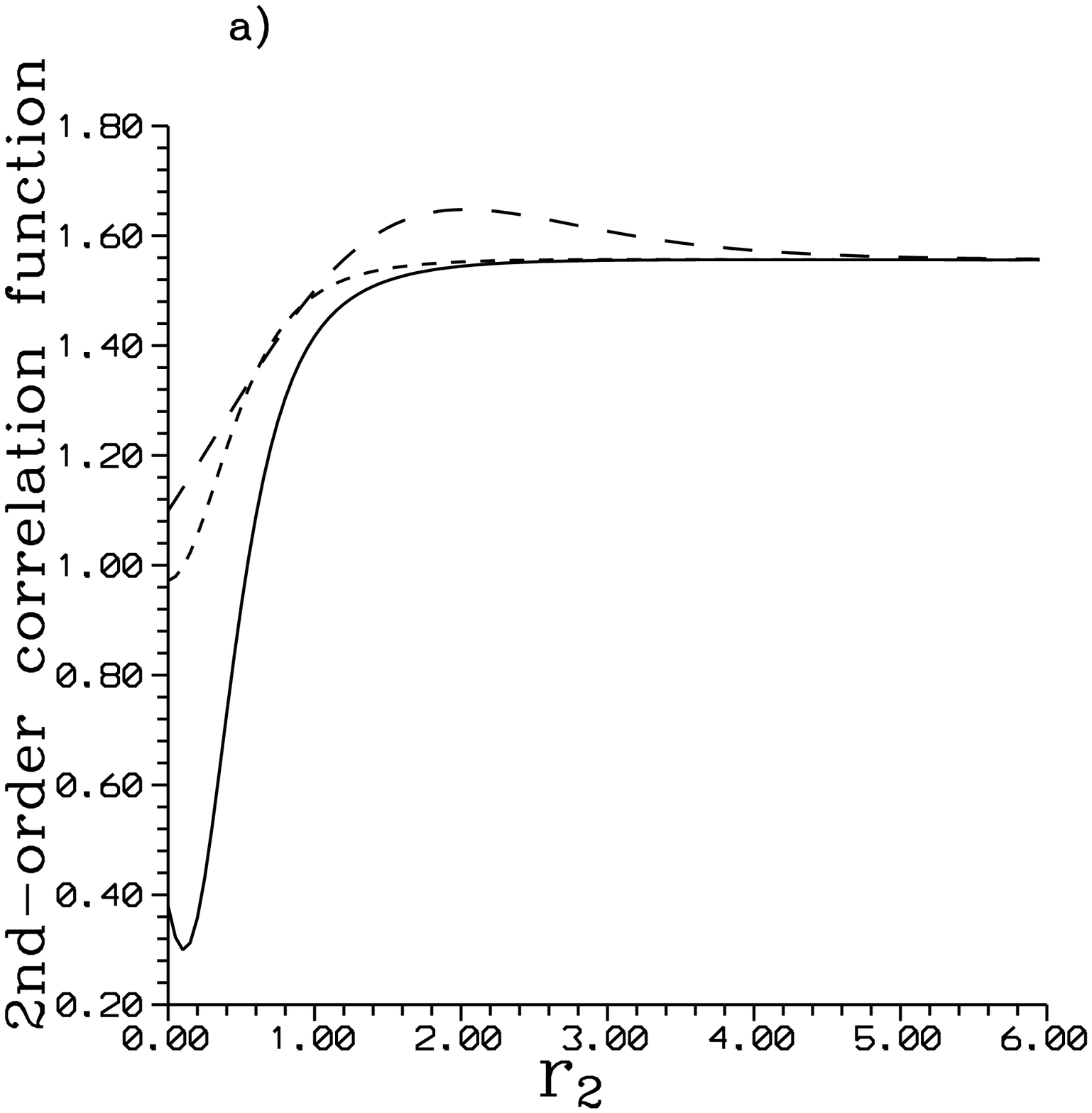}}
 \subfigure[]{\includegraphics[width=5cm]{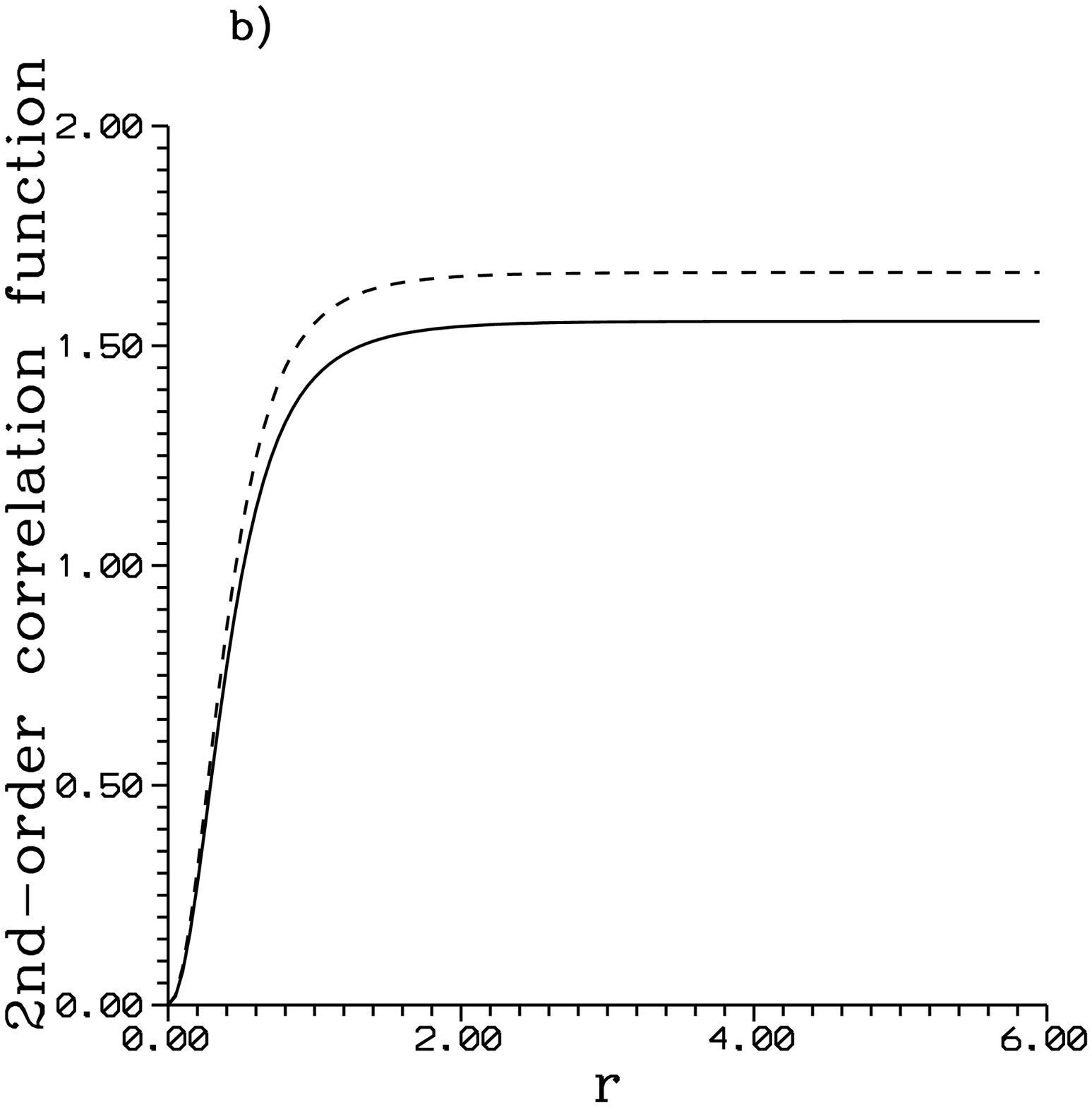}}
    \caption{
  Normalized normal second-order correlation function for:
a) three-mode squeezed number states against $r_{2}$ for different
modes,
when $%
\bar{n}_{j}=1,j=1,2,3$, $(r_{1},r_{3})=(0.5,0.3)$ for mode 1
(solid curve), mode 2 (short-dashed curve) and mode 3 (long-dashed
curve); b) for single mode squeezed number states (solid curve,
$\bar{n}_{1}=1$) and two-mode squeezed number state (dashed curve,
$\bar{n}_{j}=1,j=1,2$) against squeezing parameter $r$. }
\label{fig16}
 \end{figure}

In
Fig. 5.16a we depict normalized second-order correlation function for single
mode case
for three-mode squeezed number state against $r_{2}$ for $\bar{n}
_{j}=1,j=1,2,3$ and $(r_{1},r_{3})=(0.5,0.3)$. We notice in
general that
partially coherence behaviour is dominant again. Further we can see
sub-Poissonian behaviour for small values of $\mu $ for modes 1 and 3
with
maximum value at the third one. For sake of comparison, we displayed
the normalized second-order correlation functions for single mode
(dashed curve) and
two-mode (solid curve) squeezed number states in Fig. 5.16b for $\bar{n}
_{j}=1,j=1,2$, where we can see that both of them have sub-Poissonian
statistics for lower values of squeeze parameter,
otherwise they characterize partially coherence behaviour of light
beams.

Finally we turn our attention to discuss the effect of intermodal
correlations in terms of anticorrelations  in
three-mode squeezed coherent states. We do this by investigating the
behaviour of the Cauchy-Schwarz inequality giving by (\ref{10}).
In doing so
we have to derive  the expectation values
of cross photon-number operators between various modes,
which are given in appendix C.
In Fig. 5.17 we have plotted the quantity $I_{j,k}$ indicating the
violation of Cauchy-Schwarz inequality between the $j$th mode and the
$k$th mode against squeeze parameter
$r_{1}$, where $(r_{2},r_{3})=(0.4,0.2)$ and $\alpha_{j}=1,j=1,2,3$
(real). Further, solid, short-dashed, and long-dashed curves correspond
to the $I_{j,k}$ quantity obtained between $(1,2), (2,3),$ and $(1,3)$ modes,
respectively.
In general, photon antibunching can occur
in dependence on the values of the parameters $r_{j}$ and $\alpha_{j}$.
More precisely, in this figure we can
observe that all curves can take on negative values
reflecting the violation of the
inequality, i.e. the photons are more strongly correlated than
it is possible classically. The strongest violation of this inequality
occurs in the
$(1,3)$ mode for lower $r_{1}$ and then the curve monotonically increases
to positive values for larger $r_{1}$. The weakest
violation is in the $(2,3)$ mode for which the curve decreases
 from positive values to the negative values as $r_{1}$ increases.
 Clearly, the violation of this inequality is sensitive to the values of
 squeeze parameters and coherent amplitudes.
 As is known, the correspondence
 between quantum and classical theories can be established via
 Glauber-Sudarshan $P$-representation.
But $P$-representation does not have all the
 properties of a classical distribution function, especially for quantum
 fields. So the violation of
the  Cauchy-Schwarz inequality provides explicit evidence of the quantum
nature of intermodal correlation between modes which imply
 that the
 $P$-distribution function possesses strong quantum properties \cite{chs13}.

\begin{figure}[h]%
 \centering
    \includegraphics[width=6cm]{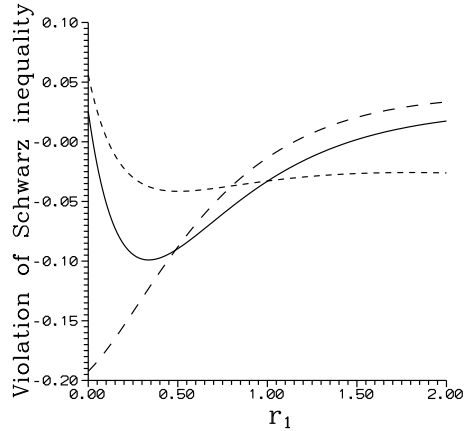}
   \caption{
The quantity $I_{j,k}$, indicating the violation of the
Cauchy-Schwarz inequality betwen the $j$th mode and $k$th mode,
against
squeeze parameter $r_{1}$, where $(r_{2},r_{3})=(0.4,0.2)$ and $%
\alpha_{j}=1,j=1,2,3$ (real). Solid, short-dashed, and long-dashed
curve correspond to the above quantity considered for modes
$(1,2), (2,3)$, and $(1,3)$, respectively. } \label{fig17}
\end{figure}

~
\subsection{ Quasiprobability distribution functions}

Here  we  extend our work to include the quasiprobability
distribution functions.  We  consider
such functions for  three-mode squeezed
coherent and  number states.  We give only one example in this thesis for
the joint
 quasiprobability functions for three-mode squeezed coherent states
 since this type of functions carries information on the intermodal
correlations  between different modes in the quantum mechanical system.
Results of section 4.3 should be used.\\
 \\
\noindent {\bf (a) Three-mode squeezed coherent states}\\
\\
Here we study quasiprobability distribution functions for three-mode
squeezed coherent state, specified by the density matrix
$\hat{\rho}_{\rm coh}= |\underline{\alpha}, \underline{r}
\rangle\langle \underline{r},\underline{\alpha}|$ and
$|\underline{\alpha}, \underline{r}
\rangle $ is given by (\ref{scf56}).
 Thus substituting $\hat{\rho}_{\rm coh}$ into (\ref{13})
and using transformations (\ref{scf49}) together with the Baker-Hausdorff
identity for normal ordering of operators and after some minor
calculations we have the $s$-parametrized joint characteristic function

\begin{eqnarray}
\begin{array}{lr}
C^{(3)}_{\rm coh}(\zeta_{1},\zeta_{2},\zeta_{3},s)=\exp[
\frac{1}{2}\sum_{j=1}^{3}(s|\zeta_{j}|^{2}-|\eta_{j}|^{2})] \\
 \times \exp[(\alpha_{1}^{*}\eta_{1}-
\alpha_{1}\eta^{*}_{1})+(\alpha_{2}^{*}\eta_{2}-
\alpha_{2}\eta_{2}^{*})+(\alpha_{3}^{*}\eta_{3}-
\alpha_{3}\eta_{3}^{*})],
\label{scf71}
\end{array}
\end{eqnarray}
 where we have used the following abbreviations
  \begin{equation}
 \eta_{1}= \zeta_{1}f_{1}-\zeta_{2}^{*}g_{3}-
\zeta_{3}^{*}h_{2},\quad
 \eta_{2}= \zeta_{2}g_{1}-\zeta_{1}^{*}f_{2}+ \zeta_{3}h_{3},
\quad
 \eta_{3}= \zeta_{3}h_{1}+\zeta_{2}g_{2}- \zeta_{1}^{*}f_{3}.
\label{scf72}
\end{equation}
Having obtained the characteristic function, we are  ready
to find the $s$-parametrized quasiprobability functions for three-mode
squeezed coherent state by inserting  (\ref{scf71}) into (\ref{14})
and evaluating the integral by using the identity
(A.4) in appendix A three times; the calculations are straightforward but
rather lengthy, thus we get

\begin{eqnarray}
\begin{array}{lr}
W^{(3)}_{\rm coh}(\beta_{1},\beta_{2},\beta_{3},s)
=\frac{2}{\pi^{3}} \frac{1}{\nu_{1}\nu_{2}(2f_{2}^{2}+2f^{2}_{3}+1-s)}
\exp\left[- \frac{
2|\beta_{1}|^{2}}{2f_{2}^{2}+2f^{2}_{3}+1-s}%
\right]\\
\\
\qquad \times \exp\left\{\frac{1}{4\nu_{1}}
\left[(A^{2}-C^{2})\cos^{2}\phi
+(B^{2}-D^{2})\sin^{2}\phi+(CD-AB)\sin (2\phi)\right]\right\} \\
 \\
 \qquad \times \exp\left\{\frac{1}{4\nu_{2}}%
\left[(B^{2}-D^{2})\cos^{2}\phi
+(A^{2}-C^{2})\sin^{2}\phi-(CD-AB)\sin (2\phi)\right]\right\},
\label{scf73}
\end{array}
\end{eqnarray}
 where

\begin{equation}
 \nu_{1}= \mu_{1}\cos^{2}\phi+ \mu_{2}\sin^{2}\phi+
\mu_{3}\sin (2\phi),\quad
 \nu_{2}= \mu_{2}\cos^{2}\phi+
\mu_{1}\sin^{2}\phi-\mu_{3}\sin (2\phi);\quad
\label{scf74}
\end{equation}
\noindent here
\begin{eqnarray}
\begin{array}{lr}
 \mu_{1}=\frac{1}{2}(2f_{2}^{2}+2f^{2}_{3}+1-s)^{-1}
\left[2f_{1}^{2}(1-s)-2g_{3}^{2}(1+s)-(1-s^{2})\right], \\
 \mu_{2}=\frac{1}{2}(2f_{2}^{2}+2f_{3}^{2}+1-s)^{-1}
\left[2f_{1}^{2}(1-s)-2h_{2}^{2}(1+s)-(1-s^{2})\right], \\
 \mu_{3}=(2f_{2}^{2}+2f_{3}^{2}+1-s)^{-1}
\left[2f_{1}^{2}g_{2}h_{1}+g_{1}h_{3}(1+s)\right].
\label{scf75}
\end{array}
\end{eqnarray}
 In equation (\ref{scf73}) we have also used the following definitions

\begin{eqnarray}
\begin{array}{lr}
A=(\tilde{\alpha_{2}^{*}}-
\tilde{\alpha_{2}})+2f_{1}g_{3}
(2f_{2}^{2}+2f_{3}^{2}+1-s)^{-1}
(\tilde{\alpha_{1}^{*}}-\tilde{\alpha_{1}}%
),\\
 B=(\tilde{\alpha_{3}^{*}}- \tilde{\alpha_{3}}%
)+2f_{1}h_{2}(2f_{2}^{2}+2f_{3}^{2}+1-s)^{-1}
(\tilde{\alpha_{1}^{*}}-\tilde{%
\alpha_{1}}), \\
 C=(\tilde{\alpha_{2}^{*}}+ \tilde{\alpha_{2}}%
)-2f_{1}g_{3}(2f_{2}^{2}+2f_{3}^{2}+1-s)^{-1}
(\tilde{\alpha_{1}^{*}}+\tilde{%
\alpha_{1}}), \\
 D=(\tilde{\alpha_{3}^{*}}+ \tilde{\alpha_{3}}%
)-2f_{1}h_{2}(2f_{2}^{2}+2f_{3}^{2}+1-s)^{-1}
(\tilde{\alpha_{1}^{*}}+\tilde{%
\alpha_{1}}),
\label{scf76}
\end{array}
\end{eqnarray}
 where
 $\phi=\frac{1}{2}\tan^{-1} \left(\frac{2\mu_{3}}{%
\mu_{1}-\mu_{2}}\right)$,
 $\tilde{\alpha_{j}}=(\bar{\alpha}_{j}- \beta_{j})$,
$j=1,2,3 $ and $\bar{\alpha_{j}}$ represents the expectation value of
the operators $\bar{A_{j}}$ given by equation (\ref{scf49}) in terms of
the coherent state. The
three-mode functions $W^{(3)}_{\rm coh}(\beta_{1},\beta_{2},\beta_{3},s)$ given
in
(\ref{scf73}) are 6-dimensional Gaussian functions and display the nonclassical
correlation nature by involving the terms $A^{2}, B^{2},$ etc. when the
cross-terms $\beta_{1}\beta_{2}, \beta_{1}\beta_{3}$, etc. are not
zero.
Furthermore, we can see that the three-mode $P$ representation does not
exist at least for some values of squeezing parameters $r_{j}$, i.e. if
$\nu_{1}=0$ or $\nu_{2}=0$ or both (see equation (\ref{scf73})), the physical
reason for non-conditional breaking down of the $P$-function lies in the
extremely strong correlation between the amplitudes of the three modes of the
system  during the evolution determined by squeeze operator \cite{[3]}.
Such tightly
correlation causes that the modes may no longer fluctuate independently
 even in small amount such as allowed in a pure coherent states \cite{entan1}.

Now we turn our attention to single mode case, using similar procedure
as
before. The $s$-parametrized quasiprobability function takes the form

\begin{equation}
W^{(1)}_{\rm coh}(\beta_{j},s)=\frac{2}{\pi(\tau_{j}-s)} \exp
\left(-2%
\frac{|\beta_{j}-\bar{\alpha}_{j}|^{2}}{\tau_{j}-s}\right),
\label{scf77}
\end{equation}
 where $j=1,2,3$ and

\begin{equation}
\tau_{1}=2f^{2}_{2}+2f^{2}_{3}+1, \quad \tau_{2}=2g^{2}_{3}+1,
\quad \tau_{3}=2h^{2}_{2}+1.
\label{scf78}
\end{equation}
 Light fields for which the $P$-representation is not a
well-behaved distribution will exhibit nonclassical features. Clearly
single mode $P$-function is well defined, because the parametric systems
evolution
broadens $P$-distribution (increases the radius of the Wigner contour
compared to the initial one) \cite{entan1} and reflects that there is no
(single mode) nonclassical behaviour, e.g. sub-Poissonian statistics and
squeezing. Furthermore, it has been shown that $P$-function for the
superposition of two fields is the convolution of the $P$-function for
each  field considered individually
\cite{chs2}, so that (\ref{scf77}) (with $s=1$)
describes the superposition of $P$-function of a coherent state with
complex
amplitude $\bar{\alpha}_{j}$ and $P$-function of a chaotic mixture with
variance $\frac{1}{2}(\tau _{j}-1)$ \cite{[3]}, i.e. displaced thermal
light.
Further, equation (\ref{scf77}) has a Gaussian form in phase space with width
and center  dependent on $r_{j},\alpha _{j}$ and with a circular symmetric
contour as a result of the fact that two quadrature variances are equal. Consequently
the single mode photon-number distribution does not exhibit oscillations, in
contrast with this for squeezed coherent state \cite{scfr6}, owing to the
noise ellipse of $W$-function which is isotropic.

Now we investigate single mode phase distribution in terms of $Q$-function,
i.e. $s=-1$. Using the same procedures as in section 5.2.5 we arrive at

\begin{eqnarray}
\begin{array}{lr}
P(\theta)=\frac{1}{2\pi\sqrt{(\tau_{j}+1)}}\exp \left[
\frac{b_{j}^{2}-4|%
\bar{\alpha}_{j}|^{2}}{2(\tau_{j}+1)}\right] \Bigl\{
\sqrt{(\tau_{j}+1)}%
\exp\left[-\frac{b_{j}^{2}}{2(\tau_{j}+1)}\right]\\
+ \frac{b_{j}\sqrt{\pi}}{2} \left[1+{\rm
erf}(\frac{\sqrt{2}b_{j}}{\sqrt{%
\tau_{j}+1}})\right]\Bigr\} ,
\label{scf79}
\end{array}
\end{eqnarray}
where

\begin{equation}
b_{j}=\bar{\alpha}_{j}\exp(-i\theta) +\bar{\alpha}^{*}_{j}\exp(i
\theta) ,
\label{scf80}
\end{equation}
 with $j=1,2,3$, $\bar{\alpha}_{j}$   have the same meaning as
before, and ${\rm erf(.)}$ is  the Gauss error function
(\ref{scf40}).

\begin{figure}[h]%
 \centering
    \includegraphics[width=6cm]{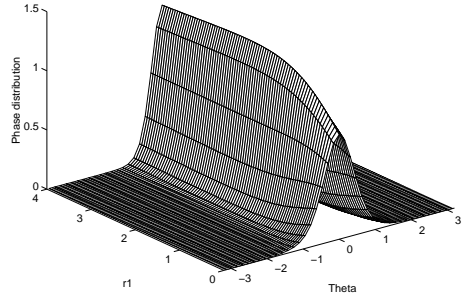}
   \caption{
Phase distribution $P(\theta ,r_{1})$ against $\theta $ and
$r_{1}$ for the first mode as output from three-mode squeezed
coherent state for $\alpha _{j}=1,j=1,2,3,r_{2}=1$ and
$r_{3}=0.5$. } \label{fig18}
\end{figure}

Here we restrict our investigation  to real $\alpha_{j}$, i.e.
$b_{j}=2\bar{\alpha}_{j}\cos\theta$, detailed examination to the formula
(\ref{scf79}) shows that it is a $2\pi$-periodic function, a symmetric
function ($P(-\theta)=P(\theta)$)
around $\theta=0$ and also it has its maximum height at $\theta=0$.
Moreover, this formula has a similar structure as that for coherent
states (which can be recovered from (\ref{scf79}) by setting $r_{j}=0$).
This means that the phase distribution of the single mode exhibits
a one-peak structure  for all values of $\alpha_{j}$
and $r_{j}$ (see Fig. 5.18 for shown values of parameters).
That is  the phase distribution of the single-mode
as output from three-mode squeezed coherent state is
insensitive to the quantum correlations between the systems.
This  situation is the same as that of
two-mode squeezed coherent state, i.e. when $r_{2}=r_{3}=0$ \cite{phas12}.
However, it has been shown that the joint phase distribution for the
two-mode squeezed vacuum depends only on the sum of the phases of the
two modes, and that the sum of the two phases is locked to a certain
value as the squeezing parameter increases \cite{phas12,{phas14}}.\\
\\
\noindent{\bf (b) Three-mode squeezed number state}\\
\\
Here we calculate single mode $s$-parametrized characteristic and
quasiprobability functions for mode 1, for three-mode squeezed number
state (\ref{scf66}). Therefore $s$-parametrized single mode
characteristic function  can be written as follows

\begin{equation}
 C_{\rm sn}^{(1)}(\zeta _{1},
,s)=\exp [\frac{1}{2}|\zeta _{1}|^{2}(s-2f^{2}_{2}-2f^{2}_{3})]
L_{n_{1}}(|\zeta _{1}|^{2}f^{2}_{1})
L_{n_{2}}(|\zeta _{1}|^{2}f^{2}_{2})L_{n_{3}}(|\zeta _{1}|^{2}f^{2}_{3}),
\label{scf81}
\end{equation}
where
(\ref{scf4}) has been used,
$L_{k}(.)$ are the Laguerre polynomials of order $k$, and subscript ${\rm
sn}$ stands for three-mode squeezed number states.

The single mode $s$-parametrized quasiprobability function for
mode 1 is given by Fourier transformation of (\ref{scf81}) as

\begin{equation}
 W_{\rm sn}^{(1)}(\beta _{1},s)=\frac{2}{\pi (\tau
_{1}-s)}%
\prod_{j=1}^{3}L_{n_{j}}[f_{j}^{2}\frac{\partial ^{2}}{\partial \beta
_{1}\partial (-\beta _{1}^{*})}]\exp \left[-\frac{2|\beta _{1}|^{2}}{\tau
_{1}-s}\right],
\label{scf82}
\end{equation}
 where $\tau _{1}$ is given in (\ref{scf78}).
 In derivation of (\ref{scf82}) we used differentiation under the sign of
 integral to perform the integration and the identity (A.4) in appendix A.
  Expressions for the second
mode and third mode can be obtained from (\ref{scf82}) by replacing
functions $f_{j}$  by  functions $g_{j}$ and $h_{j}$.
\begin{figure}[h]%
 \centering
    \includegraphics[width=6cm]{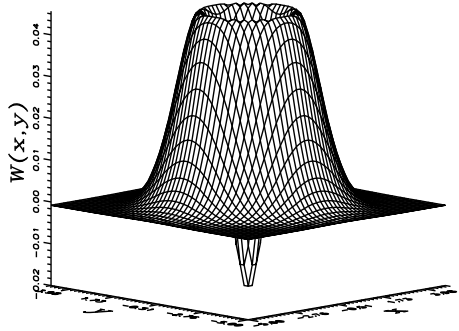}
   \caption{
$W$-function for the single mode (mode 1) as output from
three-mode squeezed number state, assuming the first mode is
$|1\rangle$ and the other modes are in vacuum for
$(r_{1},r_{2},r_{3})=(0.8,0.9,0.6)$. } \label{fig19}
\end{figure}

Here we discuss the behaviour of $W$-function for the first mode when it is
in the Fock state $|1\rangle $ while the other modes are in vacuum. In fact,
$W$-function of squeezed Fock state $|1\rangle $ is well characterized
 by inverted
hole
which is stretched as a consequence of squeezing parameter \cite{ther10}.
In the following we can use $L_{1}(x)=1-x$ to analyze the behaviour of
the model under discussion. For this case it is clear that, from equation
(\ref{scf82}), $W$-function can exhibit negative values only inside the
circle
$|\beta _{1}|^{2}<\frac{2f_{1}^{2}-1}{4f_{1}^{2}}$ with center at the
origin. However, the maximum value will be established at the circle $
|\beta _{1}|^{2}=\frac{4f_{1}^{4}-1}{4f_{1}^{2}}$. Hence, $W$-function
will not exhibit stretching since the variables ${\rm Re}\beta _{1}$ and
${\rm Im}\beta _{1}$ are not involving squeezing factor which plays an
essential role for stretched quadratures of squeezed number state. As a
result the radii of these circles are dependent on squeeze parameters
$r_{j}$, so  the negative values of $W$-function will be sensitive to
the values of squeeze parameters, i.e. the function has negative values
for a range of $|\beta_{1}|$ shorter than for individual Fock state
$|1\rangle $, where it exhibits negative values inside the circle $|\beta
_{1}|^{2}<\frac{1}{2}$. This can be recognized if one compares the well known
shape of $W$-function of Fock state $|1\rangle$ with that of the single-mode
case as output from three-mode squeezed number states. For this purpose
Fig. 5.19 is displayed for $W$-function of $|1,0,0\rangle$ against $x={\rm
Re}\beta_{1}$ and $y={\rm Im}\beta_{1}$  for $(r_{1},r_{2},r_{3})
=(0.8,0.9,0.6)$. We can see that the well known negative values of $W$-function
of Fock state $|1\rangle$ are smoothed out, in turn, the circle of negative
values is enlarged, as we have shown earlier.

The phase distribution for the $j$th mode, which is in Fock state $%
|n_{j}\rangle$ while the other modes are in vacuum states, evolves under
the action of three-mode squeezed operator (\ref{scf46}), with the aid of
$Q$-function,
to

\begin{equation}
P(\theta _{j})=\frac{1}{2\pi} ,\label{scf83}
\end{equation}
 which is uniform distribution. In other words, in spite of the
system is highly correlated the phase distribution in the single mode of
purely nonclassical state is phase insensitive. This agrees with the fact
that the  states  are represented by a density matrix which is diagonal
in the number state basis, having random phase distribution  \cite{phas9}.

\subsection{Conclusions}

In this section we have introduced new type of multidimensional squeeze
operator which is more general than usually used and which includes two
different squeezing mechanisms. This operator arises from the
time-dependent  evolution operator for the Hamiltonian representing
mutual interaction between three different modes of the field.
The origin of the nonclassical effects of
 this operator model  is the correlation  between the systems where
we have shown that the quadratures squeezing  can occur in the
combination systems rather than in the  individual systems.

The quantum statistical properties corresponding to this operator have
 been traced by means of the variances of the photon-number sum and
 difference, squeezing phenomenon, Glauber second-order correlation
  function, violation of
Cauchy-Schwarz inequality, quasiprobability distribution functions and
phase probability distribution, considered for three-mode coherent and
 number states.

For three-mode squeezed coherent state, we found that the second-order
correlation function describes partially coherent field provided that one
mechanism of squeezing is always surviving, which can be demonstrated
also by means of quasiprobability distribution function in single modes.
Nevertheless, this behaviour is in contrast with behaviour of squeezed
coherent states, where second-order correlation function can display
superthermal statistics, i.e. $g^{(2)}(0)>2$. We have found strong
violation for Cauchy-Schwarz inequality in some modes, i.e. the photons
are more strongly correlated than it is allowed classically. Concerning
the single-mode phase distribution, single peak structure is dominant
for all values of parameters provided that coherent amplitudes are real.
The signature of the correlations between the three modes appears
straightforwardly in the form of quasiprobability functions.

For three-mode squeezed number states, the second-order correlation function
is in agreement with that for the squeezed number state and in general it
exhibits partial coherence; however, sub-Poissonian behaviour is attained
for small values of $\mu$ and the maximum value is obtained only for $
r_{j}=0,j=1,2,3$. Nevertheless, it cannot exhibits superthermal statistics.
Also the range
of negative values of the $W$-function, for a single mode, are highly
sensitive to squeeze parameters. Phase distribution, for a single mode, is
insensitive to correlation between modes, and it displays a uniform form.

\chapter{Scientific results and their analysis: part II (dynamic regime)}
In this part we investigate the quantum properties of  some dynamical
systems in the framework of interaction of radiation modes with nonlinear
media described  nonlinear couplers.  We have considered three types of
such models when two, three or four modes are evolved within.
In general, all of these  models are composed from two waveguides of
the length $L$ through which the interacting modes evolve.
The interaction between the different modes in the same waveguide can be
established by strong laser  pumping, however, the interaction between the
two waveguides occurs by means of evanescent waves.
Instead of initial coherent states at the input beams, the Fock
states generated  for instance in a micromaser can  be introduced.
Outgoing fields are detected as single or compound modes by means of
homodyne, photocounting or coincidence detection.
Further, the values of the coupling constants between the different modes
in the device are chosen for the reason of simple calculations and
illustration of the physical behaviour of
the system. We mainly demonstrate the effect of their mutual relations and
their magnitudes are not substantial for such demonstrations. For an
experimental realization these values can be estimated
 as $(10^{11}-10^{12}) s^{-1}$,
provided that the pumping laser of power in mW is used,
producing about $10^{19}$
photons/s, which is sufficient power to neglect quantum noise in pumping beams.
Then all the effects shown are interpreted on the corresponding reduced time
scale. Initial state amplitudes or photon  numbers are chosen in units
because switching  properties are examined on the quantum level of
single photons.

In the following section we investigate the quantum properties of the
first model which includes two-mode case.

\section{ Quantum statistics and dynamics of nonlinear couplers with
nonlinear exchange}

In this section we discuss the quantum statistical properties
of nonlinear couplers composed of two waveguides operating by the second
subharmonic generation assuming strong coherent pumping and linear
exchange of energy between waveguides by means of evanescent waves,
however we
additionally take into account the influence of nonlinear coupling of the
parametric type of both the waveguides.
This system is
suggested to be  described by the following Hamiltonian:

\begin{eqnarray}
\begin{array}{lr}
\frac{\hat{H}}{\hbar} =
\sum_{j=1}^{2}\left\{\omega_{j}\hat{a}_{j}^{\dagger}\hat{a}
_{j}+ \lambda_{j}\left[\hat{a}_{j}^{\dagger 2}\exp (i\mu_{j}t) + {\rm
h.c.}\right]\right\}\\
\\
 +\lambda_{3}\left\{\hat{a}_{1}\hat{a}_{2}^{\dagger}
\exp [i\phi_{1}(t)] + {\rm h.c.}\right\} +\lambda_{4}\left\{\hat{a}_{1}
\hat{a_{2}}\exp [-i\phi_{2}(t)] + {\rm h.c.}\right\}, \label{scs1}
\end{array}
\end{eqnarray}
 where
$\hat{a}_{1} (\hat{a}_{1}^{\dagger})$, $\hat{a}_{2} (\hat{a}
_{2}^{\dagger} $) are annihilation (creation) operators of the fundamental
modes in the first and second waveguides having frequencies $\omega_{1}$
and $\omega_{2}$, respectively, $\mu_{j}$ are related with the frequency
of the second-harmonic modes described classically as strong coherent
fields, $\phi_{j}(t)$, $j=1,2$, are related to the difference- and
sum-frequencies of modes $1$ and $2$, respectively, $\lambda_{1}$ and
$\lambda_{2}$ are nonlinear coupling constants for the second subharmonic
generation in the first and second waveguides, respectively, $\lambda_{3}$
is the coupling constant for linear exchange between waveguides through
evanescent waves, $\lambda_{4}$ is the coupling constant for the nonlinear
exchange through simultaneous annihilation or creation of a photon in
both the subharmonic modes on expense of pumping and {\rm h.c.} means the
Hermitian conjugate terms (for further details concerning the optical
parametric processes, see \cite{perin} (Chap. 10)).

When $\mu_{j}=0$ and only the degenerate term
is considered, we have the well-known
Hamiltonian, in the interaction picture,
for squeezed light generation \cite{[6]}, where $\lambda_{1}$ (or
$\lambda_{2}$) represents the coupling constant proportional
to the quadratic susceptibility of the second-order
nonlinear process (degenerate
parametric down-conversion with classical coherent pumping),
or the coupling constant proportional to the cubic susceptibility
of the third-order nonlinear process
(degenerate four-wave mixing with classical coherent pumping) \cite{yams}.
If additionally $\phi_{1} (t)=\phi_{2}(t)=0$, the Hamiltonian (\ref{scs1})
represents a mixture of second subharmonic generation, frequency conversion
 and parametric amplification in the interaction picture \cite{[3],{mish1},
 {martin}}. Schematically, this Hamiltonian is represented in Fig. 6.1.
\newpage
\begin{picture}(120,70)(-100,6)
\put (20,5){\line(1,0){100}}
\put (20,10){\vector(1,0){80} $\chi^{(2)}$}
\put (20,15){\line(1,0){100}}
\put (20,5){\line(-1,1){35}}
\put (20,15){\line(-1,1){30}}
\put (120,5){\line(1,1){35}}
\put (120,15){\line(1,1){30}}
\put (135,25){\vector(1,1){30}}
\put (180,50){\makebox(0,0){$\hat{a}_{2}(\frac{z}{v})$}}
\put (-15,45){\vector(1,-1){30}}
\put (-27,47){\makebox(0,0){$\hat{a}_{2}(0)$}}
\put (20,-15){\line(1,0){100}}
\put (20,-25){\line(1,0){100}}

\put (20,-65){\line(1,0){100}}
\put (20,-65){\vector(0,1){40}}
\put (120,-65){\vector(0,1){40}}

\put (70,-70){\makebox(0,0){$z=vt$}}
\put (20,-20){\vector(1,0){80}$\chi^{(2)}$}
\put (25,-5){\vector(0,-1){15}$\lambda_{3}$}
\put (25,-5){\vector(0,1){15}}
\put (70,-5){\vector(0,1){15}$\lambda_{4}$}
\put (70,-5){\vector(0,-1){15}}

\put (20,-25){\line(-1,-1){30}}
\put (20,-15){\line(-1,-1){35}}
\put (120,-25){\line(1,-1){30}}
\put (120,-15){\line(1,-1){35}}
\put (175,-55){\makebox(0,0){$\hat{a}_{1}(\frac{z}{v})$}}
\put (135,-35){\vector(1,-1){30}}
\put (-15,-55){\vector(1,1){30}}
\put (-28,-52){\makebox(0,0){$\hat{a}_{1}(0)$}}
\end{picture}
\vspace{1.5in}

\noindent Figure 6.1:  Scheme of quantum nonlinear
coupler with linear and nonlinear coupling formed from two nonlinear
waveguides described by the quadratic susceptibility $\chi^{(2)}$.
The beams are described by the photon
annihilation operators as indicated; $z=vt$ is the interaction length.
Both the waveguides are pumped by strong classical coherent waves.
Outgoing fields are examined as single or compound modes by means
of homodyne detection to observe squeezing of vacuum fluctuations,
or by means a set of photodetectors to measure photon antibunchibng
and sub-Poissonian photon statistics in the standard ways.
\vspace{.5cm}

\setcounter{figure}{1}

Solution  of the equations of  motion in the Heisenberg picture for
the Hamiltonian (\ref{scs1})  are

\begin{eqnarray}
\begin{array}{lr}
 \hat{a}_{1}(t)\exp(-it\frac{\mu_{1}}{2})
=\hat{a}_{1}(0)K_{1}(t) +\hat{a}_{1}^{
\dagger}(0)L_{1}(t) +\hat{a}_{2}(0)M_{1}(t)+ \hat{a}_{2}^{
\dagger}(0)N_{1}(t),\\
\\
 \hat{a}_{2}(t)
\exp(-it\frac{\mu_{2}}{2})=\hat{a}_{2}(0)K_{2}(t) +\hat{a}_{2}^{
\dagger}(0)L_{2}(t) +\hat{a}_{1}(0)M_{2}(t)+ \hat{a}_{1}^{
\dagger}(0)N_{2}(t), \label{scs2}
\end{array}
\end{eqnarray}
 where the explicit forms of the time dependent coefficients,
 which contain all
the features of the structure, and the details about the solution
are given in appendix D.

One can see from this solution (see appendix D) that when $\bar{\Omega}_{1}$
and $\bar{\Omega}_{2}$ are real, the coupler switches the energy between the
modes which  propagate inside since the solution will include trigonometric
functions \cite{[3]}. Nevertheless, if $\bar{\Omega}_{1}$ and
$\bar{\Omega}_{2}$ are pure imaginary, the Heisenberg solutions
provide hyperbolic functions, which are growing rapidly with time, and the
coupler operates as amplifier for the input modes \cite{tuck1}. So that
 the behaviour of the coupler will be indicated essentially by the
relation between coupling constants.

In what follows, we shall employ the results obtained  here
to treat the squeezing phenomenon and quasiprobability distribution functions
for the model under consideration.

\subsection{ Squeezing phenomenon}

Here we study single-mode squeezing phenomenon when the modes are initially prepared
in thermal states with the average  thermal  photon
numbers
$\bar{n}_{j},\quad j=1,2$ as well as in the coherent states.
For this purpose (using results of section 4.2 and considering $\mu_{j} (t)$ to be the phase of the
local oscillator in the relations (\ref{12}) and (\ref{r13}),
without loss of generality, we can cancel the high frequency terms,
and $j=1,2$ stands for mode 1 and mode 2, respectively)
we calculate the quantities $S_{j}(t)$ and $Q_{j}(t)$ which are
defined in (4.9) for the initial thermal light.  For the first
mode we have the following expressions

\begin{eqnarray}
\begin{array}{lr}
S_{1}(t) = 2\bar{n}_{1}[|L_{1}(t)|^{2}+|K_{1}(t)|^{2}]
+2\bar{n}_{2}[|N_{1}(t)|^{2}+|M_{1}(t)|^{2}]
+2|L_{1}(t)|^{2}+2|N_{1}(t)|^{2}\\
\\
+2(2\bar{n}_{1}+1)[ L_{1}(t)K_{1}(t) + {\rm c.c.}]
+2(2\bar{n}_{2}+1)[ M_{1}(t)N_{1}(t) + {\rm c.c.}],
\label{scs4}
\end{array}
\end{eqnarray}

\begin{eqnarray}
\begin{array}{lr}
Q_{1}(t) = 2\bar{n}_{1}[|L_{1}(t)|^{2}+|K_{1}(t)|^{2}]
+2\bar{n}_{2}[|N_{1}(t)|^{2}+|M_{1}(t)|^{2}]
+2|L_{1}(t)|^{2}+2|N_{1}(t)|^{2}\\
\\
-2(2\bar{n}_{1}+1)[ L_{1}(t)K_{1}(t) + {\rm c.c.}]
-2(2\bar{n}_{2}+1)[ M_{1}(t)N_{1}(t) + {\rm c.c.}],\label{scs5}
\end{array}
\end{eqnarray}
where {\rm c.c.} means the complex conjugate terms.  The
corresponding expressions for the second mode can be obtained from (\ref{scs4})
and (\ref{scs5}) by using  the interchange $1\leftrightarrow 2$. Further,
the other expressions related to the injected coherent light
in the coupler are the same as (\ref{scs4}) and (\ref{scs5}), but  setting
$\bar{n}_{j}=0$.

It is known that the nonlinear coupler is a source of optical fields,
the statistical properties of which are changed
as a result of the linear and nonlinear interaction inside and between
waveguides. Consequently, one can
obtain nonclassical light from one input and, in addition,
it can be switched.
\begin{figure}[h]%
  \centering
  \subfigure[]{\includegraphics[width=5cm]{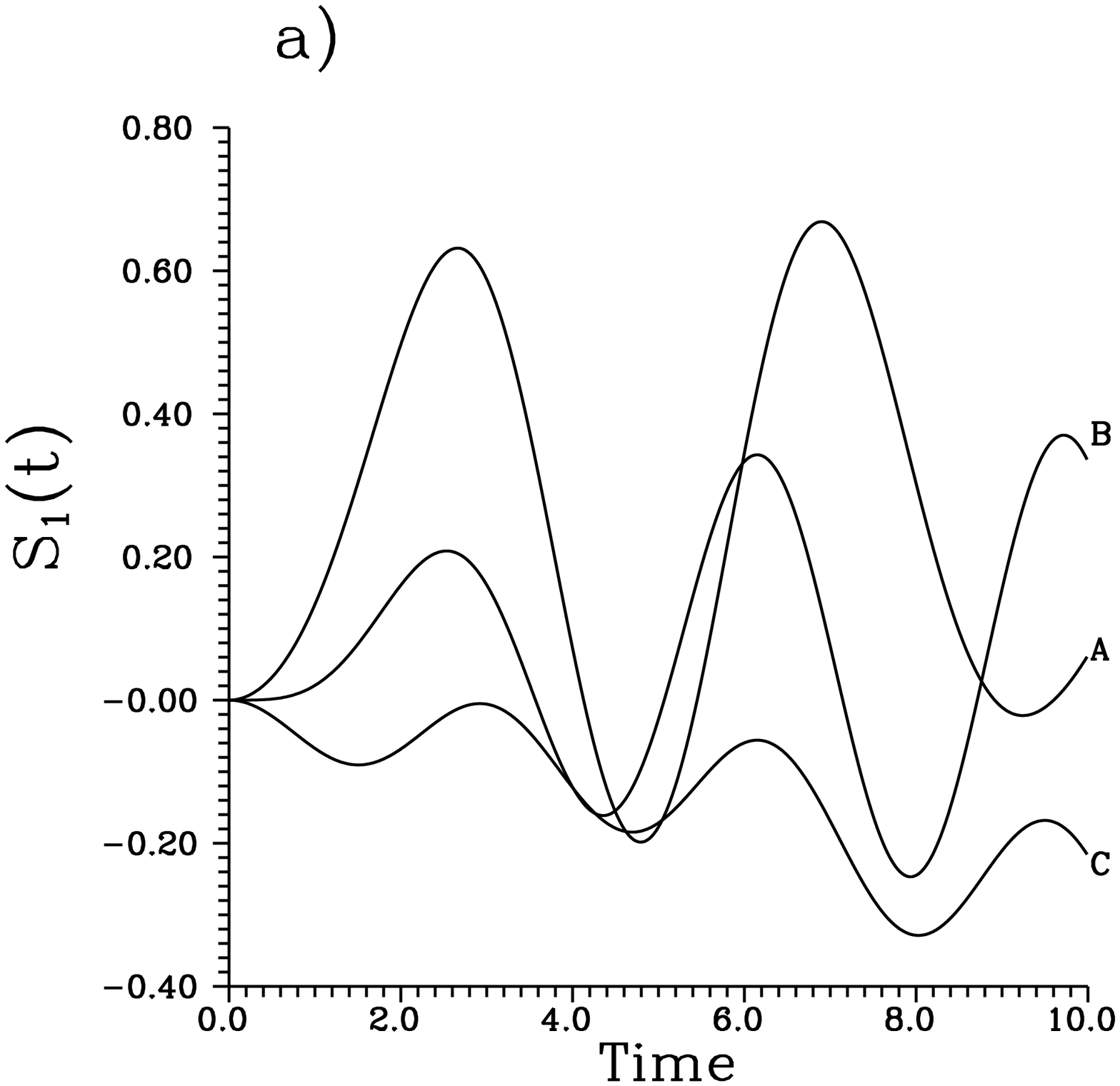}}
 \subfigure[]{\includegraphics[width=5cm]{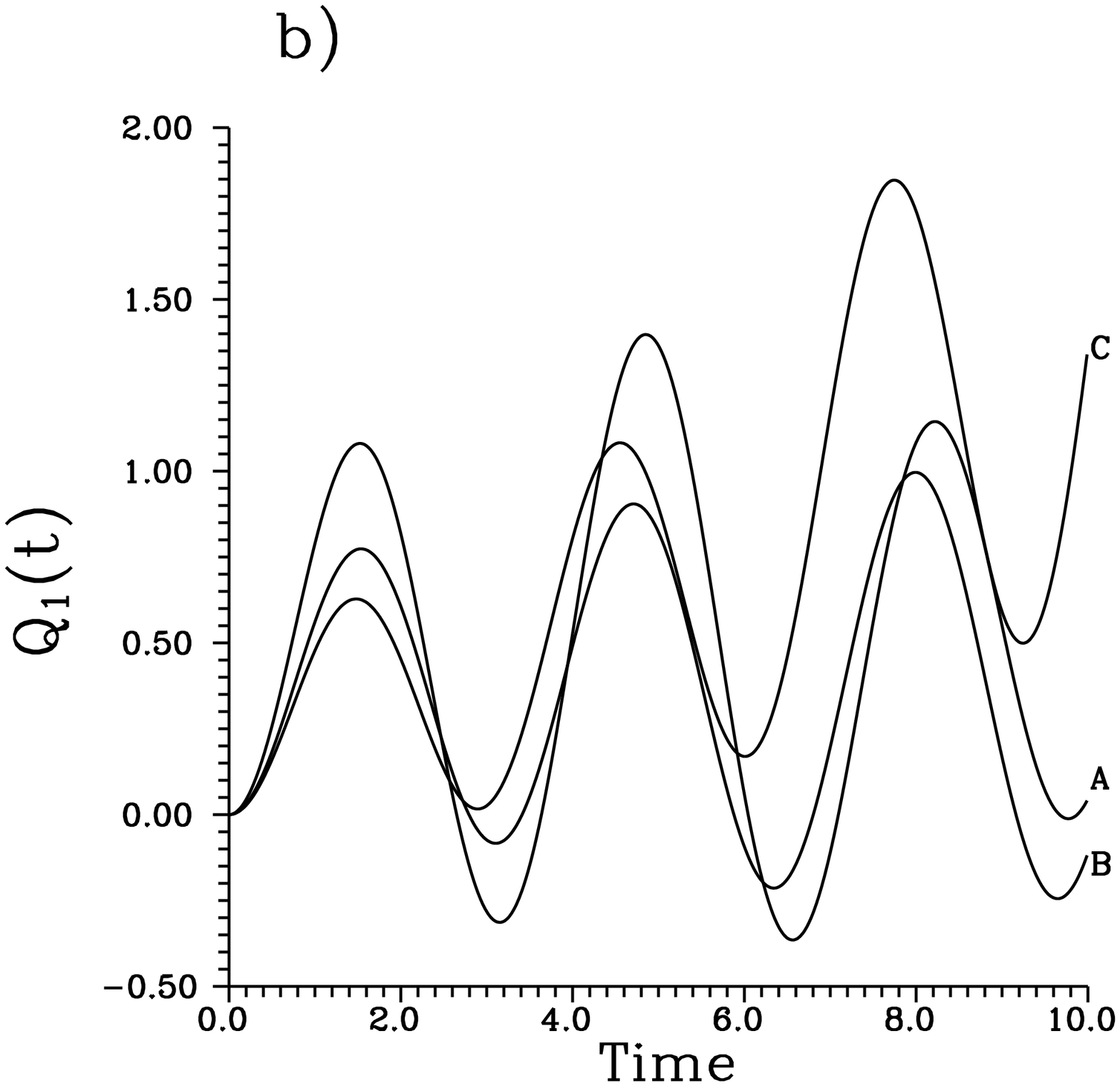}}
    \caption{
  Squeezing phenomenon for mode 1 when the modes are initially
in coherent light and: a) for the first component $S_{1}(t)$; b)
for the second component $Q_{1}(t)$; $\lambda_{3}=1$ for all
curves; curves A, B and C are corresponding to $\lambda_{1}=
\lambda_{2}=\lambda_{4}=0.25$, $\lambda_{1} =
\lambda_{2}=\lambda_{4}=0.2$ and $\lambda_{1}=0.17$,
$\lambda_{2}=\lambda_{4}=0.2$, respectively; straight line has
been put to show the bound of squeezing. } \label{fig20}
 \end{figure}

We have plotted $S_{1}(t), Q_{1}(t)$ in Figs. 6.2a,b and $S_{2}(t), Q_{2}(t)$
in Figs. 6.3a,b, when the initial light is coherent, for different values of
$\lambda_{k}$. Further we have chosen $
\lambda_{3}=1$ for all curves and for the curve $A: \lambda_{1}=\lambda_{2}
=\lambda_{4}=0.25$; for the curve $B: \lambda_{1}=\lambda_{2}
=\lambda_{4}=0.20$, and for the curve $C: \lambda_{1}=0.17,\lambda_{2}
=\lambda_{4}=0.2$. On the other hand, Fig. 6.4 gives
$S_{1}(t)$ (first mode) when the initial light is thermal light with
coupling constants as those for the curve C, where $\bar{n}_{1}=0.5$ and
$\bar{n}_{2}=0.5$ (solid curve), 1.5 (dashed curve);  straight line
 shows the bound of squeezing of the curves.
Firstly, we start our discussion by studying the case of input coherent light.
From these figures we can see how the coherent states, which are
minimum-uncertainty states, evolve in the coupler to produce squeezed light.
We can observe  the oscillatory behaviour in these curves, showing
that squeezing can be switched from one waveguide to the other in the
course of time during power transfer. Moreover, squeezing can be
interchanged between the two quadratures of the same waveguide.
More precisely, for mode 1, squeezing can occur for all selected values
of $\lambda_{k}$ in $S_{1}(t)$, but in $Q_{1}(t)$ only curves A, B can
exhibit squeezing, as shown in Figs. 6.2a,b, which reflects the dependence of
nonclassical behaviour on the strength of subharmonic generation.
\begin{figure}[h]%
  \centering
  \subfigure[]{\includegraphics[width=5cm]{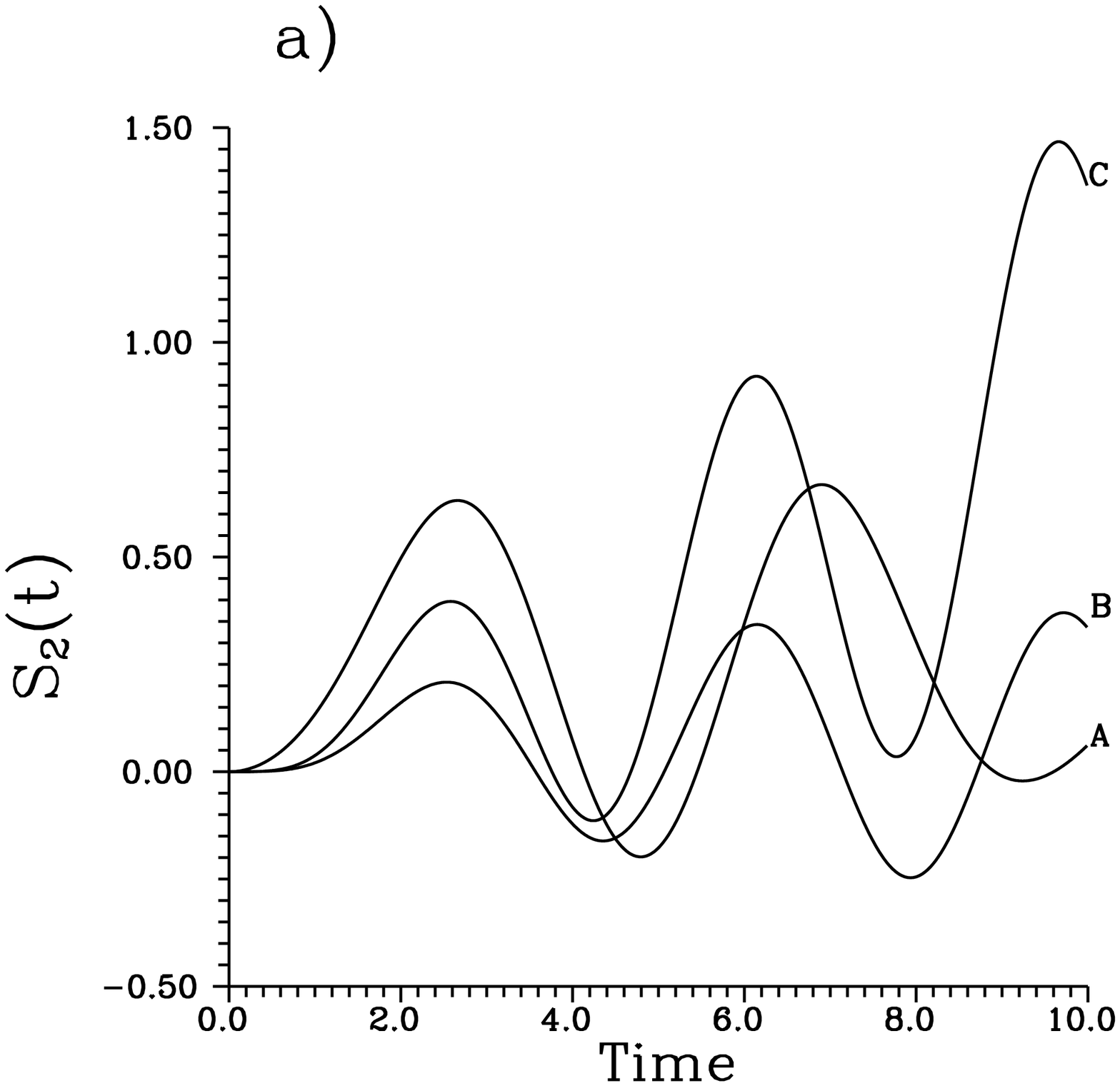}}
 \subfigure[]{\includegraphics[width=5cm]{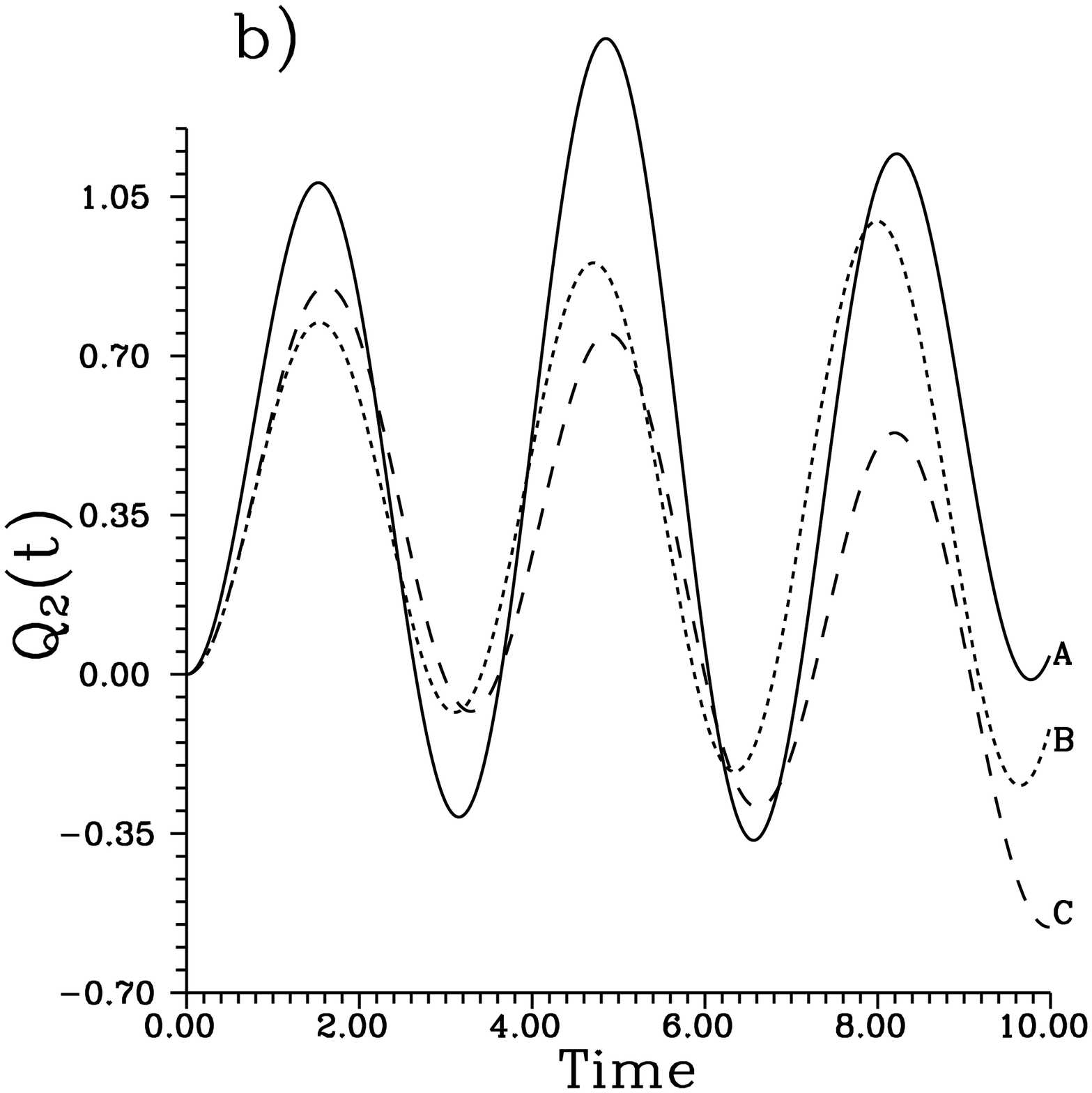}}
    \caption{
 Squeezing phenomenon for mode 2: a) for the first component $
S_{2}(t)$; b) for the second component $Q_{2}(t)$; the values of
the parameters $\lambda_{k}$ are as in Fig. 6.2. } \label{fig21}
 \end{figure}

For mode 2, we can see squeezing in all curves in both the quadratures, as shown in
Figs. 6.3a,b. It can be easily seen that the amount of squeezing is sensitive
to the strength of coupling $\lambda_{k}$ and that in general its values in
the second component are more pronounced than those in the first one.
Now if we turn our attention to the case of injected thermal light (Fig. 6.4c),
 we can observe that squeezing is available in the large
interaction time. Further, $S_{1}(t)$ exhibits oscillatory behaviour and
it evolves from unsqueezed values in the short range of interaction time,
owing to the fact that thermal states are not minimum-uncertainty states, into
squeezed values and eventually unsqueezed values can be recovered. Indeed, we
proved numerically that this behaviour is periodically recovered  with the
time.   Moreover, by comparing the dashed curve with the solid one, we can
see that increasing of the photon number in the second waveguide causes
decreasing of the amount of squeezing in the first one. This is related with
the effect of evanescent waves between waveguides and shows how one can
control light by light in the coupler. Finally, we can conclude that by
controlling  the input average thermal photon number and  the interaction
time (or  the length of the coupler), the interaction under consideration
can generate squeezed thermal light.
It is worthwhile to refere to \cite{ther9,{ther10},{ther12},{ther13}}, where more discussions related
to squeezed thermal states are given.

\begin{figure}[h]%
 \centering
    \includegraphics[width=6cm]{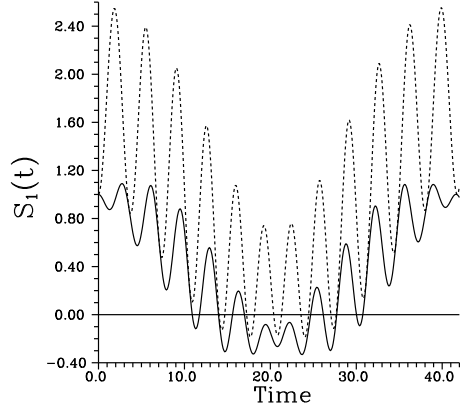}
   \caption{
Squeezing phenomenon for mode 1 when the modes are initially in
thermal light for the first component $S_{1}(t)$ with
$\bar{n}_{1}=0.5$, $\bar{n}_{2}=0.5$ (solid curve), $1.5$ (dashed
curve) and the coupling constants $\lambda_{j}$ are the same as
those for the curve C in Fig. 6.2.} \label{fig22}
\end{figure}


\subsection{Quasiprobability functions}

Here we shall continue in our investigation for the statistical
properties of the system under discussion in the basis of
quasiprobability distribution functions for single-mode case only
when both the modes are initially in number and coherent states.\\
\\
{\bf (a) Input Fock states}\\
\\
The density operator for two-mode number states is

\begin{equation}
\hat{\rho}_{\rm n}(0)={\rm |n\rangle_{1}|m\rangle_{2}}
{\rm_{2}\langle m| _{1}\langle n|}, \label{scs6}
\end{equation}
where the subscript $n$ stands for Fock states case.
The single-mode s-parametrized characteristic function
for the first mode can be obtained  using the results of section
4.3 together with the identity (\ref{scf4})
as

\begin{equation}
C^{(1)}_{n,m}(\zeta,s,t)=\exp \left\{ \frac{s}{2}|\zeta|^{2}
-\frac{1}{2}[|\nu_{1}(t)|^{2}+|\nu_{2}(t)|^{2}]\right\}
L_{n}(|\nu_{1}(t)|^{2})
L_{m}(|\nu_{2}(t)|^{2}),
\label{scs7}
\end{equation}
where
\begin{equation}
\nu_{1}(t)=\zeta K_{1}^{*}(t)-\zeta^{*}L_{1}(t), \quad
\nu_{2}(t)=\zeta M_{1}^{*}(t)-\zeta^{*}N_{1}(t),
\label{scs8}
\end{equation}
and $L_{m}(.)$ is the Laguerre polynomial of order $m$.
Inserting (\ref{scs7}) into (\ref{16}), carrying out the integration
numerically (the exact form can be found, see \cite{mfp})
and taking $s=0$ we obtain the W-function.
The result has been plotted in
\begin{figure}[h]%
  \centering
  \subfigure[]{\includegraphics[width=5cm]{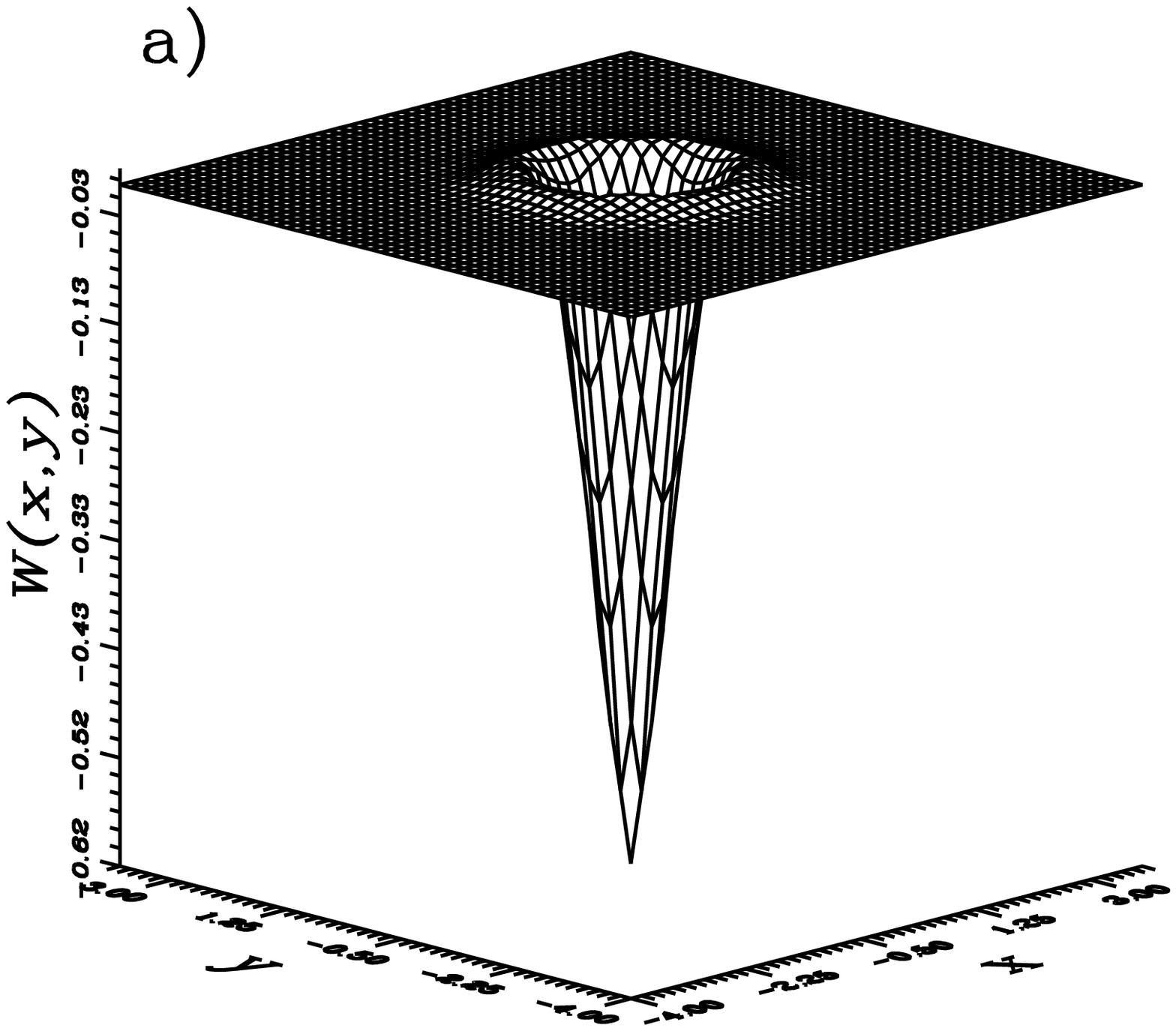}}
 \subfigure[]{\includegraphics[width=5cm]{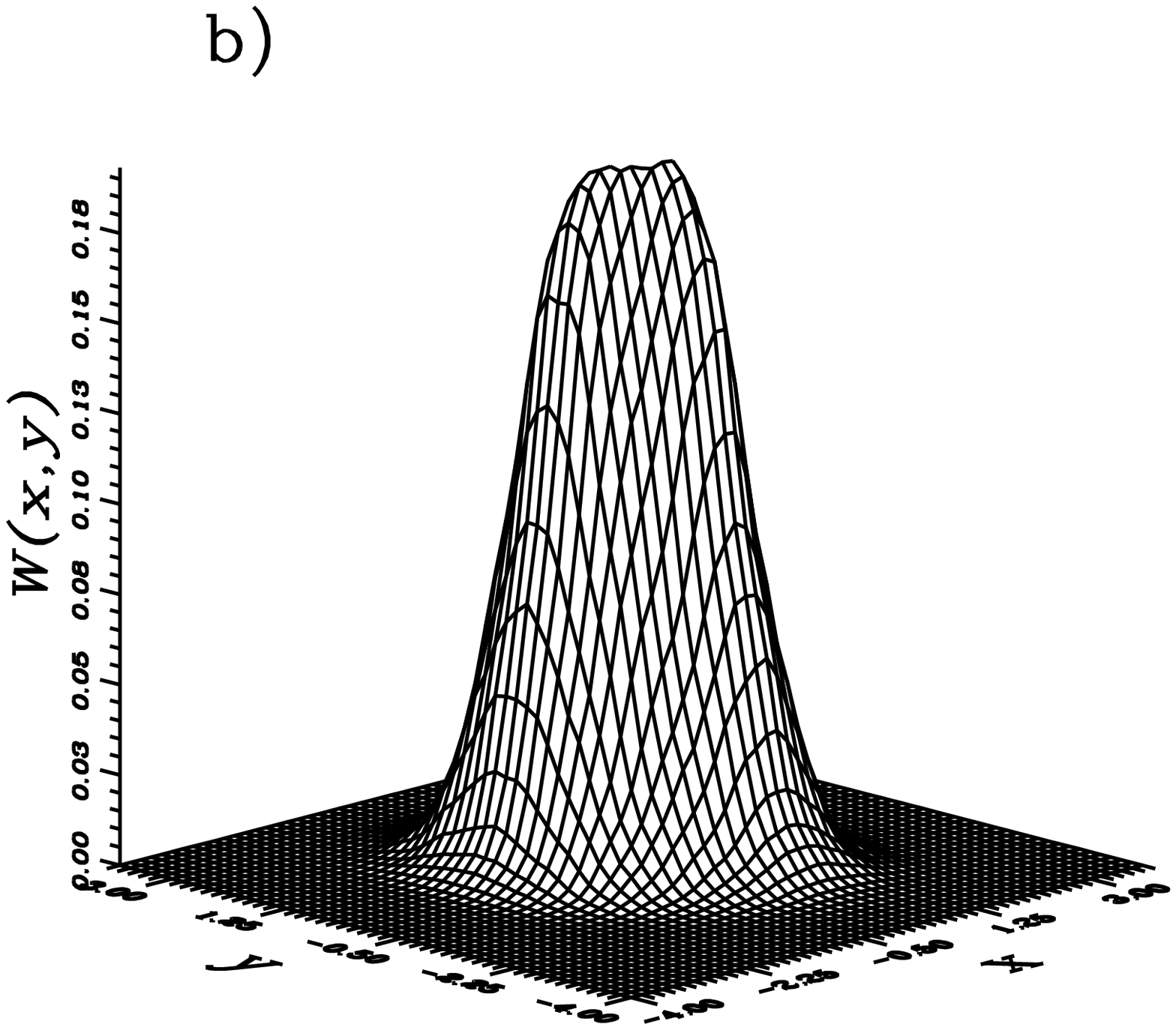}}
  \subfigure[]{\includegraphics[width=5cm]{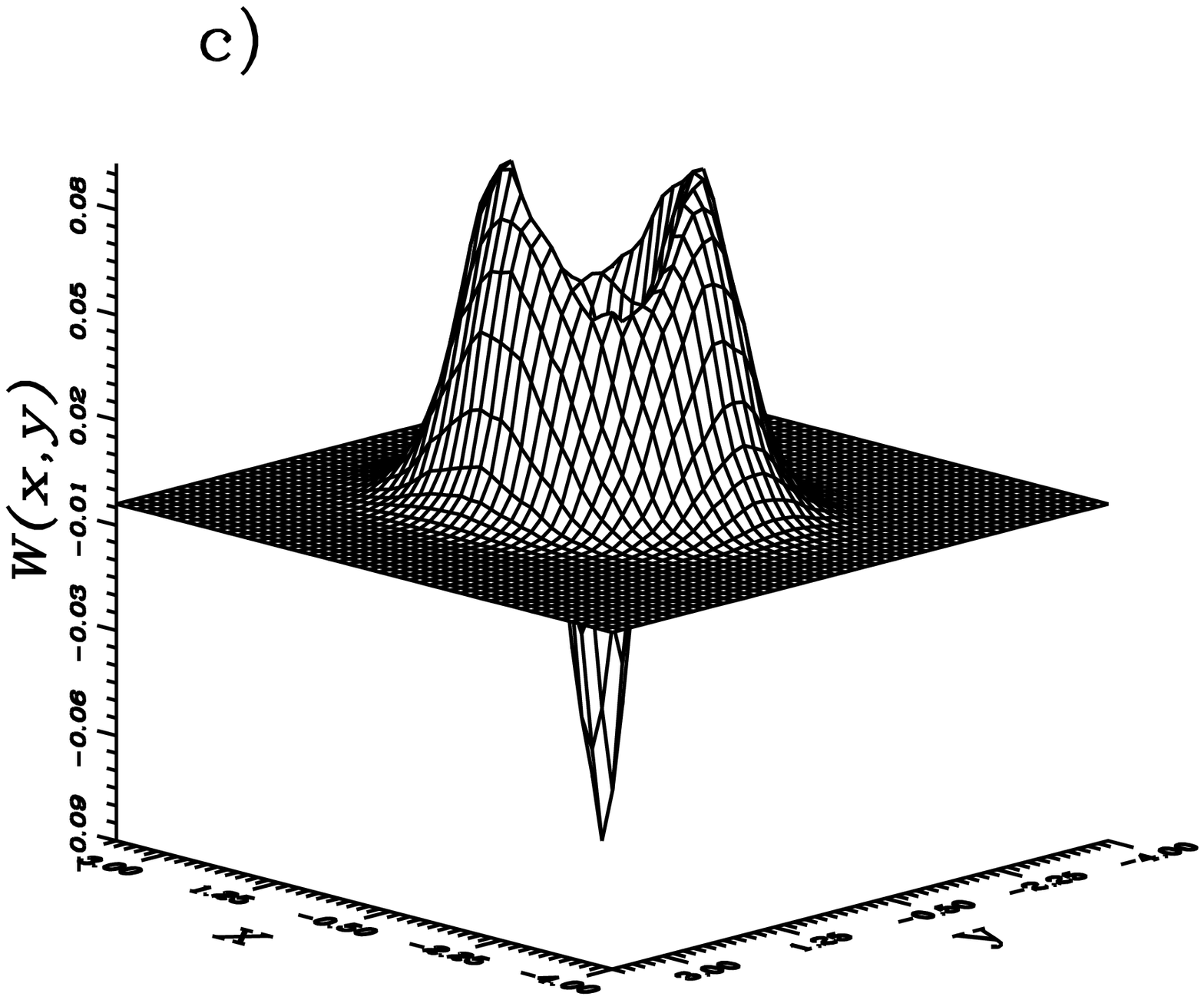}}
    \caption{
 W-function for the single mode (mode 1) for different values
of time $t$ when both the modes are initially in the number
states; the first mode is in the state $|1\rangle $ and the second
mode is in the state $|0\rangle$ and $\lambda_{3}=1$,
$\lambda_{1}=\lambda_{2}=\lambda_{4}=0.25$: a) for $t=\pi/100$; b)
for $t=\pi/2$; c) for $t=\pi$. } \label{fig23}
 \end{figure}

Figs. 6.5 against $x={\rm Re}\beta_{1}$ and
$y={\rm Im}\beta_{1}$, when the
first state is the Fock state $|1\rangle$ and the second one is the vacuum
state $|0\rangle$, i.e. $n=1, m=0$; $\lambda_{3}=1,
\lambda_{1}=\lambda_{2}=\lambda_{4}=0.25$ and for shown values of time.
In Fig. 6.5a we have the W-function for $t=\frac{\pi}{100}$, i.e.
after short time interaction between the two modes we observe
similar behaviour as for the W-function of the state $|1\rangle$,
which means that pronounced negative values are exhibited.
This
behaviour of the W-function is completely different by increasing the
time ($t=\frac{\pi}{2}$); we see disappearance of negative values of the
quasidistribution and a stretched positive peak occurs (Fig. 6.5b).
This form of $W$-function is close to that of squeezed vacuum states
\cite{[6]}, i.e. squeezed vacuum states can be generated, in principle, in this
model. It should be borne in mind that the specific direction of stretching for the
quasiprobability function of squeezed states may be achieved by choosing
a suitable
value for the  phase of squeeze parameter. Of course, in Fig. 6.5b, there is
a negligible spike at the top of the peak which can be smoothed out by
governing the coupler parameters. After larger time interaction $t=\pi$,
the negative values are reached again, but they are less pronounced and
asymmetry  can be observed due to stretching (Fig. 6.5c). So we meet a time
development of the W-function as a result of the power transfer
between the two modes inside the coupler.
This shows that the nonlinear directional coupler can be used
as a source of quantum states.
Further, we have found that the behaviour of
the Q-function is consistent with that of $W$-function
for the same shown values \cite{mfp}.\\
\\
{\bf (b) Input coherent light}\\
\\
In a similar way as we followed in the case {\bf (a)} we can study the same
quantities when both the modes are initially in coherent states
$|\alpha_{1},\alpha_{2}\rangle$.
For the single-mode case the $s$-parametrized characteristic function and
the $s$-parametrized quasiprobability function are given, respectively, as

\begin{eqnarray}
\begin{array}{lr}
C^{(1)}_{{\rm coh}}(\zeta,s,t) = \exp \left\{- \frac{1}{2}[1-s+
2|L_{1}(t)|^{2}+2|N_{1}(t)|^{2}]|\zeta|^{2} +\zeta\bar{\alpha}
_{1}^{*}(t)-\zeta^{*}\bar{\alpha}_{1}(t) \right\}\\
 \\
 \times \exp\left\{\frac{1}{2} \zeta^{2}
[N^{*}_{1}(t)M^{*}_{1}(t)+L^{*}_{1}(t)K^{*}_{1}(t)]\right\}\\
 \\
 \times \exp\left\{ \frac{1}{2}\zeta^{*2}
[N_{1}(t)M_{1}(t)+L_{1}(t)K_{1}(t)] \right\},
\label{scs9}
\end{array}
\end{eqnarray}

\begin{eqnarray}
\begin{array}{lr}
W^{(1)}_{{\rm coh}}(\beta,s,t) = \frac{1}{\pi\sqrt{ [\frac{1-s}{2}
+|L_{1}(t)|^{2}+|N_{1}(t)|^{2}]^{2}- |B_{1}(t)|^{2} }}\\
\\
 \times \exp \left\{-\frac{[\frac{1-s}{2}
+|L_{1}(t)|^{2}+|N_{1}(t)|^{2}] |\bar{\alpha}_{1}(t)-\beta|^{2}}
{[\frac{1-s}{2} +|L_{1}(t)|^{2}+|N_{1}(t)|^{2}]^{2}-|B_{1}(t)|^{2}}
\right\}\\
\\
\times \exp \left\{- \frac{\frac{1}{2}
|B_{1}(t)|[E_{1}^{2}(t)+E_{1}^{*2}(t)]} {[\frac{1-s}{2}
+|L_{1}(t)|^{2}+|N_{1}(t)|^{2}]^{2}-|B_{1}(t)|^{2}} \right\},
\label{ascs10}
\end{array}
\end{eqnarray}
 where
\begin{eqnarray}
\begin{array}{lr}
B_{1}(t)=N^{*}_{1}(t)M^{*}_{1}(t)+L^{*}_{1}(t)K^{*}_{1}(t)
=|B_{1}(t)|\exp[2i\delta_{1}(t)], \\
E_{1}(t)=[\bar{\alpha}_{1}(t)-\beta]\exp[i
\delta_{1}(t)],
\label{bscs10}
\end{array}
\end{eqnarray}
and $\bar{\alpha}_{1}(t)$ is the average of the first mode operator
$\hat{A}_{1}(t)$ with respect to input coherent light.

From (\ref{ascs10}), the $P$-function is not well defined as an ordinary function for
$|L_{1}(t)|^{2}+|N_{1}(t)|^{2}<|B_{1}(t)|$ and hence the nonclassical effects
such as squeezing of vacuum fluctuations can occur, as we have seen before.
\begin{figure}[h]%
  \centering
  \subfigure[]{\includegraphics[width=5cm]{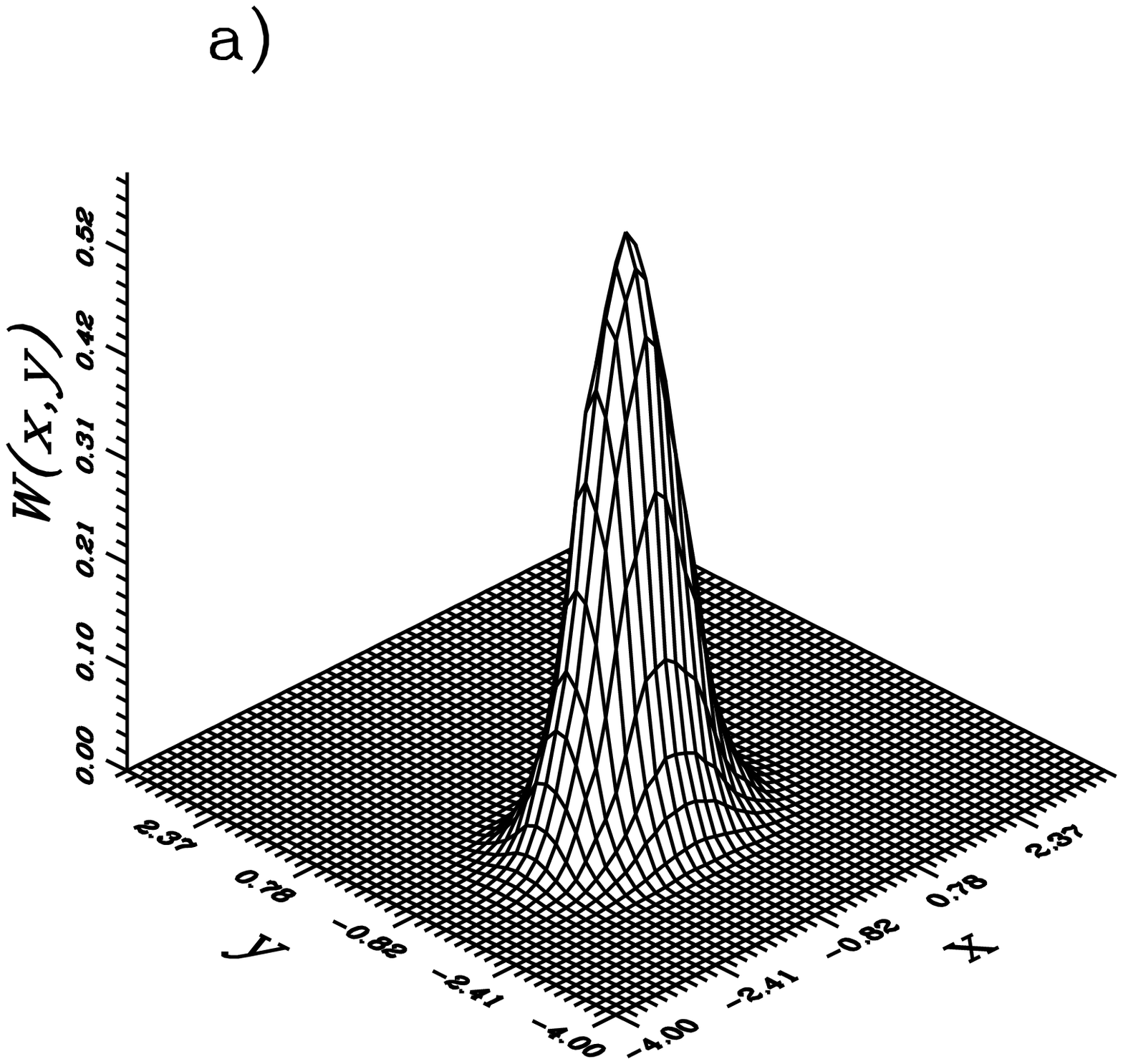}}
 \subfigure[]{\includegraphics[width=5cm]{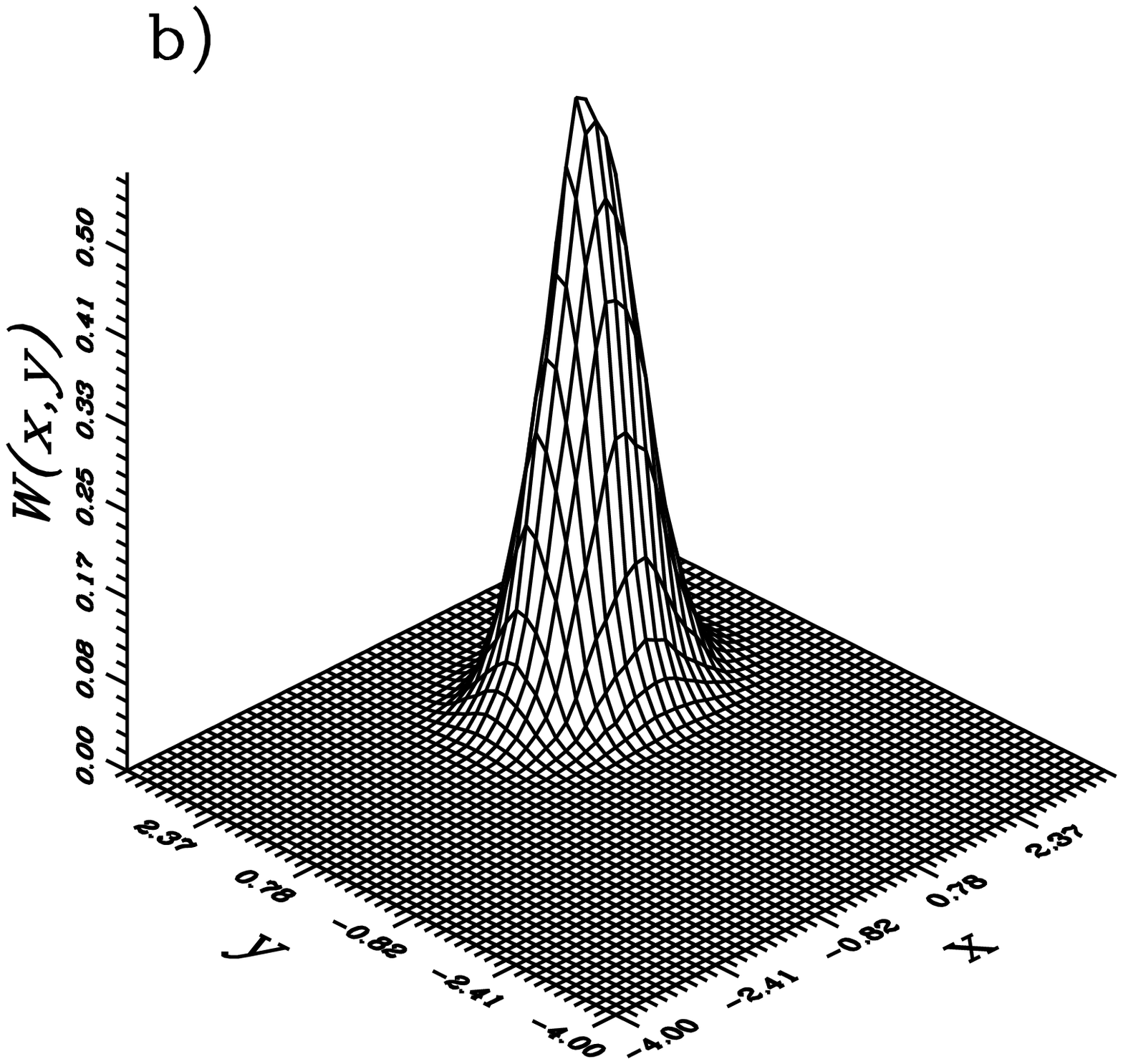}}
      \caption{
W-function for the single mode (mode 1) for different values
of time $t$ when both the modes are initially in the coherent states; $%
|\alpha_{1}|^{2}=|\alpha_{2}|^{2}=2$ and $\lambda_{k}$ are the
same as in Fig. 6.6: a) for $t=\pi$; b) for $t=2\pi$. }
\label{fig24}
 \end{figure}


Furthermore, the nonclassical effect, especially squeezing of vacuum
fluctuations in the case of our system, can be
recognized in the behaviour of $W$-function (and/or $Q$-function) in phase
space as shown in Fig. 6.6 for shown values of parameters.
For $t=0$, i.e. when there is no
interaction between the two modes, the W-function is identical with that
shown for a single mode representing a symmetric Gaussian bell in
phase space. As soon as the interaction switches on ($t > 0$), we observe
that the Gaussian centre is shifted and the rotationally
symmetric function of the initial state at $t=0$ gets to be squeezed in
various phase space directions in dependence on time, as demonstrated in
Figs. 6.6a,b. In other words, the initial symmetric contour of the
$W$-function has been stretched as the interaction switches on, i.e. noise
ellipse characterizing squeezed light appears, which
rotates in  the phase space as the interaction time  progresses.

\subsection{Conclusions}

In this section we have examined the quantum statistical properties of
radiation generated and propagated in the nonlinear optical coupler
composed
of two nonlinear waveguides operating by the second subharmonic
processes,
coupled linearly by evanescent waves and nonlinearly by nondegenerate
optical parametric process. We have demonstrated regimes for generation
and
propagation of nonclassical light exhibited by squeeezing of vacuum
fluctuations. We have also obtained quasidistribution functions
for the initial light beams which are in Fock and coherent states.
Compared to earlier results for nonlinear optical
couplers
we have shown that the nonlinear coupling increases in general quantum
noise
in the device even if in some cases it can support generation of
nonclassical light.

The motivation for examination of the system under consideration
arises from
the previous investigations of the nonlinear couplers as promising
 devices to produce nonclassical light.
When coherent light is injected initially in the system, squeezed as well
as sub-Poissonian light can be generated
\cite{mfp} (the latter  has not been  analyzed here).
For injected number states,
 squeezed vacuum states are produced.
When thermal light initially enters the coupler,  the
coupler can operate as a microwave Josephson-junction parametric
amplifier \cite{ther1}.
These effects have been recognized to result from the
competition between
linear and nonlinear properties of the system and are dependent
on the initial amplitudes of the input fields.
The crucial role plays here the mechanism of the energy exchange
between waveguids.

\section{  Generation of squeezed light in a nonlinear asymmetric directional
coupler}
In this section we study the properties of three modes interacting
in a nonlinear asymmetric directional coupler
composed of a linear waveguide and a nonlinear waveguide operating by
nondegenerate parametric amplification.
We show that such a device is a source of single-mode
squeezed light.
 This fact has been demonstrated, under certain conditions and for specific
 modes, for incident coherent beams in terms of the quasiprobability
 functions, photon-number distribution and phase distribution.

\subsection{ Equations of motion}
We consider a device (nonlinear asymmetric directional coupler)
as outlined in Fig. 6.7.  It can be described by the Hamiltonian

\begin{eqnarray}
\begin{array}{lr}
\frac{\hat{H}}{\hbar}=[\sum_{j=1}^{3}\omega_{j}\hat{a}^{\dagger}_{j}
\hat{a}_{j} ]+i\lambda_{1}
\left\{
\hat{a}_{1}\hat{a}_{2}\exp[i(\omega_{1}+\omega_{2})t]-{\rm h.c.}\right\}\\
\\
+i\lambda_{2}
\left\{\hat{a}_{1}\hat{a}^{\dagger}_{3}\exp[i(\omega_{1}-\omega_{3})t]-{\rm
h.c.}\right\}
+i\lambda_{3}
\left\{\hat{a}_{2}\hat{a}^{\dagger}_{3}\exp[i(\omega_{2}-\omega_{3})t]-{\rm
h.c.}\right\}, \label{scs10}
\end{array}
\end{eqnarray}
 where
 $\hat{a}_{j} (\hat{a}^{\dagger}_{j}),\quad j=1,2,3$ are
 the annihilation (creation) operators designated to the signal, idler
 and linear modes, respectively,
$\lambda_{1}$ and $\lambda_{j}, \quad j=2,3$ are the
corresponding nonlinear and linear coupling constants;
$\omega_{j}$ are the natural  frequencies of oscillations
of the uncoupled modes and {\rm h.c.} is the Hermitian conjugate.
The linear exchange between the two waveguides establishes through evanescent
waves provided that $\omega_{1}\simeq\omega_{2}\simeq \omega_{3}$.

  The dynamics of the system is described by the Heisenberg
  equations of motion which, using the slowly varying forms
($\hat{a}_{j}=\hat{A}_{j}\exp(-i\omega_{j}t),\quad j=1,2,3$),
read

\begin{eqnarray}
\begin{array}{lr}
\frac{d\hat{A}_{1}}{dt}=-
\lambda_{1} \hat{A}_{2}^{\dagger}-\lambda_{2} \hat{A}_{3},\\
\\
\frac{d\hat{A}_{2}}{dt}=-\lambda_{1} \hat{A}_{1}^{\dagger}
-\lambda_{3} \hat{A}_{3},\\
\\
\frac{d\hat{A}_{3}}{dt}=\lambda_{2} \hat{A}_{1} +\lambda_{3} \hat{A}_{2}.
\label{scs11}
\end{array}
\end{eqnarray}

\noindent
These are  three equations  with their Hermitian conjugates
forming a closed system which can be easily solved, restricting ourselves to
the case
$\lambda_{2}=\lambda_{3}=\frac{\lambda_{1}}{\sqrt{2}}$, to avoid the complexity
in the calculations; this means that we consider stronger  nonlinearity
in the first waveguide compared with the linear exchange between
waveguides, thereby having the solution

\begin{eqnarray}
\begin{array}{lr}
\hat{A}_{1}(t)=\hat{a}_{1}(0)f_{1}(t) +\hat{a}^{\dagger}_{1}
(0)f_{2}(t)-\hat{a}_{2}(0)f_{3}(t)\\
- \hat{a}_{2}^{\dagger}(0)f_{4}(t)-\hat{a}_{3}(0)f_{5}(t)-
\hat{a}^{\dagger}_{3}
(0)f_{6}(t),\\
\\
\hat{A}_{2}(t)=\hat{a}_{2}(0)g_{1}(t) +\hat{a}^{\dagger}_{2}
(0)g_{2}(t) -\hat{a}_{1}(0)g_{3}(t)
\\
- \hat{a}_{1}^{\dagger}(0)g_{4}(t)-\hat{a}_{3}(0)g_{5}(t)-\hat{a}^{\dagger}_{3}
(0)g_{6}(t),\\
\\
 \hat{A}_{3}(t)=\hat{a}_{3}(0)h_{1}(t) +\hat{a}^{\dagger}_{3}(0)h_{2}(t)
 +\hat{a}_{2}(0)h_{3}(t)\\
+ \hat{a}_{2}^{
\dagger}(0)h_{4}(t)+\hat{a}_{1}(0)h_{5}(t)+\hat{a}^{\dagger}_{1}(0)h_{6}(t) ,
\label{scs12}
\end{array}
\end{eqnarray}
where  the time-dependent coefficients,
i.e.  $f_{j}(t),g_{j}(t),h_{j}(t)$,   including all information about
 the system,  are

\begin{picture}(70,70)(50,20)
\put (107,37){\vector(1,0){223}$\hat{a}_{1}(\frac{L}{v})$}

\put (107,60){\vector(1,0){223}$\hat{a}_{2}(\frac{L}{v})$}
\put (120,70){\makebox(0,0){$\hat{a}_{2}(0)$}}
\put (120,45){\makebox(0,0){$\hat{a}_{1}(0)$}}
\put (107,-20){\vector(1,0){223}$\hat{a}_{3}(\frac{L}{v})$}
\put (120,-10){\makebox(0,0){$\hat{a}_{3}(0)$}}

\put (300,-14){\vector(0,1){77}}
\put (200,0){\vector(0,1){40}}

\put (300,0){\vector(0,-1){20}}
\put (200,0){\vector(0,-1){20}}

\put (310,20){\framebox(0,0){$\lambda_{3}$}}
\put (210,20){\framebox(0,0){$\lambda_{2}$}}

\put (178,-50){\framebox(140,40){$\chi^{(1)}$}}
\put (178,30){\framebox(140,40){$\chi^{(2)}\quad\lambda_{1}$}}

\put (248,-100){\vector(1,0){70}}
\put (248,-100){\vector(-1,0){70}}
\put (248,-90){\framebox(0,0){$L$}}

\put (178,30){\line(0,-1){10}}
\put (178,10){\line(0,-1){10}}
\put (178,-10){\line(0,-1){10}}
\put (178,-30){\line(0,-1){10}}
\put (178,-50){\line(0,-1){10}}
\put (178,-70){\line(0,-1){10}}
\put (178,-90){\line(0,-1){10}}

\put (318,30){\line(0,-1){10}}
\put (318,10){\line(0,-1){10}}
\put (318,-10){\line(0,-1){10}}
\put (318,-30){\line(0,-1){10}}
\put (318,-50){\line(0,-1){10}}
\put (318,-70){\line(0,-1){10}}
\put (318,-90){\line(0,-1){10}}

\put (120,10){\line(1,0){20}}
\put (150,10){\line(1,0){20}}
\put (180,10){\line(1,0){20}}
\put (210,10){\line(1,0){20}}
\put (240,10){\line(1,0){20}}
\put (270,10){\line(1,0){20}}
\put (300,10){\line(1,0){20}}
\put (330,10){\line(1,0){20}}

\end{picture}

\vspace{1.7in}

\noindent Figure 6.7: Scheme of realization of interaction in (\ref{scs10}) using
a nonlinear asymmetric directional coupler which is composed of two optical
waveguides fabricated from first-order ($\chi^{(1)}$) and second-order
($\chi^{(2)}$) materials, where $\chi$ designate susceptibility. Signal
mode 1 and idler mode 2 propagate in the first waveguide
and  linear mode 3 in the second waveguide. The interaction between
the signal and idler modes
 is established by strong pump coherent light, which is  not indicated in
  figure, with the coupling constant $\lambda_{1}$.
The interactions between the linear mode  and the signal and idler
are established linearly with the  coupling constants $\lambda_{2}$
and $\lambda_{3}$, respectively.
\vspace{.5cm}

\begin{eqnarray}
\begin{array}{lr}
f_{1,3}(t)=
\frac{1}{2} \left[
\cosh (\lambda_{1}t)
\pm \cosh (\frac{\lambda_{1}t}{2})\cos (\bar{k}t)
\pm\frac{1}{\sqrt{3}}
\sinh (\frac{\lambda_{1}t}{2})\sin (\bar{k}t)
\right],\\
\\
f_{2,4}(t)=
\frac{1}{2} \left[
\sinh (\lambda_{1}t)
\mp\sinh (\frac{\lambda_{1}}{2}t)\cos (\bar{k}t)
\mp\frac{1}{\sqrt{3}}
\cosh (\frac{\lambda_{1}t}{2})\sin (\bar{k}t)
\right],\\
\\
f_{5}(t)=
\sqrt{\frac{2}{3}}
\cosh (\frac{\lambda_{1}t}{2})\sin (\bar{k}t)
,\quad
f_{6}(t)=
-\sqrt{\frac{2}{3}}
\sinh (\frac{\lambda_{1}t}{2})\sin (\bar{k}t),\\
\\
h_{1}(t)=
 \cosh (\frac{\lambda_{1}t}{2})\cos (\bar{k}t)
-\frac{1}{\sqrt{3}}
\sinh (\frac{\lambda_{1}t}{2})\sin (\bar{k}t),\\
\\
h_{2}(t)=
- \sinh (\frac{\lambda_{1}t}{2})\cos (\bar{k}t)
+\frac{1}{\sqrt{3}}
\cosh (\frac{\lambda_{1}t}{2})\sin (\bar{k}t),\\
\\
h_{3,5}(t)= \sqrt{\frac{2}{3}}
\cosh (\frac{\lambda_{1}t}{2})\sin (\bar{k}t)
,\quad
h_{4,6}(t)= -\sqrt{\frac{2}{3}}
\sinh (\frac{\lambda_{1}t}{2})\sin (\bar{k}t),
\label{scs13}
\end{array}
\end{eqnarray}
 where $\bar{k}=\frac{\sqrt{3}}{2}\lambda_{1}$ and
 the expressions for $g_{j}(t)$  are the same  as those of
 $f_{j}(t)$.

In fact, the nature of the solution can show how the coupler does work.
To be more specific, the time-dependent coefficients  contain both
trigonometric and hyperbolic functions.
Consequently, the propagating beams inside the coupler
can be amplified as well as switched between waveguides, i.e. between
modes, in the course of time.

On the basis of the well known commutation rules for boson operators,
the following
relations can be proved for the time-dependent coefficients

\begin{eqnarray}
\begin{array}{lr}
f^{2}_{1}(t)-f^{2}_{2}(t)+f^{2}_{3}(t)-f^{2}_{4}(t)+f^{2}_{5}(t)
-f^{2}_{6}(t)=1, \quad
f_{1}(t)g_{4}(t)-f_{2}(t)g_{3}(t)\\
\\
+f_{3}(t)g_{2}(t)-f_{4}(t)g_{1}(t)-f_{5}(t)g_{6}(t)
+f_{6}(t)g_{5}(t)=0,\\
\\
f_{1}(t)g_{3}(t)
-f_{2}(t)g_{4}(t)+f_{3}(t)g_{1}(t)-f_{4}(t)g_{2}(t)-f_{5}(t)g_{5}(t)
+f_{6}(t)g_{6}(t)=0.
\label{scs14}
\end{array}
\end{eqnarray}
 The remaining relations  can be obtained from (\ref{scs14})
by means of the following transformations

\begin{eqnarray}
\begin{array}{lr}
\Bigl(f_{1}(t),f_{2}(t),f_{3}(t)
,f_{4}(t),f_{5}(t),f_{6}(t)\Bigr)
\\
\longleftrightarrow
\Bigl(-g_{3}(t),-g_{4}(t),-g_{1}(t)
,-g_{2}(t),g_{5}(t),g_{6}(t)\Bigr)
\\
\longleftrightarrow
\Bigl(h_{5}(t),h_{6}(t),-h_{3}(t)
,-h_{4}(t),-h_{1}(t),-h_{2}(t)\Bigr).
\label{scs15}
\end{array}
\end{eqnarray}

Based on these results,
we can study the quantum properties of the evolution of  different
modes in the model  when they are
initially prepared  in coherent states. This will be demonstrated
in the following parts by means of quasiprobability functions.

\subsection{Quasiprobability functions}

In the following we consider  phase
space distributions  for the single-mode
 when all modes   are initially prepared in coherent  states
 $|\alpha_{1},\alpha_{2},\alpha_{3}\rangle$
 before
 entering the coupler.
We start from knowledge of the single-mode $s$-parametrized characteristic
function which is suitable to describe the quantum statistics of light
and for the first  mode it
takes the form

\begin{equation}
C_{1}(\zeta,s,t)=\exp\left\{ \frac{1
}{2}|\zeta|^{2}[s-|\eta _{1}(t)|]+
|\eta _{2}(t)| (\zeta^{2}+\zeta^{*2})
+[\bar{\alpha}_{1}^{*}(t)\zeta
-\bar{\alpha}_{1}(t)\zeta ^{*}]\right\},
\label{scs16}
\end{equation}
where
 $\eta
_{j}(t),\quad j=1,2$, are given by

\begin{eqnarray}
\begin{array}{lr}
\eta _{1}(t)=f^{2}_{1}(t)+f^{2}_{2}(t)+f^{2}_{3}(t)
+f^{2}_{4}(t)+f^{2}_{5}(t)+f^{2}_{6}(t),\\
\\
\eta _{2}(t)=f_{1}(t)f_{2}(t) +f_{3}(t)f_{4}(t)+f_{5}(t)f_{6}(t).
\label{scs17}
\end{array}
\end{eqnarray}

Now the $s$-parametrized single-mode  quasiprobability  functions of
the signal mode, which can be derived in a similar way as we did in
the previous section, are

\begin{eqnarray}
\begin{array}{lr}
W_{1}(\beta,s,t)=\frac{2}{\pi\sqrt{[A_{+}(t)-s]
[A_{-}(t)-s]}}\\
\\
\times \exp\left\{ \frac{[l_{1}(t)-l^{*}_{1}(t)]^{2}}{2[A_{-}(t)-s]}
-\frac{[l_{1}(t)+l^{*}_{1}(t)]^{2}}{2[A_{+}(t)-s]}
\right\} ,
\label{scs18}
\end{array}
\end{eqnarray}
where $l _{1}(t)=\bar{\alpha}_{1}(t)-\beta$
($\bar{\alpha}_{1}(t)$ has the same meaning as before) and

\begin{equation}
A _{\pm}(t)=[f_{1}(t)\pm f_{2}(t)]^{2}+[f_{3}(t)\pm f_{4}(t)]^{2}
+[f_{5}(t)\pm f_{6}(t)]^{2};
 \label{scs19}
\end{equation}
the other expressions for the idler and linear modes can be
obtained from (\ref{scs16}) and (\ref{scs18}) by making use of the
transformations (\ref{scs15}); we except here the case $s=1$, which will be
discussed  shortly.
Now let us start our investigation by demonstrating the behaviour
of the $W$-function, i.e. setting $s=0$ in (\ref{scs18}).
In the language of mechanical analogy of a harmonic oscillator with dynamical
conjugate variables $x$ and $y$, i.e $\beta=x+iy$, the $W$-function can be
written in the form

\begin{equation}
W_{1}(x,y,t)=\frac{1}{2\pi\sqrt{
\langle (\triangle\hat{X}(t))^{2}\rangle
\langle (\triangle\hat{Y}(t))^{2}\rangle
}}\exp\left\{- \frac{[x-\bar{\alpha}_{x}(t)]^{2}}{
2\langle (\triangle\hat{X}(t))^{2}\rangle}
-\frac{[y-\bar{\alpha}_{y}(t)]^{2}}{2
\langle (\triangle\hat{Y}(t))^{2}\rangle}
\right\},
\label{scs20}
\end{equation}
where

\begin{equation}
\langle (\triangle\hat{X}(t))^{2}\rangle=\frac{A_{+}(t)}{4},\quad
\quad \quad
\langle (\triangle\hat{Y}(t))^{2}\rangle=\frac{A_{-}(t)}{4},
\label{scs21}
\end{equation}
and
$\bar{\alpha}_{1}(t)=\bar{\alpha}_{x}(t)+i\bar{\alpha}_{y}(t)$;
$\langle (\triangle\hat{X}(t))^{2}\rangle$
 and $\langle (\triangle\hat{Y}(t))^{2}\rangle$ are   single-mode $x$- and
 $y$-quadrature variances, respectively.

Now from (\ref{scs20}) and (\ref{scs21}) together with (\ref{scs13})
one can easily prove that the quadrature variances of the linear mode
(single-mode squeezing) are

\begin{equation}
\langle (\triangle \hat{X}(t))^{2}\rangle=
\frac{1}{4}\left\{\frac{4}{3}\sin ^{2}(\bar{k}t)+
[\cos (\bar{k}t)+\frac{1}{\sqrt{3}}\sin (\bar{k}t)]^{2}\right\}
\exp (-\lambda_{1} t),
\label{scs22}
\end{equation}

\begin{equation}
\langle (\triangle \hat{Y}(t))^{2}\rangle=
\frac{1}{4}\left\{
\frac{4}{3}\sin ^{2}(\bar{k}t)+
[\cos (\bar{k}t)-
\frac{1}{\sqrt{3}}\sin (\bar{k}t)]^{2}\right\}
\exp (\lambda_{1} t),
\label{scs23}
\end{equation}
where $\bar{k}$ has the same meaning as before.
Thus  squeezing can be achieved
in the linear waveguide in the $X$-quadrature.
Moreover, it is clear
that  squeezing values become more pronounced for a large interaction time,
i.e. for long length of the coupler $L$.
This behaviour shows that changing the power of the
linear interaction it is possible to transfer nonclassical properties,
which are generated in the nonlinear waveguide, to the linear signal mode.
For completeness, the uncertainty relation reads

\begin{equation}
\langle (\triangle\hat{X}(t))^{2}\rangle
\langle (\triangle\hat{Y}(t))^{2}\rangle=\frac{1}{16}[1
+\frac{16}{9}\sin ^{4}(\bar{k}t)].
\label{scs24}
\end{equation}
This formula reveals  that
the minimum-uncertainty relation  holds only when
$t=t_{s}=\frac{2m\pi}{\sqrt{3}\lambda_{1}}$ where $m$ is  positive
integer and we call $t_{s}$ a squeeze time.
In this
case the device provides the squeezed coherent light in the
linear mode with squeeze
parameter  $r=\frac{2m\pi}{\sqrt{3}}$.
It is important to mention that at $t=t_{s}$ the two waveguides become
completely independent (cf. equations (\ref{scs13})) and  the signal and idler modes
 can  display perfect two-mode
squeezing, which one can easily check with the help of two-mode quadrature
variances. This can be explained as follows: A portion of the energy
always remains within the guide into which the field was initially
injected. This energy grows in the course of time until $t=t_{s}$, when the
two guides are completely independent and hence the modes are trapped in
their own guides. So that  the linear mode can
yield the well known
single-mode squeezed light in the linear waveguide having its origin in
the nonlinear waveguide in which  the nonlinearity
 produces two-mode squeezing.
For later times $t\neq t_{s}$, the device generates
 single-mode squeezed light  in the linear waveguide where
the corresponding $W$-function may be broader than that of the idealized squeezed
coherent states \cite{[6]}.   This situation is close
to that of a two-photon absorber (a two-photon absorption by a reservoir of
two-level atoms from a single mode of the electromagnetic field \cite{gil1})
where a squeezed state, which is not a minimum-uncertainty state, has been
generated \cite{gil}. The origin of squeezing of initially unsqueezed light
interacting with two-photon absorbers is that the squeezing is generated
by simple quantum superposition of states of light \cite{gil1}.

We proceed in our discussion by focusing our attention on the
$P$-function.
We have shown earlier that this system
is able to provide squeezed light and this should be reflected in the
 behaviour of the $P$-representation, i.e. setting $s=1$ in (\ref{scs18}).
The significant example for this situation is the behaviour of the
linear mode. For this case, we can  show that the single-mode quadrature
squeezing is established  provided that $A_{+}(t)-1<0$ and
$A_{-}(t)-1>0$.  It is evident that the $P$-function is not
well-behaved function in this case and this is the indication of
the nonclassical fields. Indeed, if we look at the model under consideration
of a competition of parametric processes, we can find that
this result is in contrast to the  results in the
literature for simpler systems of interacting modes starting from coherent states
\cite{[3],{entan1},{entan2},{entan5},{abd1}}, where the delta function of the initial
$P$-representation of coherent light  for a single mode
becomes well-behaved distribution  during the interaction.

In the following   we  use the phase space distribution
 functions to study the photon-number distribution and phase
 distribution for the system under discussion.

\subsection{ Photon-number distribution}

Photon-number distribution, i.e. the probability of finding  $n_{1}$ photons
in the signal mode  at time $t$,  can be obtained in the photodetection
process, and can be determined by means of the relation (\ref{19})
 as

\begin{eqnarray}
\begin{array}{lr}
P(n_{1},t)=\frac{2}{\sqrt{[A_{+}(t)+1][A_{-}(t)+1] }}
\exp \left\{ \frac{[\bar{\alpha}_{1}(t)-\bar{\alpha}^{*}_{1}(t)]^{2}}{
2[A_{-}(t)+1]}
-\frac{[\bar{\alpha}_{1}(t)+\bar{\alpha}^{*}_{1}(t)]^{2}}{2[A_{+}(t)+1]}
\right\}\\
\\
\times \sum^{n_{1}}_{r=0}
\left[ \frac{A_{+}(t)-1}{A_{+}(t)+1}\right]^{r}
\left[ \frac{A_{-}(t)-1}{A_{-}(t)+1}\right]^{n_{1}-r}
\\
\\
\times
{\rm L}^{-\frac{1}{2}}_{r}\left[
-\frac{(\bar{\alpha}_{1}(t)+ \bar{\alpha}^{*}_{1}(t) )^{2}}
{(1+A_{+}(t))(A_{+}(t)-1)}\right]
 {\rm L}^{-\frac{1}{2}}_{n_{1}-r}\left[
\frac{(\bar{\alpha}_{1}(t)- \bar{\alpha}^{*}_{1}(t) )^{2}}
{(1+A_{-}(t))(A_{-}(t)-1)}\right],
\label{scs25}
\end{array}
\end{eqnarray}
where $L^{\nu}_{n_{1}}(.)$ is the associated Laguerre polynomial and the
integration has been carried out using the same technique as in appendix A.

It is convenient to
compare this distribution with the corresponding Poisson distribution

\begin{equation}
P(n_{1},t)=\frac{\langle \hat{n}_{1}(t)\rangle ^{n_{1}}}{n_{1}!}\exp
[-\langle \hat{n}_{1}(t)\rangle ],
\label{scs26}
\end{equation}
which corresponds to fully coherent field with the same mean photon number
$\langle \hat{n}_{1}(t)\rangle$.
 The  expressions for the idler and linear modes should
 be obtained from (\ref{scs25}) using the transformations (\ref{scs15}).
\setcounter{figure}{7}
\begin{figure}[h]%
 \centering
    \includegraphics[width=6cm]{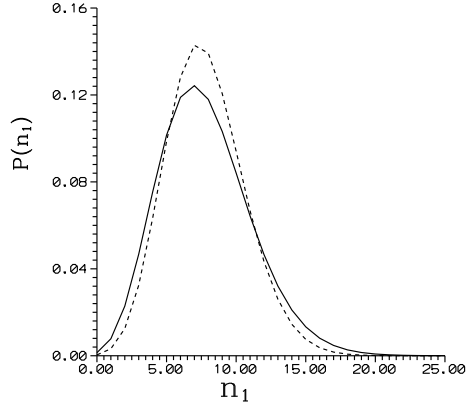}
   \caption{
Photon-number distribution $P(n_{1})$ for the signal mode  for
$\alpha_{j}=3\exp (i\frac{\pi}{3}), j=1,2,3, \quad t=1.5$ and
$\lambda_{1}=0.3$. The dashed curve is the Poisson photon-number
distribution.} \label{fig25}
\end{figure}


In Fig. 6.8, we have plotted  $P(n_{1},t=1.5)$ against
$n_{1}$ for the signal mode  (solid curve) for $\alpha_{j}=
3\exp (i\frac{\pi}{3})$ and $\lambda_{1}=0.3$.
For the sake of comparison, the corresponding
photon distribution for coherent field, (\ref{scs26}), is shown by dashed curve.
For all numerical calculations $\sum_{n_{1}=0}^{\infty}P(n_{1},t)=1$ with good
accuracy. We noted for the signal mode  that the photon-number distribution
exhibits always one-peak structure. From this figure we see that the behaviour
is rather super-Poissonian as a result of quantum fluctuations because
the solid curve is always broader than the corresponding dashed curve.
So that, in general the Poissonian light evolves in the nonlinear waveguide
as super-Poissonian light. On the other hand,  we have realized earlier
from the behaviour of $W$-function
that squeezed states can be generated. However, these states
are exhibiting oscillating photon-number distribution for certain values
of squeeze parameter.
\begin{figure}[h]%
  \centering
  \subfigure[]{\includegraphics[width=5cm]{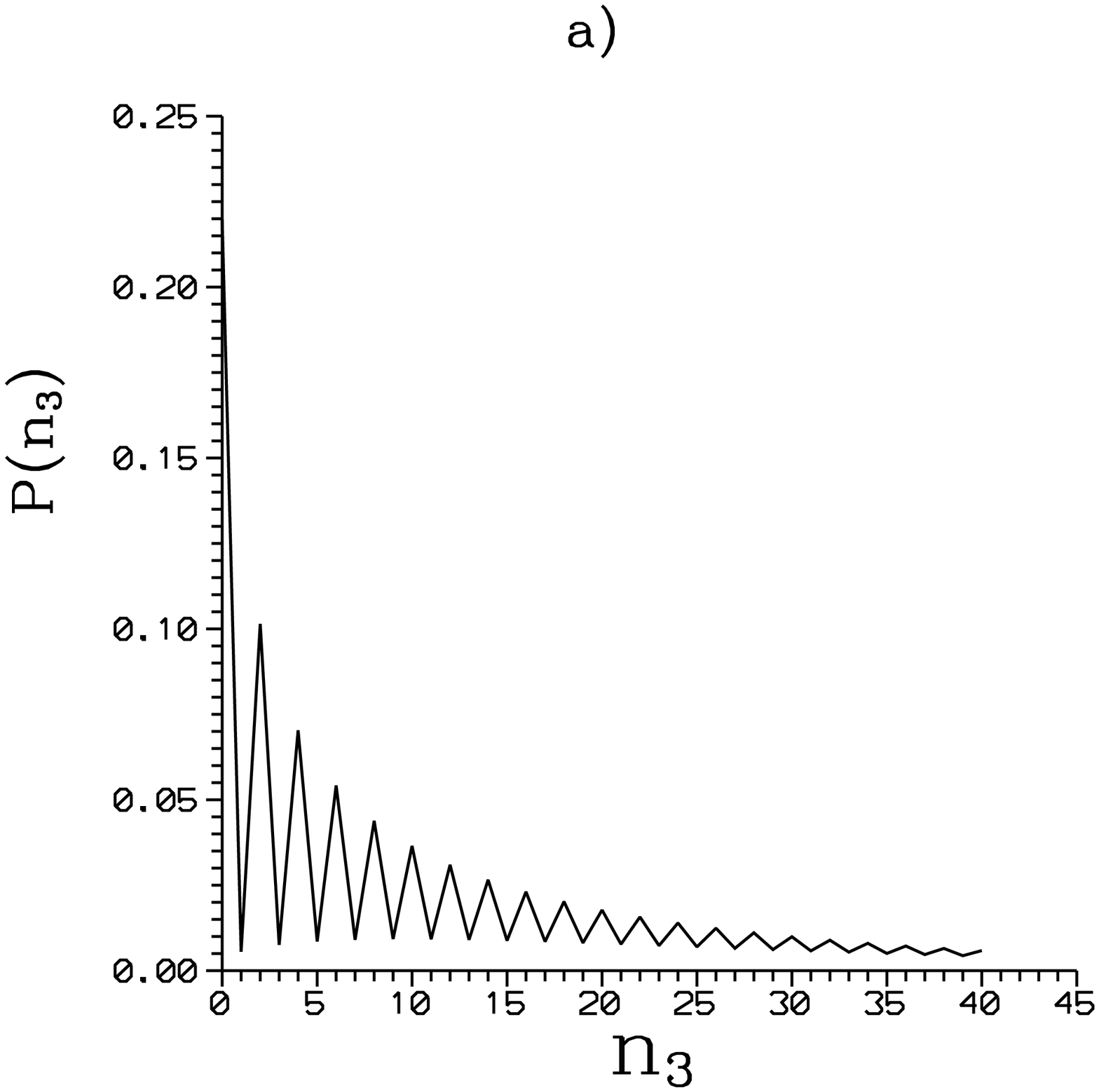}}
 \subfigure[]{\includegraphics[width=5cm]{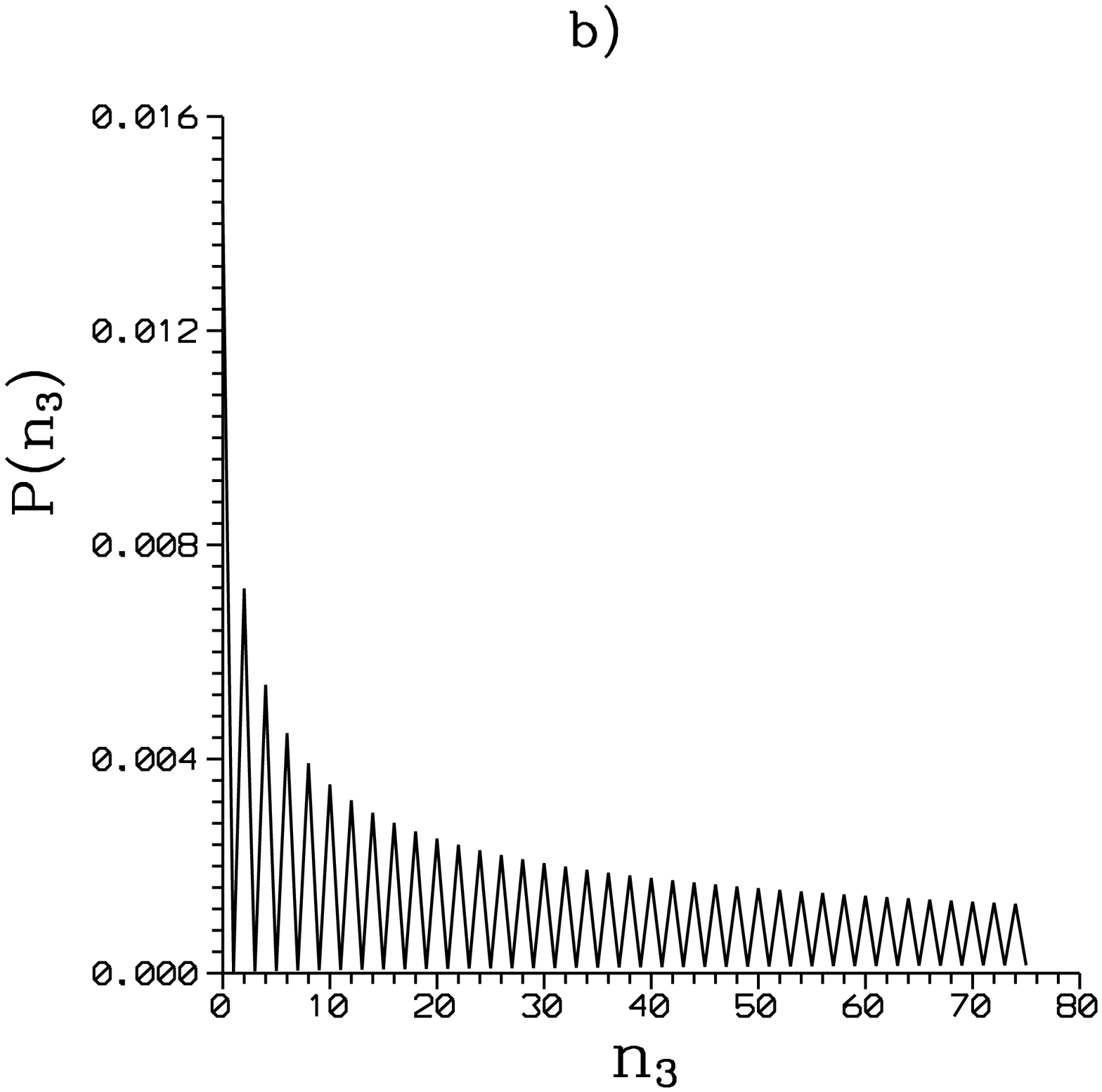}}
      \caption{
Photon-number distribution $P(n_{3})$ for the linear mode  when $
\alpha_{j}=0.5\exp(i\frac{\pi}{3}), j=1,2,3; \lambda_{1}=1$ and a)
$t=t_{s}$ ; b) $t=5$. } \label{fig26}
 \end{figure}

Indeed, these oscillations  can be recognized for the  linear mode
where this mode  can display squeezed light
(see Fig. 6.9, for shown values of parameters).  From this figure we see  the
macroscopic oscillations of the photon-number distribution for squeezed light.
Further, comparison of Figs. 6.9a and 6.9b shows that the nonclassical
oscillations of $P(n_{3})$ are faster and more pronounced when $t$ increases.
It is worthwhile mentioning that the case of Fig. 6.9a corresponds to
squeezed light with squeeze parameter
$r\simeq 3.6$.

In fact, the behaviour of the photon-number distribution in this device is
slightly different from that of the nonlinear asymmetric coupler with
strong classical stimulating light in the second harmonic mode
\cite{qu6} where oscillatory behaviour as well as sub-Poissonian
statistics for specific modes in both linear and nonlinear waveguide are
exhibited.

\subsection {Phase distribution}

We  make use of the single-mode $Q$-function to investigate the phase
distribution for the system under discussion using (4.20).
The result of the integration has the form

\begin{eqnarray}
\begin{array}{lr}
P(\theta,t)=\frac{1}{\pi C(t)\sqrt{A_{+}(t)A_{-}(t)}}\exp\left\{
\frac{[\bar{\alpha}_{1}(t)-\bar{\alpha}^{*}_{1}(t)]^{2}}{2[A_{-}(t)+1]}
-\frac{[\bar{\alpha}_{1}(t)+\bar{\alpha}^{*}_{1}(t)]^{2}}{2[A_{+}(t)+1]}\right\}
\\
\\
\times \left\{ 1+\frac{B(t)}{2}\sqrt{\frac{\pi}{C(t)}}
\exp[\frac{B(t)^{2}}{4C(t)}][1+{\rm erf}(\frac{B(t)}{2\sqrt{C(t)}})]\right\},
\label{scs27}
\end{array}
\end{eqnarray}
where

\begin{eqnarray}
\begin{array}{lr}
C(t)=\frac{2}{A_{-}(t)A_{+}(t)}[A_{+}(t)\sin ^{2}\theta+A_{-}(t)
\cos ^{2}\theta],
\\
B(t)=\frac{2}{A_{-}(t)A_{+}(t)}\left[ A_{-}(t)
\left(\bar{\alpha}_{1}(t)+\bar{\alpha}^{*}_{1}(t)
\right)\sin\theta-iA_{+}(t)
\left(\bar{\alpha}_{1}(t)-\bar{\alpha}^{*}_{1}(t)\right)
\cos\theta\right],
\label{scs28}
\end{array}
\end{eqnarray}
and ${\rm erf}(.)$ is the Gauss error function;
$A_{\pm}(t)=4\langle (\triangle\hat{X}_{\pm}(t))^{2}\rangle$
with $\langle (\triangle\hat{X}(t))^{2}\rangle$
 and $\langle (\triangle\hat{Y}(t))^{2}\rangle$ are   single-mode $x$- and
 $y$-quadrature variances, respectively;
$\bar{\alpha}_{1}(t)$ is
the expectation value of the signal mode operator in  coherent state.

\begin{figure}[h]%
  \centering
  \subfigure[]{\includegraphics[width=5cm]{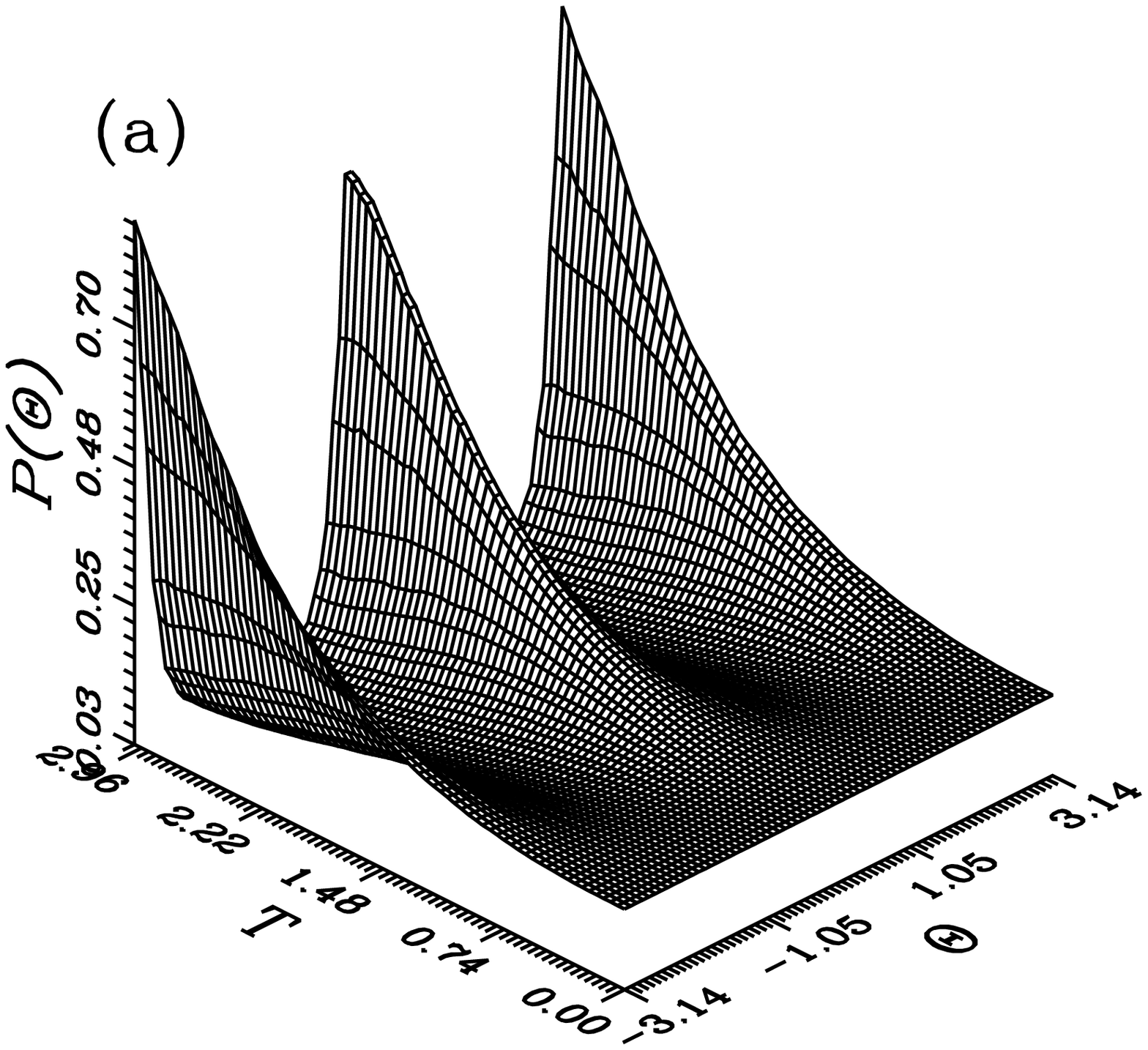}}
 \subfigure[]{\includegraphics[width=5cm]{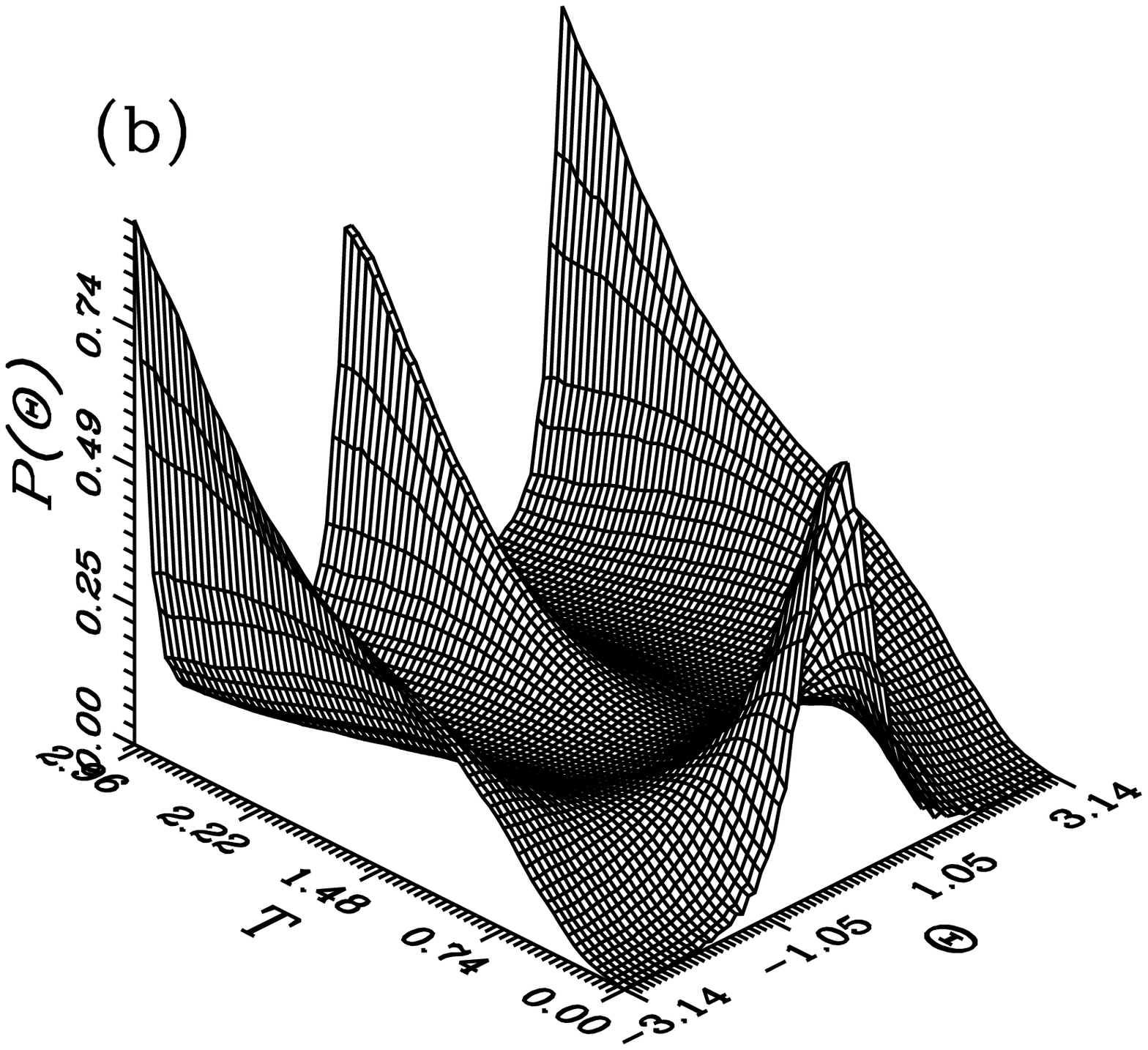}}
      \caption{
Phase distribution $P(\theta,t)$ for the signal mode  against
$\theta$ and $t$ for $\lambda_{1}=0.6$
 and a) $\alpha_{j}=0$; b) $\alpha_{j}=1,\quad j=1,2,3$.
} \label{fig27}
 \end{figure}


The behaviour of this distribution  in this device
can be understood on the basis of the competition between the two-peak
structure for the vacuum states and the single-peak structure for coherent
states provided that the coherent amplitudes are real. Such behaviour is
demonstrated in Figs. 6a,b for the signal mode and in Figs. 7a,b for
the linear mode, where the evolution of the phase distribution (\ref{scs27})
has been depicted against the phase $\theta$ and the time $t$.

\begin{figure}[h]%
  \centering
  \subfigure[]{\includegraphics[width=5cm]{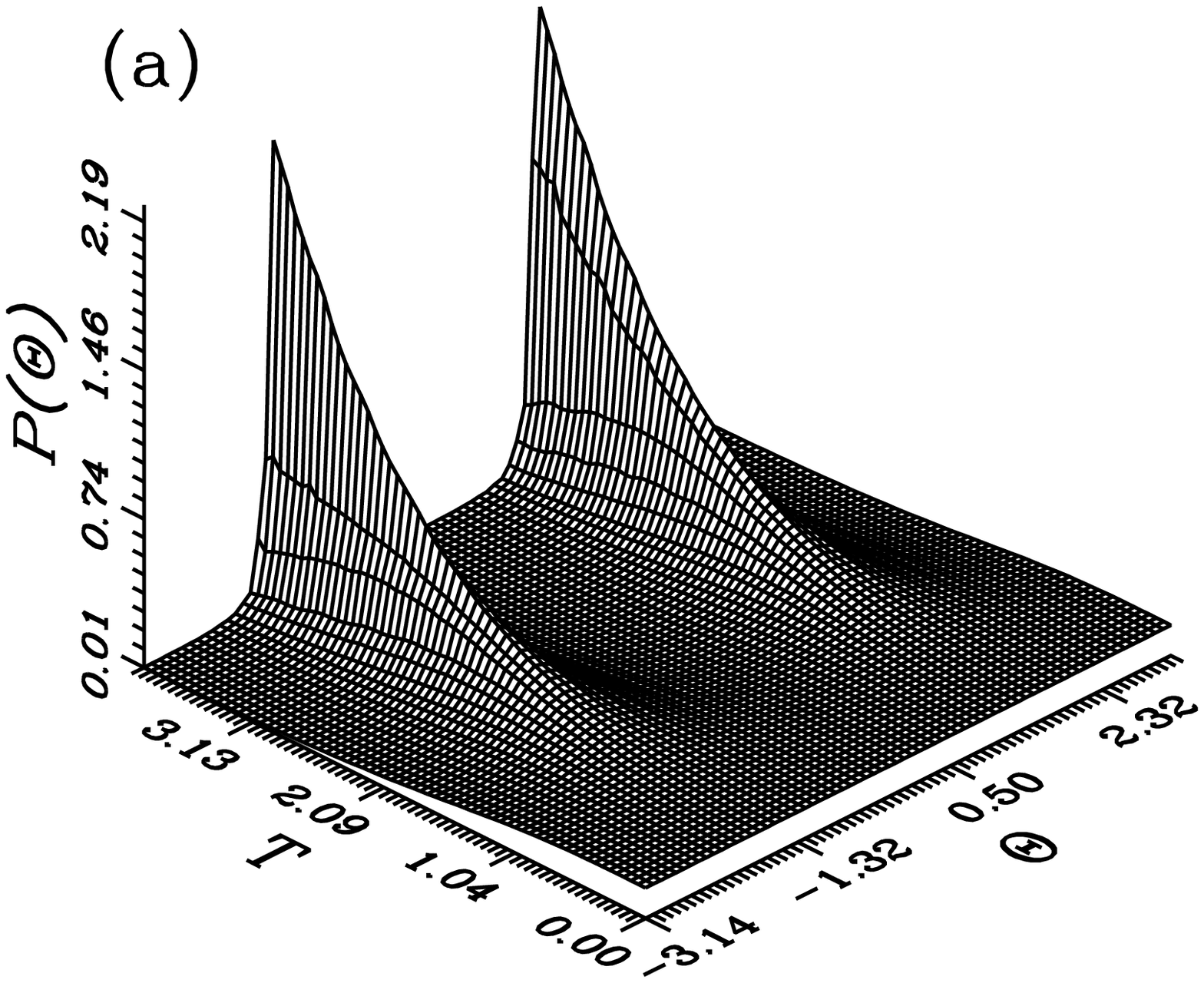}}
 \subfigure[]{\includegraphics[width=5cm]{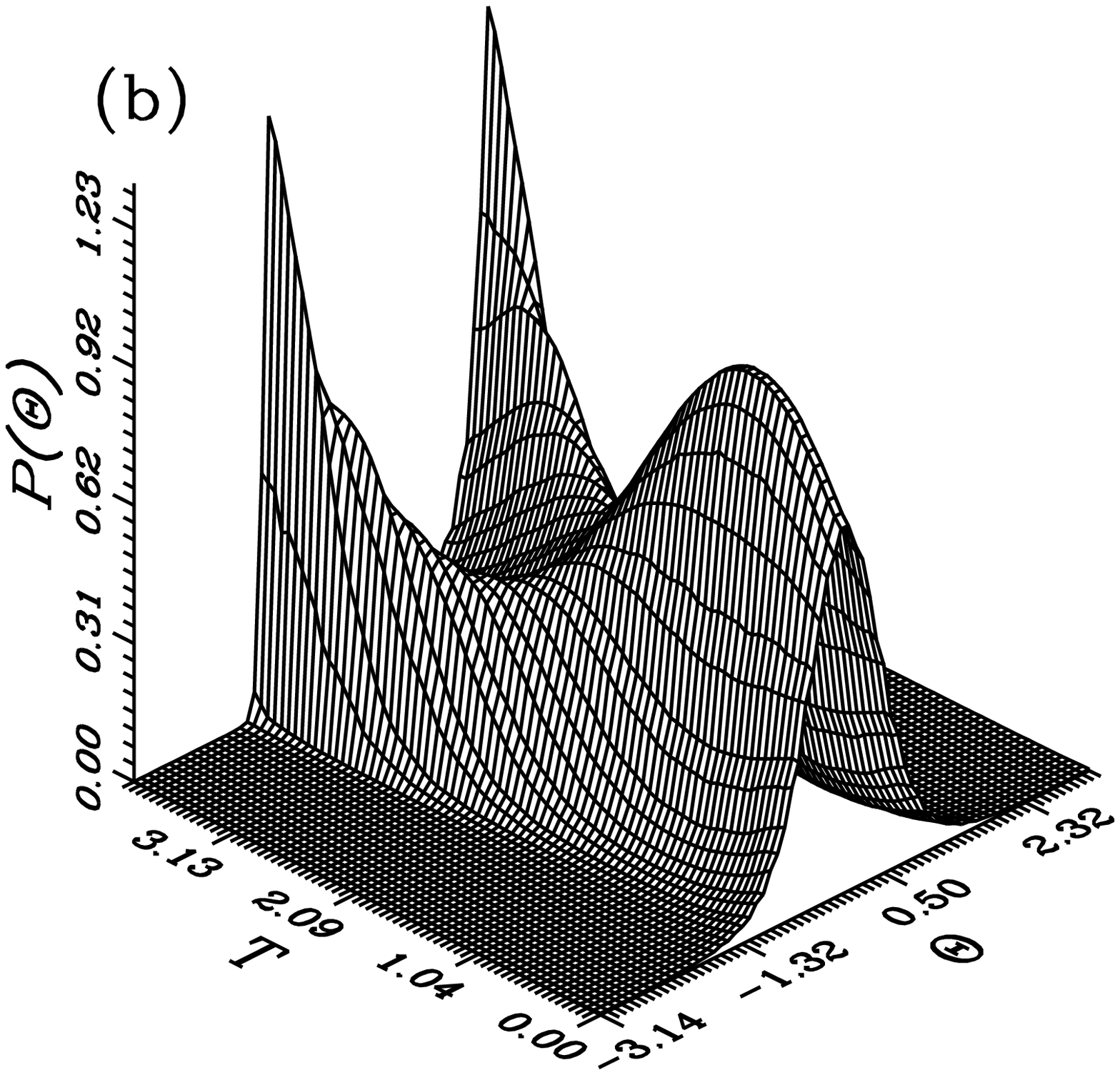}}
      \caption{
Phase distribution $P(\theta,t)$ for the linear mode  against
$\theta$ and $t$ for the same situation as in Figs. 6.10. }
\label{fig28}
 \end{figure}

On the other hand, the well-known
behaviour for the phase distribution of squeezed states may be established
when we focus our attention on the behaviour of the linear mode, as expected
(see Figs. 6.11a and b). From Fig. 6.11a one can see that the two-peak
structure is typical for squeezed vacuum states, i.e. we have two-peak
structure for $\theta=\pm\frac{\pi}{2}$ \cite{phas15,{phas16}}. Further, the
heights of these peaks are $\frac{1}{2\pi}\sqrt{\frac{A_{-}(t)}{A_{+}(t)}}$
at any time $t\quad (t>0)$. However, Fig. 6.11b displays the well-known
bifurcation shape for the phase distribution of squeezed field, i.e. the
distribution curve undergoes a transition from single- to a
double-peaked form with increasing time. Indeed this figure qualitatively is
quite similar
to that for idealized squeezed coherent states \cite{phas11}, and the distributions
differ in the behaviour of the initial peak, which is here amplified for
a while before splitting into two peaks. This is connected with the
significant influence of the development by nonlinear effects and power
transfers between waveguides. It is worthwhile  mentioning that a similar
behaviour has been obtained for a contradirectional nonlinear asymmetric
coupler with strong stimulated coherent field in the second harmonic
waveguide \cite{qu16}.

Finally, we would like to conclude this part by discussing the origin of
such behaviour of the  phase distribution. This can be easily understood by
analyzing the function $P(\theta,t)$ when $\alpha_{j}=0, \quad j=1,2,3$.
In this case the formula (\ref{scs27}) reduces to

\begin{equation}
P(\theta,t)=\frac{1}{2\pi}\frac{\sqrt{A_{+}(t)A_{-}(t)}}{
A_{+}(t)\sin ^{2}\theta+A_{-}(t)
\cos ^{2}\theta}.
\label{scs29}
\end{equation}
This distribution function is of double- or three-peak structure
according to the relation between $A_{+}(t)$ and $A_{-}(t)$.
Let us restrict our discussion to the one of the modes which are propagating
in the nonlinear waveguide, say, to the signal mode. For this mode, it can
easily be shown that
$A_{+}(t)>A_{-}(t)$ and consequently the formula (\ref{scs29}) exhibits
three-peak structure with peaks for $\theta=0,\pm \pi$. The heights of
these peaks at any time $t\neq 0$ are
$\frac{1}{2\pi}\sqrt{\frac{A_{+}(t)}{A_{-}(t)}}$. For non-zero
displacements $\alpha_{j}$, an additional factor of the form for coherent
state, but with a coherent amplitude $\bar{\alpha}_{1}(t)$ which is
the expectation value of the signal mode operator in  coherent state,
appears in the distribution.
This factor is responsible for a peak at $\theta=0$ related to coherent
component, which  competes with the
three-peak structure of vacuum to provide the previous behaviour of the
distribution. Similar argument can be adopted to explain the behaviour of the
linear mode, however, it should be borne in mind that in this case
$A_{+}(t)<A_{-}(t)$,  as we have shown earlier for squeezing phenomenon,
and hence the two-peak structure for the input vacuum case is dominant at
$\theta=\pm\frac{\pi}{2}$.

\subsection{Conclusion}

In this section we have shown that squeezed light
can be generated  in a nonlinear asymmetric directional
coupler
composed of a linear waveguide and a nonlinear waveguide operating by
nondegenerate parametric amplification.
After using the Heisenberg approach to the quantum statistics of
interacting modes, we have investigated such an effect in the linear mode,
propagating in the linear waveguide, in terms
of the quasiprobability functions, photon-number distribution and phase
distribution when the modes are initially in coherent states.
Concerning the behaviour of Glauber $P$- and Wigner $W$-functions for the
linear mode we can conclude that the former is not well-behaved function
when $t>0$, while, the latter displays the well-known behaviour for squeezed
light in which  one of the quadratures is amplified  and the other is
attenuated. For the photon-number distribution, the large scale macroscopic
oscillations related to squeezed light are established.
The phase distribution displays the bifurcation typical for squeezed field.
In fact, this behaviour shows that changing the power of the
linear interaction, it is possible to transfer nonclassical properties,
which are generated in the nonlinear waveguide, to the linear signal
mode. Further, we have shown, in general, that the system has less tendency
to generate sub-Poissonian light from  initial coherent light.

\section{ Quantum statistical properties of nondegenerate optical parametric
symmetric coupler }

In this section  we shall concentrate on studying the statistical
properties of an optical field propagating within a directional coupler
containing a parametric amplifying medium. Our starting point is the
Hamiltonian, which represents a nonlinear directional coupler composed
of two nonlinear waveguides (taking into account the case of a strong pump,
where non-depleting intensity is included into the amplifier coupling
constant). This model can be described by an effective  Hamiltonian
as

\begin{equation}
\frac{H}{\hbar }=\lambda_{1}(\hat{a}_{1}\hat{a}_{3}+\hat{a}_{1}^{\dagger }
\hat{a}_{3}^{\dagger })+\lambda _{2}(\hat{a}_{1}\hat{a}_{2}^{\dagger
}+\hat{a
}_{1}^{\dagger }\hat{a}_{2})+\lambda
_{3}(\hat{a}_{2}\hat{a}_{4}+\hat{a}
_{2}^{\dagger }\hat{a}_{4}^{\dagger }),
\label{scs30}
\end{equation}
and  the corresponding scheme is illustrated in Fig. 6.12. Two waveguides of
the length $L$ operating by nondegenerate optical
parametric processes with signal beams described by annihilation operators
$\hat{a}_{1}$ and $\hat{a}_{2}$ and idler beams described by $\hat{a}_{3}$
and $\hat{a}_{4}$ are pumped by strong laser beams of complex amplitudes
$\alpha_{1p}$ and $\alpha_{2p}$, respectively
 and  signals are coupled linearly
by evanescent waves with the strength
$\lambda_{2}$; $v$ is the speed of the waves (dispersion is neglected) and
$\chi ^{(2)}$ is quadratic susceptibility. The coupling constants
$\lambda_{1}$ and $\lambda_{3}$ include the pumping amplitudes.
It is important to mention that in Fig. 6.12
the beam splitter
 plays only auxiliary role to incline a strong classical beam
from a laser and consequently the vacuum fluctuations in the free port can be neglected.
Both the beams before entering the waveguides must be attenuated
to a low-intensity quantum level (the attenuators are not indicated in the figure)
and only these beams are involved in the described quantum interaction.

Clearly, the Hamiltonian (\ref{scs30}) has a kind of symmetry which will
be helpful to reduce the number of equations. For example, if one takes
$\hat{a}_{1}\longleftrightarrow\hat{a}_{2}$, $\hat{a}_{3} \longleftrightarrow\hat{a}_{4}
$ and $\lambda_{1} \longleftrightarrow\lambda_{3}$, the Hamiltonian
(\ref{scs30}) is still invariant.

The above Hamiltonian has been considered earlier by Janszky et al.
\cite{jansz4} (taking into consideration the coupling parameters $\lambda_{1}$
and $\lambda _{2}$ are equal), where they discussed the propagation of a
quantum field in coupler when the channels contain a parametric amplifying
medium. This situation occurs if one or both channels \cite{assa2} are made
from a second-order nonlinear material realizing a parametric process (down
conversion). They have found two operation regimes: one under threshold,
where the amplification is less effective than the coupling between
channels, and the other regime is above threshold, where the amplification
constant is greater than the coupling constant. However, the above Hamiltonian
(\ref{scs30}) will give us an advantage to see spatial effects (e.g.
switching between the channels), where the coupling parameters will play
a great role in controlling this phenomena. This will be demonstrated in
the following. Also we shall give more details for the
squeezing phenomenon and sub-Poissonian effect of this model
\cite{jansz2}.

\begin{picture}(170,170)(10,20)
\put(1,-5){\framebox(85,40){Laser pump}}
\put(86,25){\vector(4,0){42}}
\put(86,5){\vector(4,0){42}}
\put (110,15){\line(2,-1){30}}
\put (110,15){\line(2,1){30}}
\put (128,5){\vector(0,-1){30}}
\put (128,25){\vector(0,1){25}}
\put (115,42){\line(3,2){25}}
\put (115,-15){\line(3,-2){25}}
\put (128,50){\line(1,0){50}}
\put (128,-25){\line(1,0){50}}
\put (150,-50){\line(2,3){14}}
\put (140,65){\line(2,-1){17}}
\put (145,62){\vector(1,0){183}$\hat{a}_{1}(\frac{L}{v})$}
\put(145,-120){\framebox(30,50)}
\put(205,-95){\makebox(0,0){Idler laser}}
\put (160,-70) {\vector(0,1){110}}
\put (151,26){\line(2,3){14}}
\put (160,39){\vector(1,0){170}$\hat{a}_{3}(\frac{L}{v})$}
\put (160,-35){\vector(1,0){170}$\hat{a}_{4}(\frac{L}{v})$}
\put (178,-40){\framebox(140,30){$\chi^{(2)}(\lambda_{3})$}}
\put (178,35){\framebox(140,30){$\chi^{(2)}(\lambda_{1})$}}
\put (130,90){\framebox(30,50)}
\put (195,115){\makebox(0,0){Signal laser}}
\put (146,90){\vector(0,-1){107}}
\put (140,-12){\line(2,-1){17}}
\put (145,-15){\vector(1,0){183}$\hat{a}_{2}(\frac{L}{v})$}
\put(178,-60){\line(1,0){140}}
\put(178,-60){\vector(0,1){18}}
\put(318,-60){\vector(0,1){18}}
\put(250,-65){\makebox(0,0){z=vt}}
\put(200,15){\vector(0,1){47}}
\put(200,15){\vector(0,-1){30}}
\put(207,15){\makebox(0,0){$\lambda_{2}$}}
\put(120,-8){\makebox(0,0){$\alpha_{2p}$}}
\put(120,35){\makebox(0,0){$\alpha_{1p}$}}
\put(165,70){\makebox(0,0){$\hat{a}_{1}(0)$}}
\put(178,25){\makebox(0,0){$\hat{a}_{3}(0)$}}
\put(178,1){\makebox(0,0){$\hat{a}_{2}(0)$}}
\put(170,-3){\vector(1,-2){5}}
\put(170,28){\vector(1,2){5}}
\put(178,-30){\makebox(0,0){$\hat{a}_{4}(0)$}}
\put(200,-150){\makebox(0,0){Figure 6.12: Scheme of realization of interaction
in (6.31); BS are beam splitters.}}
\put(120,-45){\makebox(0,0){BS}}
\put(130,-45){\vector(1,0){23}}
\put(110,65){\makebox(0,0){BS}}
\put(120,65){\vector(1,0){20}}
\end{picture}

\vspace{2.5in}

\subsection{ Equations of motion and their solutions}

The exact solution of
the Heisenberg equation of motion for the bosonic operators
of the model are

\begin{eqnarray}
\begin{array}{lr}
\hat{a}_{1}(t)=f_{1}(t)\hat{a}_{1}(0)+if_{2}(t)\hat{a}
_{2}(0)-if_{3}(t)\hat{a}_{3}^{\dagger }(0)+f_{4}(t)\hat{a}_{4}^{\dagger
}(0),\\
\\
 \hat{a}_{2}(t)=g_{2}(t)\hat{a}_{2}(0)+ig_{1}(t)\hat{a}
_{1}(0)+g_{3}(t)\hat{a}_{3}^{\dagger }(0)-ig_{4}(t)\hat{a}_{4}^{\dagger
}(0),\\
\\
 \hat{a}_{3}(t)=h_{3}(t)\hat{a}_{3}(0)+h_{2}(t)\hat{a}
_{2}(0)-ih_{1}(t)\hat{a}_{1}^{\dagger }(0)-ih_{4}(t)\hat{a}_{4}^{\dagger
}(0),\\
\\
\hat{a}_{4}(t)=l_{4}(t)\hat{a}_{4}(0)+l_{1}(t)\hat{a}_{1}^{\dagger
}(0)-il_{2}(t)\hat{a}_{2}^{\dagger }(0)-il_{3}(t)\hat{a}_{3}(0),\hfill
\label{scs31}
\end{array}
\end{eqnarray}
where $\hat{a}_{j}(0)$ are the input operators and the expression of the time
dependent functions are given in the appendix E.

From equations (\ref{scs31}) together with the exact forms for the
time-dependent coefficients in the appendix E, it is easy to check that the well known
commutation relations between bosen operators are satisfied.

In the following  we treat some of the quantum statistical
properties of the model under discussion, using the results presented here,
related to squeezing phenomenon, sub-Poissonian and antibunching phenomena.

\subsection{ Two-mode squeezing phenomenon}

Here we study two-mode squeezing using quadrature operators $\hat{X}$
and $\hat{Y}$ which are defined in a similar sense as in section 4.2.
Since the electromagnetic field is guided inside the structure, exchange of
energy between the two waveguides is possible because of the evanescent
field between the waveguides \cite{yariv}. So we can examine several
cases of two-mode squeezing for this system using the input initial
coherent states $\prod^{4}_{j=1}|\alpha_{j}\rangle $.

In the first case, two-mode squeezing for mode 1 and mode 4, i.e. for
the signal mode in the first waveguide and the idler mode in the second
waveguide, is described by

\begin{eqnarray}
\begin{array}{lr}
2\langle (\triangle \hat{X})^{2}\rangle\exp (2\phi)= [\cosh \phi
\cos(t\Omega_{1})+\sinh \phi\cosh (t\Omega_{2})]^{2}
\\
+\left[ (\lambda _{3}\cosh \phi -\lambda _{2}\sinh \phi )\frac{\sinh
(t\Omega _{2})}{\Omega _{2}}+(\lambda _{3}\sinh \phi -\lambda _{2}\cosh
\phi)\frac{\sin (t\Omega _{1})}{\Omega _{1}}\right] ^{2},
\label{scs32}
\end{array}
\end{eqnarray}

\begin{eqnarray}
\begin{array}{lr}
2\langle (\triangle \hat{Y})^{2}\rangle\exp (-2\phi)= [\cosh \phi
\cos(t\Omega_{1})-\sinh \phi\cosh (t\Omega_{2})]^{2}
\\
+\left[ (\lambda _{3}\cosh \phi -\lambda _{2}\sinh \phi )\frac{\sinh
(t\Omega _{2})}{\Omega _{2}}-(\lambda _{3}\sinh \phi -\lambda _{2}\cosh\phi
)\frac{\sin (t\Omega _{1})}{\Omega _{1}}\right] ^{2}.
\label{scs33}
\end{array}
\end{eqnarray}

In the second case, two-mode squeezing for mode 1 and mode 2, i.e. for the
signal modes in the first and second waveguides, is described by

\begin{equation}
\langle (\triangle \hat{X})^{2}\rangle =\langle (\triangle
\hat{Y})^{2}\rangle =\frac{1}{2}[
1+f_{3}^{2}(t)+f_{4}^{2}(t)+g_{2}^{2}(t)+g_{4}^{2}(t)].
\label{scs34}
\end{equation}

In the third case, two-mode squeezing for mode 3 and mode 4, i.e. for the
idler modes in the two waveguides, is described by

\begin{equation}
\langle (\triangle \hat{X})^{2}\rangle =\langle (\triangle
\hat{Y})^{2}\rangle =\frac{1}{2}[1+h^{2}_{2}(t)+
h^{2}_{3}(t)+l^{2}_{1}(t)+l^{2}_{2}(t)].
\label{scs35}
\end{equation}
\setcounter{figure}{12}
\begin{figure}[h]%
 \centering
    \includegraphics[width=6cm]{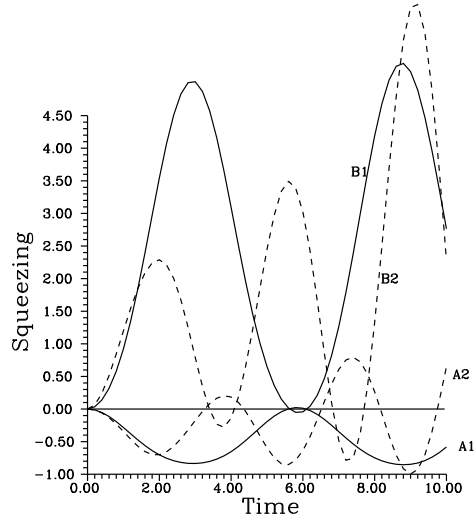}
   \caption{
Two-mode squeezing phenomenon for the first case, for solid curves
$\lambda _{1}=0.1,\lambda _{2}=1.25$ and $\lambda_{3}=0.5$; for
dashed curves $\lambda _{1}=0.6,\lambda _{2}=2$ and
$\lambda_{3}=0.5$. } \label{fig29}
\end{figure}


We can see from (\ref{scs32})--(\ref{scs35}) that this coupler can generate
squeezing, in terms of two-mode definition, only when considering the signal
mode in the first waveguide and the idler mode in the second waveguide or
vice versa.
So we concentrate on this case. In Fig. 6.13, we have plotted
squeezing component for the first case for different values of $\lambda_{j}$
against time $t$. Curves A1 and A2 are related to the $X$-component
corresponding to the two groups of values for $\lambda_{j}$
respectively,
and similarly curves B1 and B2 are related to the $Y$-component. For
$\lambda_{1}$ small, $\lambda_{2}> \lambda_{3}>\lambda_{1}$, i.e. when
linear
exchange between waveguides is stronger than nondegenerate parametric
amplification inside waveguides, we observe that squeezing is dominant
in
the first quadrature having oscillatory behaviour; nevertheless the
value of
squeezing is negligible in the second component, see curves A1 and B1.
Increasing the values of both $\lambda_{1}$ and $\lambda_{2}$ and
keeping $\lambda_{3}$ as before, we observed squeezing in both the quadratures
and  its values were more pronounced than in the earlier case with the
maximum value of squeezing in the first quadrature, see curves A2 and B2.

Thus we can conclude that this system is able to generate squeezed light in
terms of two-mode squeezing, only when the signal and idler modes are
considered in different waveguides, provided that the signal modes are
coupled. The values of squeezing are well controllable by the values of
coupling constants.

\subsection{ Second-order correlation function}

As we have mentioned earlier  sub-Poissonian photon
statistics need not be associated with antibunching, but can be
accompanied by bunching.
We  trace the nonclassical effects for the
model under discussion using the normalized second-order correlation function
for
the various modes when the initial input light modes are in number
states $
\prod^{4}_{j=1}|n\rangle_{j}$ as well as in the coherent states.
Then, we extend our discussion to demonstrate the photon antibunching
phenomenon for our model using the relation (4.4).

When the input light is in the
number states, we have

\begin{eqnarray}
\begin{array}{lr}
\langle \hat{n}_{1}(t+\tau)\rangle
_{\rm n}=f_{1}^{2}(t+\tau)\bar{n}_{1}+f_{2}^{2}(t+\tau)\bar{n
}_{2}+f_{3}^{2}(t+\tau)(\bar{n}_{3}+1)
\\
+f_{4}^{2}(t+\tau)(\bar{n}_{4}+1),
\label{scs36}
\end{array}
\end{eqnarray}

\begin{eqnarray}
\begin{array}{lr}
\langle \hat{n}_{3}(t+\tau)\rangle
_{\rm n}=h_{3}^{2}(t+\tau)\bar{n}_{3}+h_{4}^{2}(t+\tau)\bar{n
}_{4}+h_{2}^{2}(t+\tau)(\bar{n}_{2}+1)
\\
+h_{1}^{2}(t+\tau)(\bar{n}_{1}+1),
\label{scs37}
\end{array}
\end{eqnarray}
 and

\begin{eqnarray}
\begin{array}{lr}
\langle
\hat{a}^{\dagger}_{1}(t)\hat{a}^{\dagger}_{1}(t+\tau)\hat{a}_{1}(t+\tau)
\hat{a}_{1}(t)\rangle_{\rm n}
=
[f_{1}^{2}(t)\bar{n}_{1}+f_{2}^{2}(t)\bar{n
}_{2}][f_{3}^{2}(t+\tau)(\bar{n}_{3}+1)
\\
+f_{4}^{2}(t+\tau)(\bar{n}_{4}+1)]
+\bar{n}_{1}\bar{n}_{2}[f_{1}(t)
f_{1}(t+\tau)+f_{2}(t)f_{2}(t+\tau)]^{2}
\\
+f^{2}_{1}(t) f^{2}_{1}(t+\tau)\bar{n}_{1}(\bar{n}_{1}-1)+ f^{2}_{2}(t)
f^{2}_{2}(t+\tau)\bar{n}_{2}(\bar{n}_{2}+1)
\\
+2[f_{1}(t) f_{1}(t+\tau)\bar{n}_{1}+f_{2}(t)f_{2}(t+\tau)\bar{n
}_{2}][f_{3}(t) f_{3}(t+\tau)(\bar{n}_{3}+1)
\\
+ f_{4}(t)f_{4}(t+\tau)(\bar{n}_{4}+1)]
\\
+[f_{1}^{2}(t+\tau)\bar{n}_{1}+f_{2}^{2}(t+\tau)\bar{n
}_{2}][f_{3}^{2}(t)(\bar{n}_{3}+1)+f_{4}^{2}(t)(\bar{n}_{4}+1)]
\\
+(\bar{n}_{3}+1)(\bar{n}_{4}+1)[f_{3}(t) f_{4}(t+\tau)
+f_{4}(t)f_{3}(t+\tau)]^{2}
\\
+f^{2}_{3}(t) f^{2}_{3}(t+\tau)(\bar{n}_{3}+1)(\bar{n}_{3}+2)+ f^{2}_{4}(t)
f^{2}_{4}(t+\tau)(\bar{n}_{4}+1)(\bar{n}_{4}+2),
\label{scs38}
\end{array}
\end{eqnarray}

\begin{eqnarray}
\begin{array}{lr}
\langle
\hat{a}^{\dagger}_{3}(t)\hat{a}^{\dagger}_{3}(t+\tau)\hat{a}_{3}(t+\tau)
\hat{a}_{3}(t)\rangle_{\rm n}=
[h_{3}^{2}(t)\bar{n}_{3}+h_{4}^{2}(t)\bar{n
}_{4}][h_{2}^{2}(t+\tau)(\bar{n}_{2}+1)
\\
+h_{1}^{2}(t+\tau)(\bar{n}_{1}+1)]+\bar{n}_{3}\bar{n}_{4}[h_{3}(t)
h_{4}(t+\tau)+h_{4}(t)h_{3}(t+\tau)]^{2}
\\
+h^{2}_{3}(t) h^{2}_{3}(t+\tau)\bar{n}_{3}(\bar{n}_{3}-1)+ h^{2}_{4}(t)
h^{2}_{4}(t+\tau)\bar{n}_{4}(\bar{n}_{4}+1)
\\
+2[h_{3}(t) h_{3}(t+\tau)\bar{n}_{3}+h_{4}(t)h_{4}(t+\tau)\bar{n
}_{4}]
\\
\times [h_{2}(t) h_{2}(t+\tau)(\bar{n}_{1}+1)+ h_{1}(t)h_{1}(t+\tau)
(\bar{n}_{1}+1)]
\\
+[h_{3}^{2}(t+\tau)\bar{n}_{3}+h_{4}^{2}(t+\tau)\bar{n
}_{4}][h_{1}^{2}(t)(\bar{n}_{1}+1)+h_{2}^{2}(t)(\bar{n}_{2}+1)]
\\
+(\bar{n}_{2}+1)(\bar{n}_{2}+1)
[h_{1}(t) h_{2}(t+\tau)+
h_{2}(t)h_{1}(t+\tau)]^{2}
\\
+h^{2}_{1}(t) h^{2}_{1}(t+\tau)(\bar{n}_{1}+1)(\bar{n}_{1}+2)+
h^{2}_{2}(t)
h^{2}_{2}(t+\tau)(\bar{n}_{2}+1)(\bar{n}_{2}+2),
\label{scs39}
\end{array}
\end{eqnarray}
where the subscript $n$ stands for the Fock states case.

Firstly, we will discuss the sub-Poissonian statistics, i.e. we take
$\tau=0$ (see the discussion in section 4.1) in the equations
(\ref{scs36})--(\ref{scs39}). For $n_{j}=0$, i.e. for input vacuum states,
 we can get

\begin{eqnarray}
\begin{array}{lr}
\langle (\triangle \hat{n}_{1}(t))^{2}\rangle _{0}-\langle \hat{n}
_{1}(t)\rangle _{0}=(f_{3}^{2}(t)+f_{4}^{2}(t))^{2},
\\
\langle (\triangle \hat{n}_{3}(t))^{2}\rangle _{0}-\langle \hat{n}
_{3}(t)\rangle _{0}=(h_{1}^{2}(t)+h_{2}^{2}(t))^{2},
\label{scs40}
\end{array}
\end{eqnarray}
where the relation such as

\begin{equation}
\langle (\triangle \hat{n}_{j}(t))^{2}\rangle -\langle
\hat{n}_{j}(t)\rangle=\langle
\hat{a}^{\dagger 2}_{j}(t)\hat{a}^{2}_{j}(t)\rangle-\langle
\hat{a}_{j}^{\dagger}(t)\hat{a}_{j}(t)\rangle^{2},
\label{scs41}
\end{equation}
has been used. It  is clear from (\ref{scs40}) that for input vacuum states,
which have maximum pronounced
sub-Poissonian statistics, they evolve into pure super-Poissonian statistics
states for all times $t\neq 0$ inside the coupler. We display the
second-order correlation function (4.1) against time $t$ in Fig. 6.14 for
input number states with different mean photon numbers $\bar{n}_{j}$
($\bar{n}_{1}=4,\bar{n}_{j}=1,j=2,3,4$) and for different values of coupling
constants.
\begin{figure}[h]%
  \centering
  \subfigure[]{\includegraphics[width=5cm]{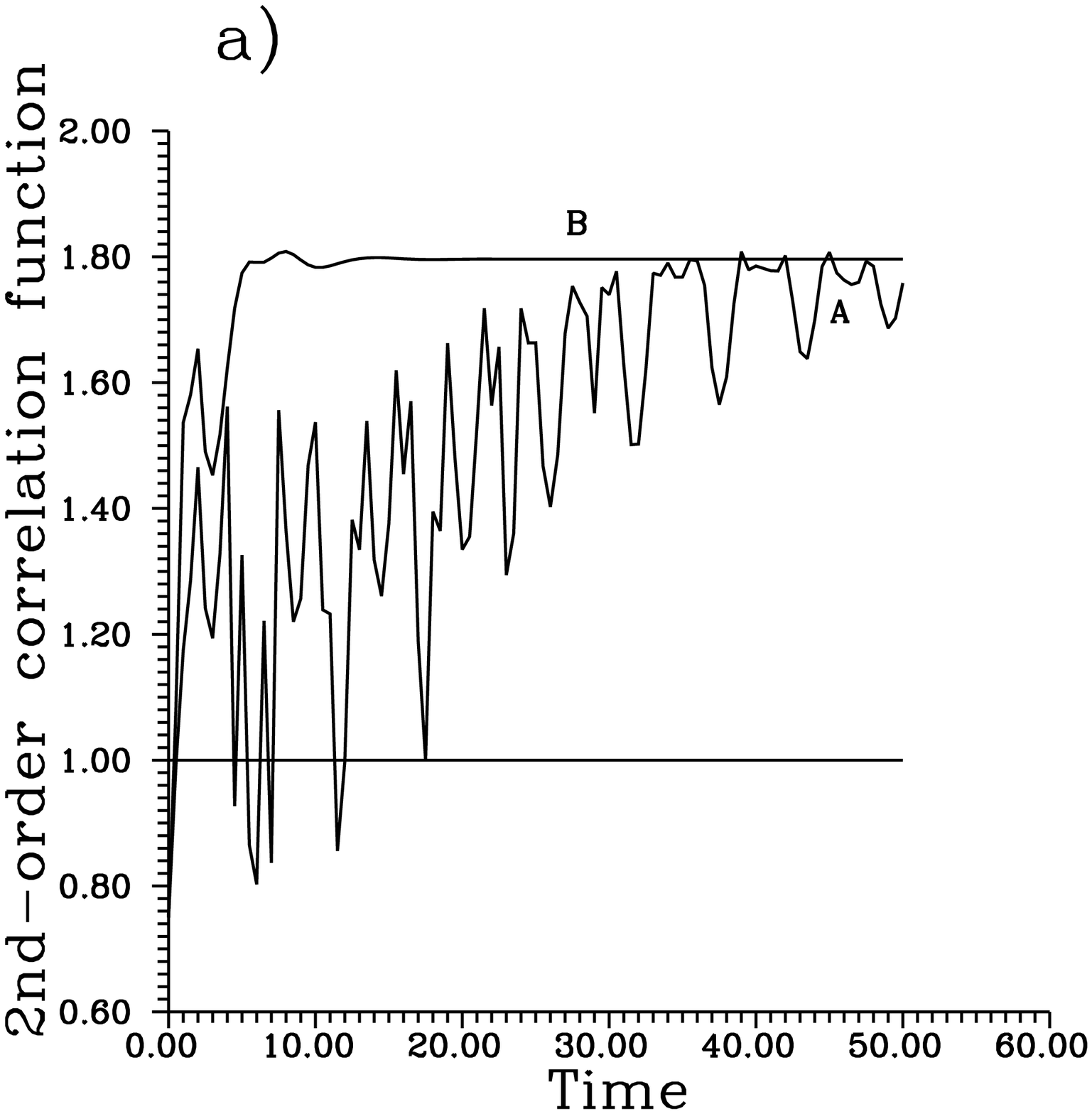}}
 \subfigure[]{\includegraphics[width=5cm]{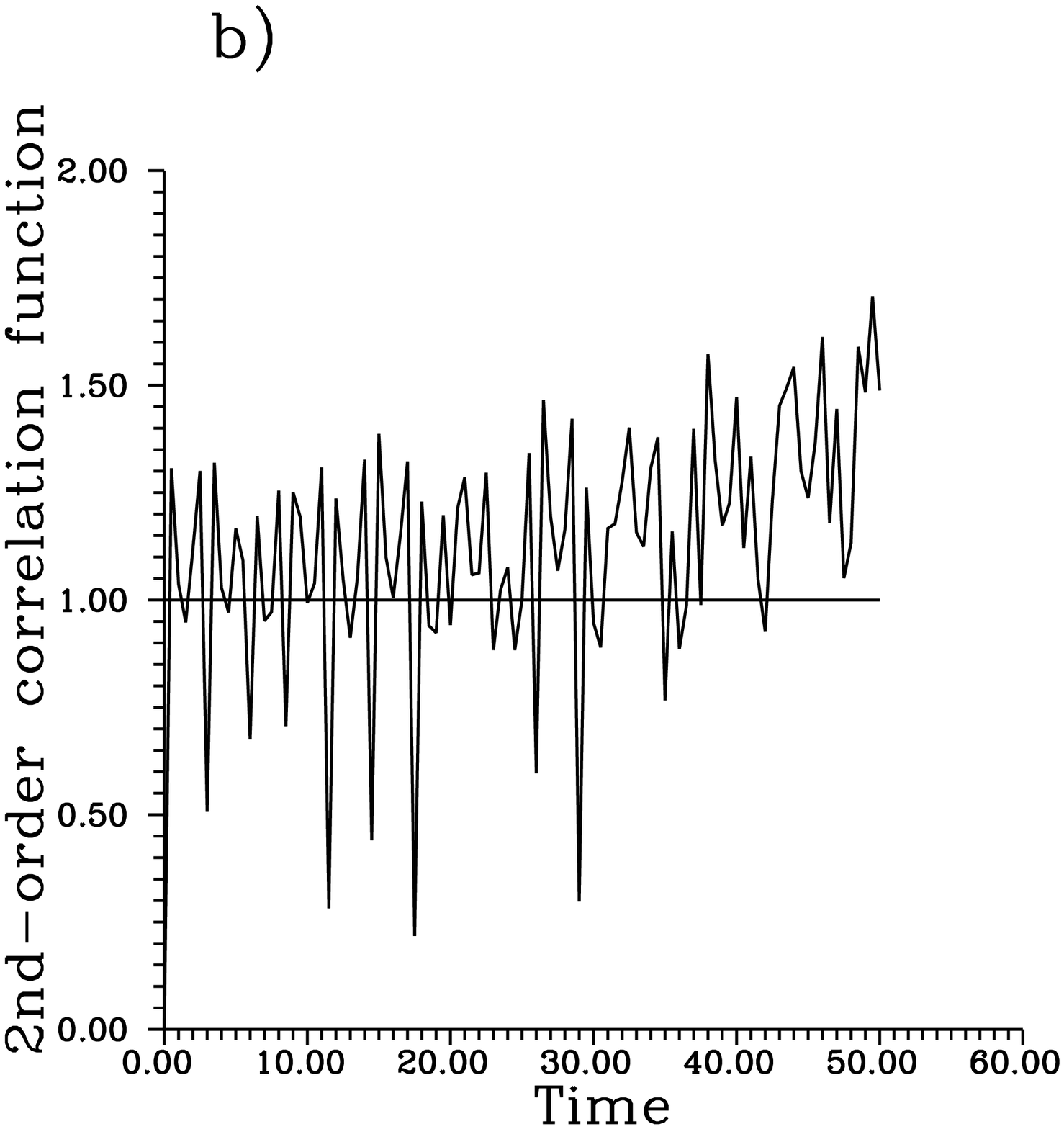}}
      \caption{
Normalized normal second-order correlation function when both the
modes are initially in the number states with mean photon numbers
$ \bar{n}_{1}=4,\bar{n}_{j}=1,j=2,3,4$, for all curves
$\lambda_{2}=1.2$, $ \lambda _{3}=0.5$ and for curve A, $\lambda
_{1}=0.1$ and for curve B, $ \lambda _{1}=0.6$: a)
$g_{1}^{(2)}(t)$ for mode 1; b) $g_{2}^{(2)}(t)$ for mode 2,
straightline is corresponding to $g^{(2)}(0)$ of the coherent
light. } \label{fig30}
 \end{figure}

In Fig. 6.14a, $g^{(2)}_{1}(t)$ for mode 1 has been shown, where
always $\lambda_{2}=1.2,$ $\lambda_{3}=0.5 $ and $\lambda_{1}=0.1,0.6$ are
corresponding to curve $A$ and curve $B$, respectively. Initially, $
g^{(2)}_{1}(0)=0.75$, which is the corresponding value for the Fock
state $|4\rangle$. Increasing the time, switching between modes starts (see
curve $A $), $g^{(2)}_{1}(t)$ has an oscillating behaviour between values for
sub-Poissonian and super-Poissonian statistics for rather short interaction
times. However, for large interaction times the oscillations are
successively washed out and super-Poissonian statistics are dominant. As
seen in curve $B$, $g^{(2)}_{1}(t)$ does no exhibit any periodic behaviour
and takes values corresponding to super-Poissonian statistics shortly after
the switching on of the interaction. On the other hand, we can see that strongly
sub-Poissonian light is generated not only for short time interaction but
also for larger times as demonstrated by the behaviour of $g^{(2)}_{2}(t)$
in the second mode for the same values of $\bar{n}_{j}$ as in the former
case and for $(\lambda_{1},\lambda_{2},\lambda_{3})=(0.1,1.2,0.5)$,
see Fig. 6.14b. Moreover, $g^{(2)}_{2}(t)$ behaves more smoothly than $g^{(2)}_{1}
(t)$ for the same values of $\bar{n}_{j}$ and $\lambda_{j}$.

We note that for $\lambda_{2}$ less than $\lambda_{1}$ and $\lambda_{3}$, $
t\neq 0$ the second-order correlation functions exhibit always
super-Poissonian statistics and there is no oscillatory behaviour regardless
of the values of input mean photon numbers, since all time dependent
functions in equations (E.1)-(E.2) of appendix E are being hyperbolic
functions, which are monotonically increasing functions.

\begin{figure}[h]%
  \centering
  \subfigure[]{\includegraphics[width=5cm]{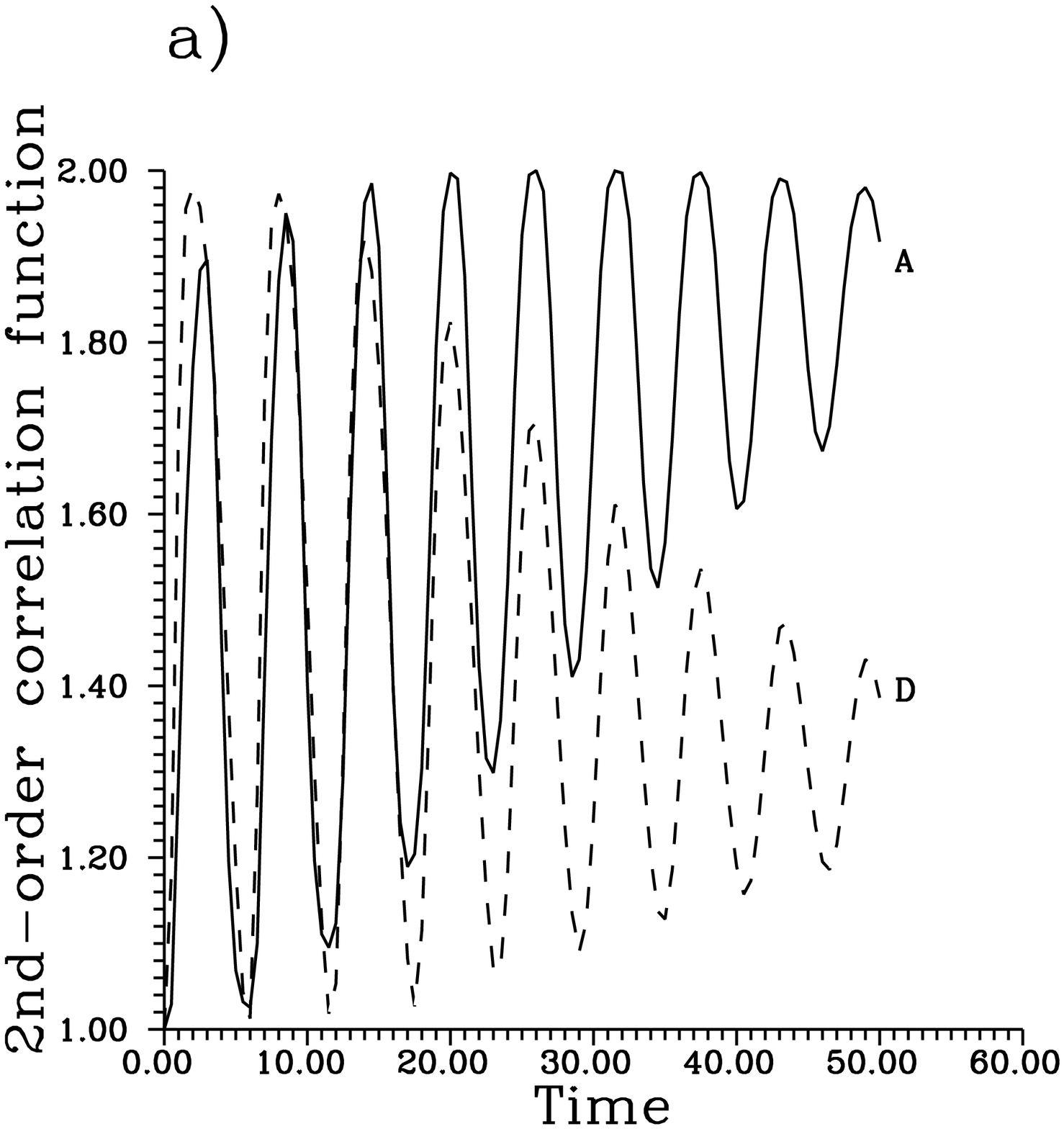}}
 \subfigure[]{\includegraphics[width=5cm]{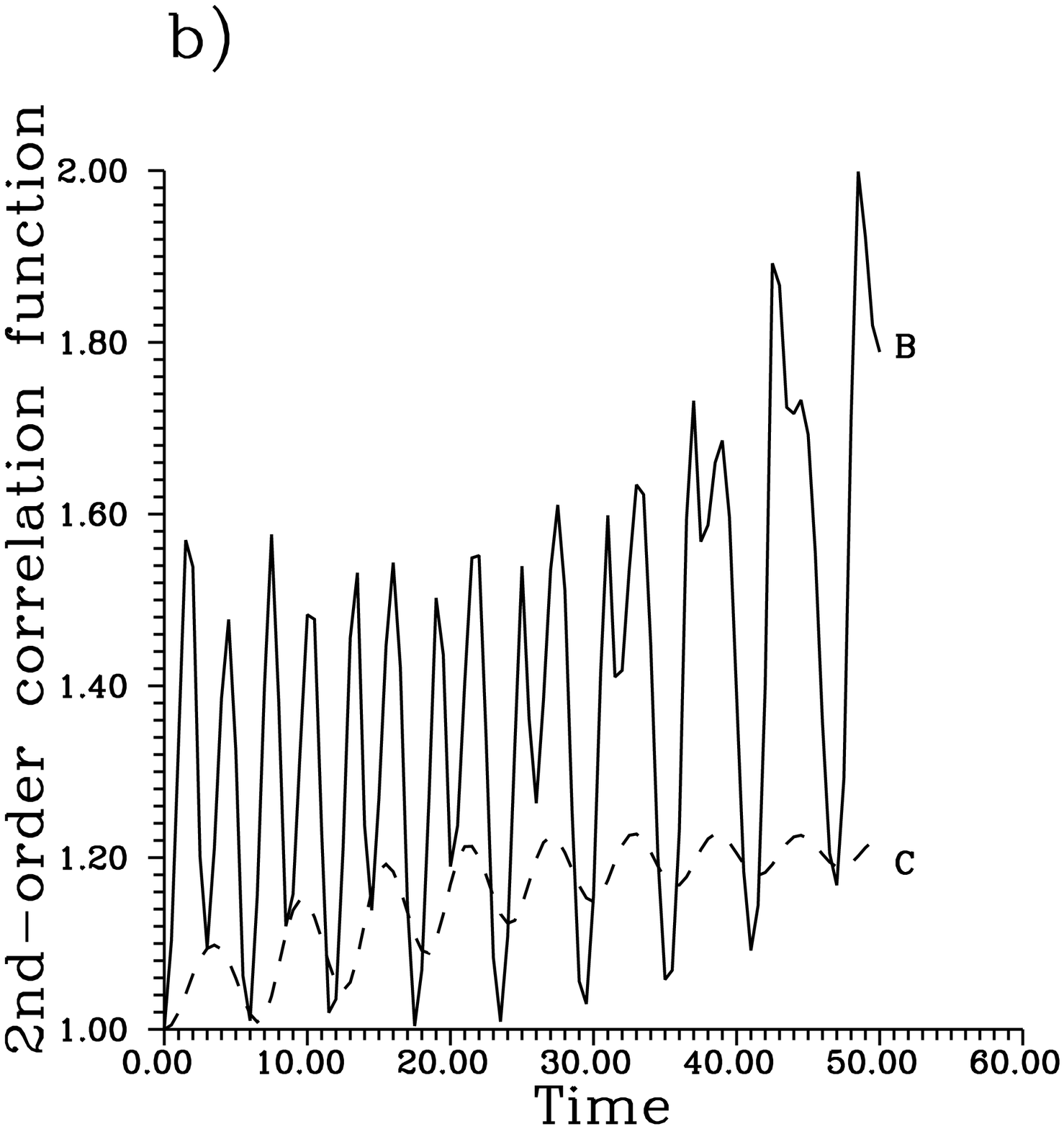}}
      \caption{
Normalized normal second-order correlation function $
g_{j}^{(2)}(t)$ for different modes when both the modes are
initially in the coherent states with $|\alpha _{j}|=1,j=1,2,3,4$,
$\psi _{1}=\frac{\pi }{2}$ and $\psi _{j}=\frac{\pi }{3},j=2,3,4$
for all curves,
 $\lambda _{1}=0.1,\lambda _{2}=1.2,\lambda _{3}=0.5$: a) first
mode-curve A and fourth mode-D; b) second mode-curve B and third
mode-C. } \label{fig31}
 \end{figure}


For input coherent light, we have
\begin{eqnarray}
\begin{array}{lr}
\langle \hat{n}_{1}(t+\tau)\rangle _{\rm coh}=|\alpha
_{1}^{*}f_{1}(t+\tau)-i\alpha
_{2}^{*}f_{2}(t+\tau)+i\alpha _{3}f_{3}(t+\tau)+\alpha
_{4}f_{4}(t+\tau)|^{2}
\\
+f_{3}^{2}(t+\tau)
+f_{4}^{2}(t+\tau),
\label{scs42}
\end{array}
\end{eqnarray}

\begin{eqnarray}
\begin{array}{lr}
\langle \hat{n}_{3}(t+\tau)\rangle _{\rm coh}=|\alpha
_{3}h_{3}(t+\tau)-i\alpha
_{1}^{*}h_{1}(t+\tau)+\alpha _{2}^{*}h_{2}(t+\tau)-i\alpha
_{4}h_{4}(t+\tau)|^{2}
\\
+h_{1}^{2}(t+\tau)+h_{2}^{2}(t+\tau),
\label{scs43}
\end{array}
\end{eqnarray}

\begin{eqnarray}
\begin{array}{lr}
\langle
\hat{a}^{\dagger}_{1}(t)\hat{a}^{\dagger}_{1}(t+\tau)\hat{a}_{1}(t+\tau)
\hat{a}_{1}(t)\rangle_{\rm coh}=
\langle \hat{n}_{1}(t)\rangle_{\rm coh}
\langle \hat{n}_{1}(t+\tau)\rangle_{\rm coh}
\\
+
[f_{3}(t) f_{3}(t+\tau)+
f_{4}(t)f_{4}(t+\tau)]^{2}
\\
+[f_{3}(t) f_{3}(t+\tau)+
f_{4}(t)f_{4}(t+\tau)][
\langle \hat{a}^{\dagger}_{1}(t)\rangle_{\rm coh}
\langle \hat{a}_{1}(t+\tau)\rangle_{\rm coh}+ {\rm c.c.}],
\label{scs44}
\end{array}
\end{eqnarray}

\begin{eqnarray}
\begin{array}{lr}
\langle
\hat{a}^{\dagger}_{3}(t)\hat{a}^{\dagger}_{3}(t+\tau)\hat{a}_{3}(t+\tau)
\hat{a}_{3}(t)\rangle_{\rm coh}=
\langle \hat{n}_{3}(t)\rangle_{\rm coh}
\langle \hat{n}_{3}(t+\tau)\rangle_{\rm coh}
\\
+
[h_{1}(t) h_{1}(t+\tau)+
h_{2}(t)h_{2}(t+\tau)]^{2}
\\
+[h_{1}(t) h_{1}(t+\tau)+
h_{2}(t)h_{2}(t+\tau)][
\langle \hat{a}^{\dagger}_{3}(t)\rangle_{\rm coh}
\langle \hat{a}_{3}(t+\tau)\rangle_{\rm coh}+ {\rm c.c.}],
\label{scs45}
\end{array}
\end{eqnarray}
where $\langle \hat{n}_{j}(t)\rangle_{\rm coh}$ can be obtained from
$\langle \hat{n}_{j}(t+\tau)\rangle_{\rm coh}$ by simply setting $\tau=0;
\langle \hat{a}_{j}(t)\rangle_{\rm coh}$ is the expectation value for
$\hat{a}_{j}(t)$ in the coherent states, {\rm c.c.} is the complex
conjugate; and {\rm coh} stands for the expressions which are related to input
coherent light.
From equations (4.1) it is clear that the condition for
sub-Poissonian statistics is that the variance $\langle (\triangle
\hat{n}
_{j}(t))^{2}\rangle $ is less that the mean photon number $\langle
\hat{n}
_{j}(t)\rangle $. Combination of (\ref{scs42})--(\ref{scs45}) into
(\ref{scs41}), after taking $\tau=0$, it is easy to  show that this
system cannot provide sub-Poissonian light for input coherent light
for $t>0$.
For instance, for mode 1, from equations (\ref{scs41}) with (\ref{scs42})
and (\ref{scs44})
it follows that sub-Poissonian light could be generated provided that

\begin{equation}
2|\alpha _{1}^{*}f_{1}(t)-i\alpha _{2}^{*}f_{2}(t)+i\alpha
_{3}f_{3}(t)+\alpha
_{4}f_{4}(t)|^{2}+f_{3}^{2}(t)+f_{4}^{2}(t)<0.
\label{scs46}
\end{equation}
It is evident that this inequality cannot be fulfilled regardless
of the values of $\alpha_{j}$ and $\lambda_{j}$. More precisely, for input
coherent light, this model can generate classical light, e.g. coherent,
partially coherent and chaotic light.
\begin{figure}[h]%
  \centering
  \subfigure[]{\includegraphics[width=5cm]{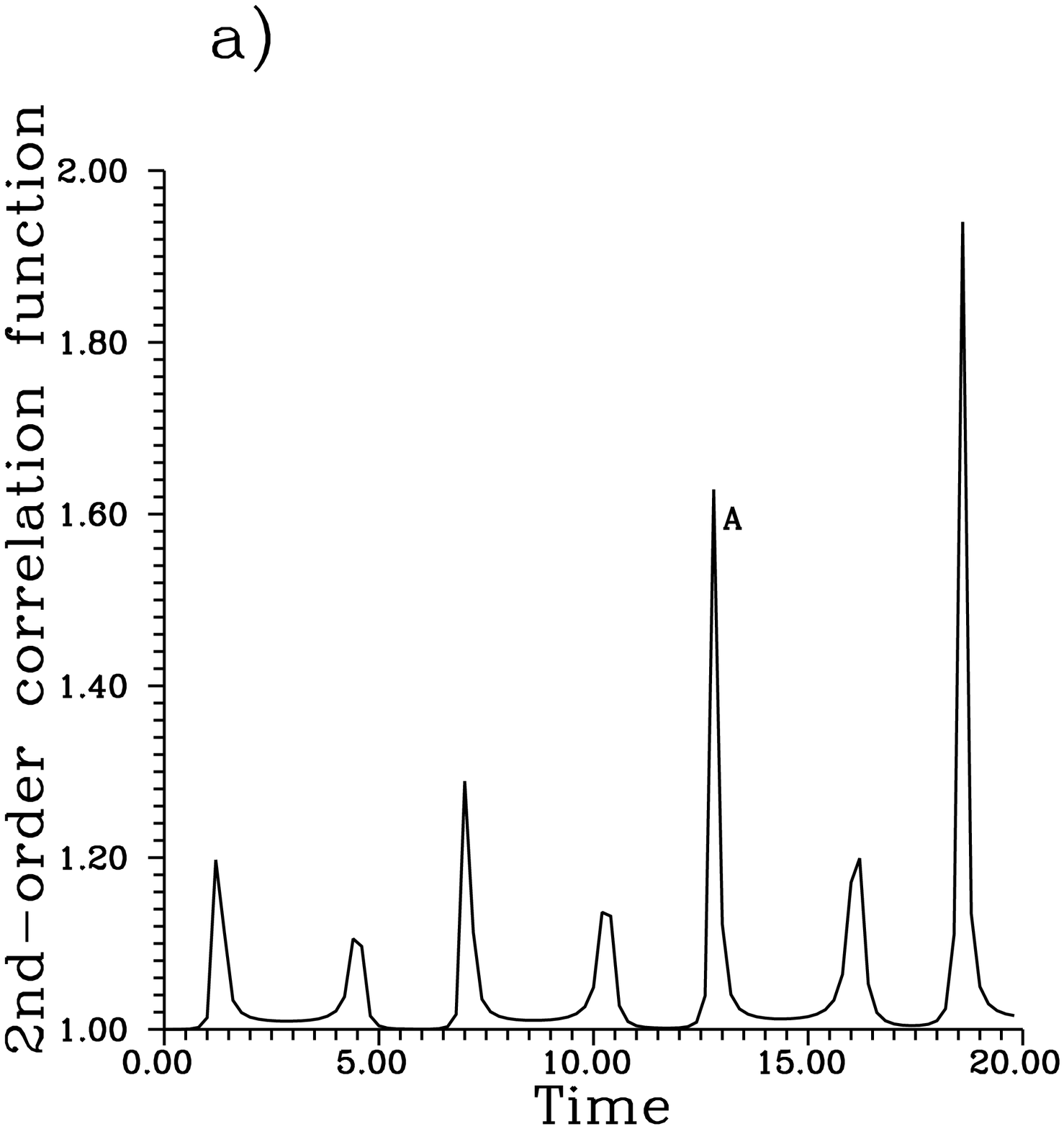}}
 \subfigure[]{\includegraphics[width=5cm]{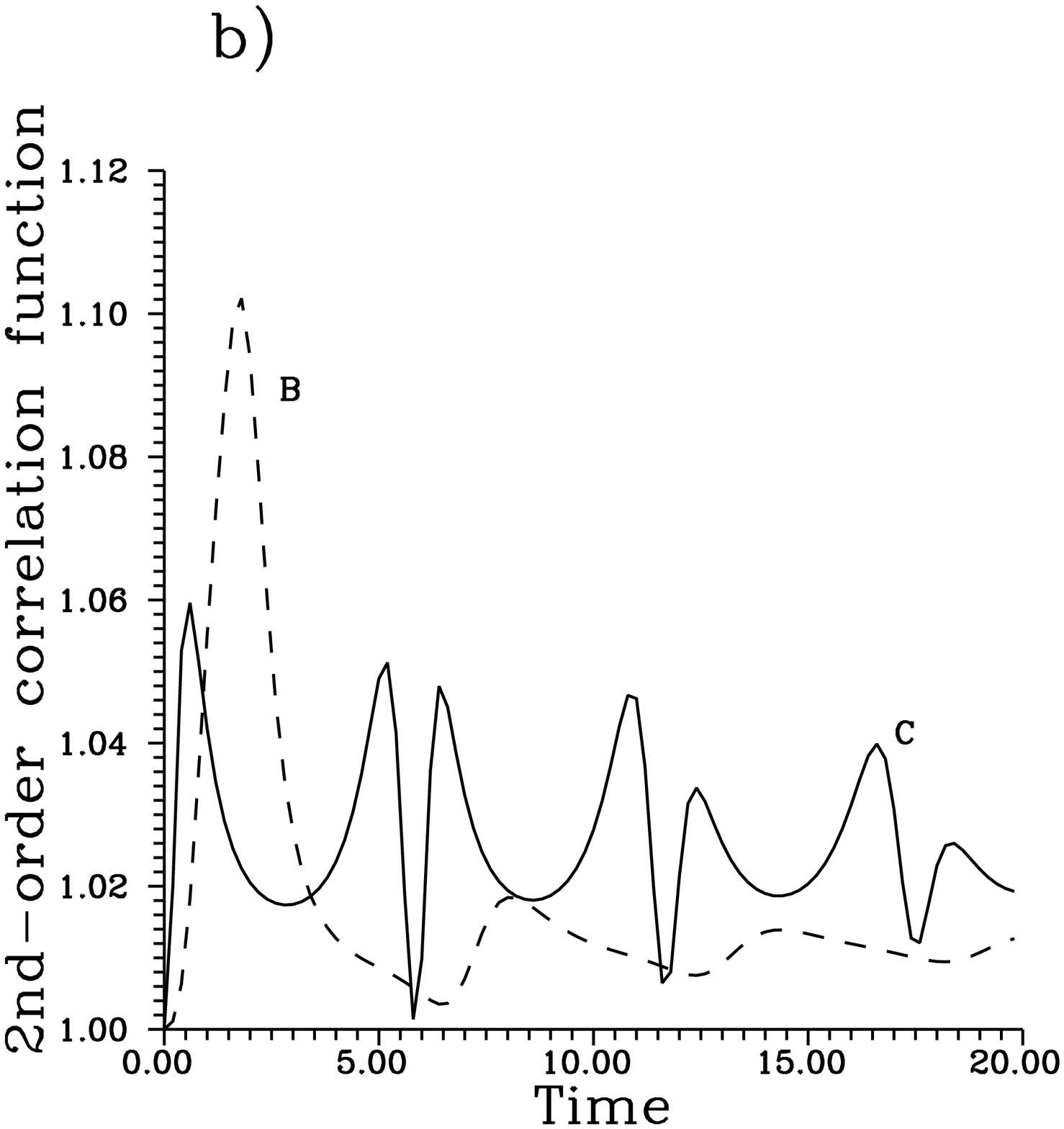}}
      \caption{
Normalized normal second-order correlation function $
g_{j}^{(2)}(t)$ for different modes when both the modes are
initially in the coherent states with $|\alpha _{1}|=10,|\alpha
_{j}|=1,j=2,3,4$, $\psi _{j}= \frac{\pi }{6},j=1,2,3,4$ where
$\lambda _{j}$ have the same values as in Fig. 6.15: a) first mode
; b) third mode-curve B and fourth mode-C. } \label{fig32}
 \end{figure}

In Figs. 6.15 and 6.16 we have plotted $g_{j}^{(2)}(t)$ against time $t$ for
$\lambda_{1}=0.1,\lambda _{2}=1.2,\lambda _{3}=0.5$ and input coherent
light with complex amplitudes $\alpha _{j}=|\alpha _{j}|e^{\psi _{j}}$. In all
these figures $g_{j}^{(2)}(0)=1$ initially. In Fig. 6.15 for $|\alpha
_{j}|=1$ and $(\psi _{1},\psi _{2},\psi _{3},\psi _{4})=(\frac{\pi }{2},\frac{\pi
}{3},\frac{\pi }{3},\frac{\pi }{3})$, we see the oscillatory behaviour in the
evolution of $g_{j}^{(2)}(t)$ in all cases showing that the photons are
transferred from one mode to the other. These oscillations are successively
disappearing for large interaction time. They can be destroyed by changing
the values of the complex amplitudes of input light, but not by changing
the values of the coupling constants (see Fig. 6.16 for shown values). This
shows how one can control light by light via nonlinear medium. In Fig. 6.16a we
can see the periodical behaviour exhibiting photon statistics between close
to Poissonian and super-Poissonian states for mode 1 with gradual increase of
values of $g^{(2)}(t)=2$ representing chaotic light for larger interaction
times. We see that the initial coherent state can be successively
approximately regenerated. From Fig. 6.16b we can see that the
precise Poisson distribution is obtained only initially and partially
coherent light is obtained for later times, with the maximal noise value
in mode 3 (curve B). For input coherent light
$2\geq g_{j}^{(2)}(t)\geq
1$ holds, i.e. one cannot obtain superchaotic light. One can also observe
some complementary behaviour of both the modes.

\begin{figure}[h]%
 \centering
    \includegraphics[width=6cm]{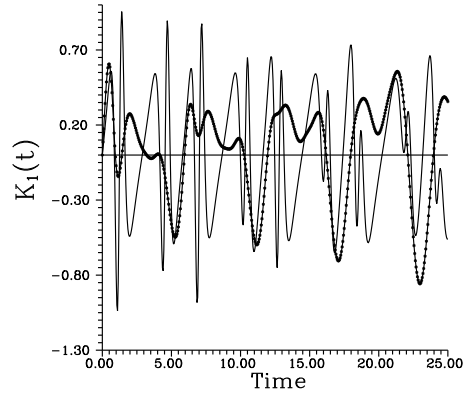}
   \caption{
The quantity $K_{1}(t)$ for mode 1 when both the modes are
initially in the number (solid curve) and coherent (star-centered
curve) input states for the same situation as in Fig. 6.14a (curve
A) and Fig. 6.15a for number  state and coherent state,
respectively. } \label{fig33}
\end{figure}


Concerning photon bunching and antibunching according to
the definition (4.4),
we have analysed the cases of input number and coherent states.
We note, in general, that the quantity $K_{j}(t)$ exhibits
oscillatory behaviour  between negative and positive values for both number
 and coherent input states, i.e. both antibunching and bunching  can
occur. The  photon antibunching is more pronounced for input number states
than for input coherent states.
Comparing these results with those for sub-Poisson statistics,
we can conclude  that there is no direct relation between
antibunching and sub-Poissonian statistics here,
in agreement with results shown in the literature
earlier \cite{chs6,{chs7}}.
We can demonstrate this graphically in Fig. 6.17,
by considering the first mode, for input number state
(solid curve) and for input coherent state (star-centered curve) for the same
values of parameters as those of Figs. 6.14a (curve A) and 6.15a
for number  state and coherent state, respectively.

\subsection{Conclusions}

The main conclusions of this section can be summarized as follows. Two-mode
squeezing, sub-Poissonian and antibunching effects
 of an optical field propagating inside a directional
coupler containing nondegenerate parametric amplification process have
been studied in the framework of Hamiltonian formalism.
Incident number states and coherent states
have been considered. We have shown that for input coherent light, i.e.
Poissonian light, the system can generate squeezed light in terms of
two-mode squeezing depending on the values of the coupling constants.
More precisely, when the linear coupling is stronger than the nonlinear one,
there is possibility to obtain squeezed light considering signal beam in
one waveguide and the idler beam in the other, provided that the signal
beams are connected by the evanescent waves. Nevertheless, this possibility is
completely smeared when both signal or idler beams are considered.
On the other hand, we have demonstrated that for input Fock
states, i.e. for sub-Poissonian light, our model can provide sub-Poissonian
light governed by coupler parameters. However, for input coherent light,
 partially coherent light and chaotic light can be generated.
This was demonstrated in terms of the second-order correlation
function.
Concerning the photon antibunching we have shown for both number and
coherent input states that the outgoing field oscillates between bunching and
antibunching giving a good indication that it need not be direct relation
between sub-Poissonian statistics and photon antibunching as shown in
the literature before \cite{chs6,{chs7}}.
\chapter{Summary of main results}
In this doctoral thesis we have studied the quantum properties of
several
 models which
have been classified as statical and dynamical systems. The first
part has been devoted to investigate the properties of the
statical models including the superposition of  squeezed displaced
number states with and without thermal noise. Also we have
developed a new type of multidimensional squeeze operator
including two different squeezing mechanisms. In the second part
the dynamical models were given  to show the interaction between
modes in the nonlinear optical coupler. The results of the
statical models can be summarized as follows:\newline 1) We
introduced general class of quantum states, as a single mode
vibration
   of electromagnetic field, sudden squeezed-plus-displaced by a collection
    of two displacements $180^{0}$ out of phase.
 For such class
the quantum statistical quantities, such as the second-order
correlation function, quasiprobability distribution functions and
the photon-number distribution function are examined analytically
and numerically. Moreover, the phase  properties  and amplitude
squeezing and phase squeezing for such class have been  examined
in the framework of Pegg-Barnett formalism. Generalizations of
some considerable results given in the literature earlier have
been reported. Also we have suggested ways for generation of such
superposition in the framework of micromaser and trapped ions.
Regimes of strongly nonclassical behaviour of such states have
been demonstrated.\newline 2) We studied the effect of thermal
noise  on the statistical behaviour of the superposition of
squeezed and displaced number states. The main tool in this
analysis is  the $s$-parametrized characteristic function. For
such a superpostion states an exact expressions for
quasiprobability distribution functions, normalized second-order
corelation function, photon-number distribution and phase space
distribution have been obtained and discussed in detail. The
behaviour of such system  reavels that the origin of the
nonclassical effects is the correlation between different
oscillators in phase space.\newline 3) We developed a new squeeze
operator, which is related to the time-dependent evolution
operator for Hamiltonian representing mutual interaction between
three different modes. Squeezing phenomenon as well as the
variances of the photon-number sum and difference have been
considered. Moreover, Glauber second-order correlation function,
the quasiprobability distribution function and phase distribution
for different states have been discussed. We have shown also that
the origin of the nonclassical effects of
 this operator model  is the correlation  between the systems.\newline

On the other hand, the dynamical models were given in the second
part to show the interaction between modes in the nonlinear
optical coupler. We have given three types of such device and the
results can be summarized as follows:\newline 4) We derived the
quantum statistical and dynamical properties of nonlinear optical
couplers composed of two nonlinear waveguides operating by the
second subharmonic generation, which are coupled linearly through
evanescent waves and nonlinearly through nondegenerate optical
parametric interaction. Main attention has been paid to generation
and transmission of nonclassical light, based on a discussion of
squeezing phenomenon and quasiprobability distribution functions.
The beams have been initially considered in  coherent and number
 states. In particular, results have been discussed in
dependence on the strength of the nonlinear coupling relatively to
the linear coupling. We have shown that if thermal fields enter
initially the waveguides the coupler plays  similar role as
 a microwave Josephson-junction parametric
amplifier to generate squeezed thermal light. \newline 5) We
showed that a nonlinear asymmetric directional coupler composed of
a linear waveguide and a nonlinear waveguide operating by
nondegenerate parametric amplification is a source of single-mode
squeezed light.
 This fact has been demonstrated, under certain conditions and for specific
 modes, for incident coherent beams in terms of the quasiprobability
 functions, photon-number distribution and phase distribution.\newline
6) We  studied the quantum statistical properties of an optical
field propagating inside a directional coupler operating by
nondegenerate parametric amplification. We have investigated the
effect of switching between the input modes and the outgoing
fields from the coupler. Particular attention has been paid to two
mode squeezing and second-order correlation function.
 Incident number and coherent states have been considered.
Furthermore, regimes for generation and transmission of squeezed
and/or
 sub-Poissonian light have been found.\newline

\addcontentsline{toc}{chapter}{Appendices}
\chapter{Appendices}

\begin{center}
{\bf Appendix A}
\end{center}
In this appendix we deduce the result of the Fourier transformation
of equation
(\ref{scf17}). This is basically depending on the generating function of
the Laguerre polynomial.  We restrict ourselves to the derivation
of the result of one term for
Wigner function, i.e. $s=0$, however, the others
can be found in a similar way.

Now let us assume we have such type of  integral

${\displaystyle
I_{n}=\int d ^{2}\zeta
{\rm L}_{n}(|k|^{2})
 \exp [-| k|^{2}/2+(\zeta -\zeta^{*})\alpha]
\exp(\beta\zeta^{*}-\beta^{*}\zeta),}\hfill (A.1) $

\noindent
where the explicit form of $k$ has been given in
(\ref{scf18}).
 Using the generating function of Laguerre polynomial \cite{scfr14}

${\displaystyle
\frac{\exp(-\frac{ty}{1-t})}{1-t}=\sum^{\infty}_{n=0} \frac{t^{n}}{n!}
{\rm L}_{n}(y),}
\hfill (A.2)$

\noindent in equation (A.1), we get

${\displaystyle
\sum^{\infty}_{n=0}
\frac{I_{n}t^{n}}{n!}=\frac{1}{1-t}
\int d ^{2}\zeta
\exp \left[-\frac{t+1}{2(1-t)} |k|^{2}+\zeta
(\alpha -\beta^{*})+\zeta^{*}(\beta-\alpha)\right] ,}\hfill  $

${\displaystyle
=\frac{1}{1-t}
\int d ^{2}\zeta
\exp \left\{-\frac{t+1}{2(1-t)}[\beta^{2}\cosh (2r)+(\beta^{2}+
\beta^{*2})C_{r}S_{r}]+\zeta
(\alpha -\beta^{*})+\zeta^{*}(\beta-\alpha)\right\},}\hfill (A.3) $

\noindent
where the notations $S_{r}$ and $C_{r}$ have the same meaning as before.
Applying the following identity \cite{perin}

${\displaystyle
\int \exp [-B|\zeta|^{2}+(c/2)\zeta^{*2}+(c_{1}/2)\zeta^{2}+
D_{1}\zeta+D\zeta^{*}]d ^{2}\zeta}\hfill $
${\displaystyle
=\frac{\pi}{\sqrt{K}}
\exp \left\{\frac{1}{K}[DD_{1}B+D^{2}(c_{1}/2)+D^{2}_{1}(c/2)\right\},
}\hfill (A.4) $

\noindent with

${\displaystyle
K=B^{2}-cc_{1}, \qquad {\rm Re}K>0, \qquad
{\rm Re}[B+(c_{1}+c)/2]>0,}\hfill  (A.5)$

\noindent to  (A.3), after minor algebra we arrive at

${\displaystyle
\sum^{\infty}_{n=0}
\frac{I_{n}t^{n}}{n!}=\frac{2\pi}{1+t}
\exp \left[-2\frac{(1-t)z}{(1+t)}\right] ,}\hfill (A.6a) $

${\displaystyle
=\frac{2\pi}{1+t} \exp (-2z)
\exp \left[4\frac{tz}{(1+t)}\right] ,}\hfill (A.6b) $

${\displaystyle
=2\pi\exp (-2z)
\sum^{\infty}_{n=0}
\frac{(-t)^{n}}{n!}{\rm L}_{n}(4z) ,}\hfill (A.6c) $

\noindent where

${\displaystyle
z=|\beta-\alpha|^{2}\cosh (2r)+[(\beta-\alpha)^{2}+
(\beta^{*}-\alpha)^{2}]S_{r}C_{r},}\hfill  $

${\displaystyle
=(x-\alpha)^{2}\exp(2r) +y^{2}\exp (-2r).}\hfill (A.7) $

\noindent In the transition from (A.6b) to (A.6c) we used the generating
function of Laguerre polynomial (A.2) again, however, in (A.7) we used
 $\beta=x+iy$ as before. Finally, (A.6c) shows that

${\displaystyle
I_{n}=2\pi (-1)^{n}\exp (-2z){\rm L}_{n}(4z) ,}\hfill (A.8) $

\noindent and this is the value of the required integral.  For $Q$-function
the right-hand side of (A.6) will be expressed in terms of
the generating function of Hermite polynomials.
\newpage
\begin{center}
{\bf Appendix B}
\end{center}
In this appendix we derive  the relations (\ref{scf49}).
Before solving the problem it is convenient to remind that
the connection between Schr\"{o}dinger's picture and Heisenberg's
picture for any  operator is

${\displaystyle
\hat{A}_{j}(t)=\hat{S}^{\dagger}(t)\hat{A}_{j}(0)\hat{S}(t) ,}\hfill (B.1) $

\noindent where $\hat{S}(t)$ is the unitary operator of time development.
So that the required  relation can be found by solving Heisenberg's
equations of motion.

The Heisenberg equation of motion for any operator \^{O} is given by

${\displaystyle
\frac{d\hat{O}}{dt}=\frac{\partial \hat{O}}{\partial
t}+\frac{1}{i\hbar }[
\hat{O},H].}\hfill (B.2)$

\noindent The equations of motion related to the effective Hamiltonian (\ref{scf44})
are

${\displaystyle
\frac{d\hat{A}_{1}}{dt} = \lambda _{1}\hat{A}_{2}^{\dagger
}+\lambda _{2}
\hat{A}^{\dagger}_{3},}\hfill (B.3a)$

${\displaystyle
\frac{d\hat{A}_{2}}{dt}=\lambda _{1}\hat{A}^{\dagger}_{1}+\lambda
_{3}\hat{A}_{3},}\hfill (B.3b)$

${\displaystyle
\frac{d\hat{A}_{3}}{dt}=\lambda _{2}\hat{A}_{1}^{\dagger}
-\lambda _{3}\hat{A}_{2}.}\hfill(B.3c)$

\noindent Equations (B.3) with their Hermitian conjugates
give a close system of
6 differential equations with time-independent coefficients  which
can be solved, for instance, by means of Laplace transformation. Nevertheless, they
can be solved straightforwardly by differentiting one of these equation
two times and subsituting from the others. Let us restrict ourselves to
(B.3a) which gives (after minor algebra)

${\displaystyle
\frac{d^{3}\hat{A}_{1}}{dt^{3}}-(\lambda^{2} _{1}+ \lambda^{2} _{2}-
\lambda^{2} _{3}) \frac{d\hat{A}_{1}}{dt}=0,}\hfill(B.4)$

\noindent and consequently the charactristic equation is

${\displaystyle
y^{3}-(\lambda^{2} _{1}+ \lambda^{2} _{2}-\lambda^{2} _{3}) y=0,
}\hfill (B.5)$

\noindent and hence  the roots of (B.5) are $y=0,
\pm z=\pm \sqrt{\lambda^{2} _{1}+ \lambda^{2} _{2}-\lambda^{2} _{3}}$.
Now considering  the condition
$\lambda^{2} _{1}+ \lambda^{2} _{2}>\lambda^{2} _{3}$,
the solution of  (B.4) is

${\displaystyle
\hat{A}_{1}(t)=c_{1}
+c_{2}\cosh (tz)
+c_{3}\sinh (tz),
}\hfill (B.6)$

\noindent where $c_{1},c_{2},c_{3}$ are constants which can be specified
from  (B.6) and its derivative.
After some simplification we arrive at

${\displaystyle
\hat{A}_{1}(t)=
\hat{A}_{1}(0)[\cosh (tz)+\frac{2\lambda^{2} _{3}}{z^{2}}
 \sinh^{2}(tz/2)]
}\hfill $

${\displaystyle
+\hat{A}^{\dagger}_{2}(0)[\frac{\lambda _{1}}{z} \sinh(tz)
-\frac{2 \lambda _{2}\lambda _{3}}{z^{2}} \sinh^{2}(tz/2)]
}\hfill $

${\displaystyle
+\hat{A}_{3}(0)[\frac{\lambda _{2}}{z} \sinh(tz)
+\frac{2 \lambda _{1}\lambda _{3}}{z^{2}} \sinh^{2}(tz/2)].
}\hfill (B.7)$

\noindent Now using the substitution $r_{j}=t\lambda_{j}, j=1,2,3$,
(B.7) reduces to the first equation  in (\ref{scf49}) where we wrote
 $\hat{A}_{j}(0)=\hat{A}_{j}$ in these relations. Similarly, the relations related to
$\hat{A}_{j}(t), j=2,3$ can be deduced.

\newpage
\begin{center}
{\bf Appendix C}
\end{center}

In this appendix we give  the explicit forms for the expectation values
of cross photon-number operators between various modes for three-mode
squeezed coherent states. The derivation is straightforward and
the relations (\ref{scf49}) should be frequently used:

\begin{math}
{\displaystyle
\langle \hat{n}_{1}\hat{n}_{2}\rangle=
f_{1}^{2}g^{2}_{3}(|\alpha_{1}|^{4}+2|\alpha_{1}|^{2})
}\hfill
\end{math}

\begin{math}
{\displaystyle
+|(\alpha_{2}f_{2}+\alpha_{3}f_{3})
(\alpha_{2}g_{1}+\alpha_{3}g_{2})|^{2}
 }\hfill
\end{math}

\begin{math}
{\displaystyle
+(f^{2}_{1}|\alpha_{1}|^{2}+f^{2}_{1}-1)|\alpha_{2}g_{1}+
\alpha_{3}g_{2})|^{2}
 }\hfill
\end{math}

\begin{math}
{\displaystyle
+(|\alpha_{1}|^{2}+1)\Bigl[
(|\alpha_{2}|^{2}+1)f_{2}g_{3}
(g_{1}f_{1}+g_{3}f_{2})+
+(|\alpha_{3}|^{2}+1)f_{3}g_{3}
(g_{2}f_{1}+g_{3}f_{3}) }\hfill
\end{math}

\begin{math}
{\displaystyle
+f_{1}g_{3}(f_{2}g_{1}
|\alpha_{2}|^{2} +f_{3}g_{2}
|\alpha_{3}|^{2})
+
(\alpha_{1}\alpha_{2}+\alpha^{*}_{1}\alpha^{*}_{2})f_{1}g_{3}
(f_{1}g_{1}+f_{2}g_{3})
 }\hfill
\end{math}

\begin{math}
{\displaystyle
+ (\alpha_{2}\alpha^{*}_{3}+\alpha^{*}_{2}\alpha_{3})g_{3}
(f_{1}f_{3}g_{1}+f_{1}f_{2}g_{2}+f_{2}f_{3}g_{3})
+(\alpha_{1}\alpha_{3}+\alpha^{*}_{1}\alpha^{*}_{3})f_{1}g_{3}
(f_{1}g_{2}+f_{3}g_{3})\Bigr] }\hfill
\end{math}

\begin{math}
{\displaystyle
+f_{1}|\alpha_{2}g_{1}+\alpha_{3}g_{2}|^{2}
[
(\alpha_{1}\alpha_{2}+\alpha^{*}_{1}\alpha^{*}_{2})f_{2}
+
(\alpha_{1}\alpha_{3}+\alpha^{*}_{1}\alpha^{*}_{3})f_{3}]
 }\hfill
\end{math}

\begin{math}
{\displaystyle
+g_{3}|\alpha_{2}f_{2}+\alpha_{3}f_{3}|^{2}
[
(\alpha_{1}\alpha_{2}+\alpha^{*}_{1}\alpha^{*}_{2})g_{1}
+
(\alpha_{1}\alpha_{3}+\alpha^{*}_{1}\alpha^{*}_{3})g_{2}]
 }\hfill
\end{math}

\begin{math}
{\displaystyle
+(\alpha_{1}\alpha_{2}+\alpha^{*}_{1}\alpha^{*}_{2})
h_{2}
[f_{1}f_{2}h_{2} + f^{2}_{1}h_{3}-h_{3}] }\hfill
\end{math}

\begin{math}
{\displaystyle
+(\alpha_{1}\alpha_{3}+\alpha^{*}_{1}\alpha^{*}_{3})
h_{2}
[f_{1}f_{2}h_{2} + f^{2}_{1}h_{1}-h_{2}] }\hfill
\end{math}

\begin{math}
{\displaystyle
+
 f_{1}h_{2}[(\alpha^{2}_{1}+\alpha^{*2}_{1})|\alpha_{2}|^{2}f_{2}h_{3}+
(\alpha^{2}_{1}+\alpha^{*2}_{1})|\alpha_{3}|^{2}f_{3}h_{1} }\hfill
\end{math}

\begin{math}
{\displaystyle
+(\alpha^{2}_{1}\alpha_{2}\alpha_{3}+
\alpha^{*2}_{1}\alpha^{*}_{2}\alpha^{*}_{3})
(h_{1}f_{2} +h_{3}f_{3})], }\hfill (C.1)
\end{math}

$\hfill $

\begin{math}
{\displaystyle
\langle \hat{n}_{2}\hat{n}_{3}\rangle=
h_{2}^{2}g^{2}_{3}[(|\alpha_{1}|^{2}+2)^{2}-3]
+(|\alpha_{1}|^{2}+1)\Bigl[
g^{2}_{3}|h_{1}\alpha_{3}+h_{3}\alpha_{2}|^{2} }\hfill
\end{math}

\begin{math}
{\displaystyle
+h^{2}_{2}|g_{1}\alpha_{2}+g_{2}\alpha_{3}|^{2}
+h_{2}h_{3}g_{1}g_{3}(2|\alpha_{2}|^{2}+1) }\hfill
\end{math}

\begin{math}
{\displaystyle
+h_{1}h_{2}g_{2}g_{3}(2|\alpha_{3}|^{2}+1)
+ h_{2}g_{3}(g_{1}h_{2}+g_{3}h_{3})(\alpha_{1}\alpha_{2}+\alpha^{*}_{1}
\alpha^{*}_{2})}\hfill
\end{math}

\begin{math}
{\displaystyle
+ h_{2}g_{3}(g_{3}h_{1}+g_{2}h_{2})(\alpha_{1}\alpha_{3}+\alpha^{*}_{1}
\alpha^{*}_{3})}\hfill
\end{math}

\begin{math}
{\displaystyle
+ h_{2}g_{3}(g_{2}h_{3}+g_{1}h_{1})
 (\alpha_{2}\alpha^{*}_{3}+\alpha^{*}_{2}\alpha_{3})
\Bigr]}\hfill
\end{math}

\begin{math}
{\displaystyle
+
|(g_{1}\alpha_{2}+g_{2}\alpha_{3})
(h_{1}\alpha_{3}+h_{3}\alpha_{2})|^{2} }\hfill
\end{math}

\begin{math}
{\displaystyle
+
g_{3}|h_{1}\alpha_{3}+h_{3}\alpha_{2}|^{2}
[g_{1}(\alpha_{1}\alpha_{2}+\alpha^{*}_{1}\alpha^{*}_{2})
+g_{2}(\alpha_{1}\alpha_{3}+\alpha^{*}_{1}\alpha^{*}_{3})]
}\hfill
\end{math}

\begin{math}
{\displaystyle
+
h_{2}|g_{1}\alpha_{2}+g_{2}\alpha_{3}|^{2}
[h_{1}(\alpha_{1}\alpha_{3}+\alpha^{*}_{1}\alpha^{*}_{3})
+h_{3}(\alpha_{1}\alpha_{2}+\alpha^{*}_{1}\alpha^{*}_{2})]
}\hfill
\end{math}

\begin{math}
{\displaystyle
+
h_{2}g_{3}(h_{1}g_{3}+h_{2}g_{2})
(\alpha_{1}\alpha_{3}+\alpha^{*}_{1}\alpha^{*}_{3})
}\hfill
\end{math}

\begin{math}
{\displaystyle
+
h_{2}g_{3}(h_{3}g_{3}+h_{2}g_{1})
(\alpha_{1}\alpha_{2}+\alpha^{*}_{1}\alpha^{*}_{2})
}\hfill
\end{math}

\begin{math}
{\displaystyle
+
h_{2}g_{3}(h_{1}g_{1}+h_{3}g_{2})
(\alpha^{2}_{1}\alpha_{2}\alpha_{3}+\alpha^{*2}_{1}\alpha^{*}_{2}
\alpha^{*}_{3})
}\hfill
\end{math}

\begin{math}
{\displaystyle
+
h_{2}h_{3}g_{1}g_{3}(\alpha^{2}_{1}\alpha^{2}_{2}+\alpha^{*2}_{1}
\alpha^{*2}_{2})
+h_{1}h_{2}g_{1}g_{3}(\alpha^{2}_{1}\alpha^{2}_{3}+\alpha^{*2}_{1}
\alpha^{*2}_{3}). }\hfill (C.2)
\end{math}

\noindent Corresponding relation between modes $1$ and $3$ can
be obtained from (C.1) using the transformation (\ref{scf61}).

\newpage
\begin{center}
{\bf Appendix D}
\end{center}
In this appendix we give  the derivation of the solution
of equations of motion (\ref{scs2}). The Heisenberg equations
of motion are

$ {\displaystyle
\frac{d\hat{a}_{1}}{dt}=-i\omega_{1}\hat{a}_{1}-2i\lambda_{1} \hat{a}
_{1}^{\dagger} \exp (it\mu_{1}) -i\lambda_{3}\hat{a}_{2}\exp
[-i\phi_{1}(t)]
-i\lambda_{4}\hat{a}_{2}^{\dagger}\exp [i\phi_{2}(t)], } \hfill (D.a1) $

$ {\displaystyle
\frac{d\hat{a}_{2}}{dt}=-i\omega_{2}\hat{a}_{2}-2i\lambda_{2} \hat{a}
_{2}^{\dagger} \exp (it\mu_{2}) - i\lambda_{3}\hat{a}_{1}\exp
[i\phi_{1}(t)]
-i\lambda_{4}\hat{a}_{1}^{\dagger}\exp [i\phi_{2}(t)]. } \hfill (D.a2) $

\noindent
 Substituting $\hat{a}_{1}=\hat{A}\exp(\frac{it}{2}\mu_{1})$
and $\hat{a}_{2}= \hat{B}\exp(\frac{it}{2}\mu_{2})$ in terms of the
slowly varying forms $\hat{A}$ and $\hat{B}$,
 equations (D.a) take the form

$ {\displaystyle
\frac{d\hat{A}}{dt}= -i(\omega_{1}+\frac{\mu_{1}}{2})\hat{A} -2i\lambda_{1}
\hat{A}^{\dagger} -i\lambda_{3}\hat{B} \exp \left[ i\frac{(\mu_{2}-\mu_{1})t
}{2}-i\phi_{1}(t)\right] } \hfill $

$ {\displaystyle \qquad -i\lambda_{4}\hat{B}^{\dagger} \exp \left[-i\frac{
(\mu_{1}+\mu_{2})t}{2}+i\phi_{2}(t)\right], }\hfill (D.b1) $

$ {\displaystyle
\frac{d\hat{B}}{dt} = -i(\omega_{2}+\frac{\mu_{2}}{2})\hat{B}
-2i\lambda_{2}
\hat{B}^{\dagger} -i\lambda_{3}\hat{A} \exp \left[ i\frac{
(\mu_{1}-\mu_{2})t
}{2}+i\phi_{1}(t)\right] } \hfill $

${\displaystyle \qquad -i\lambda_{4}\hat{A}^{\dagger} \exp \left[
-i\frac{
(\mu_{1}+\mu_{2})t}{2 }+i\phi_{2}(t)\right].}$ \hfill (D.b2)

\noindent
Now let us define

${\displaystyle
\hat{Q}_{1} =\hat{A}+\hat{A}^{\dagger},\quad
\hat{Q}_{2} =\hat{B}+\hat{B}^{\dagger},\quad
\hat{P}_{1} =\hat{A}-\hat{A}^{\dagger},\quad
\hat{P}_{2} =\hat{B}-\hat{B}^{\dagger}.
}$ \hfill (D.b3)

\noindent
Equations (D.b1)-(D.b2)
cannot be solved directly and hence some restrictions should be considered, so
that we shall consider
 $\phi_{1}(t)=\frac{1}{2}(\mu_{2}-\mu_{1})t$ and
$\phi_{2}(t)=\frac{
1 }{2}(\mu_{2}+\mu_{1})t$ and thus these equations with their
 Hermitian conjugates  lead to the
following equations:

$ {\displaystyle
\frac{d\hat{Q}_{1}}{dt}= -ik_{-}
\hat{P}_{1} -i\lambda_{-}\hat{P}_{2}, \quad
\frac{d\hat{Q}_{2}}{dt}= -iJ_{-}
\hat{P}_{2} -i\lambda_{-}\hat{P}_{1},
} \hfill $

$ {\displaystyle
\frac{d\hat{P}_{1}}{dt}= -ik_{+}
\hat{Q}_{1} -i\lambda_{+}\hat{Q}_{2},
\quad
\frac{d\hat{P}_{2}}{dt}= -iJ_{+}
\hat{Q}_{2} -i\lambda_{+}\hat{Q}_{1},
} \hfill (D.c1) $

\noindent where

$ {\displaystyle  \lambda_{\pm}=\lambda_{3}\pm \lambda_{4}, \quad
 k_{\pm}=
\omega_{1}+\frac{1}{2}\mu_{1}\pm2\lambda_{1}, \quad
 J_{\pm}=\omega_{2}+\frac{1}{2}\mu_{2}\pm2\lambda_{2}.
}\hfill (D.c2) $

The second order differential equations for
$\hat{Q}_{j}$ can be obtained from (D.c1) after some minor algebra as

$ {\displaystyle
\frac{d^{2}\hat{\bar{Q}}_{1}}{dt^{2}}+ \Omega_{1}
\hat{\bar{Q}}_{1}
= -\sqrt{g_{1}g_{2}} \hat{\bar{Q}}_{2},\quad
\frac{d^{2}\hat{\bar{Q}}_{2}}{dt^{2}}+ \Omega_{2}
\hat{\bar{Q}}_{2}
= -\sqrt{g_{1}g_{2}} \hat{\bar{Q}}_{1},
} \hfill (D.d1)$

\noindent where  $\hat{\bar{Q}}_{j}=\sqrt{g_{j}}\hat{Q}_{j}, \quad j=1,2$ and

$ {\displaystyle  g_{1}=k_{-}\lambda_{+}+\lambda_{-}J_{+}, \quad
  g_{2}=k_{+}\lambda_{-}+\lambda_{+}J_{-}, }\hfill  $

 $ {\displaystyle
 \Omega_{1}^{2}=\lambda_{-}\lambda_{+}+k_{-}k_{+}, \quad
 \Omega_{2}^{2}=\lambda_{-}\lambda_{+}+J_{-}J_{+}.} \hfill (D.d2) $

\noindent  Now using the following transformations:

 $ {\displaystyle
 \hat{\bar{Q}}_{1}  =
 \hat{X}\cos \theta +\hat{Y}\sin \theta, \quad
 \hat{\bar{Q}}_{2} =
 \hat{Y}\cos \theta -\hat{X}\sin \theta,} \hfill (D.f1) $

 \noindent
 where $\hat{X}$ and $ \hat{Y}$ are two Hermitian operators,
 equations (D.d1) can be split into two equations as

 $ {\displaystyle
\frac{d^{2}\hat{X}}{dt^{2}} +\bar{\Omega}_{1}\hat{X}=0, \quad
\frac{d^{2}\hat{Y}}{dt^{2}} +\bar{\Omega}_{1}\hat{Y}=0,} \hfill (D.f2) $

\noindent provided that
 $ \theta=\frac{1}{2}\tan
^{-1}\left(\frac{2\sqrt{g_{1}g_{2}}
} {J_{-}J_{+}-k_{-}k_{+}}\right) $, where

$\bar{\Omega}_{1}=\left[ \Omega_{1}^{2}\cos^{2}\theta +
\Omega_{2}^{2}\sin^{2}\theta -\sqrt{g_{1}g_{2}}\sin
(2\theta)\right]^{\frac{
1 }{2}}, \hfill (D.f3)  $

$\bar{\Omega}_{2}=\left[ \Omega_{2}^{2}\cos^{2}\theta +
\Omega_{1}^{2}\sin^{2}\theta +\sqrt{g_{1}g_{2}}\sin
(2\theta)\right]^{\frac{
1 }{2}}. \hfill (D.f4)  $
\noindent
Equations (D.f2) can be easily   solved and consequently
 $\hat{Q}_{j}, \quad j=1,2$  can be obtained. Similar procedures can be
 done to get $\hat{P}_{j}, \quad j=1,2$ and eventually the
 expressions  for  $\hat{a}_{j}(t), \quad j=1,2$ can be traced out
 in the form (\ref{scs2}).
The time dependent coefficients in (\ref{scs2}) are given by

$ {\displaystyle
K_{1}(t)=F_{1}(t)-\frac{i}{2}
\left[ [ k_{+} +k_{-}] G_{1}(t)+[\lambda_{+}\frac{g_{2}}{g_{1}}
+\lambda_{-}]S(t)\right],} \hfill  $

$ {\displaystyle \hfill
L_{1}(t)=-\frac{i}{2}
\left[ [ k_{+} -k_{-}] G_{1}(t)+[\lambda_{+}\frac{g_{2}}{g_{1}}
-\lambda_{-}]S(t)\right], }\hfill  $

$ {\displaystyle  M_{1}(t) =
\frac{1}{2}\left\{\left(1 +
\frac{ g_{2}}{g_{1}}\right)C(t)-
i\left[ [\lambda_{+}+\lambda_{-}]G_{1}(t)+[J_{+}\frac{
g_{2}}{g_{1}} +J_{-}]S(t)\right]\right\},}\hfill (D.g) $

$ {\displaystyle
\hfill N_{1}(t) = \frac{1}{2}\left\{\left(1
-\frac{ g_{2}}
{g_{1}}\right) C(t)-i\left[ [\lambda_{+}-\lambda_{-}]G_{1}(t)
+[ J_{+}\frac{g_{2}}{g_{1}}
-J_{-}] S(t)\right]\right\} }, \hfill  $

\noindent whereas

$ {\displaystyle
\hfill
K_{2}(t)=F_{2}(t)-\frac{i}{2}\left[ [J_{+}
+J_{-}] G_{2}(t)+[\lambda_{+}\frac{g_{2}}{g_{1}}
+\lambda_{-}]S(t)\right],}
\hfill  $

$ {\displaystyle
\hfill L_{2}(t)=-\frac{i}{2}\left[ [J_{+}
-J_{-}]G_{2}(t)+[\lambda_{+}-\lambda_{- }\frac{g_{2}}{g_{1}}]S(t)\right],
}\hfill $

$ {\displaystyle  \hfill
M_{2}(t)=\frac{1}{2}%
\left\{ \left(1 +\frac{g_{2} }{g_{1}}\right)
C(t)-i\left[ [\lambda_{+}+\lambda_{-}]G_{2}(t)+[k_{+}+k_{-}\frac{g_{2}}
{g_{1}}]S(t)\right]\right\}, } \hfill (D.h) $

$ {\displaystyle \hfill N_{2}(t) =
\frac{1}{2}
\left\{ \left( \frac{g_{2}}{g_{1}}-1\right) C(t)-i\left[ [\lambda_{+}-
\lambda_{-}]G_{2}(t)
+[k_{+}-k_{-}\frac{g_{2}}{g_{1}}]S(t)\right]\right\}. } \hfill  $

\noindent In the above equations we have defined

$ {\displaystyle  F_{1}(t)= \cos (t\bar{\Omega}_{1}) \cos ^{2}\theta +
\cos(t\bar{\Omega}_{2})\sin ^{2}\theta ,} \hfill  $

$ {\displaystyle  F_{2}(t)=\cos (t\bar{\Omega}_{2}) \cos ^{2}\theta
+\cos (t
\bar{\Omega}_{1}) \sin ^{2}\theta ,} \hfill  $

$ {\displaystyle  G_{1}(t)=\frac{\sin
(t\bar{\Omega}_{1})}{\bar{\Omega}_{1} }
\cos ^{2}\theta + \frac{\sin (t\bar{\Omega}_{2})}{\bar{\Omega}_{2}} \sin
^{2}\theta , } \hfill  $

$ {\displaystyle \hfill G_{2}(t)=\frac{\sin
(t\bar{\Omega}_{2})}{\bar{\Omega}
_{2} }\cos ^{2}\theta + \frac{\sin
(t\bar{\Omega}_{1})}{\bar{\Omega}_{1}}
\sin ^{2}\theta , }\hfill (D.k) $

$ {\displaystyle \hfill
C(t)=\frac{1}{2}\sqrt{\frac{g_{1}}{g_{2}}}\left[\cos
(t\bar{\Omega}_{2}) -\cos (t\bar{\Omega}_{1})\right]\sin (2\theta ) , }
\hfill  $

$ {\displaystyle  \hfill
S(t)=\frac{1}{2}\sqrt{\frac{g_{1}}{g_{2}}}\left[
\frac{\sin (t\bar{\Omega}_{2})}{\bar{\Omega}_{2} } - \frac{\sin
(t\bar{\Omega
}_{1})}{\bar{\Omega}_{1} } \right]\sin (2\theta ). } \hfill  $

\newpage

\begin{center}
{\bf Appendix E}
\end{center}
In this appendix we give the exact expressions for the time
dependent functions for the solutions of
the Heisenberg equations of motion for interacting modes according to
 the
Hamiltonian  (\ref{scs30}). It is reasonable to mention that the solutions
in this case
can be deduced using the similar technique as that given in
 appendix D. Thus we have

${\displaystyle
f_{1}(t)=\cos (t\Omega _{1})\cosh ^{2}\phi
         -\cosh (t\Omega _{2})\sinh^{2}\phi, }\hfill (E.1a)$

${\displaystyle
f}_{2}(t)=\frac{\lambda _{3}}{2}[\frac{\sin (t\Omega _{1})}{\Omega_{1}}
          -\frac{\sinh (t\Omega _{2})}{\Omega _{2}}]\sinh (2\phi )$

$-\lambda _{2}[\frac{\sinh (t\Omega _{1})}{\Omega _{1}}\cosh ^{2}\phi
      -\frac{\sinh (t\Omega _{2})}{\Omega _{2}}\sinh ^{2}\phi ],\hfill (E.1b)$

${\displaystyle
f_{3}(t)=\lambda _{1}[\frac{\sin (t\Omega _{1})}{\Omega _{1}}\cosh^{2}\phi
 -\frac{\sinh (t\Omega _{2})}{\Omega _{2}}\sinh ^{2}\phi ],}\hfill (E.1c)$

${\displaystyle
f_{4}(t)=\frac{1}{2}[\cos (t\Omega _{1})-\cosh (t\Omega _{2})]
 \sinh(2\phi ),}\hfill (E.1d)$

${\displaystyle
h_{1}(t)=\frac{\lambda _{2}}{2}[\frac{\sin (t\Omega _{3})}{\Omega_{3}}
         -\frac{\sinh (t\Omega _{4})}{\Omega _{4}}]\sinh (2\theta )
}\hfill $
${\displaystyle
         +\lambda_{1}[\frac{\sinh (t\Omega _{4})}{\Omega _{4}}
            \cosh ^{2}\theta -\frac{\sin (t\Omega_{3})}{\Omega _{3}}
            \sinh ^{2}\theta ],}\hfill (E.2a)$

${\displaystyle
h_{2}(t)=\frac{1}{2}[\cosh (t\Omega _{4})-\cos (t\Omega _{3})]
                     \sinh(2\theta),}\hfill (E.2b)$

${\displaystyle
h_{3}(t)=\cosh (t\Omega _{4})\cosh ^{2}\theta
         -\cos (t\Omega_{3})\sinh^{2}\theta, }\hfill (E.2c)$

${\displaystyle
h_{4}(t)=\frac{\lambda _{3}}{2}[\frac{\sin (t\Omega _{3})}{\Omega _{3}}
\frac{\sinh (t\Omega _{4})}{\Omega _{4}}]\sinh (2\theta ),}\hfill
(E.2d)$

\noindent where

${\displaystyle
\Omega _{1}=[k_{1}^{2}\cosh ^{2}\phi +\lambda _{3}^{2}\sinh ^{2}\phi
-\lambda _{2}\lambda _{3}\sinh (2\phi )]^{\frac{1}{2}},}\hfill (E.3a)$

${\displaystyle
\Omega _{2}=[k_{1}^{2}\sinh ^{2}\phi +\lambda _{3}^{2}\cosh ^{2}\phi
-\lambda _{2}\lambda _{3}\sinh (2\phi )]^{\frac{1}{2}},}\hfill (E.3b)$

\noindent and

${\displaystyle
\phi =\frac{1}{2}\tanh ^{-1}(\frac{2\lambda _{2}\lambda _{3}}{\lambda
_{2}^{2}+\lambda _{3}^{2}-\lambda _{1}^{2}}),}\hfill (E.4)$

\noindent with $k_{1}=\sqrt{\lambda _{2}^{2}-\lambda _{1}^{2}}$.

The other coefficients can be obtained with the help of the substitution $
\lambda_{1} \longleftrightarrow\lambda_{3}$ and hence

${\displaystyle
\Omega_{1} \longleftrightarrow\Omega_{3}, \Omega_{2}
\longleftrightarrow\Omega_{4}, \phi \longleftrightarrow\theta,} \hfill $

${\displaystyle
(f_{1}(t),f_{2}(t),f_{3}(t),f_{4}(t)) \longleftrightarrow\
(g_{2}(t),g_{1}(t),g_{4}(t),g_{3}(t)),}\hfill $

${\displaystyle
(h_{1}(t),h_{2}(t),h_{3}(t),h_{4}(t)) \longleftrightarrow\
(l_{2}(t),l_{1}(t),l_{4}(t),l_{3}(t)).} \hfill (E.5) $

\newpage
\addcontentsline{toc}{chapter}{Publications of the author of the
thesis}
\section*{Publications of the author of the thesis}
1- F. A. A. El-Orany: "Quantum statistical properties of
superposition of squeezed and displaced number states", Czech J.  Physics
49 (1999) 1145.\newline

\noindent 2-  F. A. A. El-Orany, J. Pe\v{r}ina  and M. Sebawe Abdalla
: "Quantum statistical properties of
superposition of squeezed and displaced  states
with  thermal noise", J. Mod. Opt.  46 (1999) 1621.
It has been presented at Workshop on Optical Properties of Microcavities,
Abdus Salam Centre, Trieste, Italy, August 6-12, 1998. \newline

\noindent 3- M. S. Abdalla, F. A. A. El-Orany and J. Pe\v{r}ina:
"Quantum statistical properties of nondegenerate optical parametric
symmetric coupler", J. Phys. A: Math. Gen.  32 (1999) 3457.\newline

\noindent 4- M. S. Abdalla, F. A. A. El-Orany and J. Pe\v{r}ina:
"Nonclassical effects in nondegenerate optical parametric
symmetric coupler ", ICO XVIII Conference, San Francisco USA,
SPIE Vol. 3749 (1999) 538.\newline

\noindent 5- M. S. Abdalla, F. A. A. El-Orany and J. Pe\v{r}ina:
"Dynamical properties of degenerate parametric amplifier with
photon-added coherent states",
Nuovo Cimento (2001) (in print).
 It has been presented at 2nd Euroconference
on Trends in Optical Nonlinear Dynamics Physical Problems and Applications
(COCOS), M\"{u}nster, Germany, October 7-10, 1999.\newline

\noindent 6-  F. A. A. El-Orany, J. Pe\v{r}ina  and M. Sebawe Abdalla:
 "Phase  properties of the superposition of squeezed
and displaced number states ", J. Opt. B: Quant. Semiclass. Opt.
2 (2000) 545. \newline

\noindent 7- M. S. Abdalla, F. A. A. El-Orany and J. Pe\v{r}ina:
"$SU(2)$- and $SU(1,1)$-squeezing of  interacting radiation modes",
Act. Phys. Slovaca 50 (2000) 613.\newline

\noindent 8- M. S. Abdalla, F. A. A. El-Orany and J. Pe\v{r}ina: "Quantum statistics
and dynamics of nonlinear couplers with nonlinear exchange", J. Mod. Opt.
47 (2000) 1055.\newline

\noindent 9- M. S. Abdalla, F. A. A. El-Orany and J. Pe\v{r}ina:
"Quantum statistical properties of highly correlated
squeezed parametric system", Act. Phys. Univ. Palacki. Olomouc 39 (2000) 25.\newline

\noindent 10- M. S. Abdalla, F. A. A. El-Orany and J. Pe\v{r}ina: "Sub-Poissonian
statistics of highly correlated squeezed parametric systems".
International Conference of Squeezed States and
 Uncertainty Relation.  The Naples 1999, NASA 2000, p. 23.\newline

\noindent 11-  F. A. A. El-Orany, J. Pe\v{r}ina  and M. Sebawe Abdalla:
"Generation of squeezed light in a nonlinear asymmetric directional coupler",
 J. Opt. B: Quant. Semiclass. Opt. 3 (2001) 67.
It has been presented at the  7th Centeral European Workshop on
 Quantum Optics, April 28-May 1, 2000, Hungary.
 Electronic Proceedings of the CEWQO Workshop
 (htt://optics.szfki.kfki/cewqo2000/).
 \newline

\noindent 12- F. A. A. El-Orany,  J. Pe\v{r}ina and M. S. Abdalla:
"Quantum properties of the parametric amplifier
with and without pumping field fluctuations",
Opt. Commun.  187 (2001) 199.\newline

\noindent 13- F. A. A. El-Orany, J. Pe\v{r}ina  and M. Sebawe Abdalla:
 "Statistical properties of three quantized interacting oscillators",
 Phys. Scripta 63 (2001) 128. \newline

\noindent 14- M. S. Abdalla, F. A. A. El-Orany and J. Pe\v{r}ina:
"Statistical properties of a solvable three-boson squeeze operator model", J.
Europ. Phys. D 13 (2001) 423.
It has been presented at XIII the International Congress on Mathematical Physics,
London, July 17-22, 2000.\newline

\noindent 15- F. A. A. El-Orany, M. Sebawe Abdalla,  A.-S. F. Obada  and
G. M. Abd-Al-Kader:  "Influence of squeezing operator on the quantum
properties of various binomial states", In. J. of Mod. Phys. B
15 (2001) 75.\newline

\newpage
\addcontentsline{toc}{chapter}{Stru\v{c}n\'e  shrnut\'{\i} v\'ysledku
diserta\v{c}n\'{\i} pr\'ace (summary in Czech)}
\section*{Stru\v{c}n\'e  shrnut\'{\i} v\'ysledku diserta\v{c}n\'{\i}
pr\'ace (summary in Czech) }
V t\'eto diserta\v{c}n\'{\i} pr\'aci jsou studov\'any kvantov\'e vlastnosti n\v{e}kter\'ych
statick\'ych a dynamick\'ych syst\'em\accent23u. Prvn\'{\i} \v{c}\'ast je v\v{e}nov\'ana v\'yzkumu
vlastnost\'{\i} statick\'ych model\accent23u, zahrnuj\'{\i}c\'{\i}m superpozice stla\v{c}en\'ych
a posunut\'ych Fockov\'ych stav\accent23u s term\'aln\'{\i}ch \v{s}umem a bez term\'aln\'{\i}ho
\v{s}umu. Je rovn\v{e}\v{z} zaveden nov\'y typ v\'{\i}cerozm\v{e}rn\'eho oper\'atoru
stla\v{c}en\'{\i}, kter\'y zahrnuje dva r\accent23uzn\'e typy mechanizm\accent23u stla\v{c}en\'{\i}.
V druh\'e \v{c}\'asti jsou vy\v{s}et\v{r}ov\'any dynamick\'e modely v souvislosti
s interakc\'{\i} m\'od\accent23u v neline\'arn\'{\i}ch optick\'ych vazebn\'{\i}ch prvc\'{\i}ch.
Motivace je d\'ana d\v{r}\'{\i}v\v{e}j\v{s}\'{\i}m v\'yzkumem t\v{e}chto prvk\accent23u jako slibn\'ych
za\v{r}\'{\i}zen\'{\i} pro generaci neklasick\'eho sv\v{e}tla. Jsou diskutov\'any
t\v{r}i typy t\v{e}chto za\v{r}\'{\i}zen\'{\i}, zalo\v{z}en\'e na interakci dvou, t\v{r}\'{\i} a \v{c}ty\v{r} m\'od\accent23u
p\v{r}i \v{s}\'{\i}\v{r}en\'{\i} prost\v{r}ed\'{\i}m. Jsou ur\v{c}eny re\v{z}imy generace a p\v{r}enosu
neklasick\'eho sv\v{e}tla.

\end{document}